\newif\ifsubmit\submitfalse
\newif\iftr\trfalse

\submittrue
\ifdefined\istr\trtrue\fi
\trtrue

\iftr
\documentclass[acmsmall,screen]{acmart}
\renewcommand\footnotetextcopyrightpermission[1]{}
\fancypagestyle{firstpagestyle}{\fancyfoot{}}
\else
\documentclass[acmsmall,screen]{acmart} 
\fi

\usepackage[T1]{fontenc}
\usepackage{microtype}
\usepackage{breakcites}

\usepackage{booktabs}   
\usepackage{subcaption} 
\usepackage{boxedminipage}
\usepackage{url}                  
\usepackage{enumitem}      
\usepackage{color}
\usepackage{xcolor}
\usepackage{xspace}
\usepackage{mathpartir}
\usepackage{listings}
\usepackage{wrapfig}

\lstdefinestyle{customcaml}{
	numbers=left,
	belowcaptionskip=1\baselineskip,
	breaklines=true,
	tabsize=4,
	xleftmargin=\parindent,
	language=Caml,
	showstringspaces=false,
  basicstyle=\linespread{0.5}\footnotesize\sffamily,
	columns=flexible,
	otherkeywords={fix, mux, xor, with, region, castP, castS, flip, rnd, castU, castNU},
	literate={{<-}{{$\leftarrow\,$}}2
		{->}{{$\rightarrow\,$}}2},
	numberstyle=\tiny\rmfamily,
	keywordstyle=\bfseries\color{green!40!black},
	commentstyle=\itshape\color{purple!40!black},
	identifierstyle=\color{blue!80!black},
	stringstyle=\color{orange},
}
\newcommand{\code}[1]{\lstinline{#1}}

\setcopyright{rightsretained}
\acmPrice{}
\acmDOI{10.1145/3371118}
\acmYear{2020}
\copyrightyear{2020}
\acmJournal{PACMPL}
\acmVolume{4}
\acmNumber{POPL}
\acmArticle{50}

\acmMonth{1}

\ifsubmit
\newcommand{\liuchang}[1]{}
\newcommand{\todo}[1]{}
\newcommand{\mwh}[1]{}
\newcommand{\aseem}[1]{}
\newcommand{\dcd}[1]{}
\newcommand{\ins}[1]{}
\else
\newcommand{\liuchang}[1]{[{\color{blue}{\textrm{Chang: #1}}}]}
\newcommand{\todo}[1]{[{\color{red}{\textrm{TODO: {#1}}}}]}
\newcommand{\mwh}[1]{\textcolor{red}{\textbf{Mike says}: #1}}
\newcommand{\aseem}[1]{\textcolor{magenta}{Aseem: #1}}
\newcommand{\dcd}[1]{\textcolor{orange!80!black}{\textbf{David says}: #1}}
\newcommand{\ins}[1]{\textcolor{purple}{Ian: #1}}
\fi

\newcommand{\lang}{\ensuremath{\lambda_\textbf{obliv}}\xspace}

\newcommand\prefigskip{\vspace*{-1em}}
\newcommand\precaptionskip{\vspace*{-1em}}
\newcommand\postcaptionskip{\vspace*{-0.5em}}

\usepackage{mathtools}
\usepackage{tikz}
\usepackage{relsize}
\usetikzlibrary{decorations.pathmorphing,shapes}
\usetikzlibrary{cd}

\newcommand{\colorTEXT}{black}

\newcommand\colorSYNTAXA{blue!45!red!50!black}

\newcommand{\colorMATH}{\colorMATHA}
\newcommand{\colorSYNTAX}{\colorSYNTAXA}

\usepackage{centernot}
\newcommand{\slashedrel}[1]{\mathrel{\centernot{#1}}}

\newcommand{\underbracketarg}[2]{\underbracket[0.5pt]{#2}_{#1}}

\newcommand{\mtext}[1]{\ifmmode\operatorname{#1}\else\textnormal{#1}\fi}
\newcommand{\mtexttt}[1]{\ifmmode\operatorname{\mathtt{#1}}\else\textnormal{\texttt{#1}}\fi}
\newcommand{\mtextit}[1]{\ifmmode\operatorname{\mathit{#1}}\else\textnormal{\textit{#1}}\fi}
\newcommand{\mtextbf}[1]{\ifmmode\operatorname{\mathbf{#1}}\else\textnormal{\textbf{#1}}\fi}
\newcommand{\mtextsc}[1]{\ifmmode\operatorname{\textsc{\smaller #1}}\else\textnormal{\textsc{\smaller #1}}\fi}
\newcommand{\mfootnotesize}[1]{\ifmmode\textnormal{\smaller\smaller\(#1\)}\else{\smaller\smaller #1}\fi}
\newcommand*\circled[1]{\text{\textcircled{\raisebox{-0.05ex}{\scalebox{0.66}{\hspace{-0.01em}#1}}}}}

\makeatletter
\def\@acmplainindent{0pt}
\def\@proofindent{\noindent}
\makeatother

\usepackage{tabularx}
\usepackage{nicefrac}
\usepackage{geometry}

\usepackage{framed}
\newtheorem{theorem}{Theorem}[section]

\newtheorem{fact}[theorem]{Fact}

\begin{document}

\title{A Language for Probabilistically Oblivious Computation}


\author{David Darais}
\authornote{Work carried out in part while at the University of Maryland}
\affiliation{
  \institution{University of Vermont}            
  \country{USA}
}
\email{david.darais@uvm.edu}          

\author{Ian Sweet}
\affiliation{
  \institution{University of Maryland}            
  \country{USA}
}
\email{ins@cs.umd.edu}          

\author{Chang Liu}
\authornotemark[1]
\affiliation{
  \institution{Citadel Securities}            
  \country{USA}
}
\email{liuchang@eecs.berkeley.edu}          

\author{Michael Hicks}
\affiliation{
  \institution{University of Maryland}            
  \country{USA}
}
\email{mwh@cs.umd.edu}          

\begin{abstract}

  An oblivious computation is one that is free of direct and indirect
  information leaks, e.g., due to observable differences in timing and
  memory access patterns.  This paper presents
  \lang, a core language whose type system enforces obliviousness.
  Prior work on type-enforced oblivious computation has focused on
  deterministic programs. \lang is new in its consideration of
  programs that implement \emph{probabilistic} algorithms, such as
  those involved in cryptography. \lang employs a substructural type
  system and a novel notion of \emph{probability region} to ensure that
  information is not leaked via the observed distribution of visible
  events. Probability regions support reasoning about
  \emph{probabilistic correlation and independence} between values, and
  our use of probability regions is motivated by a source of unsoundness
  that we discovered in the type system of ObliVM, a language for implementing
  state of the art oblivious algorithms. We prove
  that
  \lang's type system enforces obliviousness and show that it is
  expressive enough to typecheck advanced tree-based oblivious RAMs.
 \end{abstract}

%


\maketitle

\section{Introduction}
\label{sec:intro}


Cloud computing allows clients to conveniently outsource computation,
but they must trust that cloud providers do not exploit or mishandle sensitive
information. To remove the provider from the trusted computing base,
work in both industry and research has strived to produce a secure
abstract machine comprising an execution engine and protected memory:
The adversary cannot see sensitive data as it is being operated on,
nor can it observe such data at rest in memory. Such an abstract
machine can be realized by encrypting the data in memory and then
performing computations using cryptographic mechanisms (e.g., secure
multi-party computation~\cite{yao}) or secure processors~\cite{xom,aegis,sgx}.

Unfortunately, a secure abstract machine does not defend against an
adversary that can observe memory access
patterns~\cite{accesspatternleak,phantom,hide} and instruction
timing~\cite{remotetiming,kocher96timing} (as made famous by recent
Spectre and Meltdown attacks~\cite{spectre,meltdown,VanBulck:2018:FEK:3277203.3277277}),
among other ``side'' channels of information. 
For cloud computing, such an adversary is the cloud provider itself,
which has physical access to its machines, and so can observe traffic on
the memory bus.

A countermeasure against an unscrupulous provider is to store code and data in \emph{oblivious RAM}
(ORAM)~\cite{aegis,phantom}. First proposed by \citet{oram10} and \citet{oram00},
ORAM obfuscates the mapping between
addresses and data, in effect ``encrypting'' the addresses along with
the data.  Replacing RAM with ORAM solves (much of) the security
problem but incurs a substantial slowdown in practical
situations~\cite{csf13,ghostrider,phantom} as reads/writes add
overhead that is polylogarithmic in the size of the memory.

Recent work has explored methods for reducing the cost of programming
with ORAM. \citeN{csf13,ghostrider,scvm} developed a family
of type systems to check when \emph{partial} use of ORAM (alongside
normal, encrypted RAM) results in no loss of security; i.e., only when
the addresses of secret data could indirectly reveal sensitive
information must the data be stored in ORAM. This optimization
can provide order-of-magnitude asymptotic performance
improvements. \citeN{ods}
explored how to build \emph{oblivious data structures} (ODSs), such as
queues or stacks, that are more efficient than their standard
counterparts implemented on top of ORAM. In followup
work, \citeN{oblivm,oblivmsite} devised ObliVM, a programming language 
for implementing such oblivious data
structures, including ORAMs themselves.
A key feature of ObliVM is careful treatment of random numbers, which are
at the heart of state-of-the-art ORAM and ODS algorithms.
While the goal of ObliVM is that well-typed programs are secure, no formal
argument to this effect is made.

In this paper, we present \lang, a core language for oblivious
computation, inspired by ObliVM.  \lang extends a standard
language with primitives for
generating and using uniformly distributed random numbers. We prove
that \lang's type system guarantees \emph{probabilistic memory trace
  obliviousness} (PMTO), i.e., that the possible distribution of
adversary-visible execution traces is independent of the values of
secret variables. This property generalizes the deterministic MTO
property enforced by \citeN{csf13,ghostrider}, which did not consider the use of
randomness. In carrying out this work, we discovered that the ObliVM
type system is unsound, so an important contribution of \lang is a design which
achieves soundness without overly restricting or complicating the
language.

\lang's type system aims to ensure that no probabilistic correlation
forms between secrets and publicly revealed random choices. In oblivious
algorithms it is often the case that a security-sensitive random choice
is made (e.g., where to store a particular block in an ORAM), and
eventually that choice is made visible to the adversary (e.g., when a block
is accessed by the client). This transition from a hidden choice to a public
one---which we call a \emph{revelation}---is not problematic so long as the
revealed value does not communicate information about a secret. \lang
ensures that revelations do not communicate information by guaranteeing that
all revealed values are uniformly distributed.

\lang's type system, presented in Section~\ref{sec:formalism}, ensures that
revelations are uniformly distributed by treating randomly generated numbers as
\emph{affine}, meaning they cannot be freely copied. Affinity prevents
revealing the same number twice, which is problematic because
a second revelation is not uniformly distributed when conditioned on observing the first. Unfortunately, strict
affinity is too strong for implementing oblivious algorithms, which
require the ability to make copies of random numbers which are later revealed.
\lang's type system addresses this by allowing random numbers to be copied as
non-affine secret values which can never be revealed. Moreover, \lang enforces that random
numbers do not influence the choice of whether or not they are revealed, since
this could also result in a non-uniform revelation. For example, a \lang
program cannot copy a random number to a secret and then decide to
reveal the original random number based on the value of the copy. The type system
prevents such behavior by using a new mechanism we call \emph{probability
regions} to track the probabilistic (in)dependence of values in the program.
(Probability regions are missing in ObliVM, and their absence is the source of
ObliVM's unsoundness.)
Section~\ref{sec:pmto} outlines the proof that \lang enjoys
PMTO by relating its semantics to a novel \emph{mixed semantics} whose
terms operate on distributions directly, which makes it easier to
state and prove the PMTO property. Full proofs may be found in
\iftr
Appendix~\ref{sec:proofs}.
\else
the supplemental report~\cite{lobliv-tr}.
\fi

%

\lang is expressive enough to type check interesting algorithms.
Section~\ref{sec:oram-overview} presents the implementation of a
tree-based, non-recursive ORAM (NORAM) that type checks in
a straightforward extension of \lang; we have implemented a type
checker for this extension. Such an NORAM is a key component of state-of-the-art ORAM
implementations~\cite{asiacrypt11,pathoram,circuitoram} and other oblivious
data structures~\cite{ods}, and to our
knowledge ours is the first implementation automatically verified to
be oblivious. Section~\ref{sec:full-oram} shows that
\emph{recursive} ORAM, built on NORAM, is also possible but requires a
few more advanced (but standard) language features we have not
implemented, including region polymorphism, recursive and variant
types, and existential quantification. We have also
experimented with implementing oblivious data structures using our NORAM;
\iftr
Appendix~\ref{sec:ostack}
\else
the supplemental report
\fi
presents \emph{oblivious stacks} (ostacks) in detail. Unfortunately,
\lang's strict ordering on probability regions is too strong, so the
complete ostack implementation will not typecheck. An interesting
future direction would be to apply the
approach of~\citet{Zhang:2019:FTL:3352468.3341697} to integrate
\lang's type system with a general logic, such as that
by~\citet{barthe20probsep}, which can be be used to justify that
omitting the probability region order check is (locally) safe.
We elaborate in Section~\ref{sec:related} when we discuss
related work and make the case that \lang subsumes
previous work on type system design for oblivious computation.
Our type checker and all code examples are online at
\url{https://github.com/plum-umd/oblivml}.
\if

\section{Overview}
\label{sec:overview}

This section first presents the threat model. Then it discusses
\emph{deterministic} oblivious execution, considered by prior
work. Finally, it sketches our novel type system for enforcing
\emph{probabilistic} oblivious execution, which we develop in full in
the rest of the paper.

\subsection{Threat Model}
\label{sec:threat}

We assume a powerful adversary that can make fine-grained observations
about a program's execution. In particular, we use a generalization of
the \emph{program counter (PC) security model}~\cite{pcsec}: The
adversary knows the program being executed, and can observe during
execution the PC, the contents of memory, and memory access patterns.
Some \emph{secret} memory contents may be encrypted (while
\emph{public} memory is not) but all addresses used to access memory are still visible.

Consider an untrusted cloud provider
using a secure processor, like SGX~\cite{sgx}. Reads/writes from/to
memory can be directly observed, but secret memory is encrypted (using
a key kept by the processor). The pattern of accesses, timing
information, and other system features ({e.g.}, instruction cache
misses) provide information about the PC. Another setting is
secure multi-party computation (MPC) using secret
shares~\cite{gmw87}. Here, two parties simultaneously execute the same
program (and thus know the program and program counter), but certain
values---the input values from each party---are kept hidden from
both using secret sharing.

By handling such a strong adversary, our techniques can also handle
adversaries with fewer capabilities, such as those that
can observe memory traffic but not the PC, or can make timing
measurements but cannot observe the PC or memory.

\subsection{Oblivious Execution}
\label{sec:oblivexe}


Our goal is to ensure \emph{memory trace obliviousness (MTO)}, which
is a kind of noninterference
property~\cite{goguen1982security,infoflow}. This property states that
despite being able to observe each address (of instructions and data)
as it is fetched, and each public value, the adversary will not be
able to infer anything about input secret values.

We can formalize this idea as a small-step operational semantics
{{\color{\colorMATH}\(\sigma ;e \longrightarrow^{t} \sigma ^{\prime};e^{\prime}\)}}, which states that an
expression {{\color{\colorMATH}\(e\)}} in memory {{\color{\colorMATH}\(\sigma \)}} transitions to memory {{\color{\colorMATH}\(\sigma ^{\prime}\)}} and
expression {{\color{\colorMATH}\(e^{\prime}\)}} while emitting trace event {{\color{\colorMATH}\(t\)}}. Trace events include
fetched instruction addresses, public values, and addresses of
public and secret values that are read and written. (Secret \emph{values} are
not visible in the trace.) Under this model,
MTO means that running \emph{low-equivalent} input states
{{\color{\colorMATH}\(\sigma _{1};e_{1}\)}} and {{\color{\colorMATH}\(\sigma _{2};e_{2}\)}} will produce the exact same memory
trace, along with low-equivalent output states. Two states are low
equivalent if they agree on the code and public
values (but may differ on secret values). More formally, MTO states that if {{\color{\colorMATH}\(\sigma _{1};e_{1} \sim  \sigma _{2};e_{2}\)}} and
{{\color{\colorMATH}\(\sigma _{1};e_{1} \longrightarrow^{t} \sigma _{1}^{\prime};e_{1}^{\prime}\)}} then there exists {{\color{\colorMATH}\(\sigma _{2}^{\prime};e_{2}^{\prime}\)}} {s.t.}
{{\color{\colorMATH}\(\sigma _{2};e_{2} \longrightarrow^{t} \sigma _{2}^{\prime};e_{2}^{\prime}\)}} and {{\color{\colorMATH}\(\sigma _{1}^{\prime};e_{1}^{\prime} \sim  \sigma _{2}^{\prime};e_{2}^{\prime}\)}}, where {{\color{\colorMATH}\(\sim \)}} denotes low-equivalence.

\begin{figure}
\prefigskip
\begin{tabular}{lll}
\begin{minipage}{0.3\textwidth}
\begin{lstlisting}
 B[0] <- s0
 B[1] <- s1
 ...
 let s = ... // secret bit
 let r = B[s] // leaks s
    // via address trace
\end{lstlisting}
\end{minipage}
&
\begin{minipage}{0.3\textwidth}
\begin{lstlisting}
B[0] <- s0
B[1] <- s1
...
let s = ... // secret bit
let s0' = B[0]
let s1' = B[1]
let r,_ = mux(s,s1',s0')
\end{lstlisting}
\end{minipage}
&
\begin{minipage}{0.4\textwidth}
\begin{lstlisting}
let sk = flip()
let s0',s1' = mux(castS(sk),s1,s0)
B[0] <- s0'
B[1] <- s1'
...
let s = ... // secret bit
let s' = xor(s,sk)
let r = B[castP(s')]
\end{lstlisting}
\end{minipage}
\\
{\small (a) Leaky program} & {\small (b) Deterministic MTO program} & (c) {\small Probabilistic MTO program}
\\
\end{tabular}
\precaptionskip
\caption{Code examples}
\postcaptionskip
\label{fig:code-examples}
\end{figure}

To illustrate how revealing addresses can leak information, consider
the program in Figure~\ref{fig:code-examples}(a).
Here, we assume array \code{B}'s contents are secret, and thus invisible to the
adversary. Variables \code{s0}, \code{s1}, and \code{s} are secret ({i.e.}, encrypted)
inputs. The assignments on the first two lines are safe since we are just
storing secret values in the secret array. The problem is on the last
line, when the
program uses \code{s} to index \code{B}. Since the adversary is able to see which
address was used (in trace {{\color{\colorMATH}\(t\)}}), they can infer \code{s}.

The program in Figure~\ref{fig:code-examples}(b) fixes the problem.
It reads both secret values from \code{B}, and then
uses the \code{mux} to select the one indicated by \code{s}, storing it in \code{r}. The
semantics of \code{mux} is that if the first argument is 1 it pairs and
returns the second two arguments in order, otherwise it swaps them. 
To the adversary this appears as a single
program instruction, and so nothing is learned about \code{s} via
branching. Moreover, nothing is learned from the address trace: We
always unconditionally read both elements of \code{B}, no matter the value of
\code{s}.

While this approach is secure, it is inefficient: To read a
single secret value in \code{B} this code reads \emph{all} values in \code{B}, to
hide which one is being selected. If \code{B} were an array of size
{{\color{\colorMATH}\(N\)}}, this approach would turn an {{\color{\colorMATH}\(O(1)\)}} operation into an {{\color{\colorMATH}\(O(N)\)}} operation.

\subsection{Probabilistic Oblivious Execution}

To improve performance while retaining security, the key is to employ
randomness. In particular, the client can randomly generate and
hold secret a key, using it to map logical addresses used by the program to
physical addresses visible to the adversary. The program in
Figure~\ref{fig:code-examples}(c) illustrates the idea, hinting at the basic approach
to implementing an ORAM.
Rather than deterministically store \code{s0} and \code{s1} in positions 0 and 1 of
\code{B}, respectively, the program scrambles their locations according to a
coin flip, \code{sk}, generated by the call to \code{flip}, and not visible to the adversary. Using
the \code{mux} on line 2, if \code{sk} is 1 then \code{s0} and \code{s1} will be
copied to \code{s0'} and \code{s1'}, respectively, but if \code{sk} is 0 then \code{s0} and \code{s1} will be swapped,
with \code{s0} going into \code{s1'} and \code{s1} going into \code{s0'}.  (The \code{castS}
coercion on \code{sk} is a no-op, used by the type system; it will be
explained in the next subsection.) Values \code{s0'} and \code{s1'} are
then stored at positions 0 and 1, respectively, on lines 3 and 4.
When the program later wishes to look up the value at logical index \code{s},
it must consult \code{sk} to retrieve the mapping. This is done via the \code{xor}
on line 7. Then \code{s'} is used to index \code{B} and retrieve the value
logically indicated by \code{s}.

In terms of memory accesses, this program is more efficient: It
reads \code{B} only once, not twice. One can argue that more work
is done overall, but as we will see in Section~\ref{sec:case-study},
this basic idea does scale up to build recursive ORAMs with access times
of {{\color{\colorMATH}\(O(\mathit{log}_c~N)\)}} for some {{\color{\colorMATH}\(c\)}} (rather than {{\color{\colorMATH}\(O(N)\)}}).

\begin{wrapfigure}[6]{R}{1.2in}
\begin{tabular}{r||l|l}
 & \code{sk}=0 & \code{sk}=1 \\ \hline \hline
\code{s}=0 & 0,1,0 & 0,1,1 \\
\code{s}=1 & 0,1,1 & 0,1,0 \\
\end{tabular}
\precaptionskip
\caption{Possible traces}
\postcaptionskip
\label{tab:traces}
\end{wrapfigure}
This program is also secure: no
matter the value of \code{s}, the adversary learns nothing from the
address trace. Consider Figure~\ref{tab:traces} which tabulates the
four possible traces (the memory indexes used to access \code{B}) depending on
the possible values of \code{s} and \code{sk}.
This table makes plain that our program is not
\emph{deterministically} MTO. Looking at column \code{sk}=0, we can see that a
program that has \code{s}=0 may produce trace 0,1,0 while a program that uses
\code{s}=1 may produce trace 0,1,1; MTO programs may not produce
different traces when using different secrets.

But this is not actually a problem. Assuming that \code{sk}~$=0$ and
\code{sk}~$=1$ are
equally likely, we can see that address traces 0,1,0 and 0,1,1 are also
equally likely no matter whether \code{s}~$=0$ or \code{s}~$=1$. More
specifically, if we assume the adversary's expectation for secret values is
uniformly distributed, then after \emph{conditioning} on knowledge of the third
memory access, the adversary's expectation for the secret remains unchanged,
and thus nothing is learned about \code{s}.
This probabilistic model of adversary knowledge is captured
by a \emph{probabilistic}
variant of MTO. In particular, the probability of any particular trace
event {{\color{\colorMATH}\(t\)}} emitted by two low-equivalent programs should be the same
for both programs, and the resulting programs should also be
low-equivalent. More formally: If {{\color{\colorMATH}\(\sigma_1;e_1 \sim \sigma_2;e_2\)}} then
{{\color{\colorMATH}\(\mathrm{Pr}[\sigma_1;e_1 \longrightarrow^{t} \sigma_1';e_1'] = q\)}} implies
{{\color{\colorMATH}\(\mathrm{Pr}[\sigma_2;e_2 \longrightarrow^{t} \sigma_2';e_2'] = q\)}}
and {{\color{\colorMATH}\(\sigma_1';e_1' \sim \sigma_2';e_2'\)}}. \mwh{Changed the previous to
  match 4.1; old version in comments here}


\subsection{\lang: Obliviousness by Typing}
\label{sec:language-teaser}

The main contribution of this paper is \lang, an expressive language whose type
system guarantees that programs are probabilistically MTO\@.
%
\lang's type system's power derives from two key features:
\emph{affine} treatment of random values, and
\emph{probability regions} to track probabilistic (in)dependence (i.e.,
correlation) between random values that could leak information when a value is
revealed. Together, these features ensure that each time a random value is
revealed to the adversary---even if the value interacted with secrets, like the
secret memory layout of an ORAM---it is \emph{always uniformly
  distributed}, which means that its particular value communicates no
secret information.

\begin{figure}
  \centering
\prefigskip
\begin{tabular}{cc}
  \begin{minipage}{0.45\textwidth}
\begin{lstlisting}
 let sx,sy = (flip(), flip())
 let sz,_ = mux (s,sx,sy)
 output (castP(sz)) (* OK *)
 output (castP(sx)) (* Bad *)
\end{lstlisting}
\end{minipage}
&
\begin{minipage}{0.45\textwidth}
\begin{lstlisting}
 let sx,sy = (flip(),flip())
 let sk,_ = mux(castS(sx),sx,sy)
 let sz,_ = mux(s,sk,flip())
 output (castP(sz)) (* Bad *)
\end{lstlisting}
\end{minipage}
  \\
  {\small (a) Leak by multiple revelation} & {\small (b) Leak due to probabilistic dependence}
  \\
\end{tabular}
\precaptionskip
\caption{Example leaky programs (precluded by \lang type system)}
\postcaptionskip
\label{fig:ex-leaks}
\end{figure}

\paragraph*{Affinity}
In \lang, public and secret bits are given types {{\color{\colorMATH}\({\mfootnotesize{{{\color{\colorSYNTAX}\mtexttt{bit}}}}}_{{\mtext{P}}}\)}} and
{{\color{\colorMATH}\({\mfootnotesize{{{\color{\colorSYNTAX}\mtexttt{bit}}}}}_{{\mtext{S}}}\)}} respectively, and coin flips are given type
{{\color{\colorMATH}\({\mfootnotesize{{{\color{\colorSYNTAX}\mtexttt{flip}}}}}\)}}. Our formalism uses bits for
simplicity; it is easy to generalize to (random fixed-width) integers, which is
done in our implementation. Values of {{\color{\colorMATH}\({\mfootnotesize{{{\color{\colorSYNTAX}\mtexttt{flip}}}}}\)}} type are, like secret
bits of type {{\color{\colorMATH}\({\mfootnotesize{{{\color{\colorSYNTAX}\mtexttt{bit}}}}}_{{\mtext{S}}}\)}}, invisible to the adversary. But a {{\color{\colorMATH}\({\mfootnotesize{{{\color{\colorSYNTAX}\mtexttt{flip}}}}}\)}} can be
revealed by using {{\color{\colorMATH}\({\mfootnotesize{{{\color{\colorSYNTAX}\mtexttt{cast_{P}}}}}}\)}} to convert it to a public bit, as is done on line~8
of Figure~\ref{fig:code-examples}(c) to perform a (publicly visible)
array index operation.

The type system aims to ensure that a {{\color{\colorMATH}\({\mfootnotesize{{{\color{\colorSYNTAX}\mtexttt{flip}}}}}\)}} value is always
uniformly distributed when it is revealed. The uniformity requirement implies
that each flip should be revealed \emph{at most once.} Why? Because
the second time a flip is revealed, its distribution is conditioned on
prior revelations, meaning the each outcome is no longer equally
likely. To see how this situation could end up leaking secret
information,
consider the example in Figure~\ref{fig:ex-leaks}(a).
Lines 1--3 in this code are safe: we generate two coin flips that are
invisible to the adversary, and then store one of them in \code{sz}
depending on whether the secret \code{s} is 1 or not. Revealing
\code{sz} at line 3 is safe: regardless of whether \code{sz} contains
the contents of \code{sx} or \code{sy}, the fact that both are
uniformly distributed means that whatever is revealed, nothing can be
learned about \code{s}. However, revealing \code{sx} on line 4, after
having revealed \code{sz}, is not safe. This is because seeing two
ones or two zeroes in a row is more likely when \code{sz} is
\code{sx}, which happens when \code{s} is one. So this program
violates PMTO.

To prevent this problem, \lang's type system treats values of type
{{\color{\colorMATH}\({\mfootnotesize{{{\color{\colorSYNTAX}\mtexttt{flip}}}}}\)}} affinely, meaning that each can be used at most
once. The read of \code{sx} on line 2 consumes that variable, so it
cannot be used again on the problematic line 4. Likewise, flip variable
\code{sk} is consumed when passed to \code{xor} on line~7 of
Figure~\ref{fig:code-examples}(c), and \code{s'} is consumed when
revealed on line~8.


Unfortunately, a purely affine
treatment of flips would preclude useful algorithms. In particular,
notice that line~2 of Figure~\ref{fig:code-examples}(c) uses \code{sk}
as the guard of a \code{mux}. If doing so consumed \code{sk}, line~7's
use of \code{sk} would fail to type check. To avoid this
problem, \lang relaxes the affinity constraint on flips passed to
\code{castS}. In effect, programs can make many secret {{\color{\colorMATH}\({\mfootnotesize{{{\color{\colorSYNTAX}\mtexttt{bit}}}}}_{{\mtext{S}}}\)}}
copies of a flip, and compute with them, but only the original
{{\color{\colorMATH}\({\mfootnotesize{{{\color{\colorSYNTAX}\mtexttt{flip}}}}}\)}} can ultimately be revealed.

It turns out that this relaxed treatment of affinity is insufficient
to ensure PMTO\@.
The reason is that we can now use non-affine copies of
a coin to make a flip's distribution non-uniform when it is
revealed. To see how, consider the code in
Figure~\ref{fig:ex-leaks}(b).
This code flips two coins, and then uses the \code{mux} to store the
first coin flip, \code{sx}, in \code{sk} if \code{sx} is 1, else to store
the second coin flip there.  Now \code{sk} is more likely to be 1 than not:
{{\color{\colorMATH}\(\mathrm{Pr}[{{\color{\colorTEXT}\textnormal{\code{sk}}}}~=1] = \frac{3}{4}\)}} while
{{\color{\colorMATH}\(\mathrm{Pr}[{{\color{\colorTEXT}\textnormal{\code{sk}}}}~=0] = \frac{1}{4}\)}}. On line~3, the \code{mux}
will store \code{sk} in \code{sz} if secret \code{s} is 1, which means
that if the adversary observes a 1 from the output on line 4, it is
more likely than not that \code{s} is 1. The same sort of issue would
happen if we replaced line~1 from Figure~\ref{fig:code-examples}(c)
with the first two lines above: when the program looks up
\code{B[castP(s')]} on line~8, if the adversary observes 1 for the
address, it is more likely that \code{s} is 0, and vice versa if the
adversary observes 1. Notice that we have not violated affinity here:
no coin flip has been used more than once (other than uses of \code{castS}
which side-step affinity tracking). The problematic
correlation in Figure~\ref{fig:ex-leaks}(b) is incorrectly allowed by
ObliVM~\cite{oblivm}, and is the root of its unsoundness.

\paragraph*{Probability regions}
\lang's type system addresses the problem of probabilistic
correlations leading to non-uniform distributions using a novel
construct we call \emph{probability regions}, which are static names
that represent sets of coin flips, reminiscent of a points-to
location in alias analysis~\cite{Emami:1994:CIP:178243.178264}. We
have elided the region name in our examples so far, but normally
programmers should write {{\color{\colorMATH}\({\mfootnotesize{{{\color{\colorSYNTAX}\mtexttt{flip}}}}}^{\rho }()\)}} for flipping a coin in region
{{\color{\colorMATH}\(\rho \)}}, which then has type {{\color{\colorMATH}\({\mfootnotesize{{{\color{\colorSYNTAX}\mtexttt{flip}}}}}^{\rho }\)}}. Bits derived from flips via
\code{castS} carry the region of the original flip, so
\code{bit} types also include a region {{\color{\colorMATH}\(\rho \)}}.

Regions form a partial order, and the type system enforces an
invariant that each flip labeled with region {{\color{\colorMATH}\(\rho \)}} is probabilistically
independent of all bits derived from flips at regions
{{\color{\colorMATH}\(\rho ^{\prime}\)}} when {{\color{\colorMATH}\(\rho ^{\prime} \sqsubset  \rho \)}}. Then, the type system will
prevent problematic correlations arising among bits and flips, in
particular via the \code{mux} and \code{xor} operations, in a way that
could threaten uniformity.
We can see regions at work in the problematic example above: the region of the
secret bit \code{castS(sx)} is the same region as \code{sx}, since
\code{castS(sx)} was derived from \code{sx}. As such,
there is no assurance of probabilistic independence
between the guard and the branch; indeed, when conditioning on
\code{castS(sx)} to return \code{sx}, the
output will \emph{not} be uniform. On the other hand, if the guard of a
\code{mux} is a bit in region {{\color{\colorMATH}\(\rho \)}} and its branches are flips in region {{\color{\colorMATH}\(\rho ^{\prime}\)}}
where {{\color{\colorMATH}\(\rho  \sqsubset  \rho ^{\prime}\)}}, then the guard is derived from a flip that
is sure to be independent of the branches, so the uniformity of the
output is not threatened. This kind of provable independence is a
critical piece of our Tree ORAM implementation in Section~\ref{sec:case-study}.

\section{Formalism}
\label{sec:formalism}

This section presents the syntax, semantics, and type system of
\lang. The following section proves that \lang's type system is
sufficient to ensure PMTO.

\subsection{Syntax}

\begin{figure} 
\small
\prefigskip
\begingroup\color{\colorMATH}\begin{gather*}\begin{tabularx}{\linewidth}{>{\centering\arraybackslash\(}X<{\)}}\hfill\hspace{0pt} \begin{array}{rcrcl@{\hspace*{1.00em}}l
      } \ell    &{}\in {}& {\mtext{label}}   &{}\mathrel{\Coloneqq }{}& {\mtext{P}} \mathrel{|} {\mtext{S}}                       & {{\color{\colorTEXT}\textnormal{public and secret}}}
      \cr      &{} {}& \multicolumn{3}{c}{{{\color{\colorTEXT}\textnormal{({\mtextit{where}} {{\color{\colorMATH}\({\mtext{P}}\sqsubset {\mtext{S}}\)}})}}}}                   & {{\color{\colorTEXT}\textnormal{security labels}}}
      \cr  \rho    &{}\in {}& R         &{}\mathrel{\Coloneqq }{}& \ldots                                & {{\color{\colorTEXT}\textnormal{probability region}}}
      \cr  b   &{}\in {}& {\mathbb{B}}         &{}\mathrel{\Coloneqq }{}& {\mfootnotesize{{{\color{\colorSYNTAX}\mtexttt{O}}}}} \mathrel{|} {\mfootnotesize{{{\color{\colorSYNTAX}\mtexttt{I}}}}}                       & {{\color{\colorTEXT}\textnormal{bits}}}
      \cr  x,y &{}\in {}& {\mtext{var}}     &{}\mathrel{\Coloneqq }{}& \ldots                                & {{\color{\colorTEXT}\textnormal{variables}}}
      \cr  v   &{}\in {}& {\mtext{val}}     &{}\mathrel{\Coloneqq }{}& x                               & {{\color{\colorTEXT}\textnormal{variable values}}}
      \cr      &{} {}&           &{}\mathrel{|}{}& {\mfootnotesize{{{\color{\colorSYNTAX}\mtexttt{fun}}}}}_{y}(x{\mathrel{:}}\tau ).e               & {{\color{\colorTEXT}\textnormal{function values}}}
      \cr      &{} {}&           &{}\mathrel{|}{}& \langle v,v\rangle                            & {{\color{\colorTEXT}\textnormal{tuple values}}}
      \cr  \tau    &{}\in {}& {\mtext{type}}    &{}\mathrel{\Coloneqq }{}& {\mfootnotesize{{{\color{\colorSYNTAX}\mtexttt{bit}}}}}_{\ell }^{\rho }                     & {{\color{\colorTEXT}\textnormal{non-random bit}}}
      \cr      &{} {}&           &{}\mathrel{|}{}& {\mfootnotesize{{{\color{\colorSYNTAX}\mtexttt{flip}}}}}^{\rho }                       & {{\color{\colorTEXT}\textnormal{secret uniform bit}}}
      \cr      &{} {}&           &{}\mathrel{|}{}& {\mfootnotesize{{{\color{\colorSYNTAX}\mtexttt{ref}}}}}(\tau )                        & {{\color{\colorTEXT}\textnormal{reference}}}
      \cr      &{} {}&           &{}\mathrel{|}{}& \tau  \times  \tau                            & {{\color{\colorTEXT}\textnormal{tuple}}}
      \cr      &{} {}&           &{}\mathrel{|}{}& \tau  \rightarrow  \tau                            & {{\color{\colorTEXT}\textnormal{function}}}
      \end{array}
    \hfill\hspace{0pt} \begin{array}{rcl@{\hspace*{1.00em}}l
      } e \in  {\mtext{exp}}  &{}\mathrel{\Coloneqq }{}& v                               & {{\color{\colorTEXT}\textnormal{value expressions}}}
      \cr      &{}\mathrel{|}{}& b_{\ell }                            & {{\color{\colorTEXT}\textnormal{bit literal}}}
      \cr      &{}\mathrel{|}{}& {\mfootnotesize{{{\color{\colorSYNTAX}\mtexttt{flip}}}}}^{\rho }()                     & {{\color{\colorTEXT}\textnormal{coin flip in region}}}
      \cr      &{}\mathrel{|}{}& {\mfootnotesize{{{\color{\colorSYNTAX}\mtexttt{cast}}}}}_{\ell }(v)                    & {{\color{\colorTEXT}\textnormal{cast flip to bit}}}
      \cr      &{}\mathrel{|}{}& {\mfootnotesize{{{\color{\colorSYNTAX}\mtexttt{mux}}}}}(e,e,e)                    & {{\color{\colorTEXT}\textnormal{atomic conditional}}}
      \cr      &{}\mathrel{|}{}& {\mfootnotesize{{{\color{\colorSYNTAX}\mtexttt{xor}}}}}(e,e)                      & {{\color{\colorTEXT}\textnormal{bit xor}}}
      \cr      &{}\mathrel{|}{}& {\mfootnotesize{{{\color{\colorSYNTAX}\mtexttt{if}}}}}(e)\{ e\} \{ e\}                    & {{\color{\colorTEXT}\textnormal{branch conditional}}}
      \cr      &{}\mathrel{|}{}& {\mfootnotesize{{{\color{\colorSYNTAX}\mtexttt{ref}}}}}(e)                        & {{\color{\colorTEXT}\textnormal{reference creation}}}
      \cr      &{}\mathrel{|}{}& {\mfootnotesize{{{\color{\colorSYNTAX}\mtexttt{read}}}}}(e)                       & {{\color{\colorTEXT}\textnormal{reference read}}}
      \cr      &{}\mathrel{|}{}& {\mfootnotesize{{{\color{\colorSYNTAX}\mtexttt{write}}}}}(e,e)                    & {{\color{\colorTEXT}\textnormal{reference write}}}
      \cr      &{}\mathrel{|}{}& \langle e,e\rangle                            & {{\color{\colorTEXT}\textnormal{tuple creation}}}
      \cr      &{}\mathrel{|}{}& {\mfootnotesize{{{\color{\colorSYNTAX}\mtexttt{let}}}}}\hspace*{0.33em}x = e\hspace*{0.33em}{\mfootnotesize{{{\color{\colorSYNTAX}\mtexttt{in}}}}}\hspace*{0.33em}e              & {{\color{\colorTEXT}\textnormal{variable binding}}}
      \cr      &{}\mathrel{|}{}& {\mfootnotesize{{{\color{\colorSYNTAX}\mtexttt{let}}}}}\hspace*{0.33em}x,y = e\hspace*{0.33em}{\mfootnotesize{{{\color{\colorSYNTAX}\mtexttt{in}}}}}\hspace*{0.33em}e            & {{\color{\colorTEXT}\textnormal{tuple elimination}}}
      \cr      &{}\mathrel{|}{}& e(e)                            & {{\color{\colorTEXT}\textnormal{fun. application}}}
      \end{array}
    \hfill\hspace{0pt}
  \end{tabularx}
\end{gather*}\endgroup
\precaptionskip
\caption{\lang Syntax (source programs)}
\postcaptionskip
\label{fig:syntax}
\end{figure} 

Figure~\ref{fig:syntax} shows the syntax for \lang. The term language is
expressions {{\color{\colorMATH}\(e\)}}. The set of values {{\color{\colorMATH}\(v\)}} is comprised of (1) base values such as
variables {{\color{\colorMATH}\(x\)}} (included to enable a substitution-based semantics) and recursive function
definitions {{\color{\colorMATH}\({\mfootnotesize{{{\color{\colorSYNTAX}\mtexttt{fun}}}}}_{y}(x{\mathrel{:}}\tau ).e\)}} where the function body may refer to itself
using variable {{\color{\colorMATH}\(y\)}}; and (2) connectives from the expression language {{\color{\colorMATH}\(e\)}} which
identify a subset of expressions which are also values, such as pairs {{\color{\colorMATH}\(\langle v,v\rangle \)}}
with type {{\color{\colorMATH}\(\tau  \times  \tau \)}}.

Expressions also include bit literals {{\color{\colorMATH}\(b_{\ell }\)}} (of type {{\color{\colorMATH}\({\mfootnotesize{{{\color{\colorSYNTAX}\mtexttt{bit}}}}}_{\ell }^{\bot }\)}}) which are
either {{\color{\colorMATH}\({\mfootnotesize{{{\color{\colorSYNTAX}\mtexttt{O}}}}}\)}} or {{\color{\colorMATH}\({\mfootnotesize{{{\color{\colorSYNTAX}\mtexttt{I}}}}}\)}} and annotated with their security label
{{\color{\colorMATH}\(\ell \)}}.\footnote{Bit literals are not values to create symmetry with the
  alternative, \emph{mixed} semantics in the next section.}
A security label {{\color{\colorMATH}\(\ell \)}} is either
{{\color{\colorMATH}\({\mtext{S}}\)}} (secret) or {{\color{\colorMATH}\({\mtext{P}}\)}} (public). Values with the label {{\color{\colorMATH}\({\mtext{S}}\)}} are invisible to
the adversary. Bit types include this security label along with a probability
region {{\color{\colorMATH}\(\rho \)}}. The expression {{\color{\colorMATH}\({\mfootnotesize{{{\color{\colorSYNTAX}\mtexttt{flip}}}}}^{\rho }()\)}} produces a flip value, i.e.,
a uniformly random bit of type {{\color{\colorMATH}\({\mfootnotesize{{{\color{\colorSYNTAX}\mtexttt{flip}}}}}^{\rho }\)}}.
The annotation assigns the coin to region {{\color{\colorMATH}\(\rho \)}}. Coin flips are semantically secret, and have
limited use; we can compute on one using {{\color{\colorMATH}\({\mfootnotesize{{{\color{\colorSYNTAX}\mtexttt{mux}}}}}\)}} or {{\color{\colorMATH}\({\mfootnotesize{{{\color{\colorSYNTAX}\mtexttt{xor}}}}}\)}}, cast one to a
public bit via {{\color{\colorMATH}\({\mfootnotesize{{{\color{\colorSYNTAX}\mtexttt{cast}}}}}_{{\mtext{P}}}\)}}, or cast to a secret bit via {{\color{\colorMATH}\({\mfootnotesize{{{\color{\colorSYNTAX}\mtexttt{cast}}}}}_{{\mtext{S}}}\)}}. To
simplify the type system, casts only apply to values, however {{\color{\colorMATH}\({\mfootnotesize{{{\color{\colorSYNTAX}\mtexttt{cast}}}}}_{{\mtext{\ell }}}(e)\)}}
could be used as shorthand for {{\color{\colorMATH}\({\mfootnotesize{{{\color{\colorSYNTAX}\mtexttt{let}}}}}\hspace*{0.33em}x = e\hspace*{0.33em}{\mfootnotesize{{{\color{\colorSYNTAX}\mtexttt{in}}}}}\hspace*{0.33em}{\mfootnotesize{{{\color{\colorSYNTAX}\mtexttt{cast}}}}}_{{\mtext{\ell }}}(x)\)}}.

The expression {{\color{\colorMATH}\({\mfootnotesize{{{\color{\colorSYNTAX}\mtexttt{mux}}}}}(e_{1},e_{2},e_{3})\)}} unconditionally evaluates {{\color{\colorMATH}\(e_{2}\)}} and {{\color{\colorMATH}\(e_{3}\)}} and
returns their values as a pair in the given order if {{\color{\colorMATH}\(e_{1}\)}} evaluates to {{\color{\colorMATH}\({\mfootnotesize{{{\color{\colorSYNTAX}\mtexttt{I}}}}}\)}}, or in the
opposite order if it evaluates to {{\color{\colorMATH}\({\mfootnotesize{{{\color{\colorSYNTAX}\mtexttt{O}}}}}\)}}. This operation is critical for
obliviousness because it is atomic. By contrast, normal conditionals
{{\color{\colorMATH}\({\mfootnotesize{{{\color{\colorSYNTAX}\mtexttt{if}}}}}(e_{1})\{ e_{2}\} \{ e_{3}\} \)}} evaluate either {{\color{\colorMATH}\(e_{2}\)}} or {{\color{\colorMATH}\(e_{3}\)}} depending on {{\color{\colorMATH}\(e_{1}\)}}, never both,
so the branch taken is evident from the trace.  The components of tuples
{{\color{\colorMATH}\(e\)}} constructed as {{\color{\colorMATH}\(\langle e_{1},e_{2}\rangle \)}} can be accessed via {{\color{\colorMATH}\({\mfootnotesize{{{\color{\colorSYNTAX}\mtexttt{let}}}}}\hspace*{0.33em}x_{1},x_{2}=e\hspace*{0.33em}{\mfootnotesize{{{\color{\colorSYNTAX}\mtexttt{in}}}}}\hspace*{0.33em}...\)}}
\lang also has normal let binding, function application, and means to manipulate
mutable reference cells.

\lang captures the key elements that make implementing oblivious
algorithms possible, notably: random and secret bits, trace-oblivious
multiplexing, public revelation of secret random values, and general
computational support in tuples, conditionals and recursive functions.
Other features can be encoded in these, e.g., general numbers and
operators on them can be encoded as tuples of bits, and arrays can be
encoded as tuples of references (read/written using (nested) conditionals).
Our prototype interpreter implements these things directly.

\subsection{Semantics}

\begin{figure}  
\small
\prefigskip
\begingroup\color{\colorMATH}\begin{gather*}\begin{tabularx}{\linewidth}{>{\centering\arraybackslash\(}X<{\)}}\hfill\hspace{0pt} \begin{array}{rcrcl@{\hspace*{1.00em}}l
      } \iota  &{}\in {}& {\mtext{loc}} &{}\approx {}& {\mathbb{N}}            & {{\color{\colorTEXT}\textnormal{ref locations}}}
      \cr  v &{}\in {}& {\mtext{val}} &{}\mathrel{\Coloneqq }{}& \ldots             & {{\color{\colorTEXT}\textnormal{extended\ldots }}}
      \cr    &{} {}&       &{}\mathrel{|}{}& {\mfootnotesize{{{\color{\colorSYNTAX}\mtexttt{bitv}}}}}_{\ell }(b) & {{\color{\colorTEXT}\textnormal{bit value}}}
      \cr    &{} {}&       &{}\mathrel{|}{}& {\mfootnotesize{{{\color{\colorSYNTAX}\mtexttt{flipv}}}}}(b)   & {{\color{\colorTEXT}\textnormal{uniform bit value}}}
      \cr    &{} {}&       &{}\mathrel{|}{}& {\mfootnotesize{{{\color{\colorSYNTAX}\mtexttt{locv}}}}}(\iota )    & {{\color{\colorTEXT}\textnormal{location value}}}
      \end{array}
    \hfill\hspace{0pt} \begin{array}{rcrcl@{\hspace*{1.00em}}l
      } \sigma  &{}\in {}& {\mtext{store}}   &{}\triangleq {}& {\mtext{loc}} \rightharpoonup  {\mtext{val}} & {{\color{\colorTEXT}\textnormal{store}}}
      \cr  e &{}\in {}& {\mtext{exp}}     &{}\mathrel{\Coloneqq }{}& \ldots              & {{\color{\colorTEXT}\textnormal{extended\ldots }}}
      \cr  \varsigma  &{}\in {}& {\mtext{config}}  &{}\mathrel{\Coloneqq }{}& \sigma ,e           & {{\color{\colorTEXT}\textnormal{configuration}}}
      \cr  t &{}\in {}& {\mtext{trace}}   &{}\mathrel{\Coloneqq }{}& {\mfootnotesize{{{\color{\colorSYNTAX}\mtexttt{\epsilon }}}}} \mathrel{|} t\mathord{\cdotp }\varsigma      & {{\color{\colorTEXT}\textnormal{trace}}}
      \cr  E &{}\in {}& {\mtext{context}} &{}\mathrel{\Coloneqq }{}& \ldots              & {{\color{\colorTEXT}\textnormal{eval contexts\ldots }}}
      \end{array}
    \hfill\hspace{0pt}
  \\[-1.75ex]
  \cr \hfill\hspace{0pt} \begingroup\color{\colorTEXT}\boxed{\begingroup\color{\colorMATH} {\mtext{step}}_{{\mathcal{M}}} \in  {\mathbb{N}} \times  {\mtext{config}} \rightharpoonup  {\mathcal{M}} ({\mtext{config}}) \endgroup}\endgroup
  \\[-1.75ex]
  \cr  \begin{array}{lcl
     } {\mtext{step}}_{{\mathcal{M}}}(N,\sigma ,b_{\ell })
        &{}={}& {\mtext{return}}(\sigma ,{\mfootnotesize{{{\color{\colorSYNTAX}\mtexttt{bitv}}}}}_{\ell }(b))
     \cr  {\mtext{step}}_{{\mathcal{M}}}(N,\sigma ,{\mfootnotesize{{{\color{\colorSYNTAX}\mtexttt{flip}}}}}^{\rho }())
        &{}={}& {\mtext{do}}\hspace*{0.33em} b \leftarrow  {\mtext{bit}}(N) \mathrel{;} {\mtext{return}}(\sigma ,{\mfootnotesize{{{\color{\colorSYNTAX}\mtexttt{flipv}}}}}(b))
     \cr  {\mtext{step}}_{{\mathcal{M}}}(N,\sigma ,{\mfootnotesize{{{\color{\colorSYNTAX}\mtexttt{cast}}}}}_{\ell }({\mfootnotesize{{{\color{\colorSYNTAX}\mtexttt{flipv}}}}}(b)))
        &{}={}& {\mtext{return}}(\sigma ,{\mfootnotesize{{{\color{\colorSYNTAX}\mtexttt{bitv}}}}}_{\ell }(b))
     \cr  {\mtext{step}}_{{\mathcal{M}}}(N,\sigma ,{\mfootnotesize{{{\color{\colorSYNTAX}\mtexttt{mux}}}}}({\mfootnotesize{{{\color{\colorSYNTAX}\mtexttt{bitv}}}}}_{\ell _{1}}(b_{1}),{\mfootnotesize{{{\color{\colorSYNTAX}\mtexttt{bitv}}}}}_{\ell _{2}}(b_{2}),{\mfootnotesize{{{\color{\colorSYNTAX}\mtexttt{bitv}}}}}_{\ell _{3}}(b_{3})))
        &{}={}& {\mtext{return}}(\sigma ,\langle {\mfootnotesize{{{\color{\colorSYNTAX}\mtexttt{bitv}}}}}_{\ell }({\mtext{cond}}(b_{1},b_{2},b_{3})),{\mfootnotesize{{{\color{\colorSYNTAX}\mtexttt{bitv}}}}}_{\ell }({\mtext{cond}}(b_{1},b_{3},b_{2}))\rangle )
     \cr  &{} {}& \hspace*{1.00em}{{\color{\colorTEXT}\textnormal{{\mtextit{where}}}}} \hspace*{1.00em} \ell  \triangleq  \ell _{1}\sqcup \ell _{2}\sqcup \ell _{3}
     \cr  {\mtext{step}}_{{\mathcal{M}}}(N,\sigma ,{\mfootnotesize{{{\color{\colorSYNTAX}\mtexttt{mux}}}}}({\mfootnotesize{{{\color{\colorSYNTAX}\mtexttt{bitv}}}}}_{\ell }(b_{1}),{\mfootnotesize{{{\color{\colorSYNTAX}\mtexttt{flipv}}}}}(b_{2}),{\mfootnotesize{{{\color{\colorSYNTAX}\mtexttt{flipv}}}}}(b_{3})))
        &{}={}& {\mtext{return}}(\sigma ,\langle {\mfootnotesize{{{\color{\colorSYNTAX}\mtexttt{flipv}}}}}({\mtext{cond}}(b_{1},b_{2},b_{3})),{\mfootnotesize{{{\color{\colorSYNTAX}\mtexttt{flipv}}}}}({\mtext{cond}}(b_{1},b_{3},b_{2}))\rangle )
     \cr  {\mtext{step}}_{{\mathcal{M}}}(N,\sigma ,{\mfootnotesize{{{\color{\colorSYNTAX}\mtexttt{if}}}}}({\mfootnotesize{{{\color{\colorSYNTAX}\mtexttt{bitv}}}}}_{\ell }(b))\{ e_{1}\} \{ e_{2}\} )
        &{}={}& {\mtext{return}}(\sigma ,{\mtext{cond}}(b,e_{1},e_{2}))
     \cr  {\mtext{step}}_{{\mathcal{M}}}(N,\sigma ,{\mfootnotesize{{{\color{\colorSYNTAX}\mtexttt{xor}}}}}({\mfootnotesize{{{\color{\colorSYNTAX}\mtexttt{bitv}}}}}_{\ell }(b_{1}),{\mfootnotesize{{{\color{\colorSYNTAX}\mtexttt{flipv}}}}}(b_{2})))
        &{}={}& {\mtext{return}}(\sigma ,{\mfootnotesize{{{\color{\colorSYNTAX}\mtexttt{flipv}}}}}(b_{1}\oplus b_{2}))
     \cr  {\mtext{step}}_{{\mathcal{M}}}(N,\sigma ,{\mfootnotesize{{{\color{\colorSYNTAX}\mtexttt{ref}}}}}(v))
        &{}={}& {\mtext{return}}(\sigma [\iota \mapsto v],{\mfootnotesize{{{\color{\colorSYNTAX}\mtexttt{refv}}}}}(\iota )) \hspace*{1.00em} {{\color{\colorTEXT}\textnormal{{\mtextit{where {{\color{\colorMATH}\(\iota  \notin  {\mtext{dom}}(\sigma )\)}}}}}}}
     \cr  {\mtext{step}}_{{\mathcal{M}}}(N,\sigma ,{\mfootnotesize{{{\color{\colorSYNTAX}\mtexttt{read}}}}}({\mfootnotesize{{{\color{\colorSYNTAX}\mtexttt{refv}}}}}(\iota )))
        &{}={}& {\mtext{return}}(\sigma ,\sigma (\iota ))
     \cr  {\mtext{step}}_{{\mathcal{M}}}(N,\sigma ,{\mfootnotesize{{{\color{\colorSYNTAX}\mtexttt{write}}}}}({\mfootnotesize{{{\color{\colorSYNTAX}\mtexttt{refv}}}}}(\iota ),v))
        &{}={}& {\mtext{return}}(\sigma [\iota \mapsto v],\sigma (\iota ))
     \cr  {\mtext{step}}_{{\mathcal{M}}}(N,\sigma ,{\mfootnotesize{{{\color{\colorSYNTAX}\mtexttt{let}}}}}\hspace*{0.33em}x = v\hspace*{0.33em}{\mfootnotesize{{{\color{\colorSYNTAX}\mtexttt{in}}}}}\hspace*{0.33em}e)
        &{}={}& {\mtext{return}}(\sigma ,[v/x]e)
     \cr  {\mtext{step}}_{{\mathcal{M}}}(N,\sigma ,{\mfootnotesize{{{\color{\colorSYNTAX}\mtexttt{let}}}}}\hspace*{0.33em}x_{1},x_{2} = \langle v_{1},v_{2}\rangle \hspace*{0.33em}{\mfootnotesize{{{\color{\colorSYNTAX}\mtexttt{in}}}}}\hspace*{0.33em}e)
        &{}={}& {\mtext{return}}(\sigma ,[v_{1}/x_{1}][v_{2}/x_{2}]e)
     \cr  {\mtext{step}}_{{\mathcal{M}}}(N,\sigma ,(\underbracketarg {v_{1}}{{\mfootnotesize{{{\color{\colorSYNTAX}\mtexttt{fun}}}}}_{y}(x\mathrel{:}\tau ).\hspace*{0.33em}e})(v_{2}))
        &{}={}& {\mtext{return}}(\sigma ,[v_{1}/y][v_{2}/x]e)
     \\[-1.75ex] {\mtext{step}}_{{\mathcal{M}}}(N,\sigma ,E[e])
        &{}={}& {\mtext{do}}\hspace*{0.33em} \sigma ^{\prime},e^{\prime} \leftarrow  {\mtext{step}}_{{\mathcal{M}}}(N,\sigma ,e) \mathrel{;} {\mtext{return}}(\sigma ^{\prime},E[e^{\prime}])
     \cr  {\mtext{step}}_{{\mathcal{M}}}(N,\sigma ,v)
        &{}={}& {\mtext{return}}(\sigma ,v)
     \end{array}
  \\[-1.75ex]
  \cr \hfill\hspace{0pt} \begingroup\color{\colorTEXT}\boxed{\begingroup\color{\colorMATH} {\mtext{nstep}}_{{\mathcal{M}}} \in  {\mathbb{N}} \times  {\mtext{config}} \rightharpoonup  {\mathcal{M}}({\mtext{trace}}) \endgroup}\endgroup
  \cr  \begin{array}{lcl
     } {\mtext{nstep}}_{{\mathcal{M}}}(0,\varsigma )   &{}={}& {\mtext{return}}({\mfootnotesize{{{\color{\colorSYNTAX}\mtexttt{\epsilon }}}}}\mathord{\cdotp }\varsigma )
     \cr  {\mtext{nstep}}_{{\mathcal{M}}}(N+1,\varsigma ) &{}={}& {\mtext{do}}\hspace*{0.33em}t\mathord{\cdotp }\varsigma ^{\prime} \leftarrow  {\mtext{nstep}}_{{\mathcal{M}}}(N,\varsigma ) \mathrel{;} \varsigma ^{\prime \prime} \leftarrow  {\mtext{step}}_{{\mathcal{M}}}(N+1,\varsigma ^{\prime}) \mathrel{;} {\mtext{return}}(t\mathord{\cdotp }\varsigma ^{\prime}\mathord{\cdotp }\varsigma ^{\prime \prime})
     \end{array}
  \\[-1.75ex]
  \cr \hfill\hspace{0pt}  \tilde x \in  {\mathcal{D}}(A) \triangleq  \left\{  f \in  A \rightarrow  {\mathbb{R}} \mathrel{}\middle|\mathrel{} \sum \limits_{x\in A} f(x) = 1 \right\} 
    \hfill\hspace{0pt}  {\mtext{Pr}}\left[\tilde x \mathrel{\dot =} x \right] \triangleq  \tilde x(x)
    \hfill\hspace{0pt} \begingroup\color{\colorTEXT}\boxed{\begingroup\color{\colorMATH} {\mathcal{D}}(A) \in  {\mtext{set}} \endgroup}\endgroup
    \hfill\hspace{0pt}
  \\[-1.75ex]
  \cr \begin{array}{l@{\hspace*{1.00em}}l@{\hspace*{1.00em}}l
    } {\mtext{return}} \in  {\mathcal{D}}(A) & {\mtext{bind}} \in  {\mathcal{D}}(A) \times  (A \rightarrow  {\mathcal{D}}(B)) \rightarrow  {\mathcal{D}}(B) & {\mtext{bit}} \in  {\mathbb{N}} \rightarrow  {\mathcal{D}}({\mathbb{B}})
    \cr  {\mtext{return}}(x) \triangleq  \lambda x^{\prime}.\hspace*{0.33em} \left\{ \begin{array}{l@{\hspace*{1.00em}}c@{\hspace*{1.00em}}l
                           } 1 &{}{{\color{\colorTEXT}\textnormal{{\mtextit{if}}}}}{}& x = x^{\prime}
                           \cr  0 &{}{{\color{\colorTEXT}\textnormal{{\mtextit{if}}}}}{}& x \neq  x^{\prime}
                           \end{array}\right.
     & {\mtext{bind}}(\tilde x,f) \triangleq  \lambda y.\hspace*{0.33em} \sum \limits_{x}f(x)(y)\tilde x(x)
     & {\mtext{bit}}(N) \triangleq  \lambda b.\hspace*{0.33em} \nicefrac{1}{2} 
    \end{array}
  \end{tabularx}
\end{gather*}\endgroup
\precaptionskip
\caption{\lang Semantics}
\postcaptionskip
\label{fig:semantics}
\end{figure} 

Figure~\ref{fig:semantics} presents a monadic, probabilistic small-step
semantics for \lang programs. The top of the figure contains some new and
extended syntax. Values (and, by extension, expressions) are extended with
forms for bit values {{\color{\colorMATH}\({\mfootnotesize{{{\color{\colorSYNTAX}\mtexttt{bitv}}}}}_{\ell }(b)\)}}, flip values {{\color{\colorMATH}\({\mfootnotesize{{{\color{\colorSYNTAX}\mtexttt{flipv}}}}}(b)\)}}, and
reference locations {{\color{\colorMATH}\({\mfootnotesize{{{\color{\colorSYNTAX}\mtexttt{locv}}}}}(\iota )\)}}; these do not appear in source programs. Stores
{{\color{\colorMATH}\(\sigma \)}} map locations to values. Stores are paired with expressions to form
\emph{configurations} {{\color{\colorMATH}\(\varsigma \)}}. A sequence of configurations arising during an
evaluation is collected in a \emph{trace} {{\color{\colorMATH}\(t\)}}.  We define evaluation
contexts {{\color{\colorMATH}\(E\)}} (not shown) in the style of~\citet{felleisen1992revised} to enforce a
left-to-right, call-by-value evaluation strategy.

The semantics is defined using an abstract probability monad {{\color{\colorMATH}\({\mathcal{M}}\)}}. Below the
semantics we define the standard ``denotational'' discrete probability monad
{{\color{\colorMATH}\({\mathcal{D}}\)}}~\cite{10.1007/BFb0092872,Ramsey:2002:SLC:503272.503288}. The \emph{standard} semantics for our language occurs when {{\color{\colorMATH}\({\mathcal{M}} = {\mathcal{D}}\)}}, and we
leave {{\color{\colorMATH}\({\mathcal{M}}\)}} a parameter so we can instantiate the semantics to a new monad in
the next section.

In the probability monad {{\color{\colorMATH}\({\mathcal{D}}\)}}, the {{\color{\colorMATH}\({\mtext{return}}\)}} operation constructs a point
distribution, and the {{\color{\colorMATH}\({\mtext{bind}}\)}} operation encodes the law of total probability,
{i.e.}, constructs a marginal distribution from a conditional one. We only use
proper distributions in the sense that the combined mass of all elements sums
to 1. We do not denote possibly non-terminating programs directly into the
monad, and therefore do not require the use of computable
distributions~\cite{huang-computable-distributions} or
sub-probability distributions~\cite{monniaux-ai-prob}---we use the monad only to denote distributions of
configurations which occur after a finite number of small-step transitions,
which is total.

The definition of {{\color{\colorMATH}\({\mtext{step}}_{{\mathcal{M}}}\)}} describes how a single configuration advances in a
single probabilistic step, yielding a distribution of resulting configurations.
The definition uses Haskell-style {{\color{\colorMATH}\({\mtext{do}}\)}} notation as the usual notation for
{{\color{\colorMATH}\({\mtext{bind}}\)}}. Starting from the bottom, we can see that a value {{\color{\colorMATH}\(v\)}} advances to
itself (more on why, below) and evaluating a redex {{\color{\colorMATH}\(e\)}} within a context {{\color{\colorMATH}\(E\)}}
steps the former and packages its result back with the latter, as usual. The
cases for let binding, pair deconstruction, and function application are
standard, using a substitution-based semantics. Likewise, rules for creating,
reading, and writing from references operate on the store {{\color{\colorMATH}\(\sigma \)}} as
usual. 

Moving to the first case, we see that literals {{\color{\colorMATH}\(b_{\ell }\)}} evaluate in one
step to bit values. A {{\color{\colorMATH}\({\mfootnotesize{{{\color{\colorSYNTAX}\mtexttt{flip}}}}}^{\rho }()\)}} expression evaluates to either
{{\color{\colorMATH}\({\mfootnotesize{{{\color{\colorSYNTAX}\mtexttt{flipv}}}}}({\mfootnotesize{{{\color{\colorSYNTAX}\mtexttt{I}}}}})\)}} or {{\color{\colorMATH}\({\mfootnotesize{{{\color{\colorSYNTAX}\mtexttt{flipv}}}}}({\mfootnotesize{{{\color{\colorSYNTAX}\mtexttt{O}}}}})\)}} as determined by {{\color{\colorMATH}\({\mtext{bit}}(N)\)}}, which for the
monad {{\color{\colorMATH}\({\mathcal{D}}\)}} yields {{\color{\colorMATH}\(\nicefrac{1}{2} \)}} probability for each outcome. (The monad {{\color{\colorMATH}\({\mathcal{D}}\)}} does not
use the {{\color{\colorMATH}\(N\)}} parameter in its definition of {{\color{\colorMATH}\({\mtext{bit}}(N)\)}}, but a later monad
will.)
The {{\color{\colorMATH}\({\mfootnotesize{{{\color{\colorSYNTAX}\mtexttt{cast}}}}}_{\ell }\)}} case converts a flip to a similarly-labeled bit value.
The next few cases use the three-argument metafunction {{\color{\colorMATH}\({\mtext{cond}}(b,X,Y)\)}}, which
returns {{\color{\colorMATH}\(X\)}} if {{\color{\colorMATH}\(b\)}} is {{\color{\colorMATH}\({\mfootnotesize{{{\color{\colorSYNTAX}\mtexttt{I}}}}}\)}}, and {{\color{\colorMATH}\(Y\)}} otherwise.
The two \code{mux} cases operate in a similar way: they return the
second two arguments of the \code{mux} in order when the first
argument is {{\color{\colorMATH}\({\mfootnotesize{{{\color{\colorSYNTAX}\mtexttt{bitv}}}}}_{\ell }({\mfootnotesize{{{\color{\colorSYNTAX}\mtexttt{I}}}}})\)}}, and in reverse order when it is
{{\color{\colorMATH}\({\mfootnotesize{{{\color{\colorSYNTAX}\mtexttt{bitv}}}}}_{\ell }({\mfootnotesize{{{\color{\colorSYNTAX}\mtexttt{O}}}}})\)}}. The security label of the result is the join of the labels of
all elements in involved. (This is not needed for flip values, since these are
always fixed to be secret.)
The case for \code{if} also uses {{\color{\colorMATH}\({\mtext{cond}}\)}} in the expected manner. The
case for \code{xor} permits xor-ing a bit with a flip, returning a
flip.

The bottom of the figure defines function {{\color{\colorMATH}\({\mtext{nstep}}_{{\mathcal{M}}}(N,\varsigma )\)}}. It composes {{\color{\colorMATH}\(N\)}}
invocations of {{\color{\colorMATH}\({\mtext{step}}_{{\mathcal{M}}}\)}} starting at {{\color{\colorMATH}\(\varsigma \)}} to produce a distribution of traces {{\color{\colorMATH}\(t\)}}.

Both {{\color{\colorMATH}\({\mtext{step}}_{{\mathcal{M}}}\)}} and {{\color{\colorMATH}\({\mtext{nstep}}_{{\mathcal{M}}}\)}} are \emph{partial} in the usual way:
They are undefined (``stuck'') for nonsensical programs like
{{\color{\colorMATH}\({\mfootnotesize{{{\color{\colorSYNTAX}\mtexttt{locv}}}}}(\iota ) ({\mfootnotesize{{{\color{\colorSYNTAX}\mtexttt{bitv}}}}}_{\ell }(b))\)}} (treating a reference location as if it were
a function). The \lang type system, explained next, rejects
such programs while also ensuring PMTO.

\subsection{Type System}

\begin{figure} 
\small
\prefigskip
\begingroup\color{\colorMATH}\begin{gather*}
\begin{tabularx}{\linewidth}{>{\centering\arraybackslash\(}X<{\)}}\hfill\hspace{0pt} \begin{array}{rcrcl@{\hspace*{1.00em}}l
    } \vphantom{\overset {\mathord{\bullet }}x}\overset {\smash {\mathord{\bullet }}}\tau  &{}\in {}& {\mtext{t\vphantom{\overset {\mathord{\bullet }}x}\overset {\smash {\mathord{\bullet }}}{yp}e}} &{}\mathrel{\Coloneqq }{}& \tau    \mathrel{|} {\mathord{\bullet }}  \hspace*{0.33em}\hspace*{0.33em}{{\color{\colorTEXT}\textnormal{({\mtextit{where}} {{\color{\colorMATH}\(\tau  \sqsubset  {\mathord{\bullet }}\)}})}}}
    \cr  \kappa    &{}\in {}& {\mtext{kind}}     &{}\mathrel{\Coloneqq }{}& {\mfootnotesize{{{\color{\colorSYNTAX}\mtexttt{U}}}}} \mathrel{|} {\mfootnotesize{{{\color{\colorSYNTAX}\mtexttt{A}}}}}\hspace*{0.33em}\hspace*{0.33em}{{\color{\colorTEXT}\textnormal{({\mtextit{where}} {{\color{\colorMATH}\({\mfootnotesize{{{\color{\colorSYNTAX}\mtexttt{U}}}}}\sqsubset {\mfootnotesize{{{\color{\colorSYNTAX}\mtexttt{A}}}}}\)}})}}}
    \end{array}
  \hfill\hspace{0pt}
  \hfill\hspace{0pt} \begin{array}{c
    } \Gamma  \in  {\mtext{tcxt}} \triangleq  var \rightharpoonup  {\mtext{t\vphantom{\overset {\mathord{\bullet }}x}\overset {\smash {\mathord{\bullet }}}{yp}e}}
    \cr  (\Gamma _{1}\sqcup \Gamma _{2})(x)\triangleq \Gamma _{1}(x)\sqcup \Gamma _{2}(x)
    \end{array}
  \hfill\hspace{0pt}
\\[-1.75ex]
\cr \hfill\hspace{0pt} \begingroup\color{\colorTEXT}\boxed{\begingroup\color{\colorMATH} {\mathcal{K}} \in  {\mtext{type}}\!\rightarrow \!{\mtext{kind}} \endgroup}\endgroup
\\[-1.75ex]
\cr \hfill\hspace{0pt} {\mathcal{K}}({\mfootnotesize{{{\color{\colorSYNTAX}\mtexttt{bit}}}}}_{\ell }^{\rho }) \triangleq  {\mathcal{K}}(\tau _{1}\!\rightarrow \!\tau _{2}) \triangleq  {\mathcal{K}}({\mfootnotesize{{{\color{\colorSYNTAX}\mtexttt{ref}}}}}(\tau )) \triangleq  {\mfootnotesize{{{\color{\colorSYNTAX}\mtexttt{U}}}}}
  \hfill\hspace{0pt} {\mathcal{K}}({\mfootnotesize{{{\color{\colorSYNTAX}\mtexttt{flip}}}}}^{\rho }) \triangleq  {\mfootnotesize{{{\color{\colorSYNTAX}\mtexttt{A}}}}}
  \hfill\hspace{0pt} {\mathcal{K}}(\tau _{1}{\times }\tau _{2}) \triangleq  {\mathcal{K}}(\tau _{1}){\sqcup }{\mathcal{K}}(\tau _{2})
\\[-1.75ex]
\cr \hfill\hspace{0pt} \begingroup\color{\colorTEXT}\boxed{\begingroup\color{\colorMATH} \Gamma  \vdash  e \mathrel{:} \tau  \mathrel{;} \Gamma \endgroup}\endgroup
\\[-4ex]
\parbox{\linewidth}{\def\MathparLineskip{\lineskip=4pt}
\begingroup\color{\colorMATH}\begin{mathpar} \inferrule*[vcenter,lab={\mtextsc{ VarU}}
   ]{{\begin{array}{rcl
      }{\mathcal{K}}(\Gamma (x)) &{}={}& {\mfootnotesize{{{\color{\colorSYNTAX}\mtexttt{U}}}}}
      \cr \Gamma (x)    &{}={}& \tau 
      \end{array}}
      }{
      \Gamma  \vdash  x \mathrel{:} \tau  \mathrel{;} \Gamma 
   }
\and \inferrule*[vcenter,lab={\mtextsc{ VarA}}
   ]{{\begin{array}{rcl
      } {\mathcal{K}}(\Gamma (x)) &{}={}& {\mfootnotesize{{{\color{\colorSYNTAX}\mtexttt{A}}}}}
      \cr  \Gamma (x)    &{}={}& \tau 
      \end{array}}
      }{
      \Gamma  \vdash  x \mathrel{:} \tau  \mathrel{;} \Gamma [x{\mapsto }{\mathord{\bullet }}]
   }
\and \inferrule*[vcenter,lab={\mtextsc{ Bit}}
   ]{ }{
      \Gamma  \vdash  b_{\ell } \mathrel{:} {\mfootnotesize{{{\color{\colorSYNTAX}\mtexttt{bit}}}}}_{\ell }^{\bot } \mathrel{;} \Gamma 
   }
\and \inferrule*[vcenter,lab={\mtextsc{ Flip}}
   ]{
      }{
      \Gamma  \vdash  {\mfootnotesize{{{\color{\colorSYNTAX}\mtexttt{flip}}}}}^{\rho }() \mathrel{:} {\mfootnotesize{{{\color{\colorSYNTAX}\mtexttt{flip}}}}}^{\rho } \mathrel{;} \Gamma 
   }
\and \inferrule*[vcenter,lab={\mtextsc{ Cast-S}}
   ]{ \Gamma  \vdash  x \mathrel{:} {\mfootnotesize{{{\color{\colorSYNTAX}\mtexttt{flip}}}}}^{\rho } \mathrel{;} \underline{\hspace{0.66em}}
      }{
      \Gamma  \vdash  {\mfootnotesize{{{\color{\colorSYNTAX}\mtexttt{cast}}}}}_{{\mtext{S}}}(x) \mathrel{:} {\mfootnotesize{{{\color{\colorSYNTAX}\mtexttt{bit}}}}}_{{\mtext{S}}}^{\rho } \mathrel{;} \Gamma 
   }
\and \inferrule*[vcenter,lab={\mtextsc{ Cast-P}}
   ]{ \Gamma  \vdash  x \mathrel{:} {\mfootnotesize{{{\color{\colorSYNTAX}\mtexttt{flip}}}}}^{\rho } \mathrel{;} \Gamma ^{\prime}
      }{
      \Gamma  \vdash  {\mfootnotesize{{{\color{\colorSYNTAX}\mtexttt{cast}}}}}_{{\mtext{P}}}(x) \mathrel{:} {\mfootnotesize{{{\color{\colorSYNTAX}\mtexttt{bit}}}}}_{{\mtext{P}}}^{\bot } \mathrel{;} \Gamma ^{\prime}
   }
\and \inferrule*[vcenter,lab={\mtextsc{ If}}
   ]{{\begin{array}{c@{\hspace*{0.33em}\hspace*{0.33em}}rcl
      }                             & \Gamma ^{\prime} &{}\vdash {}& e_{1} \mathrel{:} \tau  \mathrel{;} \Gamma _{1}^{\prime \prime}
      \cr  \Gamma   \vdash  e \mathrel{:} {\mfootnotesize{{{\color{\colorSYNTAX}\mtexttt{bit}}}}}_{{\mtext{P}}}^{\bot } \mathrel{;} \Gamma ^{\prime} & \Gamma ^{\prime} &{}\vdash {}& e_{2} \mathrel{:} \tau  \mathrel{;} \Gamma _{2}^{\prime \prime}
      \end{array}}
      }{
      \Gamma  \vdash  {\mfootnotesize{{{\color{\colorSYNTAX}\mtexttt{if}}}}}(e)\{ e_{1}\} \{ e_{2}\}  \mathrel{:} \tau  \mathrel{;} \Gamma _{1}^{\prime \prime} \sqcup  \Gamma _{2}^{\prime \prime}
   }
\and \inferrule*[vcenter,lab={\mtextsc{ Mux-Bit}}
   ]{{\begin{array}{rcl@{\hspace*{0.33em}\hspace*{0.33em}}rcl
      } \Gamma   &{}\vdash {}& e_{1} \mathrel{:} {\mfootnotesize{{{\color{\colorSYNTAX}\mtexttt{bit}}}}}_{\ell _{1}}^{\rho _{1}} \mathrel{;} \Gamma ^{\prime} &
      \cr  \Gamma ^{\prime} &{}\vdash {}& e_{2} \mathrel{:} {\mfootnotesize{{{\color{\colorSYNTAX}\mtexttt{bit}}}}}_{\ell _{2}}^{\rho _{2}} \mathrel{;} \Gamma ^{\prime \prime} & \ell  &{}={}& \ell _{1}\!\sqcup \!\ell _{2}\!\sqcup \!\ell _{3}
      \cr  \Gamma ^{\prime \prime} &{}\vdash {}& e_{3} \mathrel{:} {\mfootnotesize{{{\color{\colorSYNTAX}\mtexttt{bit}}}}}_{\ell _{3}}^{\rho _{3}} \mathrel{;} \Gamma ^{\prime \prime \prime} & \rho  &{}={}& \rho _{1}\!\sqcup \!\rho _{2}\!\sqcup \!\rho _{3}
      \end{array}}
      }{
      \Gamma  \vdash  {\mfootnotesize{{{\color{\colorSYNTAX}\mtexttt{mux}}}}}(e_{1},e_{2},e_{3}) \mathrel{:} {\mfootnotesize{{{\color{\colorSYNTAX}\mtexttt{bit}}}}}_{\ell }^{\rho }\!\times \!{\mfootnotesize{{{\color{\colorSYNTAX}\mtexttt{bit}}}}}_{\ell }^{\rho } \mathrel{;} \Gamma ^{\prime \prime \prime}
   }
\and \inferrule*[vcenter,lab={\mtextsc{ Mux-Flip}}
   ]{{\begin{array}{rcl@{\hspace*{0.33em}\hspace*{0.33em}}rcl
      } \Gamma   &{}\vdash {}& e_{1} \mathrel{:} {\mfootnotesize{{{\color{\colorSYNTAX}\mtexttt{bit}}}}}_{\ell _{1}}^{\rho _{1}} \mathrel{;} \Gamma ^{\prime} & \rho _{1} &{}\sqsubset {}& \rho _{2}
      \cr  \Gamma ^{\prime} &{}\vdash {}& e_{2} \mathrel{:} {\mfootnotesize{{{\color{\colorSYNTAX}\mtexttt{flip}}}}}^{\rho _{2}} \mathrel{;} \Gamma ^{\prime \prime}    & \rho _{1} &{}\sqsubset {}& \rho _{3}
      \cr  \Gamma ^{\prime \prime} &{}\vdash {}& e_{3} \mathrel{:} {\mfootnotesize{{{\color{\colorSYNTAX}\mtexttt{flip}}}}}^{\rho _{3}} \mathrel{;} \Gamma ^{\prime \prime \prime}    & \multicolumn{3}{c}{\rho  = \rho _{1}\!\sqcup \rho _{2}\!\sqcup \!\rho _{3}}
      \end{array}}
      }{
      \Gamma  \vdash  {\mfootnotesize{{{\color{\colorSYNTAX}\mtexttt{mux}}}}}(e_{1},e_{2},e_{3}) \mathrel{:} {\mfootnotesize{{{\color{\colorSYNTAX}\mtexttt{flip}}}}}^{\rho }\! \times \! {\mfootnotesize{{{\color{\colorSYNTAX}\mtexttt{flip}}}}}^{\rho } \mathrel{;} \Gamma ^{\prime \prime \prime}
   }
\and \inferrule*[vcenter,lab={\mtextsc{ Xor-Flip}}
   ]{{\begin{array}{rcl@{\hspace*{0.33em}\hspace*{0.33em}}c
      } \Gamma   &{}\vdash {}& e_{1} \mathrel{:} {\mfootnotesize{{{\color{\colorSYNTAX}\mtexttt{bit}}}}}_{\ell _{1}}^{\rho _{1}} \mathrel{;} \Gamma ^{\prime} &
      \cr  \Gamma ^{\prime} &{}\vdash {}& e_{2} \mathrel{:} {\mfootnotesize{{{\color{\colorSYNTAX}\mtexttt{flip}}}}}^{\rho _{2}}    \mathrel{;} \Gamma ^{\prime \prime} & \rho _{1} \sqsubset  \rho _{2}
      \end{array}}
      }{
      \Gamma  \vdash  {\mfootnotesize{{{\color{\colorSYNTAX}\mtexttt{xor}}}}}(e_{1},e_{2}) \mathrel{:} {\mfootnotesize{{{\color{\colorSYNTAX}\mtexttt{flip}}}}}^{\rho _{2}} \mathrel{;} \Gamma ^{\prime \prime}
   }
\and \inferrule*[vcenter,lab={\mtextsc{ Ref}}
   ]{ \Gamma  \vdash  e \mathrel{:} \tau  \mathrel{;} \Gamma ^{\prime}
      }{
      \Gamma  \vdash  {\mfootnotesize{{{\color{\colorSYNTAX}\mtexttt{ref}}}}}(e) \mathrel{:} {\mfootnotesize{{{\color{\colorSYNTAX}\mtexttt{ref}}}}}(\tau ) \mathrel{;} \Gamma ^{\prime}
   }
\and \inferrule*[vcenter,lab={\mtextsc{ Read}}
   ]{ {\mathcal{K}}(\tau ) = {\mfootnotesize{{{\color{\colorSYNTAX}\mtexttt{U}}}}}
   \\\\ \Gamma  \vdash  e \mathrel{:} {\mfootnotesize{{{\color{\colorSYNTAX}\mtexttt{ref}}}}}(\tau ) \mathrel{;} \Gamma ^{\prime}
      }{
      \Gamma  \vdash  {\mfootnotesize{{{\color{\colorSYNTAX}\mtexttt{read}}}}}(e) \mathrel{:} \tau  \mathrel{;} \Gamma ^{\prime}
   }
\and \inferrule*[vcenter,lab={\mtextsc{ Write}}
   ]{ \Gamma   \vdash  e_{1} \mathrel{:} {\mfootnotesize{{{\color{\colorSYNTAX}\mtexttt{ref}}}}}(\tau ) \mathrel{;} \Gamma ^{\prime}
   \\ \Gamma ^{\prime} \vdash  e_{2} \mathrel{:} \tau  \mathrel{;} \Gamma ^{\prime \prime}
      }{
      \Gamma  \vdash  {\mfootnotesize{{{\color{\colorSYNTAX}\mtexttt{write}}}}}(e_{1},e_{2}) \mathrel{:} \tau  \mathrel{;} \Gamma ^{\prime \prime}
   }
\and \inferrule*[vcenter,lab={\mtextsc{ Tup}}
   ]{ \Gamma   \vdash  e_{1} \mathrel{:} \tau _{1} \mathrel{;} \Gamma ^{\prime}
   \\ \Gamma ^{\prime} \vdash  e_{2} \mathrel{:} \tau _{2} \mathrel{;} \Gamma ^{\prime \prime}
      }{
      \Gamma  \vdash  \langle e_{1},e_{2}\rangle  \mathrel{:} \tau _{1} \times  \tau _{2} \mathrel{;} \Gamma ^{\prime \prime}
   }
\and \inferrule*[vcenter,lab={\mtextsc{ Fun}}
   ]{{\begin{array}{rcl@{\hspace*{0.33em}\hspace*{0.33em}}rcl
      }    &{} {}&              & \Gamma ^{+} &{}={}& \Gamma \uplus [x {\mapsto } \tau _{1},y {\mapsto } (\tau _{1} {\rightarrow } \tau _{2})]
      \cr  \Gamma ^{+} &{}\vdash {}& e \mathrel{:} \tau _{2} \mathrel{;} \Gamma ^{+ \prime} & \Gamma ^{+ \prime}&{}={}& \Gamma \uplus [x {\mapsto } \underline{\hspace{0.66em}},y {\mapsto } \underline{\hspace{0.66em}}]
      \end{array}}
      }{
      \Gamma  \vdash  {\mfootnotesize{{{\color{\colorSYNTAX}\mtexttt{fun}}}}}_{y}(x\mathrel{:}\tau _{1}).\hspace*{0.33em} e \mathrel{:} \tau _{1} \rightarrow  \tau _{2} \mathrel{;} \Gamma 
   }
\and \inferrule*[vcenter,lab={\mtextsc{ App}}
   ]{{\begin{array}{rcl
      } \Gamma   &{}\vdash {}& e_{1} \mathrel{:} \tau _{1} \rightarrow  \tau _{2} \mathrel{;} \Gamma ^{\prime}
      \cr  \Gamma ^{\prime} &{}\vdash {}& e_{2} \mathrel{:} \tau _{1}      \mathrel{;} \Gamma ^{\prime \prime}
      \end{array}}
      }{
      \Gamma   \vdash  e_{1}(e_{2}) \mathrel{:} \tau _{2} \mathrel{;} \Gamma ^{\prime \prime}
   }
\and \inferrule*[vcenter,lab={\mtextsc{ Let}}
   ]{{\begin{array}{rcl@{\hspace*{0.33em}\hspace*{0.33em}}rcl
      } \Gamma    &{}\vdash {}& e_{1} \mathrel{:} \tau _{1} \mathrel{;} \Gamma ^{\prime}  & \Gamma ^{\prime +} &{}={}& \Gamma ^{\prime}\uplus [x{\mapsto }\tau _{1}]
      \cr  \Gamma ^{\prime +} &{}\vdash {}& e_{2} \mathrel{:} \tau _{2} \mathrel{;} \Gamma ^{\prime \prime +} & \Gamma ^{\prime \prime +} &{}={}& \Gamma ^{\prime \prime}\uplus [x{\mapsto }\underline{\hspace{0.66em}} ]
      \end{array}}
      }{
      \Gamma  \vdash  {\mfootnotesize{{{\color{\colorSYNTAX}\mtexttt{let}}}}}\hspace*{0.33em}x = e_{1}\hspace*{0.33em}{\mfootnotesize{{{\color{\colorSYNTAX}\mtexttt{in}}}}}\hspace*{0.33em}e_{2} \mathrel{:} \tau _{2} \mathrel{;} \Gamma ^{\prime \prime}
   }
\and \inferrule*[vcenter,lab={\mtextsc{ Let-Tup}}
   ]{{\begin{array}{rcl@{\hspace*{0.33em}\hspace*{0.33em}}rcl
      } \Gamma    &{}\vdash {}& e_{1} \mathrel{:} \tau _{1} \times  \tau _{2} \mathrel{;} \Gamma ^{\prime} & \Gamma ^{\prime +} &{}={}& \Gamma ^{\prime}\uplus [x_{1}{\mapsto }\tau _{1},x_{2}{\mapsto }\tau _{2}]
      \cr  \Gamma ^{\prime +} &{}\vdash {}& e_{2} \mathrel{:} \tau _{3} \mathrel{;} \Gamma ^{\prime \prime +}     & \Gamma ^{\prime \prime +} &{}={}& \Gamma ^{\prime \prime}\uplus [x_{1}{\mapsto }\underline{\hspace{0.66em}} ,x_{2}{\mapsto }\underline{\hspace{0.66em}} ]
      \end{array}}
      }{
      \Gamma  \vdash  {\mfootnotesize{{{\color{\colorSYNTAX}\mtexttt{let}}}}}\hspace*{0.33em}x_{1},x_{2} = e_{1} \hspace*{0.33em}{\mfootnotesize{{{\color{\colorSYNTAX}\mtexttt{in}}}}}\hspace*{0.33em} e_{2} \mathrel{:} \tau _{3} \mathrel{;} \Gamma ^{\prime \prime}
   }
\end{mathpar}\endgroup
}
\end{tabularx}
\end{gather*}\endgroup
\precaptionskip
\caption{\lang Type System (source programs)}
\postcaptionskip
\label{fig:typing-rules}
\end{figure} 

Figure~\ref{fig:typing-rules} defines the type system for \lang source programs
as rules for judgment {{\color{\colorMATH}\(\Gamma  \vdash  e \mathrel{:} \tau  \mathrel{;} \Gamma '\)}}, which states that under type
environment {{\color{\colorMATH}\(\Gamma \)}} expression {{\color{\colorMATH}\(e\)}} has type {{\color{\colorMATH}\(\tau \)}}, and yields residual type
environment {{\color{\colorMATH}\(\Gamma ^{\prime}\)}}. We discuss typing configurations, including non-source
program values, in the next section. Type environments map variables to either
types {{\color{\colorMATH}\(\tau \)}} or inaccessibility tags {{\color{\colorMATH}\({\mathord{\bullet }}\)}}, which are used to enforce
affinity of flips. We discuss the three key features of the type
system---affinity, probability regions, and information flow
control---in turn.

\paragraph*{Affinity}
To enforce non-duplicability, when an affine variable is used by the
program, its type is removed from the residual
environment. Figure~\ref{fig:typing-rules} defines kinding metafunction {{\color{\colorMATH}\({\mathcal{K}}\)}}
that assigns a type either the kind universal {{\color{\colorMATH}\({\mfootnotesize{{{\color{\colorSYNTAX}\mtexttt{U}}}}}\)}} (freely duplicatable) or
affine {{\color{\colorMATH}\({\mfootnotesize{{{\color{\colorSYNTAX}\mtexttt{A}}}}}\)}} (non-duplicatable). Bits, functions, and references (but not
their contents, necessarily) are always universal,
and flips are always affine. A pair is considered affine if either of
its components is. Rule {\mtextsc{ VarU}} in Figure~\ref{fig:typing-rules} types
universally-kinded variables; the output environment {{\color{\colorMATH}\(\Gamma \)}} is the same
as the input environment. Rule {\mtextsc{ VarA}} types an affine variable by
marking it {{\color{\colorMATH}\({\mathord{\bullet }}\)}} in the output environment. This rule is sufficient to
rule out the first problematic example in Section~\ref{sec:language-teaser}.

Rules {\mtextsc{ Cast-S}} and {\mtextsc{ Cast-P}} permit converting flips to bits
via the {{\color{\colorMATH}\({\mfootnotesize{{{\color{\colorSYNTAX}\mtexttt{cast}}}}}_{{\mtext{S}}}\)}} and {{\color{\colorMATH}\({\mfootnotesize{{{\color{\colorSYNTAX}\mtexttt{cast}}}}}_{{\mtext{P}}}\)}} coercions, respectively. The
first converts a {{\color{\colorMATH}\({\mfootnotesize{{{\color{\colorSYNTAX}\mtexttt{flip}}}}}^{\rho }\)}} to a {{\color{\colorMATH}\({\mfootnotesize{{{\color{\colorSYNTAX}\mtexttt{bit}}}}}_{{\mtext{S}}}^{\rho }\)}} and does \emph{not} make
its argument inaccessible (it returns the original {{\color{\colorMATH}\(\Gamma \)}}) while the
second converts to a {{\color{\colorMATH}\({\mfootnotesize{{{\color{\colorSYNTAX}\mtexttt{bit}}}}}_{{\mtext{P}}}^{\bot }\)}} and does make it inaccessible
(returning {{\color{\colorMATH}\(\Gamma '\)}}). The type system is enforcing that any
random number is made adversary-visible at most once; secret copies
are allowed because they are never revealed.

References may contain affine values, but references themselves are
universal. Rather than track the affinity of aliased contents
specifically, the {\mtextsc{ Read}} rule disallows reading out of a reference
cell whose contents are affine. Since the write operation returns the
\emph{old} contents of the cell, programs can see the existing
contents of any reference by first writing in a valid
replacement~\cite{Baker:1992:LLL:142137.142162}.

The {\mtextsc{ Fun}} rule ensures that no affine variables in the defining
context are consumed within the body of the function, i.e., they are
not captured by its closure. We write {{\color{\colorMATH}\(\Gamma \uplus [x\mapsto \underline{\hspace{0.66em}}, y\mapsto \underline{\hspace{0.66em}} ]\)}} to split a
context into a part that binds {{\color{\colorMATH}\(x\)}} and {{\color{\colorMATH}\(y\)}} and a part {{\color{\colorMATH}\(\Gamma \)}} that binds the rest;
the {{\color{\colorMATH}\(\Gamma \)}} part is returned, dropping the {{\color{\colorMATH}\(x\)}} and {{\color{\colorMATH}\(y\)}}
bindings. Both {\mtextsc{ Let}} and {\mtextsc{ Let-Tup}} similarly remove their bound
variables.

Finally, note that different variables could be made inaccessible in
different branches of a conditional, so {\mtextsc{ If}} types each branch in the
same initial context, but then joins their the output contexts; if a
variable is made inaccessible by one branch, it will be inaccessible
in the joined environment. Contexts are joined pointwise, and the join of two
pointed types {{\color{\colorMATH}\(\vphantom{\overset {\mathord{\bullet }}x}\overset {\smash {\mathord{\bullet }}}\tau _{1} \sqcup  \vphantom{\overset {\mathord{\bullet }}x}\overset {\smash {\mathord{\bullet }}}\tau _{2}\)}} is {{\color{\colorMATH}\({\mathord{\bullet }}\)}} when either {{\color{\colorMATH}\(\vphantom{\overset {\mathord{\bullet }}x}\overset {\smash {\mathord{\bullet }}}\tau _{i}\)}} is {{\color{\colorMATH}\({\mathord{\bullet }}\)}}, the same as
{{\color{\colorMATH}\(\vphantom{\overset {\mathord{\bullet }}x}\overset {\smash {\mathord{\bullet }}}\tau _{i}\)}} when both {{\color{\colorMATH}\(\vphantom{\overset {\mathord{\bullet }}x}\overset {\smash {\mathord{\bullet }}}\tau _{i}\)}} are equal and not {{\color{\colorMATH}\({\mathord{\bullet }}\)}}, and undefined otherwise.

\paragraph*{Information flow}
The type system aims to ensure that bits {{\color{\colorMATH}\(b_{\ell }\)}} whose security label
{{\color{\colorMATH}\(\ell \)}} is secret {{\color{\colorMATH}\({\mtext{S}}\)}} cannot be learned by an adversary. Bit types
{{\color{\colorMATH}\({\mfootnotesize{{{\color{\colorSYNTAX}\mtexttt{bit}}}}}_{\ell }^{\rho }\)}} include the security label {{\color{\colorMATH}\(\ell \)}}. The rules treat types
with different labels as distinct, preventing so-called \emph{explicit} flows. For
example, the {\mtextsc{ Write}} rule prevents assigning a secret bit (of type
{{\color{\colorMATH}\({\mfootnotesize{{{\color{\colorSYNTAX}\mtexttt{bit}}}}}_{{\mtext{S}}}^{\rho }\)}}) to a reference whose type is
{{\color{\colorMATH}\({\mfootnotesize{{{\color{\colorSYNTAX}\mtexttt{ref}}}}}({\mfootnotesize{{{\color{\colorSYNTAX}\mtexttt{bit}}}}}_{{\mtext{P}}}^{\rho })\)}}. Likewise, a function of type {{\color{\colorMATH}\({\mfootnotesize{{{\color{\colorSYNTAX}\mtexttt{bit}}}}}_{{\mtext{P}}}^{\rho } \rightarrow 
\tau \)}} cannot be called with an argument of type {{\color{\colorMATH}\({\mfootnotesize{{{\color{\colorSYNTAX}\mtexttt{bit}}}}}_{{\mtext{S}}}^{\rho }\)}}, per the
{\mtextsc{ App}} rule. In our implementation we relax {\mtextsc{ App}} (but not {\mtextsc{ Write}}, due to
the invariance of reference types) to allow public bits
when secrets are expected; this is not done here just to keep things simpler.

The rules also aim to prevent \emph{implicit} information flows. A
typical static information flow type system~\cite{infoflow} would
require the type of the conditional's guard
to be less secret than the type of what it returns; e.g., the guard's
type could be {{\color{\colorMATH}\({\mfootnotesize{{{\color{\colorSYNTAX}\mtexttt{bit}}}}}_{{\mtext{S}}}^{\rho }\)}} but only if the final type {{\color{\colorMATH}\(\tau \)}} is secret
too. However, in \lang we must be more restrictive: rule {\mtextsc{ If}}
requires the guard to be public since the
adversary-visible execution trace reveals which branch is taken, and
thus the truth of the guard. Branching on secrets must be done via
\code{mux}. Notice that rule {\mtextsc{ Mux-Bit}} sets the label {{\color{\colorMATH}\(\ell \)}} of the
each element of the returned pair to be the join of the labels on the
guard and the remaining components. As such, if the guard was secret,
then the returned results will be.  The {\mtextsc{ Mux-Flip}} rule
always returns flips, which are invisible to the adversary, so the
guard can be secret or public.

\paragraph*{Probability regions.}
A probability region {{\color{\colorMATH}\(\rho \)}} appears on both {{\color{\colorMATH}\({\mfootnotesize{{{\color{\colorSYNTAX}\mtexttt{bit}}}}}\)}} and {{\color{\colorMATH}\({\mfootnotesize{{{\color{\colorSYNTAX}\mtexttt{flip}}}}}\)}}
types. The region is a static name for a collection of flip values and
secret bit values that may be derived from them. A flip value is
associated with a region {{\color{\colorMATH}\(\rho \)}} when it is created, per rule {\mtextsc{ Flip}}. Rule
{\mtextsc{ Cast-S}} ascribes the region {{\color{\colorMATH}\(\rho \)}} from the input {{\color{\colorMATH}\({\mfootnotesize{{{\color{\colorSYNTAX}\mtexttt{flip}}}}}^{\rho }\)}} to the
output type {{\color{\colorMATH}\({\mfootnotesize{{{\color{\colorSYNTAX}\mtexttt{bit}}}}}_{{\mtext{S}}}^{\rho }\)}}, tracking the flip value(s) from which the
secret bit value was possibly derived. Per rule {\mtextsc{ Bit}}, bit literals
have probability region {{\color{\colorMATH}\(\bot \)}}, as do public bits produced by
{{\color{\colorMATH}\({\mfootnotesize{{{\color{\colorSYNTAX}\mtexttt{cast}}}}}_{{\mtext{P}}}\)}}, per rule {\mtextsc{ Cast-P}}.

Regions form a join semi-lattice.
The type system maintains the invariant that flips at region {{\color{\colorMATH}\(\rho \)}} are
probabilistically independent of all secret bits in regions {{\color{\colorMATH}\(\rho ^{\prime}\)}} when
strictly ordered {{\color{\colorMATH}\(\rho ^{\prime} \sqsubset  \rho \)}}. Strict ordering is used because it is
{\mtextit{irreflexive}} and {\mtextit{asymmetric}}. The semantic property of
interest---probabilistic independence---is likewise irreflexive (except
for point distributions), and asymmetry restricts future mux operations
between values in one direction only; we say more below.

Consider the {\mtextsc{ Mux-Flip}} rule. If a secret
bit is typed at region {{\color{\colorMATH}\(\rho _{1}\)}} and a flip value at region {{\color{\colorMATH}\(\rho _{2}\)}}, and {{\color{\colorMATH}\(\rho _{1} \slashedrel\sqsubset  \rho _{2}\)}},
then it may be that the values are correlated, and a {{\color{\colorMATH}\({\mfootnotesize{{{\color{\colorSYNTAX}\mtexttt{mux}}}}}\)}} involving the
values may produce flips that are non-uniform. 
Both the {\mtextsc{ Mux-Flip}} and {\mtextsc{ Mux-Bit}} rules return outputs whose region is the join
of the regions of all inputs, indicating that the result of the {{\color{\colorMATH}\({\mfootnotesize{{{\color{\colorSYNTAX}\mtexttt{mux}}}}}\)}} is only
independent of values that were jointly independent of each of its components.

Because freshly generated random bits are always independent of each other, the
programmer is free to choose any regions when generating them via {{\color{\colorMATH}\({\mfootnotesize{{{\color{\colorSYNTAX}\mtexttt{flip}}}}}^{\rho }()\)}}
expressions. However, once chosen, the ordering establishes an invariant which
constrains the order in which mux operations can occur subsequently in the
program.
Requiring strict region ordering for mux operations is enough to reject the example from
the end of Section~\ref{sec:language-teaser}, as it could produce a
non-uniform coin \code{sk}. We recast the example below, labeled (a), using
regions {{\color{\colorMATH}\(\rho _{1} \sqsubset  \rho _{2}\)}}.
\begin{center}
\begin{tabular}{cc}
\begin{minipage}{0.35\textwidth}
\begin{lstlisting}[escapeinside={([}{])}]
let sx,sy = (flip([$^{\rho_1}$])(),flip([$^{\rho_2}$])())
let sk,_ = mux(castS(sx),sx,sy)
\end{lstlisting}
{\small (a) Incorrect example}
\end{minipage}
&
\begin{minipage}{0.45\textwidth}
\begin{lstlisting}[escapeinside={([}{])}]
 let sx = flip([$^{\rho_1}$])() in
 let sy,sz = mux(castS(sx),flip([$^{\rho_2}$])(),flip([$^{\rho_2}$])())
\end{lstlisting}
{\small (b) Correct example}
\end{minipage}
\end{tabular}
\end{center}
The type checker first ascribes types {{\color{\colorMATH}\({\mfootnotesize{{{\color{\colorSYNTAX}\mtexttt{flip}}}}}^{{\rho _{1}}}\)}} and
{{\color{\colorMATH}\({\mfootnotesize{{{\color{\colorSYNTAX}\mtexttt{flip}}}}}^{{\rho _{2}}}\)}} to \code{sx} and \code{sy}, respectively, according to
rules {\mtextsc{ Let-Tup}}, {\mtextsc{ Flip}}, and {\mtextsc{ Tup}}. It uses {\mtextsc{ Cast-S}} to give
\code{castS(sx)} type {{\color{\colorMATH}\({\mfootnotesize{{{\color{\colorSYNTAX}\mtexttt{bit}}}}}_{{\mtext{S}}}^{{\rho _{1}}}\)}} and leaves \code{sx}
accessible so that {\mtextsc{ VarA}} can be used to give it and \code{sy} types
{{\color{\colorMATH}\({\mfootnotesize{{{\color{\colorSYNTAX}\mtexttt{flip}}}}}^{{\rho _{1}}}\)}} and {{\color{\colorMATH}\({\mfootnotesize{{{\color{\colorSYNTAX}\mtexttt{flip}}}}}^{{\rho _{2}}}\)}}, respectively (then making them
inaccessible). Rule {\mtextsc{ Mux-Flip}} will now fail because the independence
conditions do not hold. In particular, the region {{\color{\colorMATH}\(\rho _{1}\)}} of the
guard is not strictly less than the region {{\color{\colorMATH}\(\rho _{1}\)}} of the second
argument, i.e., {{\color{\colorMATH}\(\rho _{1} \slashedrel\sqsubset  \rho _{1}\)}}.
The program labeled (b) above is well-typed.
Here, the bit in
the guard has region {{\color{\colorMATH}\(\rho _{1}\)}}, the region of the two flips is {{\color{\colorMATH}\(\rho _{2}\)}} and {{\color{\colorMATH}\(\rho _{1}
\sqsubset  \rho _{2}\)}} as required by {\mtextsc{ Mux-Flip}}. It is easy
to see that both \code{sy} and
\code{sz} are uniformly distributed and independent of \code{sx}.

Rule {\mtextsc{ Xor-Flip}} permits xor'ing a secret with a flip, returning a
flip, as long as the secret's region and the flip's region are
well ordered, which preserves uniformity.

We might be tempted not to order regions but instead
maintain an invariant that flips and bits
in distinct regions are independent. This turns out to not work.
While at the outset a fresh flip value is independent of all other
values in the context of the program, the region ordering is needed to
ensure that {\mfootnotesize{{{\color{\colorSYNTAX}\mtexttt{mux}}}}} operations will only occur in ``one direction.''
{E.g.}, if two fresh flip values are created {{\color{\colorMATH}\(x = {\mfootnotesize{{{\color{\colorSYNTAX}\mtexttt{flip}}}}}^{\rho _{1}}()\)}} and {{\color{\colorMATH}\(y
= {\mfootnotesize{{{\color{\colorSYNTAX}\mtexttt{flip}}}}}^{\rho _{2}}\)}}, it is true that {{\color{\colorMATH}\(x\)}} and {{\color{\colorMATH}\(y\)}} are mutually independent. Thus it would
seem reasonable that {{\color{\colorMATH}\({\mfootnotesize{{{\color{\colorSYNTAX}\mtexttt{mux}}}}}({\mfootnotesize{{{\color{\colorSYNTAX}\mtexttt{cast}}}}}_{S}(x),y,\ldots )\)}} and
{{\color{\colorMATH}\({\mfootnotesize{{{\color{\colorSYNTAX}\mtexttt{mux}}}}}({\mfootnotesize{{{\color{\colorSYNTAX}\mtexttt{cast}}}}}_{S}(y),x,\ldots )\)}} should both be well typed. While they are
both safe in isolation, the combination is problematic. Consider the
results of each {\mfootnotesize{{{\color{\colorSYNTAX}\mtexttt{mux}}}}}---they are both flip values, and they are both
valid to reveal using {{\color{\colorMATH}\({\mfootnotesize{{{\color{\colorSYNTAX}\mtexttt{cast}}}}}_{P}\)}} individually. However, the resulting
values are correlated (revealing one tells you information about the
distribution of the other), which violates the uniformity guarantee of
all {{\color{\colorMATH}\({\mfootnotesize{{{\color{\colorSYNTAX}\mtexttt{cast}}}}}_{P}\)}} results. By ordering the regions, we are essentially
promising to only allow {\mfootnotesize{{{\color{\colorSYNTAX}\mtexttt{mux}}}}} operations like this in one direction
but not the other, and therefore uniformity is never violated for
revealed flip values. For example, by requiring {{\color{\colorMATH}\(\rho _{1} \sqsubset  \rho _{2}\)}} we allow the
first {\mfootnotesize{{{\color{\colorSYNTAX}\mtexttt{mux}}}}} above but not the second.

\paragraph*{Type safety}

\lang is type safe in the traditional sense, i.e., that a well-typed
program will not get stuck. However, our interest is in the stronger
property that type-safe \lang programs do not reveal secret
information via inferences an adversary can draw from observing their
execution. We state and prove this stronger property in the next section.

\section{Probabilistic Memory Trace Obliviousness}
\label{sec:pmto}

\begin{figure} 
  \small
\prefigskip
\begingroup\color{\colorMATH}\begin{gather*}\begin{tabularx}{\linewidth}{>{\centering\arraybackslash\(}X<{\)}}\hfill\hspace{0pt} \begin{array}{rcrcl
      } \vphantom{\overset {\mathord{\bullet }}x}\overset {\smash {\mathord{\bullet }}}v &{}\in {}& {\mtext{v\vphantom{\overset {\mathord{\bullet }}x}\overset {\smash {\mathord{\bullet }}}alue}}  &{}\mathrel{\Coloneqq }{}& \ldots  \mathrel{|} {\mathord{\bullet }}
      \cr  \vphantom{\overset {\mathord{\bullet }}x}\overset {\smash {\mathord{\bullet }}}e &{}\in {}& {\mtext{e\vphantom{\overset {\mathord{\bullet }}x}\overset {\smash {\mathord{\bullet }}}xp}}    &{}\mathrel{\Coloneqq }{}& \ldots  \mathrel{|} {\mathord{\bullet }}
      \end{array}
    \hfill\hspace{0pt} \begin{array}{rcrcl
      } \vphantom{\overset {\mathord{\bullet }}x}\overset {\smash {\mathord{\bullet }}}\sigma  &{}\in {}& {\mtext{st\vphantom{\overset {\mathord{\bullet }}x}\overset {\smash {\mathord{\bullet }}}ore}}  &{}\triangleq {}&  {\mtext{loc}} \rightharpoonup  {\mtext{v\vphantom{\overset {\mathord{\bullet }}x}\overset {\smash {\mathord{\bullet }}}alue}}
      \cr  \vphantom{\overset {\mathord{\bullet }}x}\overset {\smash {\mathord{\bullet }}}\varsigma  &{}\in {}& {\mtext{co\vphantom{\overset {\mathord{\bullet }}x}\overset {\smash {\mathord{\bullet }}}nfig}} &{}\mathrel{\Coloneqq } {}&  \vphantom{\overset {\mathord{\bullet }}x}\overset {\smash {\mathord{\bullet }}}\sigma ,\vphantom{\overset {\mathord{\bullet }}x}\overset {\smash {\mathord{\bullet }}}e
      \end{array}
    \hfill\hspace{0pt} \vphantom{\overset {\mathord{\bullet }}x}\overset {\smash {\mathord{\bullet }}}t\in {\mtext{tr\vphantom{\overset {\mathord{\bullet }}x}\overset {\smash {\mathord{\bullet }}}ace}}  \mathrel{\Coloneqq } {\mfootnotesize{{{\color{\colorSYNTAX}\mtexttt{\epsilon }}}}} \mathrel{|} \vphantom{\overset {\mathord{\bullet }}x}\overset {\smash {\mathord{\bullet }}}t\mathord{\cdotp }\vphantom{\overset {\mathord{\bullet }}x}\overset {\smash {\mathord{\bullet }}}\varsigma 
    \hfill\hspace{0pt}
  \\[-1.75ex]
  \cr \hfill\hspace{0pt} \begingroup\color{\colorTEXT}\boxed{\begingroup\color{\colorMATH} {\mtext{obs}} \in  ({\mtext{exp}} \rightarrow  {\mtext{e\vphantom{\overset {\mathord{\bullet }}x}\overset {\smash {\mathord{\bullet }}}xp}}) \times  ({\mtext{store}} \rightarrow  {\mtext{st\vphantom{\overset {\mathord{\bullet }}x}\overset {\smash {\mathord{\bullet }}}ore}}) \times  ({\mtext{config}} \rightarrow  {\mtext{co\vphantom{\overset {\mathord{\bullet }}x}\overset {\smash {\mathord{\bullet }}}nfig}}) \times  ({\mtext{trace}} \rightarrow  {\mtext{tr\vphantom{\overset {\mathord{\bullet }}x}\overset {\smash {\mathord{\bullet }}}ace}}) \endgroup}\endgroup
  \\[-1.75ex]
  \cr \hfill\hspace{0pt} \begin{array}{lcl
      } {\mtext{obs}}(x)                    &{}\triangleq {}& x
      \cr  {\mtext{obs}}({\mfootnotesize{{{\color{\colorSYNTAX}\mtexttt{fun}}}}}_{y}(x\mathrel{:}\tau ).\hspace*{0.33em}e)      &{}\triangleq {}& {\mfootnotesize{{{\color{\colorSYNTAX}\mtexttt{fun}}}}}_{y}(x\mathrel{:}\tau ).\hspace*{0.33em}{\mtext{obs}}(e)
      \cr  {\mtext{obs}}({\mfootnotesize{{{\color{\colorSYNTAX}\mtexttt{bitv}}}}}_{P}(b))         &{}\triangleq {}& {\mfootnotesize{{{\color{\colorSYNTAX}\mtexttt{bitv}}}}}_{P}(b)
      \cr  {\mtext{obs}}({\mfootnotesize{{{\color{\colorSYNTAX}\mtexttt{bitv}}}}}_{S}(b))         &{}\triangleq {}& {\mathord{\bullet }}
      \cr  {\mtext{obs}}({\mfootnotesize{{{\color{\colorSYNTAX}\mtexttt{flipv}}}}}(b))           &{}\triangleq {}& {\mathord{\bullet }}
      \cr  {\mtext{obs}}({\mfootnotesize{{{\color{\colorSYNTAX}\mtexttt{locv}}}}}(\iota ))            &{}\triangleq {}& {\mathord{\bullet }}
      \cr  {\mtext{obs}}(b_{P})                 &{}\triangleq {}& b_{P}
      \cr  {\mtext{obs}}(b_{S})                 &{}\triangleq {}& {\mathord{\bullet }}
      \cr  {\mtext{obs}}({\mfootnotesize{{{\color{\colorSYNTAX}\mtexttt{flip}}}}}^{\rho }())          &{}\triangleq {}& {\mfootnotesize{{{\color{\colorSYNTAX}\mtexttt{flip}}}}}^{\rho }()
      \cr  {\mtext{obs}}({\mfootnotesize{{{\color{\colorSYNTAX}\mtexttt{cast}}}}}_{\ell }(v))         &{}\triangleq {}& {\mfootnotesize{{{\color{\colorSYNTAX}\mtexttt{cast}}}}}_{\ell }({\mtext{obs}}(v))
      \end{array}
    \hfill\hspace{0pt} \begin{array}{lcl
      }
      \cr  {\mtext{obs}}({\mfootnotesize{{{\color{\colorSYNTAX}\mtexttt{mux}}}}}(e_{1},e_{2},e_{3}))      &{}\triangleq {}& {\mfootnotesize{{{\color{\colorSYNTAX}\mtexttt{mux}}}}}({\mtext{obs}}(e_{1}),{\mtext{obs}}(e_{2}),{\mtext{obs}}(e_{3}))
      \cr  {\mtext{obs}}({\mfootnotesize{{{\color{\colorSYNTAX}\mtexttt{xor}}}}}(e_{1},e_{2}))         &{}\triangleq {}& {\mfootnotesize{{{\color{\colorSYNTAX}\mtexttt{xor}}}}}({\mtext{obs}}(e_{1}),{\mtext{obs}}(e_{2}))
      \cr  {\mtext{obs}}({\mfootnotesize{{{\color{\colorSYNTAX}\mtexttt{if}}}}}(e_{1})\{ e_{2}\} \{ e_{3}\} )     &{}\triangleq {}& {\mfootnotesize{{{\color{\colorSYNTAX}\mtexttt{if}}}}}({\mtext{obs}}(e_{1}))\{ {\mtext{obs}}(e_{2})\} \{ {\mtext{obs}}(e_{3})\} 
      \cr  {\mtext{obs}}({\mfootnotesize{{{\color{\colorSYNTAX}\mtexttt{ref}}}}}(e))             &{}\triangleq {}& {\mfootnotesize{{{\color{\colorSYNTAX}\mtexttt{ref}}}}}({\mtext{obs}}(e))
      \cr  {\mtext{obs}}({\mfootnotesize{{{\color{\colorSYNTAX}\mtexttt{read}}}}}(e))            &{}\triangleq {}& {\mfootnotesize{{{\color{\colorSYNTAX}\mtexttt{read}}}}}({\mtext{obs}}(e))
      \cr  {\mtext{obs}}({\mfootnotesize{{{\color{\colorSYNTAX}\mtexttt{write}}}}}(e_{1},e_{2}))       &{}\triangleq {}& {\mfootnotesize{{{\color{\colorSYNTAX}\mtexttt{write}}}}}({\mtext{obs}}(e_{1}),{\mtext{obs}}(e_{2}))
      \cr  {\mtext{obs}}(\langle e_{1},e_{2}\rangle )              &{}\triangleq {}& \langle {\mtext{obs}}(e_{1}),{\mtext{obs}}(e_{2})\rangle 
      \cr  {\mtext{obs}}({\mfootnotesize{{{\color{\colorSYNTAX}\mtexttt{let}}}}}\hspace*{0.33em}x=e_{1}\hspace*{0.33em}{\mfootnotesize{{{\color{\colorSYNTAX}\mtexttt{in}}}}}\hspace*{0.33em}e_{2})   &{}\triangleq {}& {\mfootnotesize{{{\color{\colorSYNTAX}\mtexttt{let}}}}}\hspace*{0.33em}x={\mtext{obs}}(e_{1})\hspace*{0.33em}{\mfootnotesize{{{\color{\colorSYNTAX}\mtexttt{in}}}}}\hspace*{0.33em}{\mtext{obs}}(e_{2})
      \cr  {\mtext{obs}}({\mfootnotesize{{{\color{\colorSYNTAX}\mtexttt{let}}}}}\hspace*{0.33em}x,y=e_{1}\hspace*{0.33em}{\mfootnotesize{{{\color{\colorSYNTAX}\mtexttt{in}}}}}\hspace*{0.33em}e_{2}) &{}\triangleq {}& {\mfootnotesize{{{\color{\colorSYNTAX}\mtexttt{let}}}}}\hspace*{0.33em}x,y={\mtext{obs}}(e_{1})\hspace*{0.33em}{\mfootnotesize{{{\color{\colorSYNTAX}\mtexttt{in}}}}}\hspace*{0.33em}{\mtext{obs}}(e_{2})
      \cr  {\mtext{obs}}(e_{1}(e_{2}))               &{}\triangleq {}& {\mtext{obs}}(e_{1})({\mtext{obs}}(e_{2}))
      \end{array}
    \hfill\hspace{0pt}
  \\[-1.75ex]
  \cr  \hfill\hspace{0pt}  \begin{array}{rcl
        } {\mtext{obs}}(\sigma ) &{}\triangleq {}& \{ \iota  {\mapsto } {\mtext{obs}}(v) \mathrel{|} \iota  {\mapsto } v {\in } \sigma \} 
        \cr  {\mtext{obs}}(\sigma ,e) &{}\triangleq {}& {\mtext{obs}}(\sigma ),{\mtext{obs}}(e)
        \end{array}
     \hfill\hspace{0pt}  \begin{array}{rcl
        } {\mtext{obs}}({\mfootnotesize{{{\color{\colorSYNTAX}\mtexttt{\epsilon }}}}}) &{}\triangleq {}& {\mfootnotesize{{{\color{\colorSYNTAX}\mtexttt{\epsilon }}}}}
        \cr  {\mtext{obs}}(t\mathord{\cdotp }\varsigma ) &{}\triangleq {}& {\mtext{obs}}(t)\mathord{\cdotp }{\mtext{obs}}(\varsigma )
        \end{array}
     \hfill\hspace{0pt}
  \\[-1.75ex]
  \cr  \hfill\hspace{0pt} \widetilde {\mtext{obs}}(\tilde t) \triangleq  {\mtext{do}}\hspace*{0.33em}t\leftarrow \tilde t \mathrel{;} {\mtext{return}}({\mtext{obs}}(t))
     \hfill\hspace{0pt} \begingroup\color{\colorTEXT}\boxed{\begingroup\color{\colorMATH} \widetilde {\mtext{obs}} \in  {\mathcal{D}}({\mtext{trace}}) \rightarrow  {\mathcal{D}}({\mtext{tr\vphantom{\overset {\mathord{\bullet }}x}\overset {\smash {\mathord{\bullet }}}ace}}) \endgroup}\endgroup
  \end{tabularx}
\end{gather*}\endgroup
\precaptionskip
\caption{Adversary observability}
\postcaptionskip
\label{fig:adversary}
\end{figure} 

The main metatheoretic result of this paper is that \lang's type
system ensures probabilistic memory trace obliviousness (PMTO). This
section defines this property, and then walks through its proof.

\subsection{What is PMTO?}

Figure~\ref{fig:adversary} presents a model {{\color{\colorMATH}\({\mtext{obs}}\)}} of the adversary's view of
a computation as a new class of values,
expressions and traces that ``hide'' sub-expressions
considered to be secret (written {{\color{\colorMATH}\({\mathord{\bullet }}\)}}).  Secret bit expressions, secret bit values, and secret flip
values all map to {{\color{\colorMATH}\({\mathord{\bullet }}\)}}. Compound values, expressions, stores, traces etc. call {{\color{\colorMATH}\({\mtext{obs}}\)}} in recursive
positions as expected.

Probabilistic memory trace obliviousness (PMTO), stated formally
below, holds when observationally equivalent configurations induce
distributions of traces that are themselves observationally equivalent
after $N$ steps, for any $N$.\footnote{Noninterference
  properties are often stated with a non-empty store.
  Our notion of expression equivalence is simpler, and supports
  low-equivalent expressions that pre-populate such a
  store, so there is no loss of generality.}
\begin{proposition}[Probabilistic Memory Trace Obliviousness (PMTO)]\label{prop:pmto}\
  \begin{itemize}[label={},leftmargin=0pt]\item  If: \hspace*{0.16em}{{\color{\colorMATH}\(e_{1}\)}}\hspace*{0.16em} and \hspace*{0.16em}{{\color{\colorMATH}\(e_{2}\)}}\hspace*{0.16em} are closed source expressions, \hspace*{0.16em}{{\color{\colorMATH}\(\vdash  e_{1} \mathrel{:} \tau \)}}\hspace*{0.16em}, \hspace*{0.16em}{{\color{\colorMATH}\(\vdash  e_{2} \mathrel{:} \tau \)}}\hspace*{0.16em} and \hspace*{0.16em}{{\color{\colorMATH}\({\mtext{obs}}(e_{1}) = {\mtext{obs}}(e_{2})\)}}
  \item  Then: (1) \hspace*{0.16em}{{\color{\colorMATH}\({\mtext{nstep}}_{{\mathcal{D}}}(N,\varnothing ,e_{1})\)}}\hspace*{0.16em} and \hspace*{0.16em}{{\color{\colorMATH}\({\mtext{nstep}}_{{\mathcal{D}}}(N,\varnothing ,e_{2})\)}}\hspace*{0.16em} are defined
  \item  And: (2) \hspace*{0.16em}{{\color{\colorMATH}\(\widetilde {{\mtext{obs}}}({\mtext{nstep}}_{{\mathcal{D}}}(N,\varnothing ,e_{1})) = \widetilde {{\mtext{obs}}}({\mtext{nstep}}_{{\mathcal{D}}}(N,\varnothing ,e_{2}))\)}}.
  \end{itemize}
\end{proposition}
\noindent
(1) ensures that information is not leaked due to lack of progress, {i.e.}, if
either program gets ``stuck,'' and that the main property (2) applies to all
related, well-typed source expressions {{\color{\colorMATH}\(e_{1}\)}} and {{\color{\colorMATH}\(e_{2}\)}}.

\subsection{Proof Approach}

\newcommand\shortLRef[1]{\hyperref[#1]{L\ref*{#1}}}
\begin{wrapfigure}{R}{0.5\textwidth}
\vspace*{-3ex}
\begin{framed} 
\vspace*{-2ex}
\begin{center}
\hspace*{-2.5em}
\begingroup
\renewcommand\colorMATH{black}
\begin{tikzcd}%
  [ampersand replacement=\&
  ,row sep=large
  ,column sep=large
  ,decoration={snake,amplitude=2pt}]
  \&\&\& \underline {\hat t_{2}} \ar[dd,"{=_{\hat \lceil \mathord{\cdotp }\hat \rceil }\mathrlap{\ \textit{(\shortLRef{thm:simulation-mixed})}}}" description,dash,decorate]
\\
\&     e_{2}       \ar[drr,"{{\mtext{nstep}}_{{\mathcal{I}}}}" description,near end]
                \ar[urr,"{{\mtext{\underline {nste\hspace{-1pt}}\hspace{1pt}p}}_{{\mathcal{I}}}}" description]
\&     \underline {\hat t_{1}} \ar[dd,"{=_{\hat \lceil \mathord{\cdotp }\hat \rceil }\mathrlap{\ \textit{(\shortLRef{thm:simulation-mixed})}}}" description,dash,decorate,crossing over]
                \ar[ur,"{\approx _{\sim }\mathrlap{\ \textit{(\shortLRef{thm:pmto-mixed})}}}" description,dash,decorate]
\\
       e_{1}       \ar[drr,"{{\mtext{nstep}}_{{\mathcal{I}}}}" description]
                \ar[urr,"{{\mtext{\underline {nste\hspace{-1pt}}\hspace{1pt}p}}_{{\mathcal{I}}}}" description,crossing over]
                \ar[ur,"{=_{{\mtext{obs}}}}" description,dash,decorate]
\&\&\& \hat t_{2}
\\
\&\&   \hat t_{1}     \ar[ur,"{\approx _{=_{{\mtext{obs}}}}\mathrlap{\ \textit{(\shortLRef{thm:low-equivalence-soundness})}}}" description,dash,decorate]
\end{tikzcd}
\endgroup
\end{center}
\vspace*{-2ex}
\end{framed}
\vspace*{-2ex}
\caption{Proof Approach as a Diagram}
\label{fig:proof-approach}
\vspace*{-2ex}
\end{wrapfigure} 

The remainder of this section works through our proof of PMTO
(Theorem~\ref{thm:pmto}) which we complete in the following steps: (1) we
develop a new probability monad called ``intensional distributions'' which
simplifies reasoning about conditional independence between probabilistic
values (\S \ref{sec:intensional-distributions}); (2) we define an alternative
syntax, semantics and type system for \lang programs called the ``mixed
semantics'' which uses intensional distributions to simplify inductive reasoning
about the adversary's view of probabilistic secret values
(\S \ref{sec:mixed-sem}, \S \ref{sec:mixed-sem-typing}); (3) we show that
evaluation in the mixed semantics corresponds exactly with the ground truth
semantics through simulation lemmas; (4) we prove that key invariants about
probabilistic values are ensured by well-typed mixed terms, and that terms
remain well-typed throughout evaluation---this establishes PMTO for the mixed
semantics; and (5) we demonstrate PMTO for the ground truth semantics as a
consequence of lemmas established in steps (3--4) and a soundness lemma relating
equivalent distributions of mixed terms to adversary-equivalent distributions
of standard terms.

In Figure~\ref{fig:proof-approach} we summarize the structure of this proof
approach in a diagram. On the left are two programs {{\color{\colorMATH}\(e_{1}\)}} and {{\color{\colorMATH}\(e_{2}\)}} which are
equal modulo adversary observation {{\color{\colorMATH}\(=_{{\mtext{obs}}}\)}}, which translates to
{{\color{\colorMATH}\({\mtext{obs}}(e_{1}) = {\mtext{obs}}(e_{2})\)}} as sketched in Proposition~\ref{prop:pmto}, and
means {{\color{\colorMATH}\(e_{1}\)}} and {{\color{\colorMATH}\(e_{2}\)}} agree on public values and program structure but may differ in secrets. 
The rightward moving arrows represent running each program
in either the ground truth semantics {{\color{\colorMATH}\({\mtext{step}}_{{\mathcal{I}}}\)}}---the same semantics from
Figure~\ref{fig:semantics} but instantiated with the intensional distribution
monad {{\color{\colorMATH}\({\mathcal{I}}\)}}---and the mixed semantics {{\color{\colorMATH}\({\mtext{\underline {ste\hspace{-1pt}}\hspace{1pt}p}}_{{\mathcal{I}}}\)}}. Each of these executions
result in intensional distributions of standard and mixed traces,
respectively. In step (3) above we prove Lemma~\ref{thm:simulation-mixed}
to show these distributions are equivalent according to {{\color{\colorMATH}\(=_{\hat \lceil \mathord{\cdotp }\hat \rceil }\)}} which uses
{{\color{\colorMATH}\(\hat \lceil \mathord{\cdotp }\hat \rceil \)}} to project distributions of mixed traces to distributions of standard
traces. In step (4) above we prove Lemma~\ref{thm:pmto-mixed} to
establish PMTO for the mixed semantics; i.e., that the
resulting distributions of mixed traces are equivalent modulo an
underlying low-equivalence relation {{\color{\colorMATH}\(\approx _{\sim }\)}}. In step (5) we prove
Lemma~\ref{thm:low-equivalence-soundness}, which combines results
from (3--4) to establish PMTO for the standard semantics (instantiated
with {{\color{\colorMATH}\({\mathcal{I}}\)}})---the resulting distributions of standard traces
are equivalent modulo equality of adversary observations, notated
{{\color{\colorMATH}\(\approx _{=_{{\mtext{obs}}}}\)}}. The last step of PMTO
(Theorem~\ref{thm:pmto}) is not shown:
Lemma~\ref{thm:simulation-intensional} proves via simulation that the
intensional distribution monad {{\color{\colorMATH}\({\mathcal{I}}\)}} corresponds with the usual denotational probability
monad presented in Section~\ref{sec:formalism}.

\subsection{Mixed Semantics}
\label{sec:mixed-sem}

An intuitive approach to proving Proposition~\ref{prop:pmto} is to
prove that a single-step version of it holds for {{\color{\colorMATH}\({\mtext{step}}_{{\mathcal{D}}}\)}}, and then
use that fact in an inductive proof over {{\color{\colorMATH}\({\mtext{nstep}}_{{\mathcal{D}}}\)}}. Unfortunately,
proving the single-step version quickly runs into trouble. Consider
a source program {{\color{\colorMATH}\({\mfootnotesize{{{\color{\colorSYNTAX}\mtexttt{cast}}}}}_{{\mtext{P}}}({\mfootnotesize{{{\color{\colorSYNTAX}\mtexttt{flip}}}}}^{\rho }())\)}} which steps to each of
the expressions {{\color{\colorMATH}\({\mfootnotesize{{{\color{\colorSYNTAX}\mtexttt{cast}}}}}_{{\mtext{P}}}({\mfootnotesize{{{\color{\colorSYNTAX}\mtexttt{flipv}}}}}({\mfootnotesize{{{\color{\colorSYNTAX}\mtexttt{I}}}}}))\)}} and {{\color{\colorMATH}\({\mfootnotesize{{{\color{\colorSYNTAX}\mtexttt{cast}}}}}_{{\mtext{P}}}({\mfootnotesize{{{\color{\colorSYNTAX}\mtexttt{flipv}}}}}({\mfootnotesize{{{\color{\colorSYNTAX}\mtexttt{O}}}}}))\)}}
with probability {{\color{\colorMATH}\(\nicefrac{1}{2} \)}}. These expressions are observationally equivalent---the adversary's
view of each is {{\color{\colorMATH}\({\mfootnotesize{{{\color{\colorSYNTAX}\mtexttt{cast}}}}}_{{\mtext{P}}}({\mathord{\bullet }})\)}}. For single-step PMTO to be satisfied, each of
these terms must {{\color{\colorMATH}\({\mtext{step}}\)}} to an equivalent distribution. Unfortunately, they do
not: The first produces a point distribution of the expression
{{\color{\colorMATH}\({\mfootnotesize{{{\color{\colorSYNTAX}\mtexttt{bitv}}}}}_{{\mtext{P}}}({\mfootnotesize{{{\color{\colorSYNTAX}\mtexttt{I}}}}})\)}} and the second produces a point distribution of the
expression {{\color{\colorMATH}\({\mfootnotesize{{{\color{\colorSYNTAX}\mtexttt{bitv}}}}}_{{\mtext{P}}}({\mfootnotesize{{{\color{\colorSYNTAX}\mtexttt{O}}}}})\)}}, which are not observationally the same.

\begin{figure} 
  \small
\prefigskip
\begingroup\color{\colorMATH}\begin{gather*}\begin{tabularx}{\linewidth}{>{\centering\arraybackslash\(}X<{\)}}\hfill\hspace{0pt} \begingroup\color{\colorTEXT}\boxed{\begingroup\color{\colorMATH} {\mtext{\underline {ste\hspace{-1pt}}\hspace{1pt}p}} \in  {\mathbb{N}} \times  {\mtext{\underline {confi\hspace{-1pt}}\hspace{1pt}g}} \rightharpoonup  {\mathcal{I}} ({\mtext{\underline {confi\hspace{-1pt}}\hspace{1pt}g}}) \endgroup}\endgroup
  \\[-1.75ex]
  \cr \begin{array}{lcl
    } {\mtext{\underline {ste\hspace{-1pt}}\hspace{1pt}p}}(N,\underline \sigma ,b_{\ell })
       &{}\triangleq {}& {\mtext{return}}(\underline \sigma ,{\mfootnotesize{{{\color{\colorSYNTAX}\mtexttt{bitv}}}}}_{\ell }({\mtext{return}}(b)))
    \cr  {\mtext{\underline {ste\hspace{-1pt}}\hspace{1pt}p}}(N,\underline \sigma ,{\mfootnotesize{{{\color{\colorSYNTAX}\mtexttt{flip}}}}}^{\rho }())
       &{}\triangleq {}& {\mtext{return}}(\underline \sigma ,{\mfootnotesize{{{\color{\colorSYNTAX}\mtexttt{flipv}}}}}({\mtext{bit}}(N)))
    \cr  {\mtext{\underline {ste\hspace{-1pt}}\hspace{1pt}p}}(N,\underline \sigma ,{\mfootnotesize{{{\color{\colorSYNTAX}\mtexttt{cast}}}}}_{S}({\mfootnotesize{{{\color{\colorSYNTAX}\mtexttt{flipv}}}}}(\hat b)))
       &{}\triangleq {}& {\mtext{return}}(\underline \sigma ,{\mfootnotesize{{{\color{\colorSYNTAX}\mtexttt{bitv}}}}}_{S}(\hat b))
    \cr  {\mtext{\underline {ste\hspace{-1pt}}\hspace{1pt}p}}(N,\underline \sigma ,{\mfootnotesize{{{\color{\colorSYNTAX}\mtexttt{cast}}}}}_{P}({\mfootnotesize{{{\color{\colorSYNTAX}\mtexttt{flipv}}}}}(\hat b)))
       &{}\triangleq {}& {\mtext{do}}\hspace*{0.33em}b \leftarrow  \hat b \mathrel{;} {\mtext{return}}(\underline \sigma ,{\mfootnotesize{{{\color{\colorSYNTAX}\mtexttt{bitv}}}}}_{P}({\mtext{return}}(b)))
    \cr  {\mtext{\underline {ste\hspace{-1pt}}\hspace{1pt}p}}(N,\underline \sigma ,{\mfootnotesize{{{\color{\colorSYNTAX}\mtexttt{mux}}}}}({\mfootnotesize{{{\color{\colorSYNTAX}\mtexttt{bitv}}}}}_{\ell _{1}}(\hat b_{1}),{\mfootnotesize{{{\color{\colorSYNTAX}\mtexttt{bitv}}}}}_{\ell _{2}}(\hat b_{2}),{\mfootnotesize{{{\color{\colorSYNTAX}\mtexttt{bitv}}}}}_{\ell _{3}}(\hat b_{3})))
       &{}\triangleq {}&  {\mtext{return}}(\underline \sigma ,\langle {\mfootnotesize{{{\color{\colorSYNTAX}\mtexttt{bitv}}}}}_{\ell }(\widehat {\mtext{cond}}(\hat b_{1},\hat b_{2},\hat b_{3})),{\mfootnotesize{{{\color{\colorSYNTAX}\mtexttt{bitv}}}}}_{\ell }(\widehat {\mtext{cond}}(\hat b_{1},\hat b_{3},\hat b_{2}))\rangle )
    \cr  &{} {}& \hspace*{1.00em}{{\color{\colorTEXT}\textnormal{{\mtextit{where {{\color{\colorMATH}\(\ell  \triangleq  \ell _{1}\sqcup \ell _{2}\sqcup \ell _{3}\)}}}}}}}
    \cr  {\mtext{\underline {ste\hspace{-1pt}}\hspace{1pt}p}}(N,\underline \sigma ,{\mfootnotesize{{{\color{\colorSYNTAX}\mtexttt{mux}}}}}({\mfootnotesize{{{\color{\colorSYNTAX}\mtexttt{bitv}}}}}_{\ell }(\hat b_{1}),{\mfootnotesize{{{\color{\colorSYNTAX}\mtexttt{flipv}}}}}(\hat b_{2}),{\mfootnotesize{{{\color{\colorSYNTAX}\mtexttt{flipv}}}}}(\hat b_{3})))
       &{}\triangleq {}&   {\mtext{return}}(\underline \sigma ,\langle {\mfootnotesize{{{\color{\colorSYNTAX}\mtexttt{flipv}}}}}(\widehat {\mtext{cond}}(\widehat {b_{1}},\hat b_{2},\hat b_{3})),{\mfootnotesize{{{\color{\colorSYNTAX}\mtexttt{flipv}}}}}(\widehat {\mtext{cond}}(\hat b_{1},\hat b_{3},\hat b_{2}))\rangle )
    \cr  {\mtext{\underline {ste\hspace{-1pt}}\hspace{1pt}p}}(N,\underline \sigma ,{\mfootnotesize{{{\color{\colorSYNTAX}\mtexttt{xor}}}}}({\mfootnotesize{{{\color{\colorSYNTAX}\mtexttt{bitv}}}}}_{\ell _{1}}(\hat b_{1}),{\mfootnotesize{{{\color{\colorSYNTAX}\mtexttt{flipv}}}}}(\hat b_{2})))
       &{}\triangleq {}& {\mtext{return}}(\underline \sigma ,{\mfootnotesize{{{\color{\colorSYNTAX}\mtexttt{flipv}}}}}(\hat b_{1}\mathrel{\hat \oplus }\hat b_{2}))
    \cr  {\mtext{\underline {ste\hspace{-1pt}}\hspace{1pt}p}}(N,\underline \sigma ,{\mfootnotesize{{{\color{\colorSYNTAX}\mtexttt{if}}}}}({\mfootnotesize{{{\color{\colorSYNTAX}\mtexttt{bitv}}}}}_{\ell }(\hat b))\{ \underline e_{1}\} \{ \underline e_{2}\} )
       &{}\triangleq {}& {\mtext{do}}\hspace*{0.33em}b\leftarrow \hat b \mathrel{;} {\mtext{return}}(\underline \sigma ,{\mtext{cond}}(b,\underline e_{1},\underline e_{2}))
    \cr  {\mtext{\underline {ste\hspace{-1pt}}\hspace{1pt}p}}(N,\underline \sigma ,{\mfootnotesize{{{\color{\colorSYNTAX}\mtexttt{ref}}}}}(\underline v))
       &{}\triangleq {}& {\mtext{return}}(\underline \sigma [\iota \mapsto \underline v],{\mfootnotesize{{{\color{\colorSYNTAX}\mtexttt{refv}}}}}(\iota )) \hspace*{1.00em} {{\color{\colorTEXT}\textnormal{{\mtextit{where {{\color{\colorMATH}\(\iota  \notin  {\mtext{dom}}(\underline \sigma )\)}}}}}}}
    \cr  {\mtext{\underline {ste\hspace{-1pt}}\hspace{1pt}p}}(N,\underline \sigma ,{\mfootnotesize{{{\color{\colorSYNTAX}\mtexttt{read}}}}}({\mfootnotesize{{{\color{\colorSYNTAX}\mtexttt{refv}}}}}(\iota )))
       &{}\triangleq {}& {\mtext{return}}(\underline \sigma ,\underline \sigma (\iota ))
    \cr  {\mtext{\underline {ste\hspace{-1pt}}\hspace{1pt}p}}(N,\underline \sigma ,{\mfootnotesize{{{\color{\colorSYNTAX}\mtexttt{write}}}}}({\mfootnotesize{{{\color{\colorSYNTAX}\mtexttt{refv}}}}}(\iota ),\underline v))
       &{}\triangleq {}& {\mtext{return}}(\underline \sigma [\iota \mapsto \underline v],\underline \sigma (\iota ))
    \cr  {\mtext{\underline {ste\hspace{-1pt}}\hspace{1pt}p}}(N,\underline \sigma ,{\mfootnotesize{{{\color{\colorSYNTAX}\mtexttt{let}}}}}\hspace*{0.33em}x = \underline v\hspace*{0.33em}{\mfootnotesize{{{\color{\colorSYNTAX}\mtexttt{in}}}}}\hspace*{0.33em}\underline e)
       &{}\triangleq {}& {\mtext{return}}(\underline \sigma ,\underline e[\underline v/x])
    \cr  {\mtext{\underline {ste\hspace{-1pt}}\hspace{1pt}p}}(N,\underline \sigma ,{\mfootnotesize{{{\color{\colorSYNTAX}\mtexttt{let}}}}}\hspace*{0.33em}x_{1},x_{2} = \langle \underline v_{1},\underline v_{2}\rangle \hspace*{0.33em}{\mfootnotesize{{{\color{\colorSYNTAX}\mtexttt{in}}}}}\hspace*{0.33em}\underline e)
       &{}\triangleq {}& {\mtext{return}}(\underline \sigma ,\underline e[\underline v_{1}/x_{1}][\underline v_{2}/x_{2}])
    \cr  {\mtext{\underline {ste\hspace{-1pt}}\hspace{1pt}p}}(N,\underline \sigma ,(\underbracket[0.14 ex][0.35 ex]{{\mfootnotesize{{{\color{\colorSYNTAX}\mtexttt{fun}}}}}_{y}(x\mathrel{:}\tau ).\hspace*{0.33em}\underline e}_{\underline v_{1}})(\underline v_{2}))
       &{}\triangleq {}& {\mtext{return}}(\underline \sigma ,\underline e[\underline v_{1}/y][\underline v_{2}/x])
    \\[-1.75ex] {\mtext{\underline {ste\hspace{-1pt}}\hspace{1pt}p}}(N,\underline \sigma ,\underline E[\underline e])
       &{}\triangleq {}& {\mtext{do}}\hspace*{0.33em} \underline \sigma ^{\prime},\underline e^{\prime} \leftarrow  {\mtext{\underline {ste\hspace{-1pt}}\hspace{1pt}p}}(N,\underline \sigma ,\underline e) \mathrel{;} {\mtext{return}}(\underline \sigma ^{\prime},\underline E[\underline e^{\prime}])
    \cr  {\mtext{\underline {ste\hspace{-1pt}}\hspace{1pt}p}}(N,\underline \sigma ,\underline v)
       &{}\triangleq {}& {\mtext{return}}(\underline \sigma ,\underline v)
    \end{array}
  \\[-1.75ex]
  \cr \hfill\hspace{0pt} \begingroup\color{\colorTEXT}\boxed{\begingroup\color{\colorMATH} {\mtext{\underline {nste\hspace{-1pt}}\hspace{1pt}p}} \in  {\mathbb{N}} \times  {\mtext{\underline {confi\hspace{-1pt}}\hspace{1pt}g}} \rightharpoonup  {\mathcal{I}}(\underline {\mtext{trace}}) \endgroup}\endgroup
  \cr \begin{array}{lcl
    } {\mtext{\underline {nste\hspace{-1pt}}\hspace{1pt}p}}(0,\underline \varsigma )   &{}\triangleq {}& {\mtext{return}}({\mfootnotesize{{{\color{\colorSYNTAX}\mtexttt{\epsilon }}}}}\mathord{\cdotp }\underline \varsigma )
    \cr  {\mtext{\underline {nste\hspace{-1pt}}\hspace{1pt}p}}(N+1,\underline \varsigma ) &{}\triangleq {}& {\mtext{do}}\hspace*{0.33em}\underline t\mathord{\cdotp }\underline \varsigma ^{\prime} \leftarrow  {\mtext{\underline {nste\hspace{-1pt}}\hspace{1pt}p}}(N,\underline \varsigma ) \mathrel{;} \underline \varsigma ^{\prime \prime} \leftarrow  {\mtext{\underline {ste\hspace{-1pt}}\hspace{1pt}p}}(N+1,\underline \varsigma ^{\prime}) \mathrel{;} {\mtext{return}}(\underline t\mathord{\cdotp }\underline \varsigma ^{\prime}\mathord{\cdotp }\underline \varsigma ^{\prime \prime})
    \end{array}
  \end{tabularx}
\end{gather*}\endgroup
\precaptionskip
\caption{Mixed Language Semantics, where {{\color{\colorMATH}\(\hat b \in  {\mathcal{I}}({\mathbb{B}})\)}} is a
  distributional bit value (see text)}
\postcaptionskip
\label{fig:mixed-sem}
\end{figure} 

To address this problem, we define an alternative {\mtextit{mixed}} semantics which
embeds \emph{distributional bit values} directly into (single)
traces. Instead of the semantics of {{\color{\colorMATH}\({\mfootnotesize{{{\color{\colorSYNTAX}\mtexttt{flip}}}}}^{\rho }()\)}} producing two possible
outcomes, in the mixed semantics it produces just one: a single
distributional value {{\color{\colorMATH}\({\mfootnotesize{{{\color{\colorSYNTAX}\mtexttt{flipv}}}}}(\hat b)\)}} where the {{\color{\colorMATH}\(\hat b\)}} represents either
{{\color{\colorMATH}\({\mfootnotesize{{{\color{\colorSYNTAX}\mtexttt{I}}}}}\)}} or {{\color{\colorMATH}\({\mfootnotesize{{{\color{\colorSYNTAX}\mtexttt{O}}}}}\)}} with equal probability. Doing this is like treating
{{\color{\colorMATH}\({\mfootnotesize{{{\color{\colorSYNTAX}\mtexttt{flip}}}}}^{\rho }()\)}} expressions lazily, and lines up (mixed) traces with the adversary's view
{{\color{\colorMATH}\({\mathord{\bullet }}\)}}.

The mixed semantics amends the syntax of {{\color{\colorMATH}\({\mfootnotesize{{{\color{\colorSYNTAX}\mtexttt{flipv}}}}}\)}} and {{\color{\colorMATH}\({\mfootnotesize{{{\color{\colorSYNTAX}\mtexttt{bitv}}}}}_{\ell }\)}}
to be distributional (i.e., they contain {{\color{\colorMATH}\(\hat b\)}} rather than just
{{\color{\colorMATH}\(b\)}}). Other values from the standard semantics' syntax (top of
Figure~\ref{fig:semantics}) are unchanged. As such, a distribution
of pairs of bit values (say) is represented as pair of
distributional bit values. To allow values inside the pair to be
correlated, we represent them using what we call \emph{intensional
  distributions}---intensional distributions are written
{{\color{\colorMATH}\({\mathcal{I}}(A)\)}} and discussed in the next subsection.

The mixed semantics is shown in Figure~\ref{fig:mixed-sem}.
The mixed semantics step function {{\color{\colorMATH}\({\mtext{\underline {ste\hspace{-1pt}}\hspace{1pt}p}}(N,\underline \sigma ,\underline e)\)}} maps a
configuration, {{\color{\colorMATH}\(\underline \varsigma  \triangleq  \underline \sigma , \underline e\)}} to an intensional distribution of
configurations {{\color{\colorMATH}\({\mathcal{I}}({\mtext{\underline {confi\hspace{-1pt}}\hspace{1pt}g}})\)}}. Mixed semantics expressions (and
values, etc.) are underlined to distinguish them from the standard
semantics, and operations on distributional values are hatted.

Most of the cases for the mixed semantics are structurally the same as
the standard semantics. The key
differences are the handling of {{\color{\colorMATH}\({\mfootnotesize{{{\color{\colorSYNTAX}\mtexttt{flip}}}}}^{\rho }()\)}} and {{\color{\colorMATH}\({\mfootnotesize{{{\color{\colorSYNTAX}\mtexttt{cast}}}}}_{\ell }(\underline v)\)}}. For the
first, the standard semantics samples from the fresh uniform
distribution immediately, while the mixed semantics produces a single
uniform distributional value.
This distributional value is sampled at
the evaluation of {{\color{\colorMATH}\({\mfootnotesize{{{\color{\colorSYNTAX}\mtexttt{cast}}}}}_{P}\)}}, which matches the adversary's view.

A secret literal will produce a point distribution
on that literal.
The semantic operations for {{\color{\colorMATH}\({\mfootnotesize{{{\color{\colorSYNTAX}\mtexttt{if}}}}}\)}}, {{\color{\colorMATH}\({\mfootnotesize{{{\color{\colorSYNTAX}\mtexttt{mux}}}}}\)}} and {{\color{\colorMATH}\({\mfootnotesize{{{\color{\colorSYNTAX}\mtexttt{xor}}}}}\)}} are lifted
monadically to operate over distributions of secrets, e.g.,
{{\color{\colorMATH}\(\hat b_{1}\mathrel{\hat \oplus }\hat b_{2} \triangleq  {\mtext{do}}\hspace*{0.33em}b_{1} \leftarrow  \hat b_{1} \mathrel{;} b_{2} \leftarrow  \hat b_{2} \mathrel{;} {\mtext{return}} (b_{1}\oplus b_{2})\)}}. Other
operations are as usual, e.g., let expressions and tuple elimination
reduce via substitution and are not lifted to distributions.

\subsection{Capturing Correlations with Intensional Distributions}
\label{sec:intensional-distributions}

As mentioned, a distributional bit value {{\color{\colorMATH}\(\hat b\)}} can be viewed as a lazy
interpretation of a call {{\color{\colorMATH}\({\mfootnotesize{{{\color{\colorSYNTAX}\mtexttt{flip}}}}}^{\rho }()\)}}. To be sound, this interpretation must
properly model conditional probabilities between variables.

\paragraph*{Example}
Consider the program {{\color{\colorMATH}\({\mfootnotesize{{{\color{\colorSYNTAX}\mtexttt{let}}}}}\hspace*{0.33em}x =
{\mfootnotesize{{{\color{\colorSYNTAX}\mtexttt{flip}}}}}^{\rho }()\hspace*{0.33em}{\mfootnotesize{{{\color{\colorSYNTAX}\mtexttt{in}}}}}\hspace*{0.33em}\langle {\mfootnotesize{{{\color{\colorSYNTAX}\mtexttt{cast}}}}}_{{\mtext{P}}}(x),{\mfootnotesize{{{\color{\colorSYNTAX}\mtexttt{cast}}}}}_{{\mtext{P}}}(x)\rangle \)}}.%
\footnote{%
  Although this program violates affinity and would be rejected for
  that reason by our type system, its runtime semantics is
  well-defined and serves as a helpful demonstration.}%
After two evaluation steps in the standard semantics, the program will
be reduced to either
{{\color{\colorMATH}\(\langle {\mfootnotesize{{{\color{\colorSYNTAX}\mtexttt{cast}}}}}_{{\mtext{P}}}({\mfootnotesize{{{\color{\colorSYNTAX}\mtexttt{flipv}}}}}({\mfootnotesize{{{\color{\colorSYNTAX}\mtexttt{I}}}}})),{\mfootnotesize{{{\color{\colorSYNTAX}\mtexttt{cast}}}}}_{{\mtext{P}}}({\mfootnotesize{{{\color{\colorSYNTAX}\mtexttt{flipv}}}}}({\mfootnotesize{{{\color{\colorSYNTAX}\mtexttt{I}}}}}))\rangle \)}} or
{{\color{\colorMATH}\(\langle {\mfootnotesize{{{\color{\colorSYNTAX}\mtexttt{cast}}}}}_{{\mtext{P}}}({\mfootnotesize{{{\color{\colorSYNTAX}\mtexttt{flipv}}}}}({\mfootnotesize{{{\color{\colorSYNTAX}\mtexttt{O}}}}})),{\mfootnotesize{{{\color{\colorSYNTAX}\mtexttt{cast}}}}}_{{\mtext{P}}}({\mfootnotesize{{{\color{\colorSYNTAX}\mtexttt{flipv}}}}}({\mfootnotesize{{{\color{\colorSYNTAX}\mtexttt{O}}}}}))\rangle \)}}, with equal
probability. The standard rules for {{\color{\colorMATH}\({\mfootnotesize{{{\color{\colorSYNTAX}\mtexttt{cast}}}}}_{{\mtext{P}}}\)}} would then yield
(equally likely) {{\color{\colorMATH}\(\langle {\mfootnotesize{{{\color{\colorSYNTAX}\mtexttt{bitv}}}}}_{{\mtext{P}}}({\mfootnotesize{{{\color{\colorSYNTAX}\mtexttt{I}}}}}),{\mfootnotesize{{{\color{\colorSYNTAX}\mtexttt{bitv}}}}}_{{\mtext{P}}}({\mfootnotesize{{{\color{\colorSYNTAX}\mtexttt{I}}}}})\rangle \)}} and
{{\color{\colorMATH}\(\langle {\mfootnotesize{{{\color{\colorSYNTAX}\mtexttt{bitv}}}}}_{{\mtext{P}}}({\mfootnotesize{{{\color{\colorSYNTAX}\mtexttt{O}}}}}),{\mfootnotesize{{{\color{\colorSYNTAX}\mtexttt{bitv}}}}}_{{\mtext{P}}}({\mfootnotesize{{{\color{\colorSYNTAX}\mtexttt{O}}}}})\rangle \)}}.
In the mixed semantics this program will evaluate in two steps to
{{\color{\colorMATH}\(\langle {\mfootnotesize{{{\color{\colorSYNTAX}\mtexttt{cast}}}}}_{{\mtext{P}}}({\mfootnotesize{{{\color{\colorSYNTAX}\mtexttt{flipv}}}}}(\hat b)),{\mfootnotesize{{{\color{\colorSYNTAX}\mtexttt{cast}}}}}_{{\mtext{P}}}({\mfootnotesize{{{\color{\colorSYNTAX}\mtexttt{flipv}}}}}(\hat b))\rangle \)}} where {{\color{\colorMATH}\(\hat b\)}} is a
distributional value. At this point, the mixed semantics rule for
{{\color{\colorMATH}\({\mfootnotesize{{{\color{\colorSYNTAX}\mtexttt{cast}}}}}_{{\mtext{P}}}\)}} uses monadic bind to sample {{\color{\colorMATH}\(\hat b\)}} to yield some {{\color{\colorMATH}\(b\)}} (which is
either {{\color{\colorMATH}\({\mfootnotesize{{{\color{\colorSYNTAX}\mtexttt{I}}}}}\)}} or {{\color{\colorMATH}\({\mfootnotesize{{{\color{\colorSYNTAX}\mtexttt{O}}}}}\)}}) and return it as a point distribution. The
semantics needs to ``remember'' the bit chosen for the first {{\color{\colorMATH}\({\mfootnotesize{{{\color{\colorSYNTAX}\mtexttt{cast}}}}}_{{\mtext{P}}}\)}}
so that when it samples the second, the same bit is returned. Sampling
independently would yield incorrect outcomes such as
{{\color{\colorMATH}\(\langle {\mfootnotesize{{{\color{\colorSYNTAX}\mtexttt{bitv}}}}}_{{\mtext{P}}}({\mfootnotesize{{{\color{\colorSYNTAX}\mtexttt{O}}}}}),{\mfootnotesize{{{\color{\colorSYNTAX}\mtexttt{bitv}}}}}_{{\mtext{P}}}({\mfootnotesize{{{\color{\colorSYNTAX}\mtexttt{I}}}}})\rangle \)}}.

\begin{figure} 
  \small
\prefigskip
\begingroup\color{\colorMATH}\begin{gather*}\begin{tabularx}{\linewidth}{>{\centering\arraybackslash\(}X<{\)}}\hfill\hspace{0pt} \begin{array}{c
      } \begin{array}{rcrcl
         } a   &{}\in {}& A
         \cr  \hat x &{}\in {}& {\mathcal{I}}(A)    &{}\mathrel{\Coloneqq } {}& a \mathrel{|} \text{\guilsinglleft}\hat x\hspace*{0.33em}\hat x\text{\guilsinglright}
         \cr  p   &{}\in {}& {\mtext{rpath}} &{}\mathrel{\Coloneqq } {}& \mathord{\cdotp } \mathrel{|} \circled{H}\mathrel{:: }p \mathrel{|} \circled{T}\mathrel{:: }p
         \end{array}
      \\[-1.75ex]
      \cr  \begin{array}{lcl
         } \underline{\hspace{0.66em}}[\underline{\hspace{0.66em}}] &{}\in {}& {\mathcal{I}}(A) \times  {\mtext{rpath}} \rightharpoonup  A
         \cr  a[p]           &{}\triangleq {}& a
         \cr  \text{\guilsinglleft}\hat x_{1}\hspace*{0.33em}\hat x_{2}\text{\guilsinglright}[\circled{H} \mathrel{:: } p] &{}\triangleq {}& \hat x_{1}[p]
         \cr  \text{\guilsinglleft}\hat x_{1}\hspace*{0.33em}\hat x_{2}\text{\guilsinglright}[\circled{T} \mathrel{:: } p] &{}\triangleq {}& \hat x_{2}[p]
         \\[-1.75ex]
         \cr  {\mtext{support}} &{}\in {}& {\mathcal{I}}(A) \rightarrow  \wp (A)
         \cr  {\mtext{support}}(\hat x) &{}\triangleq {}& \{ a \mathrel{|} \hat x[p] = a\} 
         \\[-1.75ex]
         \cr  \pi _{1}              &{}\in {}& {\mathcal{I}}(A) \rightarrow  {\mathcal{I}}(A)
         \cr  \pi _{1}(a)           &{}\triangleq {}& a
         \cr  \pi _{1}(\text{\guilsinglleft}\hat x_{1}\hspace*{0.33em}\hat x_{2}\text{\guilsinglright}) &{}\triangleq {}& \hat x_{1}
         \\[-1.75ex]
         \cr  \pi _{2}              &{}\in {}& {\mathcal{I}}(A) \rightarrow  {\mathcal{I}}(A)
         \cr  \pi _{2}(a)           &{}\triangleq {}& a
         \cr  \pi _{2}(\text{\guilsinglleft}\hat x_{1}\hspace*{0.33em}\hat x_{2}\text{\guilsinglright}) &{}\triangleq {}& \hat x_{2}
         \\[-1.75ex]
         \cr  {\mtext{Pr}}\left[\overline {\hat x \mathrel{\dot =} x} \mathrel{}\middle|\mathrel{} \overline {\hat y \mathrel{\dot =} y}\right]
            &{}\triangleq {}&
            \frac{{\mtext{Pr}}\left[\overline {\hat x\mathrel{\dot =}x},\overline {\hat y\mathrel{\dot =}y}\right]}
                 {{\mtext{Pr}}\left[\overline {\hat y\mathrel{\dot =}y}\right]}
         \end{array}
      \end{array}
    \hfill\hspace{0pt} \begin{array}{lcl
      } {\mtext{height}}              &{}\in {}& {\mathcal{I}}(A) \rightarrow  {\mathbb{N}}
      \cr  {\mtext{height}}(a)           &{}\triangleq {}& 0
      \cr  {\mtext{height}}(\text{\guilsinglleft}\hat x_{1}\hspace*{0.33em}\hat x_{2}\text{\guilsinglright}) &{}\triangleq {}& 1+ {\mtext{max}}({\mtext{height}}(\hat x_{1}),{\mtext{height}}(\hat x_{2}))
      \\[-1.75ex]
      \cr  {\mtext{length}}      &{}\in {}& {\mtext{rpath}} \rightarrow  {\mathbb{B}}
      \cr  {\mtext{length}}(\mathord{\cdotp })   &{}\triangleq {}& 0
      \cr  {\mtext{length}}(\underline{\hspace{0.66em}}\mathrel{:: }p) &{}\triangleq {}& 1 + {\mtext{length}}(p)
      \\[-1.75ex]
      \cr  {\mtext{bit}}      &{}\in {}& {\mathbb{N}} \rightarrow  {\mathcal{I}}({\mathbb{B}})
      \cr  {\mtext{bit}}(0)   &{}\triangleq {}& \text{\guilsinglleft}{\mfootnotesize{{{\color{\colorSYNTAX}\mtexttt{I}}}}}\hspace*{0.33em}{\mfootnotesize{{{\color{\colorSYNTAX}\mtexttt{O}}}}}\text{\guilsinglright}
      \cr  {\mtext{bit}}(N+1) &{}\triangleq {}& \text{\guilsinglleft}{\mtext{bit}}(N)\hspace*{0.33em}{\mtext{bit}}(N)\text{\guilsinglright}
      \\[-1.75ex]
      \cr  {\mtext{return}}    &{}\in {}& A \rightarrow  {\mathcal{I}}(A)
      \cr  {\mtext{return}}(a) &{}\triangleq {}& a
      \\[-1.75ex]
      \cr  {\mtext{bind}}                &{}\in {}& {\mathcal{I}}(A) \times  (A \rightarrow  {\mathcal{I}}(B)) \rightarrow  {\mathcal{I}}(B)
      \cr  {\mtext{bind}}(a,f)           &{}\triangleq {}& f(a)
      \cr  {\mtext{bind}}(\text{\guilsinglleft}\hat x_{1}\hspace*{0.33em}\hat x_{2}\text{\guilsinglright},f) &{}\triangleq {}& \text{\guilsinglleft}{\mtext{bind}}(\hat x_{1},\pi _{1}{\circ }f)\hspace*{0.33em}{\mtext{bind}}(\hat x_{2},\pi _{2}{\circ }f)\text{\guilsinglright}
      \\[-1.75ex]
      \cr  {\mtext{Pr}}\left[\overline {\hat x \mathrel{\dot =} x}\right]
         &{}\triangleq {}&
         \frac{\left|\{ p \hspace*{0.33em}\mathrel{|}\hspace*{0.33em} {\mtext{length}}(p) = h, \overline {\hat x[p] = x}\} \right|}
              {2^{h}}
      \\[-1.75ex]
      \cr  \multicolumn{3}{l}{\hspace*{1.00em} {{\color{\colorTEXT}\textnormal{{\mtextit{where}}}}}\hspace*{0.33em} h \triangleq  {\mtext{max}}(\overline {{\mtext{height}}(\hat x)})}
      \end{array}
    \hfill\hspace{0pt}
  \end{tabularx}
\end{gather*}\endgroup
\precaptionskip
\caption{Intensional Distributions}
\postcaptionskip
\label{fig:intensional-dist}
\end{figure} 

\paragraph*{Intensional distributions}
%
%
As shown in the upper left of Figure~\ref{fig:intensional-dist}, an
intensional distribution {{\color{\colorMATH}\({\mathcal{I}}(A)\)}} over a set {{\color{\colorMATH}\(A\)}} is a binary tree
with elements {{\color{\colorMATH}\(a\)}} of {{\color{\colorMATH}\(A\)}} at the leaves. It represents a distribution as a
function from input entropy---a sequence of coin flips---to a result
in {{\color{\colorMATH}\(A\)}}.  Each node {{\color{\colorMATH}\(\text{\guilsinglleft}\hat x_{1}\hspace*{0.33em}\hat x_{2}\text{\guilsinglright}\)}} in the tree represents two sets of worlds
determined by the result of a coin flip: the left side {{\color{\colorMATH}\(\hat x_{1}\)}} defines the
worlds in which the coin was heads, and the right side {{\color{\colorMATH}\(\hat x_{2}\)}} defines
those in which it was tails. Each level of the tree represents a
distinct coin flip, with the earliest coin flip at the root, and later
coin flips at lower levels. The height of a tree represents an upper bound on
the number of coin flips upon which a distribution's values depends. Each path
through the tree is a possible world.

For example, {{\color{\colorMATH}\(\text{\guilsinglleft}\text{\guilsinglleft}3\hspace*{0.33em}4\text{\guilsinglright}\hspace*{0.33em}\text{\guilsinglleft}3\hspace*{0.33em}5\text{\guilsinglright}\text{\guilsinglright}\)}} is an intensional distribution of numbers in a
scenario where two coins have been flipped. There are four
possible worlds. {{\color{\colorMATH}\(\text{\guilsinglleft}3\hspace*{0.33em}4\text{\guilsinglright}\)}} is the world where the 0th coin
came up heads. {{\color{\colorMATH}\(3\)}} is the outcome in the world where both
coins came up heads, while {{\color{\colorMATH}\(4\)}} is the outcome where the 0th
coin was heads but the 1th coin was tails. {{\color{\colorMATH}\(\text{\guilsinglleft}3\hspace*{0.33em}5\text{\guilsinglright}\)}} is the
world where the 0th coin came up tails, with {{\color{\colorMATH}\(3\)}} the outcome when
the 1th coin was heads, and {{\color{\colorMATH}\(5\)}} when it was tails.

We can derive the probabilities of particular outcomes by counting the
number of paths that reach them. In the example, 3 has
probability {{\color{\colorMATH}\(\frac{1}{2}\)}}, while 4 has probability {{\color{\colorMATH}\(\frac{1}{4}\)}}, and 5 has
probability {{\color{\colorMATH}\(\frac{1}{4}\)}}. Importantly, intensional distributions have
enough structure to represent correlations: We can see that we always
get a 3 when the 1th coin flip is heads, regardless of whether the 0th
coin flip was heads or tails. Conversely, the distribution
{{\color{\colorMATH}\(\text{\guilsinglleft}\text{\guilsinglleft}3\hspace*{0.33em}3\text{\guilsinglright}\hspace*{0.33em}\text{\guilsinglleft}4\hspace*{0.33em}5\text{\guilsinglright}\text{\guilsinglright}\)}} ascribes outcomes 3, 4, and 5 the same probabilities
as {{\color{\colorMATH}\(\text{\guilsinglleft}\text{\guilsinglleft}3\hspace*{0.33em}4\text{\guilsinglright}\hspace*{0.33em}\text{\guilsinglleft}3\hspace*{0.33em}5\text{\guilsinglright}\text{\guilsinglright}\)}}, but represents the situation in which the we
always get 3 when 0th coin flip is heads. An equivalent representation
of {{\color{\colorMATH}\(\text{\guilsinglleft}\text{\guilsinglleft}3\hspace*{0.33em}3\text{\guilsinglright}\hspace*{0.33em}\text{\guilsinglleft}4\hspace*{0.33em}5\text{\guilsinglright}\text{\guilsinglright}\)}} is {{\color{\colorMATH}\(\text{\guilsinglleft}3\hspace*{0.33em}\text{\guilsinglleft}4\hspace*{0.33em}5\text{\guilsinglright}\text{\guilsinglright}\)}}.
Although the {{\color{\colorMATH}\(3\)}} only appears once, it is logically
extended to the larger sub-tree {{\color{\colorMATH}\(\text{\guilsinglleft}3\hspace*{0.33em}3\text{\guilsinglright}\)}} for the purposes of counting.
To compute a probability, all paths are considered of a
fixed length equal to the height of the tree, and shorter sub-trees are
extended to copy leaves that appear at shorter height. Trees are equal {{\color{\colorMATH}\(=\)}} when
they are syntactically equal modulo these extensions.

In the figure, a path {{\color{\colorMATH}\(p\)}} through the tree is a sequence of
coin flip outcomes, either {{\color{\colorMATH}\(\circled{H}\)}} or {{\color{\colorMATH}\(\circled{T}\)}}. The operation {{\color{\colorMATH}\(\hat x[p]\)}} follows
a path {{\color{\colorMATH}\(p\)}} through the tree {{\color{\colorMATH}\(\hat x\)}} going left on {{\color{\colorMATH}\(\circled{H}\)}} and right on
{{\color{\colorMATH}\(\circled{T}\)}}. When a leaf {{\color{\colorMATH}\(a\)}} is reached, it is simply returned, per the case
{{\color{\colorMATH}\(a[p]\)}}; if {{\color{\colorMATH}\(p\)}} happens to not be $\cdot$, returning {{\color{\colorMATH}\(a\)}} is tantamount to
extending the tree logically, as mentioned above.
Computing the probability of an outcome {{\color{\colorMATH}\(x\)}} for intensional
distribution {{\color{\colorMATH}\(\hat x\)}} is shown at the bottom of the figure. As with the
example above, it counts the number of paths that have outcome {{\color{\colorMATH}\(x\)}},
scaled by the total possible worlds. The probability of an event
involving multiple distributions is similar. Conditional
probability works as usual.

Finally, looking at the middle right of the figure, consider the
monadic operations used by the semantics in
Figure~\ref{fig:mixed-sem}. The {{\color{\colorMATH}\({\mtext{bit}}(N)\)}} operation produces a
uniform distribution of bits following the {{\color{\colorMATH}\(N\)}}th coin flip, where the
outcomes are entirely determined by the {{\color{\colorMATH}\(N\)}}th flip, i.e., independent
of the flips that preceded it, which appear higher in the tree. {{\color{\colorMATH}\({\mtext{return}}(a)\)}}
simply returns {{\color{\colorMATH}\(a\)}}---this corresponds to a point distribution of {{\color{\colorMATH}\(a\)}}
since it is the outcome in all possible worlds (recall {{\color{\colorMATH}\(a[p] = a\)}} for
all {{\color{\colorMATH}\(p\)}}). Lastly, {{\color{\colorMATH}\({\mtext{bind}}(\hat x,f)\)}} applies {{\color{\colorMATH}\(f\)}} to each possible world
in {{\color{\colorMATH}\(\hat x\)}}, gathering up the results in an intensional distribution tree
that is of equal or greater height to that of {{\color{\colorMATH}\(\hat x\)}}; the height could grow if
{{\color{\colorMATH}\(f\)}} returns a tree larger than {{\color{\colorMATH}\(\hat x\)}}, and {{\color{\colorMATH}\({\mtext{bind}}(\hat x,f)[p] = f(\hat x[p])[p]\)}} for
all paths {{\color{\colorMATH}\(p\)}}.


\paragraph*{Example revisited}
Reconsider the example {{\color{\colorMATH}\({\mfootnotesize{{{\color{\colorSYNTAX}\mtexttt{let}}}}}\hspace*{0.33em}x =
{\mfootnotesize{{{\color{\colorSYNTAX}\mtexttt{flip}}}}}^{\rho }()\hspace*{0.33em}{\mfootnotesize{{{\color{\colorSYNTAX}\mtexttt{in}}}}}\hspace*{0.33em}\langle {\mfootnotesize{{{\color{\colorSYNTAX}\mtexttt{cast}}}}}_{{\mtext{P}}}(x),{\mfootnotesize{{{\color{\colorSYNTAX}\mtexttt{cast}}}}}_{{\mtext{P}}}(x)\rangle \)}}.
According to the mixed semantics starting with $N=0$,
{{\color{\colorMATH}\({\mfootnotesize{{{\color{\colorSYNTAX}\mtexttt{flip}}}}}^{\rho }()\)}} evaluates to {{\color{\colorMATH}\({\mfootnotesize{{{\color{\colorSYNTAX}\mtexttt{flipv}}}}}(\text{\guilsinglleft}{\mfootnotesize{{{\color{\colorSYNTAX}\mtexttt{I}}}}}\hspace*{0.33em}{\mfootnotesize{{{\color{\colorSYNTAX}\mtexttt{O}}}}}\text{\guilsinglright})\)}}, which is then (as
precipitated by {{\color{\colorMATH}\({\mtext{\underline {nste\hspace{-1pt}}\hspace{1pt}p}}\)}}) substituted for {{\color{\colorMATH}\(x\)}} in the body of the
{{\color{\colorMATH}\({\mfootnotesize{{{\color{\colorSYNTAX}\mtexttt{let}}}}}\)}}, producing
{{\color{\colorMATH}\(\langle {\mfootnotesize{{{\color{\colorSYNTAX}\mtexttt{cast}}}}}_{{\mtext{P}}}({\mfootnotesize{{{\color{\colorSYNTAX}\mtexttt{flipv}}}}}(\text{\guilsinglleft}{\mfootnotesize{{{\color{\colorSYNTAX}\mtexttt{I}}}}}\hspace*{0.33em}{\mfootnotesize{{{\color{\colorSYNTAX}\mtexttt{O}}}}}\text{\guilsinglright})),{\mfootnotesize{{{\color{\colorSYNTAX}\mtexttt{cast}}}}}_{{\mtext{P}}}({\mfootnotesize{{{\color{\colorSYNTAX}\mtexttt{flipv}}}}}(\text{\guilsinglleft}{\mfootnotesize{{{\color{\colorSYNTAX}\mtexttt{I}}}}}\hspace*{0.33em}{\mfootnotesize{{{\color{\colorSYNTAX}\mtexttt{O}}}}}\text{\guilsinglright}))\rangle \)}}.
Now we apply the context rule for {{\color{\colorMATH}\(\underline E[\underline e]\)}} where {{\color{\colorMATH}\(\underline E\)}} is
{{\color{\colorMATH}\(\langle [],{\mfootnotesize{{{\color{\colorSYNTAX}\mtexttt{cast}}}}}_{{\mtext{P}}}({\mfootnotesize{{{\color{\colorSYNTAX}\mtexttt{flipv}}}}}(\text{\guilsinglleft}{\mfootnotesize{{{\color{\colorSYNTAX}\mtexttt{I}}}}}\hspace*{0.33em}{\mfootnotesize{{{\color{\colorSYNTAX}\mtexttt{O}}}}}\text{\guilsinglright}))\rangle \)}} and {{\color{\colorMATH}\(\underline e\)}} is
{{\color{\colorMATH}\({\mfootnotesize{{{\color{\colorSYNTAX}\mtexttt{cast}}}}}_{{\mtext{P}}}({\mfootnotesize{{{\color{\colorSYNTAX}\mtexttt{flipv}}}}}(\text{\guilsinglleft}{\mfootnotesize{{{\color{\colorSYNTAX}\mtexttt{I}}}}}\hspace*{0.33em}{\mfootnotesize{{{\color{\colorSYNTAX}\mtexttt{O}}}}}\text{\guilsinglright}))\)}}. The rule invokes {{\color{\colorMATH}\({\mtext{\underline {ste\hspace{-1pt}}\hspace{1pt}p}}\)}} on
the latter, which performs {{\color{\colorMATH}\({\mtext{do}}\hspace*{0.33em}b \leftarrow  \text{\guilsinglleft}{\mfootnotesize{{{\color{\colorSYNTAX}\mtexttt{I}}}}}\hspace*{0.33em}{\mfootnotesize{{{\color{\colorSYNTAX}\mtexttt{O}}}}}\text{\guilsinglright} \mathrel{;}
{\mtext{return}}(\underline \sigma ,{\mfootnotesize{{{\color{\colorSYNTAX}\mtexttt{bitv}}}}}_{P}({\mtext{return}}(b)))\)}} per the rule for {{\color{\colorMATH}\({\mfootnotesize{{{\color{\colorSYNTAX}\mtexttt{cast}}}}}_{{\mtext{P}}}\)}}. Per the
definitions of {{\color{\colorMATH}\({\mtext{bind}}\)}} and {{\color{\colorMATH}\({\mtext{return}}\)}}, this will return the intensional
{\mtextit{distribution of configurations}} {{\color{\colorMATH}\(\text{\guilsinglleft}(\underline \sigma ,{\mfootnotesize{{{\color{\colorSYNTAX}\mtexttt{bitv}}}}}_{P}({\mfootnotesize{{{\color{\colorSYNTAX}\mtexttt{I}}}}}))\hspace*{0.33em}(\underline \sigma ,{\mfootnotesize{{{\color{\colorSYNTAX}\mtexttt{bitv}}}}}_{P}({\mfootnotesize{{{\color{\colorSYNTAX}\mtexttt{O}}}}}))\text{\guilsinglright}\)}}.
Back to the context rule, its use of {{\color{\colorMATH}\({\mtext{bind}}\)}} will re-package up these
possibilities with {{\color{\colorMATH}\(\underline E\)}}:
\vspace{-1ex}
\begingroup\color{\colorMATH}\begin{gather*}
\begin{array}{c
} \text{\guilsinglleft} (\underline \sigma ,\langle {\mfootnotesize{{{\color{\colorSYNTAX}\mtexttt{bitv}}}}}_{P}({\mfootnotesize{{{\color{\colorSYNTAX}\mtexttt{I}}}}}),{\mfootnotesize{{{\color{\colorSYNTAX}\mtexttt{cast}}}}}_{{\mtext{P}}}({\mfootnotesize{{{\color{\colorSYNTAX}\mtexttt{flipv}}}}}(\text{\guilsinglleft}{\mfootnotesize{{{\color{\colorSYNTAX}\mtexttt{I}}}}}\hspace*{0.33em}{\mfootnotesize{{{\color{\colorSYNTAX}\mtexttt{O}}}}}\text{\guilsinglright}))\rangle )
   \hspace*{0.33em} (\underline \sigma ,\langle {\mfootnotesize{{{\color{\colorSYNTAX}\mtexttt{bitv}}}}}_{P}({\mfootnotesize{{{\color{\colorSYNTAX}\mtexttt{O}}}}}),{\mfootnotesize{{{\color{\colorSYNTAX}\mtexttt{cast}}}}}_{{\mtext{P}}}({\mfootnotesize{{{\color{\colorSYNTAX}\mtexttt{flipv}}}}}(\text{\guilsinglleft}{\mfootnotesize{{{\color{\colorSYNTAX}\mtexttt{I}}}}}\hspace*{0.33em}{\mfootnotesize{{{\color{\colorSYNTAX}\mtexttt{O}}}}}\text{\guilsinglright}))\rangle )
   \text{\guilsinglright}
\vspace{-1ex}
\end{array}
\end{gather*}\endgroup
In this distribution of configurations there are two worlds---the
left configuration occurs when the 0th coin flip is heads, and
right when it is tails. Inside of each of these
configurations is a distributional value {{\color{\colorMATH}\({\mfootnotesize{{{\color{\colorSYNTAX}\mtexttt{flipv}}}}}(\text{\guilsinglleft}{\mfootnotesize{{{\color{\colorSYNTAX}\mtexttt{I}}}}}\hspace*{0.33em}{\mfootnotesize{{{\color{\colorSYNTAX}\mtexttt{O}}}}}\text{\guilsinglright})\)}}, where
once again the left side is due to the coin flip being heads, and the
right side being tails. \emph{Both are relative to the same coin flip}.
As such, there are two ``unreachable'' paths in the inner trees: the right-branch
of the left distributional value, and the left branch of the right
distributional value, shown here with bullets:
\vspace{-1ex}
\begingroup\color{\colorMATH}\begin{gather*}
\begin{array}{c
} \text{\guilsinglleft} (\underline \sigma ,\langle {\mfootnotesize{{{\color{\colorSYNTAX}\mtexttt{bitv}}}}}_{P}({\mfootnotesize{{{\color{\colorSYNTAX}\mtexttt{I}}}}}),{\mfootnotesize{{{\color{\colorSYNTAX}\mtexttt{cast}}}}}_{{\mtext{P}}}({\mfootnotesize{{{\color{\colorSYNTAX}\mtexttt{flipv}}}}}(\text{\guilsinglleft}{\mfootnotesize{{{\color{\colorSYNTAX}\mtexttt{I}}}}}\hspace*{0.33em}{\mathord{\bullet }}\text{\guilsinglright}))\rangle )
   \hspace*{0.33em}  (\underline \sigma ,\langle {\mfootnotesize{{{\color{\colorSYNTAX}\mtexttt{bitv}}}}}_{P}({\mfootnotesize{{{\color{\colorSYNTAX}\mtexttt{O}}}}}),{\mfootnotesize{{{\color{\colorSYNTAX}\mtexttt{cast}}}}}_{{\mtext{P}}}({\mfootnotesize{{{\color{\colorSYNTAX}\mtexttt{flipv}}}}}(\text{\guilsinglleft}{\mathord{\bullet }}\hspace*{0.33em}{\mfootnotesize{{{\color{\colorSYNTAX}\mtexttt{O}}}}}\text{\guilsinglright}))\rangle )
   \text{\guilsinglright}
\vspace{-1ex}
\end{array}
\end{gather*}\endgroup
The next step of the computation will force the distributional value
to be {{\color{\colorMATH}\({\mfootnotesize{{{\color{\colorSYNTAX}\mtexttt{I}}}}}\)}} in the left branch and {{\color{\colorMATH}\({\mfootnotesize{{{\color{\colorSYNTAX}\mtexttt{O}}}}}\)}} in the right
branch. Here's how. First, the definition of {{\color{\colorMATH}\({\mtext{\underline {nste\hspace{-1pt}}\hspace{1pt}p}}\)}} is a
{{\color{\colorMATH}\({\mtext{bind}}\)}} on the above distribution of configurations with {{\color{\colorMATH}\({\mtext{\underline {ste\hspace{-1pt}}\hspace{1pt}p}}\)}} as the
function {{\color{\colorMATH}\(f\)}} passed to {{\color{\colorMATH}\({\mtext{bind}}\)}}. The definition of {{\color{\colorMATH}\({\mtext{bind}}\)}} constructs a new
distribution tree which calls {{\color{\colorMATH}\({\mtext{\underline {ste\hspace{-1pt}}\hspace{1pt}p}}\)}} on the left configuration, and then
takes the left branch ({{\color{\colorMATH}\(\pi _{1}\)}}) of the tree that comes back, and likewise for the
right configuration and the right branch that comes back
({{\color{\colorMATH}\(\pi _{2}\)}}). Here {{\color{\colorMATH}\({\mtext{\underline {ste\hspace{-1pt}}\hspace{1pt}p}}\)}} will invoke cast and context rules similarly
as before, returning a two-element tree with {{\color{\colorMATH}\({\mfootnotesize{{{\color{\colorSYNTAX}\mtexttt{bitv}}}}}_{P}({\mfootnotesize{{{\color{\colorSYNTAX}\mtexttt{I}}}}})\)}} on the
left and {{\color{\colorMATH}\({\mfootnotesize{{{\color{\colorSYNTAX}\mtexttt{bitv}}}}}_{P}({\mfootnotesize{{{\color{\colorSYNTAX}\mtexttt{O}}}}})\)}} on the right. These
occurrences of {{\color{\colorMATH}\(\pi _{1}\)}} and {{\color{\colorMATH}\(\pi _{2}\)}} ``pick'' the left ({{\color{\colorMATH}\({\mfootnotesize{{{\color{\colorSYNTAX}\mtexttt{I}}}}}\)}} case) and right
({{\color{\colorMATH}\({\mfootnotesize{{{\color{\colorSYNTAX}\mtexttt{O}}}}}\)}} case), respectively, resulting in the final configuration
{{\color{\colorMATH}\(
\begin{array}{c
} \text{\guilsinglleft} (\underline \sigma ,\langle {\mfootnotesize{{{\color{\colorSYNTAX}\mtexttt{bitv}}}}}_{P}({\mfootnotesize{{{\color{\colorSYNTAX}\mtexttt{I}}}}}),{\mfootnotesize{{{\color{\colorSYNTAX}\mtexttt{bitv}}}}}_{P}({\mfootnotesize{{{\color{\colorSYNTAX}\mtexttt{I}}}}})\rangle )
   \hspace*{0.33em} (\underline \sigma ,\langle {\mfootnotesize{{{\color{\colorSYNTAX}\mtexttt{bitv}}}}}_{P}({\mfootnotesize{{{\color{\colorSYNTAX}\mtexttt{O}}}}}),{\mfootnotesize{{{\color{\colorSYNTAX}\mtexttt{bitv}}}}}_{P}({\mfootnotesize{{{\color{\colorSYNTAX}\mtexttt{O}}}}})\rangle )
   \text{\guilsinglright}
\end{array}
\)}}
\paragraph*{Simulation}
The concept of ``unreachable'' paths in a distributional value is
captured by a projection operation which ``flattens'' a distribution of mixed
terms (which have distributional values) into a distribution of standard terms
(which do not have distributional values). This projection will (1) discard
unreachable paths of distributional values, and (2) corresponds to evaluation
in the standard semantics instantiated with the intensional distribution monad.

\begin{figure} 
  \small
\prefigskip
\begingroup\color{\colorMATH}\begin{gather*}\begin{tabularx}{\linewidth}{>{\centering\arraybackslash\(}X<{\)}}\hfill\hspace{0pt} \begingroup\color{\colorTEXT}\boxed{\begingroup\color{\colorMATH} \lceil \underline{\hspace{0.66em}}\rceil  \in  ({\mtext{\underline {ex\hspace{-1pt}}\hspace{1pt}p}} \rightarrow  {\mathcal{I}}({\mtext{exp}})) \times  (\underline {\mtext{store}} \rightarrow  {\mathcal{I}}({\mtext{store}})) \times  ({\mtext{\underline {confi\hspace{-1pt}}\hspace{1pt}g}} \rightarrow  {\mathcal{I}}({\mtext{config}})) \times  (\underline {\mtext{trace}} \rightarrow  {\mathcal{I}}({\mtext{trace}})) \endgroup}\endgroup
  \\[-1.75ex]
  \cr \hfill\hspace{0pt} \begin{array}{lcl
      } \lceil x\rceil                         &{}\triangleq {}& {\mtext{return}}(x)
      \cr  \lceil {\mfootnotesize{{{\color{\colorSYNTAX}\mtexttt{locv}}}}}(\iota )\rceil                 &{}\triangleq {}& {\mtext{return}}({\mfootnotesize{{{\color{\colorSYNTAX}\mtexttt{locv}}}}}(\iota ))
      \cr  \lceil b_{\ell }\rceil                      &{}\triangleq {}& {\mtext{return}}(b_{\ell })
      \cr  \lceil {\mfootnotesize{{{\color{\colorSYNTAX}\mtexttt{flip}}}}}^{\rho }()\rceil               &{}\triangleq {}& {\mtext{return}}({\mfootnotesize{{{\color{\colorSYNTAX}\mtexttt{flip}}}}}^{\rho }())
      \end{array}
    \hfill\hspace{0pt} \begin{array}{lcl
      } \lceil {\mfootnotesize{{{\color{\colorSYNTAX}\mtexttt{fun}}}}}_{y}(x\mathrel{:}\tau ).\hspace*{0.33em}\underline e\rceil         &{}\triangleq {}& {\mtext{do}}\hspace*{0.33em}e \leftarrow  \lceil \underline e\rceil  \mathrel{;} {\mtext{return}}({\mfootnotesize{{{\color{\colorSYNTAX}\mtexttt{fun}}}}}_{y}(x\mathrel{:}\tau ).\hspace*{0.33em}e)
      \cr  \lceil {\mfootnotesize{{{\color{\colorSYNTAX}\mtexttt{bitv}}}}}_{\ell }(\hat b)\rceil            &{}\triangleq {}& {\mtext{do}}\hspace*{0.33em}b \leftarrow  \hat b \mathrel{;} {\mtext{return}}({\mfootnotesize{{{\color{\colorSYNTAX}\mtexttt{bitv}}}}}_{\ell }(b))
      \cr  \lceil {\mfootnotesize{{{\color{\colorSYNTAX}\mtexttt{flipv}}}}}(\hat b)\rceil              &{}\triangleq {}& {\mtext{do}}\hspace*{0.33em}b \leftarrow  \hat b \mathrel{;} {\mtext{return}}({\mfootnotesize{{{\color{\colorSYNTAX}\mtexttt{flipv}}}}}(b))
      \cr  \lceil {\mfootnotesize{{{\color{\colorSYNTAX}\mtexttt{cast}}}}}_{\ell }(\underline v)\rceil            &{}\triangleq {}& {\mtext{do}}\hspace*{0.33em}v \leftarrow  \lceil \underline v\rceil  \mathrel{;} {\mtext{return}}({\mfootnotesize{{{\color{\colorSYNTAX}\mtexttt{cast}}}}}_{\ell }(v))
      \end{array}
    \hfill\hspace{0pt}
  \\[-1.75ex]
  \cr \begin{array}{lcl
    } \lceil {\mfootnotesize{{{\color{\colorSYNTAX}\mtexttt{mux}}}}}(\underline e_{1},\underline e_{2},\underline e_{3})\rceil     &{}\triangleq {}& {\mtext{do}}\hspace*{0.33em}e_{1} \leftarrow  \lceil \underline e_{1}\rceil  \mathrel{;} e_{2} \leftarrow  \lceil \underline e_{2}\rceil  \mathrel{;} e_{3} \leftarrow  \lceil \underline e_{3}\rceil  \mathrel{;} {\mtext{return}}({\mfootnotesize{{{\color{\colorSYNTAX}\mtexttt{mux}}}}}(e_{1},e_{2},e_{3}))
    \cr  \lceil {\mfootnotesize{{{\color{\colorSYNTAX}\mtexttt{xor}}}}}(\underline e_{1},\underline e_{2})\rceil          &{}\triangleq {}& {\mtext{do}}\hspace*{0.33em}e_{1} \leftarrow  \lceil \underline e_{1}\rceil  \mathrel{;} e_{2} \leftarrow  \lceil \underline e_{2}\rceil  \mathrel{;} {\mtext{return}}({\mfootnotesize{{{\color{\colorSYNTAX}\mtexttt{xor}}}}}(e_{1},e_{2}))
    \cr  \lceil {\mfootnotesize{{{\color{\colorSYNTAX}\mtexttt{if}}}}}(\underline e_{1})\{ \underline e_{2}\} \{ \underline e_{3}\} \rceil    &{}\triangleq {}& {\mtext{do}}\hspace*{0.33em}e_{1} \leftarrow  \lceil \underline e_{1}\rceil  \mathrel{;} e_{2} \leftarrow  \lceil \underline e_{2}\rceil  \mathrel{;} e_{3} \leftarrow  \lceil \underline e_{3}\rceil  \mathrel{;} {\mtext{return}}({\mfootnotesize{{{\color{\colorSYNTAX}\mtexttt{if}}}}}(e_{1})\{ e_{2}\} \{ e_{3}\} )
    \cr  \lceil {\mfootnotesize{{{\color{\colorSYNTAX}\mtexttt{ref}}}}}(\underline e_{1})\rceil               &{}\triangleq {}& {\mtext{do}}\hspace*{0.33em}e_{1} \leftarrow  \lceil \underline e_{1}\rceil  \mathrel{;} {\mtext{return}}({\mfootnotesize{{{\color{\colorSYNTAX}\mtexttt{ref}}}}}(e_{1}))
    \cr  \lceil {\mfootnotesize{{{\color{\colorSYNTAX}\mtexttt{read}}}}}(\underline e_{1})\rceil              &{}\triangleq {}& {\mtext{do}}\hspace*{0.33em}e_{1} \leftarrow  \lceil \underline e_{1}\rceil  \mathrel{;} {\mtext{return}}({\mfootnotesize{{{\color{\colorSYNTAX}\mtexttt{read}}}}}(e_{1}))
    \cr  \lceil {\mfootnotesize{{{\color{\colorSYNTAX}\mtexttt{write}}}}}(\underline e_{1},\underline e_{2})\rceil        &{}\triangleq {}& {\mtext{do}}\hspace*{0.33em}e_{1} \leftarrow  \lceil \underline e_{1}\rceil  \mathrel{;} e_{2} \leftarrow  \lceil \underline e_{2}\rceil  \mathrel{;} {\mtext{return}}({\mfootnotesize{{{\color{\colorSYNTAX}\mtexttt{write}}}}}(e_{1},e_{2}))
    \cr  \lceil \langle \underline e_{1},\underline e_{2}\rangle \rceil               &{}\triangleq {}& {\mtext{do}}\hspace*{0.33em}e_{1} \leftarrow  \lceil \underline e_{1}\rceil  \mathrel{;} e_{2} \leftarrow  \lceil \underline e_{2}\rceil  \mathrel{;} {\mtext{return}}(\langle e_{1},e_{2}\rangle )
    \cr  \lceil {\mfootnotesize{{{\color{\colorSYNTAX}\mtexttt{let}}}}}\hspace*{0.33em}x=\underline e_{1}\hspace*{0.33em}{\mfootnotesize{{{\color{\colorSYNTAX}\mtexttt{in}}}}}\hspace*{0.33em}\underline e_{2}\rceil    &{}\triangleq {}& {\mtext{do}}\hspace*{0.33em}e_{1} \leftarrow  \lceil \underline e_{1}\rceil  \mathrel{;} e_{2} \leftarrow  \lceil \underline e_{2}\rceil  \mathrel{;} {\mtext{return}}({\mfootnotesize{{{\color{\colorSYNTAX}\mtexttt{let}}}}}\hspace*{0.33em}x=e_{1}\hspace*{0.33em}{\mfootnotesize{{{\color{\colorSYNTAX}\mtexttt{in}}}}}\hspace*{0.33em}e_{2})
    \cr  \lceil {\mfootnotesize{{{\color{\colorSYNTAX}\mtexttt{let}}}}}\hspace*{0.33em}x,y=\underline e_{1}\hspace*{0.33em}{\mfootnotesize{{{\color{\colorSYNTAX}\mtexttt{in}}}}}\hspace*{0.33em}\underline e_{2}\rceil  &{}\triangleq {}& {\mtext{do}}\hspace*{0.33em}e_{1} \leftarrow  \lceil \underline e_{1}\rceil  \mathrel{;} e_{2} \leftarrow  \lceil \underline e_{2}\rceil  \mathrel{;} {\mtext{return}}({\mfootnotesize{{{\color{\colorSYNTAX}\mtexttt{let}}}}}\hspace*{0.33em}x,y=e_{1}\hspace*{0.33em}{\mfootnotesize{{{\color{\colorSYNTAX}\mtexttt{in}}}}}\hspace*{0.33em}e_{2})
    \cr  \lceil \underline e_{1}(\underline e_{2})\rceil                &{}\triangleq {}& {\mtext{do}}\hspace*{0.33em}e_{1} \leftarrow  \lceil \underline e_{1}\rceil  \mathrel{;} e_{2} \leftarrow  \lceil \underline e_{2}\rceil  \mathrel{;} {\mtext{return}}(e_{1}(e_{2}))
    \end{array}
  \\[-1.75ex]
  \cr \hfill\hspace{0pt} \lceil \varnothing \rceil  \triangleq  {\mtext{return}}(\varnothing )
    \hfill\hspace{0pt} \lceil \{ \iota  \mapsto  \underline v\} \uplus \underline \sigma \rceil  \triangleq  {\mtext{do}}\hspace*{0.33em}v \leftarrow  \lceil \underline v\rceil  \mathrel{;} \sigma  \leftarrow  \lceil \underline \sigma \rceil  \mathrel{;} {\mtext{return}}(\{ \iota \mapsto v\} \uplus \sigma )
    \hfill\hspace{0pt}
  \\[-1.75ex]
  \cr \hfill\hspace{0pt} \lceil \underline \sigma ,\underline e\rceil  \triangleq  {\mtext{do}}\hspace*{0.33em}\sigma  \leftarrow  \underline \sigma  \mathrel{;} e \leftarrow  \underline e \mathrel{;} {\mtext{return}}(\sigma ,e)
    \hfill\hspace{0pt} \lceil {\mfootnotesize{{{\color{\colorSYNTAX}\mtexttt{\epsilon }}}}}\rceil      \triangleq  {\mtext{return}}({\mfootnotesize{{{\color{\colorSYNTAX}\mtexttt{\epsilon }}}}})
    \hfill\hspace{0pt} \lceil \underline t\mathord{\cdotp }\underline \varsigma \rceil  \triangleq  {\mtext{do}}\hspace*{0.33em}t \leftarrow  \underline t \mathrel{;} \varsigma  \leftarrow  \underline \varsigma  \mathrel{;} {\mtext{return}}(t\mathord{\cdotp }\varsigma )
    \hfill\hspace{0pt}
  \\[-1.75ex]
  \cr \hfill\hspace{0pt} \hat \lceil \underline {\hat t}\hat \rceil  \triangleq  {\mtext{do}}\hspace*{0.33em}\underline t \leftarrow  \underline {\hat t} \mathrel{;} \lceil \underline t\rceil 
    \hfill\hspace{0pt} \begingroup\color{\colorTEXT}\boxed{\begingroup\color{\colorMATH} \hat \lceil \underline{\hspace{0.66em}}\hat \rceil  \in  {\mathcal{I}}(\underline {\mtext{trace}}) \rightarrow  {\mathcal{I}}({\mtext{trace}}) \endgroup}\endgroup
  \end{tabularx}
\end{gather*}\endgroup
\precaptionskip
\caption{Mixed Semantics Projection}
\postcaptionskip
\label{fig:projection}
\end{figure} 

Projection is defined in Figure~\ref{fig:projection}. The definition is a
straightforward use of bind to recursively flatten embedded
distributional values.
In our example,
the projection of the mixed term before the step shows what is left after
discarding the unreachable distribution elements:
\vspace{-1ex}
\begingroup\color{\colorMATH}\begin{gather*}
\begin{array}{c
} \hat \lceil 
   \text{\guilsinglleft}(\underline \sigma ,\langle {\mfootnotesize{{{\color{\colorSYNTAX}\mtexttt{bitv}}}}}_{P}({\mfootnotesize{{{\color{\colorSYNTAX}\mtexttt{I}}}}}),{\mfootnotesize{{{\color{\colorSYNTAX}\mtexttt{cast}}}}}_{{\mtext{P}}}({\mfootnotesize{{{\color{\colorSYNTAX}\mtexttt{flipv}}}}}(\text{\guilsinglleft}{\mfootnotesize{{{\color{\colorSYNTAX}\mtexttt{I}}}}}\hspace*{0.33em}{\mfootnotesize{{{\color{\colorSYNTAX}\mtexttt{O}}}}}\text{\guilsinglright}))\rangle )
   \hspace*{0.33em}(\underline \sigma ,\langle {\mfootnotesize{{{\color{\colorSYNTAX}\mtexttt{bitv}}}}}_{P}({\mfootnotesize{{{\color{\colorSYNTAX}\mtexttt{O}}}}}),{\mfootnotesize{{{\color{\colorSYNTAX}\mtexttt{cast}}}}}_{{\mtext{P}}}({\mfootnotesize{{{\color{\colorSYNTAX}\mtexttt{flipv}}}}}(\text{\guilsinglleft}{\mfootnotesize{{{\color{\colorSYNTAX}\mtexttt{I}}}}}\hspace*{0.33em}{\mfootnotesize{{{\color{\colorSYNTAX}\mtexttt{O}}}}}\text{\guilsinglright}))\rangle )
   \text{\guilsinglright}
   \hat \rceil 
\cr  =
   \text{\guilsinglleft}(\underline \sigma ,\langle {\mfootnotesize{{{\color{\colorSYNTAX}\mtexttt{bitv}}}}}_{P}({\mfootnotesize{{{\color{\colorSYNTAX}\mtexttt{I}}}}}),{\mfootnotesize{{{\color{\colorSYNTAX}\mtexttt{cast}}}}}_{{\mtext{P}}}({\mfootnotesize{{{\color{\colorSYNTAX}\mtexttt{flipv}}}}}({\mfootnotesize{{{\color{\colorSYNTAX}\mtexttt{I}}}}}))\rangle )
   \hspace*{0.33em}(\underline \sigma ,\langle {\mfootnotesize{{{\color{\colorSYNTAX}\mtexttt{bitv}}}}}_{P}({\mfootnotesize{{{\color{\colorSYNTAX}\mtexttt{O}}}}}),{\mfootnotesize{{{\color{\colorSYNTAX}\mtexttt{cast}}}}}_{{\mtext{P}}}({\mfootnotesize{{{\color{\colorSYNTAX}\mtexttt{flipv}}}}}({\mfootnotesize{{{\color{\colorSYNTAX}\mtexttt{O}}}}}))\rangle )
   \text{\guilsinglright}
\vspace{-1ex}
\end{array}
\end{gather*}\endgroup
and where the RHS corresponds exactly to the step of computation using the
standard semantics.

We prove that the projected, mixed semantics simulates the standard semantics.
\begin{lemma}[Simulation (Mixed)]
\label{thm:simulation-mixed}
  If \hspace*{0.16em}{{\color{\colorMATH}\(e\)}}\hspace*{0.16em} is a source expression, then \hspace*{0.16em}{{\color{\colorMATH}\(\lceil {\mtext{\underline {nste\hspace{-1pt}}\hspace{1pt}p}}(N,\varnothing ,e)\rceil  = {\mtext{nstep}}_{{\mathcal{I}}}(N,\varnothing ,e)\)}}.
\end{lemma}
To relate to ``ground truth'', we also prove that the standard semantics
using intensional distributions {{\color{\colorMATH}\({\mathcal{I}}\)}} simulates the standard semantics
using the denotational probability monad {{\color{\colorMATH}\({\mathcal{D}}\)}}.
\begin{lemma}[Simulation (Intensional)]
\label{thm:simulation-intensional}
  {{\color{\colorMATH}\({\mtext{Pr}}\left[{\mtext{nstep}}_{{\mathcal{I}}}(N,\varnothing ,e) \mathrel{\dot =} t\right] = {\mtext{Pr}}\left[{\mtext{nstep}}_{{\mathcal{D}}}(N,\varnothing ,e) \mathrel{\dot =} t\right]\)}}.
\end{lemma}

\subsection{Mixed Semantics Typing}
\label{sec:mixed-sem-typing}

\begin{figure} 
  \small
\prefigskip
\begingroup\color{\colorMATH}\begin{gather*}
\begin{tabularx}{\linewidth}{>{\centering\arraybackslash\(}X<{\)}}\hfill\hspace{0pt} \Psi ^{F} {\in } {\mtext{flipset}} \triangleq  \wp ({\mathcal{I}}({\mathbb{B}}))
  \hfill\hspace{0pt} \Psi ^{B} {\in } {\mtext{bitset}}  \triangleq  R {\rightarrow } \wp ({\mathcal{I}}({\mathbb{B}}))
  \hfill\hspace{0pt} \Psi     {\in } {\mtext{fbset}}   \mathrel{\Coloneqq } \Psi ^{F}{,}\Psi ^{B}
  \hfill\hspace{0pt} \Phi     {\in } {\mtext{history}}  \mathrel{\Coloneqq } \overline {\underline {\hat \varsigma } \mathrel{\dot =} \underline \varsigma }
  \hfill\hspace{0pt}
\\[-1.75ex]
\cr  (\Psi _{1}^{F},\Psi _{1}^{B}) \uplus  (\Psi _{2}^{F},\Psi _{2}^{B}) \triangleq  (\Psi _{1}^{F} \uplus  \Psi _{2}^{F}),(\Psi _{1}^{B} \cup  \Psi _{2}^{B})
\\[-1.75ex]
\cr  \left[\overline {\hat x} \mathrel{\bot \!\!\!\bot } \overline {\hat y} \mathrel{}\middle|\mathrel{} \overline {\hat z \mathrel{\dot =} z} \right] \mathrel{\overset \vartriangle {\iff   }} \forall  \overline x,\overline y.\hspace*{0.33em}{\mtext{Pr}}\left[\overline {\hat x \mathrel{\dot =} x},\overline {\hat y\mathrel{\dot =}y} \mathrel{}\middle|\mathrel{} \overline {\hat z \mathrel{\dot =} z}\right] = {\mtext{Pr}}\left[\overline {\hat x\mathrel{\dot =}x} \mathrel{}\middle|\mathrel{} \overline {\hat z \mathrel{\dot =} z}\right]{\mtext{Pr}}\left[\overline {\hat y\mathrel{\dot =}y} \mathrel{}\middle|\mathrel{} \overline {\hat z \mathrel{\dot =} z}\right]
\\[-1.75ex]
\cr  \hline
\\[-1.75ex] \hfill\hspace{0pt} \inferrule*[vcenter,lab={\mtextsc{ Flip-Value}}
     ]{ {\mtext{Pr}}\left[\hat b \mathrel{\dot =} {\mfootnotesize{{{\color{\colorSYNTAX}\mtexttt{I}}}}} \mathrel{}\middle|\mathrel{} \Phi  \right] = \nicefrac{1}{2} 
     \\ \left[ \hat b \mathrel{\bot \!\!\!\bot } \Psi ^{F},\Psi ^{B}(\{ \rho ^{\prime} \mathrel{|} \rho ^{\prime} \sqsubset  \rho \} ) \mathrel{}\middle|\mathrel{} \Phi  \right]
        }{
        (\Psi ^{F},\Psi ^{B}),\Phi  \vdash  \hat b \mathrel{:} {\mfootnotesize{{{\color{\colorSYNTAX}\mtexttt{flip}}}}}^{\rho }
     }
  \hfill\hspace{0pt} \begingroup\color{\colorTEXT}\boxed{\begingroup\color{\colorMATH}\Psi ,\Phi \vdash \hat b\mathrel{:}{\mfootnotesize{{{\color{\colorSYNTAX}\mtexttt{flip}}}}}^{\rho }\endgroup}\endgroup
\\[-1.75ex]
\parbox{\linewidth}{\def\MathparLineskip{\lineskip=4pt}
\begingroup\color{\colorMATH}\begin{mathpar} \inferrule*[vcenter,lab={\mtextsc{ BitV-P}}
   ]{ }{
      \Psi ,\Phi ,\Sigma ,\Gamma  \vdash  {\mfootnotesize{{{\color{\colorSYNTAX}\mtexttt{bitv}}}}}_{P}({\mtext{return}}(b)) \mathrel{:} {\mfootnotesize{{{\color{\colorSYNTAX}\mtexttt{bit}}}}}_{P}^{\bot } \mathrel{;} \Gamma ,\varnothing ,\varnothing 
   }
\and \inferrule*[vcenter,lab={\mtextsc{ BitV-S}}
   ]{ }{
      \Psi ,\Phi ,\Sigma ,\Gamma  \vdash  {\mfootnotesize{{{\color{\colorSYNTAX}\mtexttt{bitv}}}}}_{S}(\hat b) \mathrel{:} {\mfootnotesize{{{\color{\colorSYNTAX}\mtexttt{bit}}}}}_{S}^{\rho } \mathrel{;} \Gamma ,\varnothing ,\{ \rho \mapsto \{ \hat b\} \} 
   }
\and\inferrule*[vcenter,lab={\mtextsc{ FlipV}}
  ]{ \Psi ,\Phi  \vdash  \hat b \mathrel{:}{\mfootnotesize{{{\color{\colorSYNTAX}\mtexttt{flip}}}}}^{\rho }
     }{
     \Psi ,\Phi ,\Sigma ,\Gamma  \vdash  {\mfootnotesize{{{\color{\colorSYNTAX}\mtexttt{flipv}}}}}(\hat b) \mathrel{:} {\mfootnotesize{{{\color{\colorSYNTAX}\mtexttt{flip}}}}}^{\rho } \mathrel{;} \Gamma ,\{ \hat b\} ,\varnothing 
  }
\and\inferrule*[vcenter,lab={\mtextsc{ LocV}}
  ]{ \Sigma (\iota ) = \tau 
     }{
     \Psi ,\Phi ,\Sigma ,\Gamma  \vdash  {\mfootnotesize{{{\color{\colorSYNTAX}\mtexttt{locv}}}}}(\iota ) \mathrel{:} \tau  \mathrel{;} \Gamma  , \varnothing ,\varnothing 
  }
\\ \cdots 
\and \inferrule*[vcenter,lab={\mtextsc{ Ref}}
   ]{ \Psi ,\Phi ,\Sigma ,\Gamma  \vdash  \underline e \mathrel{:} \tau  \mathrel{;} \Gamma ^{\prime},\Psi ^{\prime}
      }{
      \Psi ,\Phi ,\Sigma ,\Gamma  \vdash  {\mfootnotesize{{{\color{\colorSYNTAX}\mtexttt{ref}}}}}(\underline e) \mathrel{:} {\mfootnotesize{{{\color{\colorSYNTAX}\mtexttt{ref}}}}}(\tau ) \mathrel{;} \Gamma ^{\prime},\Psi ^{\prime}
   }
\and\inferrule*[vcenter,lab={\mtextsc{ Tup}}
  ]{{\begin{array}{rcl
     } \Psi \uplus \Psi _{2},\Phi ,\Sigma ,\Gamma   \vdash  \underline e_{1} \mathrel{:} \tau _{1} \mathrel{;} \Gamma ^{\prime},\Psi _{1}
     \cr  \Psi \uplus \Psi _{1},\Phi ,\Sigma ,\Gamma ^{\prime} \vdash  \underline e_{2} \mathrel{:} \tau _{2} \mathrel{;} \Gamma ^{\prime \prime},\Psi _{2}
     \end{array}}
     }{
     \Psi ,\Phi ,\Sigma ,\Gamma  \vdash  \langle \underline e_{1},\underline e_{2}\rangle  \mathrel{:} \tau _{1} \times  \tau _{2} \mathrel{;} \Gamma ^{\prime \prime},\Psi _{1}\uplus \Psi _{2}
  }
\and \cdots 
\end{mathpar}\endgroup
}
\\[-1.75ex]
\cr \hfill\hspace{0pt} \inferrule*[vcenter,lab={\mtextsc{ Store-Empty}}
    ]{ }{
       \Psi  , \Phi  , \Sigma  \vdash  \varnothing  \mathrel{;} \varnothing ,\varnothing 
    }
  \hfill\hspace{0pt} \inferrule*[vcenter,lab={\mtextsc{ Store-Cons}}
    ]{ \Psi  \uplus  \Psi _{\sigma } , \Phi  , \Sigma  , \varnothing  \vdash  \underline v \mathrel{:} \Sigma (\iota ) \mathrel{;} \varnothing  , \Psi _{v}
    \\\\ \Psi  \uplus  \Psi _{v}   , \Phi  , \Sigma  , \varnothing  \vdash  \underline \sigma  \mathrel{;} \Psi _{\sigma }
       }{
       \Psi  , \Phi  , \Sigma  \vdash  \{ \iota  \mapsto  \underline v\} \uplus \underline \sigma  \mathrel{;} \Psi _{v} \uplus  \Psi _{\sigma }
    }
  \hfill\hspace{0pt} \begingroup\color{\colorTEXT}\boxed{\begingroup\color{\colorMATH}\Psi ,\Phi ,\Sigma \vdash \underline \sigma \mathrel{;}\Psi \endgroup}\endgroup
\\[-1.75ex]
\cr  \hfill\hspace{0pt}
   \inferrule*[vcenter,lab={\mtextsc{ Config}}
   ]{ \Psi  \uplus  \Psi _{e}   , \Phi  , \Sigma  \vdash  \underline \sigma  \mathrel{;} \Psi _{\sigma }
   \\ \Psi  \uplus  \Psi _{\sigma } , \Phi  , \Sigma  , \varnothing  \vdash  \underline e \mathrel{:} \tau  \mathrel{;} \varnothing  , \Psi _{e}
      }{
      \Psi  , \Phi  , \Sigma  \vdash  \underline \sigma ,\underline e \mathrel{:} \tau  \mathrel{;} \Psi _{\sigma } \uplus  \Psi _{e}
   }
  \hfill\hspace{0pt} \begingroup\color{\colorTEXT}\boxed{\begingroup\color{\colorMATH}\Phi ,\Sigma \vdash \underline \varsigma \mathrel{:}\tau ,\Psi \endgroup}\endgroup
\end{tabularx}
\end{gather*}\endgroup
\precaptionskip
\caption{Mixed Semantics Typing}
\postcaptionskip
\label{fig:mixed-typing-rules}
\end{figure} 

Our type system aims to ensure that {{\color{\colorMATH}\({\mfootnotesize{{{\color{\colorSYNTAX}\mtexttt{cast}}}}}_{P}\)}} will
produce {\mfootnotesize{{{\color{\colorSYNTAX}\mtexttt{I}}}}} and {\mfootnotesize{{{\color{\colorSYNTAX}\mtexttt{O}}}}} with equal probability, meaning neither
outcome leaks information. We establish this invariant in the PMTO
proof as a
consequence of type preservation for mixed terms. The mixed term typing
judgment extends typing of source-program expressions
(Figure~\ref{fig:typing-rules}) with some additional elements, and
considers non-source values.

The judgment has the form {{\color{\colorMATH}\(\Psi ,\Phi ,\Sigma \vdash \underline \varsigma \mathrel{:}\tau ,\Psi \)}}, and is shown at the bottom
of Figure~\ref{fig:mixed-typing-rules}. Here, {{\color{\colorMATH}\(\Sigma \)}} is a \emph{store
  context}, which maps store locations to types; it is used
to type the store {{\color{\colorMATH}\(\underline \sigma \)}} in rules {\mtextsc{ Store-Cons}} and {\mtextsc{ LocV}} as usual. {{\color{\colorMATH}\(\Phi \)}}
represents \emph{trace history} which encodes the exact sequence of
evaluation steps taken to reach the present one. The type
system reasons about the probability of distributional
values conditioned on this trace history having occurred.
The {{\color{\colorMATH}\(\Psi \)}} is an \emph{fbset}, which is a technical device used to collect all
distributional bit values {{\color{\colorMATH}\(\hat b\)}} that appear in {{\color{\colorMATH}\(\smash{\underline \varsigma }\)}}. Per the top of the figure,
the fbset is a pair {{\color{\colorMATH}\((\Psi ^{F},\Psi ^{B})\)}},
where {{\color{\colorMATH}\(\Psi ^{F}\)}} is a \emph{flipset} containing those {{\color{\colorMATH}\(\hat b\)}} that appear
inside of flip values, and {{\color{\colorMATH}\(\Psi ^{B}\)}} is a \emph{bitset} containing those
{{\color{\colorMATH}\(\hat b\)}} inside bit values. The latter is a map from a
region {{\color{\colorMATH}\(\rho \)}} to a set of bit values in that region. The {{\color{\colorMATH}\(\Psi \)}} to the right
of the turnstile contains all of the flip and secret bit values in the
configuration itself, while the {{\color{\colorMATH}\(\Psi \)}} to the left of it captures those
in the evaluation context and store.

The expression typing judgment {{\color{\colorMATH}\(\Psi ,\Phi ,\Sigma ,\Gamma  \vdash  \underline e \mathrel{:} \tau  \mathrel{;}
\Gamma ,\Psi \)}} is similar but includes variable contexts {{\color{\colorMATH}\(\Gamma \)}} as in the
source-program type rules.
We can see secret bit values being added to {{\color{\colorMATH}\(\Psi ^{B}\)}} in the {\mtextsc{ BitV-S}} rule, where {{\color{\colorMATH}\(\Psi ^{B}\)}} is
the singleton map from {{\color{\colorMATH}\(\rho \)}}, the region of the bit value, to {{\color{\colorMATH}\(\{ \hat b\} \)}},
while {{\color{\colorMATH}\(\Psi ^{F}\)}} is empty. Conversely, in the {\mtextsc{ FlipV}} rule {{\color{\colorMATH}\(\Psi ^{B}\)}} is
empty while {{\color{\colorMATH}\(\Psi ^{F}\)}} is the singleton set {{\color{\colorMATH}\(\{ \hat b\} \)}}. We can see the
maintenance of {{\color{\colorMATH}\(\Psi \)}} to the left of the turnstile in the {\mtextsc{ Tup}}
rule. Recursively typing the pair's left component {{\color{\colorMATH}\(\underline e_{1}\)}} yields fbset
{{\color{\colorMATH}\(\Psi _{1}\)}} to the right of the turnstile, which is used when typing {{\color{\colorMATH}\(\underline e_{2}\)}},
and vice versa; the {\mtextsc{ Store-Cons}} rule similarly handles
the store and the expression. The rules combine two fbsets
using the {{\color{\colorMATH}\(\uplus \)}} operator. Per the top of the figure, it acts as
disjoint union for flipsets but normal union for bitsets, mirroring
the handling of affine and universal variables.

The key invariants ensured by typing are defined
by the judgment {{\color{\colorMATH}\(\Psi ,\Phi \vdash \hat b\mathrel{:}{\mfootnotesize{{{\color{\colorSYNTAX}\mtexttt{flip}}}}}^{\rho }\)}}, which is invoked by
expression-typing rule {\mtextsc{ FlipV}} and defined in the {\mtextsc{ Flip-Value}}
rule. This judgment establishes that in a configuration reached by an
execution path {{\color{\colorMATH}\(\Phi \)}} the flip value {{\color{\colorMATH}\(\hat b\)}} is uniformly distributed (first
premise), and that it can be typed at region {{\color{\colorMATH}\(\rho \)}} because it is
properly independent of the other secret bit values in smaller regions
{{\color{\colorMATH}\(\Psi ^{B}(\{ \rho ^{\prime} \mathrel{|} \rho ^{\prime}\sqsubset \rho \} )\)}} and flip values {{\color{\colorMATH}\(\Psi ^{F}\)}} (second
premise). Conditional independence is defined in the figure in the usual
way---the overbar notation represents some sequence of random variables and/or
condition events.

We prove a type preservation lemma to establish that these
invariants are preserved.
\begin{lemma}[Type Preservation]
\label{thm:type-preservation}
  If \hspace*{0.16em}{{\color{\colorMATH}\(e\)}}\hspace*{0.16em} is a closed source expression, \hspace*{0.16em}{{\color{\colorMATH}\(\underline t\mathord{\cdotp }\underline \varsigma  \in 
  {\mtext{support}}({\mtext{\underline {nste\hspace{-1pt}}\hspace{1pt}p}}(N,\varnothing ,e))\)}}\hspace*{0.16em} and \hspace*{0.16em}{{\color{\colorMATH}\(\vdash  e \mathrel{:} \tau \)}}\hspace*{0.16em}, then there exists \hspace*{0.16em}{{\color{\colorMATH}\(\Sigma \)}}\hspace*{0.16em} and
  \hspace*{0.16em}{{\color{\colorMATH}\(\Psi \)}}\hspace*{0.16em} {s.t.} \hspace*{0.16em}{{\color{\colorMATH}\(\Phi ,\Sigma  \vdash  \underline \varsigma  \mathrel{:} \tau ,\Psi \)}}\hspace*{0.16em} where \hspace*{0.16em}{{\color{\colorMATH}\(\Phi  \triangleq  \left[{\mtext{\underline {nste\hspace{-1pt}}\hspace{1pt}p}}(N,\varnothing ,e) \mathrel{\dot =} \underline t\mathord{\cdotp }\underline \varsigma \right]\)}}.
\end{lemma}
When a configuration takes any number of steps, the resulting
configuration is well-typed under new trace history {{\color{\colorMATH}\(\Phi \)}}. Updating
{{\color{\colorMATH}\(\Phi \)}} is not arbitrary---it is necessary to
satisfy a proof obligation as used in a later lemma (\nameref{thm:pmto-mixed}). The new {{\color{\colorMATH}\(\Sigma \)}}
and {{\color{\colorMATH}\(\Psi \)}} are new store typings (in case new references were allocated), and the
new fbset (in case flip values were either created or consumed).
The proof of preservation uses a sublemma which shows typesafe substitution;
this lemma makes crucial use of affinity to ensure that aggregated {{\color{\colorMATH}\(\Psi _{1} \uplus  \Psi _{2}\)}} in
contexts for compound expressions ({e.g.}, pairs) are truly disjoint, which
will be true only because the substitution is guaranteed to only occur in {{\color{\colorMATH}\(\Psi _{1}\)}},
{{\color{\colorMATH}\(\Psi _{2}\)}}, or neither, but not both. 

The key property established by type preservation is that flip values
remain well-typed.
Recall that the first premise of {\mtextsc{ Flip-Value}}---uniformity---is crucial in establishing that it
is safe to reveal the flip via the {{\color{\colorMATH}\({\mfootnotesize{{{\color{\colorSYNTAX}\mtexttt{cast}}}}}_{P}\)}} coercion to a public bit. The
second premise is crucial in re-establishing the first premise after some
{\mtextit{other}} flip has been revealed. When another flip is revealed, this information
will be added to trace history, and it is not true that uniformity
conditioned on the current history {{\color{\colorMATH}\(\Phi \)}} automatically implies uniformity in the new
history {{\color{\colorMATH}\(\Phi ^{\prime}\)}}; this must be proved. Because the second premise
establishes independence from all other flips, we are able reestablish the
first premise via the second after some other flip is revealed to complete the proof.

Note that we also prove a progress lemma to ensure that no
well-typed evaluation reaches a stuck state; along with preservation,
this lemma establishes standard type soundness for \lang under the
mixed semantics.

\subsection{Proving PMTO}

To prove PMTO (Proposition~\ref{prop:pmto}) we first prove a variant
of it for the mixed semantics, and then apply a few more lemmas to
show that PMTO holds for the standard semantics too.
\begin{lemma}[PMTO (Mixed)]
\label{thm:pmto-mixed}
  If \hspace*{0.16em}{{\color{\colorMATH}\(\underline e_{1}\)}}\hspace*{0.16em} and \hspace*{0.16em}{{\color{\colorMATH}\(\underline e_{2}\)}}\hspace*{0.16em} are closed source expressions, \hspace*{0.16em}{{\color{\colorMATH}\(\vdash  \underline e_{1} \mathrel{:} \tau \)}}\hspace*{0.16em}, \hspace*{0.16em}{{\color{\colorMATH}\(\vdash  \underline e_{2} \mathrel{:} \tau \)}}\hspace*{0.16em} and \hspace*{0.16em}{{\color{\colorMATH}\(\underline e_{1} \sim 
  \underline e_{2}\)}}\hspace*{0.16em}, then (1) \hspace*{0.16em}{{\color{\colorMATH}\({\mtext{\underline {nste\hspace{-1pt}}\hspace{1pt}p}}(N,\varnothing ,\underline e_{1})\)}}\hspace*{0.16em} and \hspace*{0.16em}{{\color{\colorMATH}\({\mtext{\underline {nste\hspace{-1pt}}\hspace{1pt}p}}(N,\varnothing ,\underline e_{2})\)}}\hspace*{0.16em} are defined,
  and (2) \hspace*{0.16em}{{\color{\colorMATH}\({\mtext{\underline {nste\hspace{-1pt}}\hspace{1pt}p}}(N,\varnothing ,\underline e_{1}) \approx _{\sim } {\mtext{\underline {nste\hspace{-1pt}}\hspace{1pt}p}}(N,\varnothing ,\underline e_{2})\)}}.
\end{lemma}
The judgment {{\color{\colorMATH}\(\underline e_{1} \sim  \underline e_{2}\)}} in the premise indicates that the two expressions are
\emph{low equivalent}, meaning that the adversary cannot tell them
apart. The definition of this judgment is basically standard (given in
\iftr
the Appendix)
\else
the supplemental report~\cite{lobliv-tr})
\fi
and we can easily prove that it is implied by {{\color{\colorMATH}\({\mtext{obs}}(e_{1}) =
{\mtext{obs}}(e_{2})\)}} for source expressions. Mixed
PMTO establishes equivalence of the distributions of mixed configurations modulo
low-equivalence. We define two distributions as
equivalent modulo an underlying equivalence relation as follows:
\begingroup\color{\colorMATH}\begin{gather*}\begin{tabularx}{\linewidth}{>{\centering\arraybackslash\(}X<{\)}}\begin{array}{lcl
    } \hat x_{1} \approx _{\sim _{A}} \hat x_{2} &{}\mathrel{\overset \vartriangle {\iff   }}{}& \forall x.\hspace*{0.33em} \left(\sum \limits_{x^{\prime} \mathrel{|} x^{\prime} \sim _{A} x} {\mtext{Pr}}\left[ \hat x_{1} \mathrel{\dot =} x^{\prime} \right]\right) = \left(\sum \limits_{x^{\prime} \mathrel{|} x^{\prime} \sim _{A} x} {\mtext{Pr}}\left[ \hat x_{2} \mathrel{\dot =} x^{\prime} \right]\right)
    \end{array}
  \end{tabularx}
\end{gather*}\endgroup
This definition captures the idea that two distributions are equivalent
when, for any equivalence class within the relation (represented by element
{{\color{\colorMATH}\(x\)}}), each distribution assigns equal mass to the whole class.
For Mixed PMTO, the relation {{\color{\colorMATH}\(\sim _{A}\)}} is instantiated to low equivalence, which
we write just as {{\color{\colorMATH}\(\sim \)}}. When the underlying relation is equality, we recover the
usual notion of distribution equivalence: equality of probability mass
functions.

We prove \nameref{thm:pmto-mixed} by induction over steps {{\color{\colorMATH}\(N\)}} and then unfolding the monadic
definition of {{\color{\colorMATH}\({\mtext{\underline {nste\hspace{-1pt}}\hspace{1pt}p}}(N+1)\)}}. The induction appeals to a single-step PMTO
sublemma. (As mentioned in Section~\ref{sec:mixed-sem}, such a proof would not have
been possible in the standard semantics.) To use this one-step PMTO sublemma,
it must be that the configuration at {{\color{\colorMATH}\(N\)}} steps is well-typed {w.r.t.} current
trace history {{\color{\colorMATH}\(\Phi \)}}; we get
this well-typing {w.r.t.} {{\color{\colorMATH}\(\Phi \)}} from \nameref{thm:type-preservation}, discussed
earlier.

A final major lemma in our PMTO proof is a notion of soundness for
low-equivalence on mixed terms, in particular, that equivalence modulo {{\color{\colorMATH}\(\sim \)}} for
distributions of mixed traces implies equality of adversary-observable traces
in the standard semantics:
\begin{lemma}[Low-equivalence Soundness]
\label{thm:low-equivalence-soundness}
   If \hspace*{0.16em}{{\color{\colorMATH}\(\underline {\hat t_{1}} \approx _{\sim } \underline {\hat t_{2}}\)}}\hspace*{0.16em} then \hspace*{0.16em}{{\color{\colorMATH}\(\widehat {\mtext{obs}}(\hat \lceil \underline {\hat t_{1}}\hat \rceil ) \approx _{=}
   \widehat {\mtext{obs}}(\hat \lceil \underline {\hat t_{2}}\hat \rceil )\)}}.
\end{lemma}
In this lemma we use a lifting of {{\color{\colorMATH}\({\mtext{obs}}\)}} for intensional
distributions, written {{\color{\colorMATH}\(\widehat {\mtext{obs}}\)}}; its definition is identical to {{\color{\colorMATH}\(\widetilde {\mtext{obs}}\)}}
in Figure~\ref{fig:adversary} but with the intensional distribution monad
{{\color{\colorMATH}\({\mathcal{I}}\)}} instead of {{\color{\colorMATH}\({\mathcal{D}}\)}}.



We now complete the full proof of PMTO. The general strategy is to
first consider two well-typed source programs which are equal modulo adversary
observation. Next, these programs are transported to the mixed language, where
low-equivalence is established. The programs are
executed in the mixed semantics, and PMTO for mixed terms is applied, which
appeals to type preservation. Due to PMTO for mixed terms, the results will be
low-equivalent, and via soundness of low-equivalence, we conclude equality of
distributions modulo adversary observation after projection. The final steps are
via simulation lemmas, showing that this final projection lines up with
executions of the initial programs in the standard
semantics. 

\noindent
\begin{minipage}{\linewidth}
\begin{theorem}[PMTO]\label{thm:pmto}\
  \begin{itemize}[label={},leftmargin=0pt]\item  If: \hspace*{0.16em}{{\color{\colorMATH}\(e_{1}\)}}\hspace*{0.16em} and \hspace*{0.16em}{{\color{\colorMATH}\(e_{2}\)}}\hspace*{0.16em} are closed source expressions, \hspace*{0.16em}{{\color{\colorMATH}\(\vdash  e_{1} \mathrel{:} \tau \)}}\hspace*{0.16em}, \hspace*{0.16em}{{\color{\colorMATH}\(\vdash  e_{2} \mathrel{:} \tau \)}}\hspace*{0.16em} and \hspace*{0.16em}{{\color{\colorMATH}\({\mtext{obs}}(e_{1}) = {\mtext{obs}}(e_{2})\)}}
  \item  Then: (1) \hspace*{0.16em}{{\color{\colorMATH}\({\mtext{nstep}}_{{\mathcal{D}}}(N,\varnothing ,e_{1})\)}}\hspace*{0.16em} and \hspace*{0.16em}{{\color{\colorMATH}\({\mtext{nstep}}_{{\mathcal{D}}}(N,\varnothing ,e_{2})\)}}\hspace*{0.16em} are defined
  \item  And: (2) \hspace*{0.16em}{{\color{\colorMATH}\(\widetilde {{\mtext{obs}}}({\mtext{nstep}}_{{\mathcal{D}}}(N,\varnothing ,e_{1})) = \widetilde {{\mtext{obs}}}({\mtext{nstep}}_{{\mathcal{D}}}(N,\varnothing ,e_{2}))\)}}.
  \end{itemize}
\end{theorem}
\begin{proof}\
  \begin{itemize}[label={},leftmargin=0pt]\item  (1) is by Progress (see \iftr appendix). \else supplemental report). \fi (2) is by the following:
  \item  {{\color{\colorMATH}\(\begin{array}[t]{rcl@{\hspace*{1.00em}}l
      }   &{} {}&   {\mtext{obs}}(e_{1}) = {\mtext{obs}}(e_{2})
      \cr    &{}\implies   {}& e_{1} \sim  e_{2}                                                                            & \lbag {{\color{\colorTEXT}\textnormal{\hspace*{0.33em} by simple induction \hspace*{0.33em}}}}\rbag 
      \cr    &{}\implies   {}& {\mtext{\underline {nste\hspace{-1pt}}\hspace{1pt}p}}(N,\varnothing ,e_{1}) \approx _{\sim } {\mtext{\underline {nste\hspace{-1pt}}\hspace{1pt}p}}(N,\varnothing ,e_{2})                                           & \lbag {{\color{\colorTEXT}\textnormal{\hspace*{0.33em} by \nameref{thm:pmto-mixed} \hspace*{0.33em}}}}\rbag 
      \cr    &{}\implies   {}& \widehat {{\mtext{obs}}}(\hat \lceil {\mtext{\underline {nste\hspace{-1pt}}\hspace{1pt}p}}_{{\mathcal{I}}}(N,\varnothing ,e_{1})\hat \rceil ) \approx _{=} \widehat {{\mtext{obs}}}(\hat \lceil {\mtext{\underline {nste\hspace{-1pt}}\hspace{1pt}p}}_{{\mathcal{I}}}(N,\varnothing ,e_{2})\hat \rceil ) & \lbag {{\color{\colorTEXT}\textnormal{\hspace*{0.33em} by \nameref{thm:low-equivalence-soundness} \hspace*{0.33em}}}}\rbag 
      \cr    &{}\implies   {}& \widehat {{\mtext{obs}}}({\mtext{nstep}}_{{\mathcal{I}}}(N,\varnothing ,e_{1})) \approx _{=} \widehat {{\mtext{obs}}}({\mtext{nstep}}_{{\mathcal{I}}}(N,\varnothing ,e_{2}))                 & \lbag {{\color{\colorTEXT}\textnormal{\hspace*{0.33em} by \nameref{thm:simulation-mixed} \hspace*{0.33em}}}}\rbag 
      \cr    &{}\implies   {}& \widetilde {{\mtext{obs}}}({\mtext{nstep}}_{{\mathcal{D}}}(N,\varnothing ,e_{1})) = \widetilde {{\mtext{obs}}}({\mtext{nstep}}_{{\mathcal{D}}}(N,\varnothing ,e_{2}))                    & \lbag {{\color{\colorTEXT}\textnormal{\hspace*{0.33em} by \nameref{thm:simulation-intensional} \hspace*{0.33em}}}}\rbag 
      \end{array}\)}}
  \end{itemize}
\end{proof}
A detailed proof is given in the
\iftr
Appendix~\ref{sec:proofs}.
\else
supplemental report~\cite{lobliv-tr}.
\fi
\end{minipage}

\section{Implementation and Tree-based ORAM Case Study}
\label{sec:case-study}

We have
implemented an interpreter and type checker for a language that extends \lang in
several (straightforward) ways.
First, we add natural number literals and random values; these can be
  encoded in \lang as fixed-width tuples of {{\color{\colorMATH}\({\mfootnotesize{{{\color{\colorSYNTAX}\mtexttt{bitv}}}}}\)}} and {{\color{\colorMATH}\({\mfootnotesize{{{\color{\colorSYNTAX}\mtexttt{flipv}}}}}\)}}
  respectively. We write them annotated with a security level, e.g.,
  \code{2 S} or \code{2 P}, and write \code{rnd R ()} to generate a
  random number at region \code{R}. We write \code{natS} to be the
  type of a secret number in region {{\color{\colorMATH}\(\bot \)}}; \code{natP} for the type of a
  public number; \code{R natS} for the type of a secret number in the
  region \code{R}.  We also write \code{R rnd} to be the
  type of a random natural number in the region \code{R}.
Second, we add arrays; in our code examples, we write
  \code{a[n]} and \code{a[n] <- e} to read and write array
  elements. An array of length $N$ can be encoded in \lang as
  an $N$-tuple of references, using nested conditional expressions to
  access the correct (public) index and swapping out affine
  contents, as must be done with references.
Finally, we add records, which are like tuples but permit field accessor
  notation, \code{r.x}; if \code{x} is affine, doing so only consumes the field \code{x}
rather than consuming all of \code{r}. 

To demonstrate the expressiveness of \lang, we have used our extended language to 
program (and type check) a series of interesting oblivious
algorithms. Section~\ref{sec:oram-overview} presents a modern
\emph{non-recursive, tree-based ORAM} (NORAM), which is a key component of
state-of-the-art ORAM
implementations~\cite{asiacrypt11,pathoram,circuitoram}. To our
knowledge, ours is the first implementation automatically verified to
be oblivious. Building on this NORAM, Section~\ref{sec:full-oram}
presents a full \emph{recursive} ORAM. Type checking it requires some
advanced (but standard) language features we have not implemented,
including region polymorphism, recursive and variant types, and
existential quantification. Finally, the
\iftr
appendix
\else
supplemental report~\cite{lobliv-tr}
\fi
presents a mostly complete implementation of \emph{oblivious stacks} (ostacks),
a kind of oblivious data structure~\cite{ods} that builds on top of NORAM.
The \lang type system is not powerful enough to reason that ostacks'
use of NORAM is safe; the region ordering requirement is too strong.
Sections~\ref{sec:related} and~\ref{sec:conc} discuss
integrating \lang's type system with a general-purpose logic as a way
to potentially overcome this limitation.
Our type checker and all the examples 
are online at \url{https://github.com/plum-umd/oblivml}.

\subsection{Tree-based ORAM: Overview}
\label{sec:treeoram-overview}

A complete ORAM implements the same API as a standard array: A
\code{read} operation takes an ORAM \code{oram} and index \code{i} as
arguments, and returns data \code{d} stored at that index; a
\code{write} operation updates \code{oram} at \code{i} with a given
\code{d}. We assume that the ORAM contents and the indexes are
not visible to the adversary (i.e., they are encrypted). A simple 
implementation is a \emph{Trivial ORAM}. It consists of an array of
$N$ ``buckets,'' each of which consists of an index \code{i} and 
data \code{d}. A \code{read} at index \code{j} iterates over the
entire array and retrieves the data associated with \code{j}, if
present. The data is returned when the iteration is complete (or a
default value is returned, if \code{j} is not present). Since each
\code{read} touches every bucket, nothing is leaked about \code{i}. Of
course, this is very inefficient---the read takes time $O(N)$ where
$N$ is the size of the array. (The code example in
Figure~\ref{fig:code-examples}(b) does something similar.)

A tree-based ORAM~\cite{asiacrypt11,pathoram,circuitoram} offers
better performance. It breaks its implementation into two
parts. The first is a tree-like structure
\code{noram} for storing the actual data blocks; this is called a
\emph{non-recursive ORAM} (or NORAM) for reasons that will be clear in
the next subsection. The second part is the \emph{position map}
\code{pm} that maps logical data block indexes to \emph{position tags}
that indicate the block's position in the tree.

NORAMs do not implement \code{read} and \code{write} operations
directly; instead they implement two more-primitive operations called
\code{noram_readAndRemove} (or \code{noram_rr}, for short) and
\code{noram_add}. The former reads the designated data block from
\code{noram} and also removes it, while the latter adds the given
data. Putting it all together, a Tree ORAM \code{read} from index
\code{i} works in four steps: (1) retrieve tag \code{t} from
\code{pm[i]}; (2) call %
\code{noram_rr noram i t} to remove the data \code{d} at \code{i}
using \code{t} to assist the lookup; (3) update \code{pm[i]} with a
randomly generated tag \code{t2}; and (4) call \code{noram_add noram i
  t2 d} to add back data \code{d}, but with the new tag, before
returning it. An ORAM \code{write} has the same four steps, but in
step (4) we add the provided data, rather than the original. (A fifth
step in both cases, \emph{eviction}, will be explained later.)
As with the example in Figure~\ref{fig:code-examples}(c),
non-recursive ORAM combines randomness (and its tree structure) to
avoid having $O(N)$ cost for the entire map: Under
the right assumptions, these operations take time $O(\log(N))$. 

The position tags mask the relationship between a logical index and
the location of its corresponding data block in the tree. As blocks are
read and written, they are shuffled around in the tree, and their new
locations are recorded in the position map.  As such, two ORAM
\code{read} operations to the same index \code{i} will involve
different access patterns in a way that leaks nothing about the index
\emph{assuming} lookups and updates to the position map itself leak no
information. This assumption could be satisfied by making the position
map a Trivial ORAM, but then we would lose our performance
benefits. In the next subsection we simply assume we have a leak-free
position map and in Section~\ref{sec:full-oram} we show how one can be
obtained by efficiently storing the position map \emph{recursively} in
the NORAM tree structure itself.

\subsection{Tree-based Non-recursive ORAM}
\label{sec:oram-overview}

Now we present the details of our implementation of tree-based NORAM
in \lang.

\paragraph{Data definition}
The type of a tree-based NORAM is defined as follows:

\begin{lstlisting}[escapeinside={([}{])},numbers=none]
type block  = { is_dummy : R bitS ; idx : R natS ; tag : R natS ; data : (R ([{{\color{\colorMATH}\(\vee \)}}]) R' rnd) * (R ([{{\color{\colorMATH}\(\vee \)}}]) R' rnd) }
type bucket = block array
type noram  = bucket array
\end{lstlisting}

A \code{noram} is an array of $2N-1$ \code{buckets} which represents a complete tree
in the style of a heap data structure: for the node at index
$i\in\{0,...,2N-2\}$, its parents, left child, and right child
correspond to the nodes at index $(i-1)/2$, $2i+1$, and $2i+2$,
respectively. Each \code{bucket} is an array of \code{blocks}, each of which is a
record where the \code{data} field contains the data stored in that bucket.
The other three components of the block are secret; they are (1) the
\code{is_dummy} bit indicating if the block is dummy (empty) or not; (2) the
index (\code{idx}) of the block; and (3) the position \code{tag} of the block.
Note that the \code{bucket} type, ignoring the position \code{tag},
is essentially a Trivial ORAM. In the operations
discussed below, all functions prefixed with \code{trivial} are operations over buckets.

The region \lstinline[mathescape]{R $\vee$ R'} should be read as
``\code{R} join \code{R'}'' 
and corresponds to the join operation, {{\color{\colorMATH}\(\sqcup \)}}, over regions {{\color{\colorMATH}\(\rho \)}} in Section~\ref{sec:formalism}.
Notice that we have \lstinline[mathescape]{R $\sqsubset $ R $\vee $ R'}, which will be important when discussing well-typedness of \code{mux}
in the discussion that follows. We choose type \lstinline[mathescape]{(R $\vee$ R' rnd) * (R $\vee$ R' rnd)}
for the data portion to illustrate that affine values can be stored in the
NORAM, and to set up our implementation of full, recursive ORAM, next.


\paragraph{Operations}
The code for \code{noram_rr} is given below; we explain it just afterward.

\begin{lstlisting}[escapeinside={([}{])}]
let rec trivial_rr_h (troram : bucket) (idx : R natS) (i : natP) (acc : block) : block =
  if i = length(troram) then acc
  else
    (* read out the current block, replace with dummy *)
    let curr = bucket[i] <- (dummy_block ()) in
    (* check if the current block is non-dummy, and its index matches the queried one *)
    let swap : R bitS = !curr.is_dummy && curr.idx = idx in
    let (curr, acc) = mux(swap, acc, curr) in
    (* when swap is false, this equivalent to writing the data back; otherwise, acc
       stores the found block and is passed into the next iteration *)
    let _ = bucket[i] <- curr in
    trivial_rr_h troram idx (i + 1) acc

let trivial_rr (troram : bucket) (idx : R natS) : (R ([{{\color{\colorMATH}\(\vee \)}}]) R' rnd) * (R ([{{\color{\colorMATH}\(\vee \)}}]) R' rnd) =
  let ret: block = trivial_rr_h troram idx 0 (dummy_block ()) in
  ret.data

let rec noram_rr_h (noram : noram) (idx : R natS) (tag : natP) (level : natP) (acc : block) : block =
  (* compute the first index into the bucket array at depth level *)
  let base : natP = (pow 2 level) - 1 in
  if base >= length(noram) then acc
  else
    let bucket_loc : natP = base + (tag & base) in (* the bucket on the path to access *)
    let bucket = noram[bucket_loc] in
    let acc = trivial_rr_h bucket idx 0 acc in
    noram_rr_h noram idx tag (level + 1) acc

let noram_rr (noram : noram) (idx : R natS) (tag : natP) : (R ([{{\color{\colorMATH}\(\vee \)}}]) R' rnd) * (R ([{{\color{\colorMATH}\(\vee \)}}]) R' rnd) =
  let ret = noram_rr_h noram idx tag 0 (dummy_block ()) in
  ret.data
\end{lstlisting}

\code{noram_rr} takes the NORAM \code{noram}
and the index \code{idx} of the desired element as arguments. The
\code{tag} argument is the position tag, which identifies a path
through the \code{noram} binary tree along which the indexed value
will be stored, if present. This tag's type \code{natP} means it
is publicly visible. Initially it is stored, secretly, in the position
map, but prior to passing it to this function it must be revealed (via \code{castP}) 
because it (or derivatives of it) will be used to index the arrays that
make up the NORAM, and array indexes are always adversary-visible.

\code{noram_rr} works by calling \code{noram_rr_h} which
recursively works its way down the identified path. It maintains an
accumulator, \code{acc : block}, over the course of the
traversal. Initially, \code{acc} is a dummy block. The \code{dummy_block ()}
is a function call rather than a constant because the block record contains
\lstinline[mathescape]{data: (R $\vee$ R' rnd) * (R $\vee$ R' rnd)}. This member of the record must be generated
fresh for each new block, since its contents are treated affinely.
Each recursive call to \code{noram_rr_h} moves to a node the next level down in
the tree, as determined by the tag. At each node, it reads out the
bucket array, which as mentioned earlier is essentially a Trivial
ORAM. The \code{trivial_rr} function calls \code{trivial_rr_h} to
iterate through the entire bucket, to obliviously read out the desired
block, if present.

Notice that we are using arrays with both affine and non-affine (universal) contents in this code.
The \code{noram} type has contents which are kind {{\color{\colorMATH}\({\mfootnotesize{{{\color{\colorSYNTAX}\mtexttt{U}}}}}\)}}, since the
type of its contents is an array.
As such, we can read from \code{noram} without writing a new value (line 24). However,
the \code{bucket} type has contents which are kind {{\color{\colorMATH}\({\mfootnotesize{{{\color{\colorSYNTAX}\mtexttt{A}}}}}\)}}, since the type of its contents are tuples which contain
type \lstinline[mathescape]{R $\vee$ R' rnd}. So, when we index into members of values of type \code{bucket} we must write a
dummy block (line 5).

This algorithm for \code{noram_rr} will access $\log{N}$ buckets (where $N$ is the number
of buckets in the \code{noram}), and each bucket access causes a \code{trivial_rr}
which takes time $b$ where $b$ is the size of each bucket. Therefore, the
\code{noram_rr} operation above takes time $O(b \log{N})$. In the state-of-the-art ORAM constructions, such
as Circuit ORAM~\cite{circuitoram}, $b$ can be parameterized as a
constant (e.g., $4$), which renders the overall time complexity of
\code{noram_rr} to be $O(\log{N})$. This is asymptotically faster than
implementing the entire ORAM as a Trivial ORAM, which takes time $O(N)$.

The \code{noram_add} routine has the following signature:
\begin{lstlisting}[escapeinside={([}{])},numbers=none]
val noram_add : noram -> (idx : R natS) -> (tag : R natS) -> (data : (R ([{{\color{\colorMATH}\(\vee \)}}]) R' rnd) * (R ([{{\color{\colorMATH}\(\vee \)}}]) R' rnd)) -> unit
\end{lstlisting}
Like the \code{noram_rr} operation, it takes an index and a position
tag, but here the position tag is secret, since it will not be
examined by the algorithm. In particular, \code{noram_add} simply stores
a block consisting of the dummy bit, index, position tag, and data
into the root bucket of the \code{noram}. It does this as a Trivial
ORAM operation: It iterates down the root bucket's array similarly to
\code{trivial_rr} above, but stores the new block in the first
available slot.

To avoid overflowing the root's bucket due to repeated \code{noram_add}s,
our NORAM employs an additional \code{eviction} routine.
It is called after both \code{noram_add} and \code{noram_rr}, to move
blocks closer to the leaf buckets. This
routine maintains the key invariant that each data block should
reside on the path from the root to the leaf corresponding to its position tag.
Different tree-based ORAM implementations differ only in their choices
of $b$ and the eviction strategies. The simple eviction strategy we
implement (due to~\citet{asiacrypt11}) picks
two random nodes at each level of the tree, reads a single non-empty block
from each chosen node's bucket, and then writes that block one level further down
either to the left or right according to the position tag; a dummy block is written in the opposite
direction to make the operation oblivious.

\subsection{Recursive ORAM}
\label{sec:full-oram}

\newcommand{\pmp}{\ensuremath{\mathrm{PM}}\xspace}

As described in Section~\ref{sec:treeoram-overview}, a complete ORAM
combines a non-recursive ORAM with a position map. So far, we have not
said where the position map should be stored, and how.
One approach is to implement it
as just a regular array stored in hidden memory, e.g.,
on-chip (invisible to the adversary) in a secure processor deployment
of ORAM (see Section~\ref{sec:threat}). However, this is not possible for MPC-based deployments, in
which both parties secret-share the map, and thus the adversary can observe the
access pattern on the map itself. To block this side channel, we could
implement the position map itself as an ORAM, e.g., a Trivial ORAM. But to do
so would ruin the efficiency gain of our tree-based NORAM, since the
position map lookup would have time $O(N)$, as compared to
$O(\log(N))$ time for \code{noram_rr} and \code{noram_add}.

We could implement the position map in a NORAM in an attempt to get
back logarithmic-time efficiency, but doing so seems to ``kick the can
down the road'' because we now need another position map for our
position map! We can close this cycle by having each recursively
defined position map be smaller than the previous. In particular, to
implement a map with $N$ integer keys we can use a map of $N/c$ keys,
each of which maps to $c$ values, for a small constant $c$. Lookup of key
$k$ translates to looking up key $k/c$ in the smaller map, and then
returning the $(k \% c)$th value (which takes time $c$ to do
obliviously). We can apply this idea recursively, 
ultimately yielding $\log_c(N)$ maps numbered $i = 1 ... \log_c(N)$,
where map $i$ has $\frac{N}{c^i}$ keys (and each key maps to $c$
values). We can implement each map at level $i$ as a NORAM
until $i$ is large enough that we can use a Trivial ORAM to tie it off
(e.g., when $\frac{N}{c^i}$ is 4). The
complexity of looking up a key will thus be
$\sum_{i=1}^{\log_c(N)} O(\log(\frac{N}{c^i}) + c)$. Setting $c$ to be a
constant $2$ means that the complexity 
of the lookup procedure is $O(\log(N)^2)$. This construction is called
a \emph{recursive ORAM}.


\paragraph{Data Definition and Operations}
A recursive ORAM thus has the type \code{oram}, given below.
\begin{lstlisting}[escapeinside={([}{])},numbers=none]
type oram = (noram array) * bucket
\end{lstlisting}
The data blocks are stored in the \code{noram} at index $0$ in the
first component, an \code{noram array}; the remaining \code{noram}s in
that array consist of progressively smaller position maps, finally
ending in a trival ORAM, the second component (a \code{bucket}). 

We implement the \code{tree_rr} as a call to the function
\code{tree_rr_h}, which takes an additional public \code{level}
argument, to indicate at which point in the list of \code{oram}{s} to start
its work (initially, 0).
\begin{lstlisting}[escapeinside={([}{])}]
let rec tree_rr_h (oram : oram) (idx : natS) (level : natP): (R ([{{\color{\colorMATH}\(\vee \)}}]) R' rnd) * (R ([{{\color{\colorMATH}\(\vee \)}}]) R' rnd) =
  let (norams, troram) = oram in
  let levels : natP = length(norams) in
  if level >= levels then trivial_rr troram idx
  else
    let (r0, r1) : (R ([{{\color{\colorMATH}\(\vee \)}}]) R' rnd) * (R ([{{\color{\colorMATH}\(\vee \)}}]) R' rnd) = tree_rr_h oram (idx / 2) (level + 1) in
    let (r0', tag) = mux(idx % 2 = 0, rnd (R ([{{\color{\colorMATH}\(\vee \)}}]) R') (), r0) in
    let (r1', tag) = mux(idx % 2 = 1, tag, r1) in
    let _ = tree_add_h oram (idx / 2) (level + 1) (r0', r1') in
    noram_rr norams[level] idx (castP tag)

let tree_rr (oram : oram) (idx : natS): (R ([{{\color{\colorMATH}\(\vee \)}}]) R' rnd) * (R ([{{\color{\colorMATH}\(\vee \)}}]) R' rnd) =
  tree_rr_h oram idx 0
\end{lstlisting}

In the code above, the \code{level} indicates the embedded NORAM from
which to read.  For example, when \code{level} is 0, the data NORAM
should be read. For any other \code{level} $> 0$, the NORAM will be
one of the embedded position maps. Recall that each NORAM at level $i$
has its position map at level $i + 1$, with the exception of the very
last NORAM which uses a Trivial ORAM for its position map.  The
recursive call to \code{tree_rr_h} on line 6 reads out of the next
level's map, returning the pair \code{(r0, r1)}. These are the two
possible position tags for \code{nrorams[level]}---we should return
\code{r0} if \code{idx \% 2 = 0} and \code{r1} if \code{idx \% 2 =
  1}. The muxes on lines 7 and 8 obliviously achieve this, reading the
proper result into \code{tag}, replacing it with a freshly generated
tag, to satisfy the affinity requirement. Line 9 writes the updated
block \code{(r0', r1')} for \code{idx / 2} back, using an analogous
\code{tree_add_h} routine, for which a level can be
specified. Finally, line 10 reveals the retrieved position tag for
index \code{idx}, so that it can be passed to \code{noram_rr}. Since
level 0 corresponds to the actual data of the ORAM, that is what will
finally be returned to the client.



The \code{tree_add} routine is similar so we do not show it all. As
with \code{tree_rr} it recursively adds the corresponding bits of the
position tag into the array of \code{noram}{s}. At each level of the recursion
there is a snippet like the following:
\begin{lstlisting}[escapeinside={([}{])}]
let new_tag : R ([{{\color{\colorMATH}\(\vee \)}}]) R' rnd = rnd R ([{{\color{\colorMATH}\(\vee \)}}]) R' () in
let sec_tag = castS new_tag in (* does NOT consume new_tag *)
let (r0, r1) : (R ([{{\color{\colorMATH}\(\vee \)}}]) R' rnd) * (R ([{{\color{\colorMATH}\(\vee \)}}]) R' rnd) = tree_rr_h oram (idx / 2) (level + 1) in
let r0', tag = mux (idx % 2 = 0, new_tag, r0) in  (* replaces with new tag *)
let r1', tag = mux (idx % 2 = 1, tag, r1) in
let _ = tree_add_h oram (idx / 2) (level + 1) (r0', r1') in
noram_add norams[level] idx sec_tag data (* adds to Tree ORAM *)
\end{lstlisting}
Lines 1 and 2 generate a new tag, and make a secret copy of it. The
new tag is then stored in the recursive ORAM---lines 3--5 are similar
to \code{tree_add_h} but replace the found tag with \code{new_tag},
not some garbage value, at the appropriate level of the position map
(line 6). Finally, \code{sec_tag} is used to store the data in the
appropriate level of the \code{noram}.

We note that neither \code{tree_rr} nor \code{tree_add} are complete
ORAM operations on their own: to implement a full ORAM \code{read},
for example, we would need to call \code{tree_rr} with a call to
\code{tree_add}. 

\paragraph{Discussion}
Unfortunately (as astute readers may have noticed), the code snippet for
\code{add} will not type check. In particular, the \code{sec_tag}
argument has type \lstinline[mathescape]{R $\vee$ R' natS} but \code{noram_add} requires it to have
type \code{R natS}. This is because the position tags for the \code{noram} at
\code{level} are stored as the data of the \code{noram} at \code{level + 1},
and these are in different regions. We cannot put them in the same region because
we require a single \code{noram}'s metadata to have a strictly smaller region
than its data (i.e., \code{R} {{\color{\colorMATH}\(\sqsubset \)}} \code{R} {{\color{\colorMATH}\(\vee \)}} \code{R'}).

We can solve this problem by extending the language to support variant
and recursive types, existential quantification, and \emph{region
  polymorphism}, where region-polymorphic variables may have ordering
constraints. With these changes, the type of \code{oram} would be the
following:
 \begin{lstlisting}[escapeinside={([}{])},numbers=none]
type (R1,R2) block = { is_dummy : R1 bitS ; idx : R1 natS ; tag : R1 natS ; data: (R2 rnd) * (R2 rnd) } ([\text{where }]) R1 ([{{\color{\colorMATH}\(\sqsubset \)}}]) R2
type (R1,R2) bucket = (R1,R2) block array
type (R1,R2) noram  = (R1,R2) bucket array
 type (R1, R2) oram = 
  ([\hspace{0.25em}])Trivial of (R1,R2) bucket
 | Recursive of ([{{\color{\colorMATH}\(\exists \)}}])R. (R, R1) noram * (R1,R2) oram ([\text{where }]) R1 ([{{\color{\colorMATH}\(\sqsubset \)}}]) R2
\end{lstlisting}
We re-present the definitions for the elements of \code{noram}, which
we now parameterize with polymorphic region variables. For
\code{block}, we add the constraint that \code{R1} {{\color{\colorMATH}\(\sqsubset \)}} \code{R2}. When
originally presenting NORAM, this wasn't needed because we were using
concrete regions---notice that \code{R} and \code{R} {{\color{\colorMATH}\(\vee \)}} \code{R'}
from our previous \code{noram} definition satisfy the constraint on
\code{R1} and \code{R2}, respectively, in the new definition.
Type \code{oram} is also parameterized by region 
variables, and is now a recursive variant: it can be either a trivial
ORAM or a recursive ORAM. The latter is an NORAM paired with an ORAM,
which acts as its position map. Importantly, the region \code{R2} of
the ORAM data is properly ordered with the region of the position map
\code{R1}. The code would be roughly the same as the code given above,
except that rather than indexing the \code{norams} array at each
recursive \code{level}, it simply recurses down the \code{oram}
datastructure. Constructing such a datastructure would require
satisfying the region constraints at each level, which is easy to do
by simply using distinct regions for each region variable.  Along with
our other code examples at \url{https://github.com/plum-umd/oblivml}, we
show how this could work using OCaml-style functors.



\paragraph{Oblivious Stacks}
Other oblivious data structures~\cite{ods} can be
built in \lang, and on top of \code{noram} in particular.
\iftr
 Appendix~\ref{sec:ostack}
\else
The supplemental report
\fi
presents a development of probabilistic oblivious stacks
(\emph{ostacks}). As explained there, the strict ordering of probability
regions imposes a similar problem on ostacks as on recursive ORAMs,
but for ostacks the problem cannot be addressed with straightforward
language extensions. Instead, different reasoning 
principles are required. It's possible these can be integrated into
\lang via inclusion of a general-purpose logic.

\section{Related work}
\label{sec:related}


Lampson first pointed out various covert, or ``side,'' channels of information leakage
during a program's execution~\cite{confinementproblem}.
%
Defending against side-channel leakage is challenging. Previous works
have attempted to thwart such leakage
from various angles:
processor architectures that mitigate
leakage through timing~\cite{pret,rubylee}, power consumption~\cite{rubylee},
or memory-traces~\cite{phantom,ghostrider,ascend00,ascend01};
program analysis techniques that formally
ensure that a program
has bounded or no leakage through
instruction traces~\cite{pcsec},
timing channels~\cite{pcsec,agat00transforming,time-mitigation,sabelfeld06,secverilog},
or memory traces~\cite{csf13,scvm,ghostrider};
algorithmic techniques that
transform programs and algorithms
to their side-channel-mitigating or side-channel-free counterparts
while introducing only mild costs---e.g.,
works on mitigating timing channel leakage~\cite{askarov10predictive,barthe2010security,Zhang:2011:PMT:2046707.2046772},
and on preventing memory-trace
leakage~\cite{oram00,oram10,asiacrypt11,pathoram,circuitoram,ods,circuitstruct,blantongraph,oblivialg00,oblivialg01,Chan:2019:FDO:3310435.3310585}.
Often, the most effective and efficient
is through a comprehensive co-design approach combining these
areas of advances---in fact, several aforementioned works
indeed combine (a subset of) algorithms, architecture, and programming language
techniques~\cite{ghostrider,ascend00,ascend01,time-mitigation,secverilog}.

Our work belongs to a large category of work that aims to
statically enforce \emph{noninterference}, e.g., by
typing~\cite{Volpano:1996:STS:353629.353648,infoflow}.
\citeN{ghostrider,scvm,csf13} developed a type system that
ensures programs are MTO, generalizing a
line of prior works 
on (language-enforced) timing channel security~\cite{agat00transforming},
program counter security~\cite{pcsec}. In Liu et al's work, types are extended to
indicate where values are allocated; as per our above example data can
be public or secret, but can also reside in ORAM. Trace events are
extended to model ORAM accesses as opaque to the
adversary (similar to the Dolev-Yao modeling of encrypted
messages~\cite{Dolev:1981:SPK:1382435.1382728}): the adversary knows
that an access occurred, but not the address or whether it was a read
or a write. Liu et al's type
system enforces obliviousness of deterministic programs that use
(assumed-to-be-correct) ORAM. \lang's key
advance is that it applies to \emph{probabilistic} programs. It 
need not assume the existence of ORAM as a primitive; rather, \lang's
probabilistic nature is sufficient to allow us to
\emph{program} ORAM, per Section~\ref{sec:case-study}.
Thus we can express state-of-the-art algorithmic
results and formally reason about the security of their
implementations,
building a bridge between algorithmic and programming language
techniques. 


ObliVM~\cite{oblivm} is a language for programming probabilistically oblivious
algorithms intended to be run as secure multiparty
computations~\cite{yao}. Its type system also employs
affine types to ensure random numbers are used at most once. However,
it provides no mechanism to disallow constructing a non-uniformly
distributed random number. When such random numbers are generated,
they can be distinguished by an attacker from uniformly distributed
random numbers when being revealed. Therefore, the type system in
ObliVM does not guarantee obliviousness. \lang's use of
probability regions enforces that all random numbers are
uniformly random, and thus eliminates this channel of information
leakage. Moreover, we prove that this mechanism (and the others in
\lang) are sufficient to prove PMTO. 

Our
probabilistic memory trace obliviousness property bears some
resemblance to probabilistic notions of
noninterference. Much prior
work~\cite{sabelfeld2000probabilistic,smith2003probabilistic,russo2006securing,ngo2014effective}
is concerned with how random choices made by a thread scheduler could
cause the distribution of visible events to differ due to the values
of secrets. Here, the source of nondeterminism is the (external)
scheduler, rather than the program itself, as in our
case. \citeN{Smith:2006:SIF:1180337.1180341,Smith:2007:FPS:1255329.1255341}
consider how the influence of random numbers may affect the likelihood of certain outcomes,
mostly being concerned with termination channels. Their programming
model is not as rich as ours, as a secret random number is never
permitted to be made public; such an ability is the main source of
complexity in \lang, and is crucial for supporting oblivious
algorithms.


Some prior work aims to quantify the information released by a
(possibly randomized) program (e.g.,
\citeN{kopf2013automation,mu2009abstraction}) according to
entropy-based measures. Work on verifying the correctness of
differentially private
algorithms~\cite{DBLP:journals/toplas/BartheKOB13,Zhang:2017:LTA:3009837.3009884,DBLP:journals/corr/abs-1905-12594},
essentially aims to bound possible leakage; by contrast, we enforce
that \emph{no} information leaks due to a program's execution.












Our intensional distributions---while a novel syntactic device instrumental to
our proof approach---are readily interpretable as measurable sets over infinite
streams of bits, and there is prior work which has considered such models such
as Kozen's seminal treatment~\cite{Kozen:1979:SPP:1382433.1382610} among
others~\cite{huang-computable-distributions,park-sampling-functions,scibior-prob-monads,ramsey-stochastic-lambda,BARKER201647}.
A novelty in our model is support for conditional probabilistic
reasoning. This reasoning is enabled by our interpretation of monadic bind as
conditioning on outcomes, and performing sampling of new bits via operations
external to monad operations; doing so is in contrast to prior work which
interprets monadic bind directly as (effectively) sampling new random bits.

There is a rich history for {\mtextit{reasoning}} about probabilistic
programs~\cite{Sato:2019:FVH:3302515.3290351}, in particular relational
properties~\cite{hsu-thesis,Barthe:2014:PRV:2535838.2535847,Barthe:2017:CPP:3009837.3009896}
and program
logics~\cite{10.1007/978-3-319-89884-1_5,Rand:2015:VPH:2875516.2875608},
including trace properties~\cite{Smith:2019:TAM:3302515.3290352}, privacy
properties~\cite{barthe2015higher,reed2010distance,dfuzz}, obliviousness
properties~\cite{Ohrimenko:2016:OMM:3241094.3241143}, and uniformity and
independence~\cite{LPAR-21:Proving_uniformity_and_independence}. Much of this
work is focused on verification techniques for some program of
interest, and not on proof techniques for establishing metatheoric properties of
entire languages (e.g., via a type system).

Perhaps the most closely related program logic to our setting is
Probabilistic Separation Logic (PSL)~\cite{barthe20probsep}.  PSL is a
variant of separation logic in which separating conjunction models
probabilistic independence. It supports reasoning about (conditional)
independence and uniformity, which are both also key ideas in
\lang. There is a similar connection between some of PSL's proof rules
and \lang's type rules; e.g., \lang's Mux-Flip rule and PSL's RCond
rule both reason about conditional independence. It would be
interesting to explore how to embed \lang's type system in PSL's
logic, which might simplify reasoning about security for PSL, and open
up reasoning about correctness for \lang programs. It might also
permit proofs of uniformity that \lang's strict region ordering
currently forbid. How to combine these two is not obvious, though, as
PSL works on an imperative ``while'' language with a fixed set of
(global) variables, while \lang is functional, and supports
dynamically-sized data structures. Interesting future work!

\section{Conclusions}
\label{sec:conc}

This paper has presented \lang, a core language suitable for
expressing computations whose execution should be oblivious to a
powerful adversary who can observe an execution's trace of
instructions and memory accesses, but not see private
values. Unlike prior formalisms, \lang can be used to express
probabilistic algorithms whose security depends crucially on the use
of randomness. To do so, \lang tracks the use of randomly generated
numbers via a substructural (affine) type system, and employs a novel
concept called \emph{probability regions}. The latter are used to
track a random number's probabilistic (in)dependence on other random
numbers. We have proved that together these mechanisms ensure that a
random number's revelation in the visible trace does not perturb the
distribution of possible events so as to make secrets more likely. We
have demonstrated that \lang's type system is powerful enough to
accept sophisticated algorithms, including forms of oblivious
RAMs. To the best of our knowledge, by type checking an implementation of
tree-based ORAM in \lang we have carried out the first automated proof that
this algorithm is secure.

While \lang advances the state of the art in security type systems,
there are still oblivious algorithms it is not powerful enough to
check. As noted at the end of Section~\ref{sec:case-study}
\iftr
(and the appendix),
\else
(and the supplemental report),
\fi
the strict ordering on probability regions is sound but cannot handle
some idioms. More precise reasoning about probabilities is needed.
We believe that a promising way forward is to integrate \lang's
type-level mechanisms with richer systems for formal reasoning. For
example, we could adopt the approach of \emph{semantic typing},
embedding \lang's type rules as lemmas in a richer logic,
as done in RustBelt~\cite{Jung:2017:RSF:3177123.3158154} or
Fuzzi~\cite{DBLP:journals/corr/abs-1905-12594}. The
logic of~\citet{barthe20probsep} is a good candidate, but it needs
further extensions too. Another benefit of
embedding \lang's type system into a full logic is that we can use the
logic to reason about algorithm correctness, something \lang
does not do.

\begin{acks}                            
  We thank Aseem Rastogi, Kesha Heitala, Joe Near, and the anonymous reviewers for comments on earlier
  drafts of this paper, and Elaine Shi for helpful discussions throughout our process.
  This material is based upon work supported by the
  \grantsponsor{GS100000001}{National Science
    Foundation}{http://dx.doi.org/10.13039/100000001} under Grant
  Nos.~\grantnum{GS100000001}{CNS-1563722},
  \grantnum{GS100000001}{CNS-1314857}, 
  \grantnum{GS100000001}{CNS-1111599} and \grantnum{GS100000001}{CCF-1901278}; by
  \grantsponsor{GS100000002}{DARPA}{} under contracts
  \grantnum{GS100000002}{FA8750-15-2-0104} and
  \grantnum{GS100000002}{FA8750-16-C-0022}; and by ODNI/IARPA via 2019-1902070008.
  Any opinions, findings,
  conclusions or recommendations expressed in this material are those
  of the author and do not necessarily reflect the views of the
  NSF, DARPA, ODNI, IARPA, or the U.S. Government.
\end{acks}

\nocite{infoflow,ods,scvm}
\bibliography{refs,crypto,ref}

\iftr

\clearpage

\appendix
\section{Case Study: Oblivious Stacks}
\label{sec:ostack}

This section considers implementing an oblivious data structure, an
\emph{oblivious stack}, in \lang, building on top of the non-recursive
ORAM presented in Section~\ref{sec:oram-overview}.  The type system is
not powerful enough to completely check the algorithm, however; we
explain why and suggest solutions.

\subsection{Algorithm}

The goal of an oblivious stack is to hide both its data and
which operations (pushes or pops) are taking place---only the total
number of operations should be revealed. 
To do this, we could implement the stack using an ORAM rather than a
normal (encrypted) array, and we could merge the code for push and pop
so as to mask which operation is taking place (despite knowledge of
the PC). Code to do this is shown in the \code{stackop} function in
Figure~\ref{fig:det-ostack}.\dcd{TODO: from reviews: make it more explicit what
typechecks and what doesn't up front in this section.}

\begin{figure}
\prefigskip
\centering
\begin{boxedminipage}{0.7\textwidth}
\begin{lstlisting}[escapeinside={([}{])}]
 type block  = { is_dummy : R bit ; idx : R natS ; tag : R natS ; data : natS * (R ([{{\color{\colorMATH}\(\vee \)}}]) R' rnd) }
 type bucket = block array
 type noram  = bucket array
 type oram   = (noram array) * bucket
 type stack  = oram * natS ref

 let stackop ((oram, rid_r) : stack) (ispush : bitS) (d : natS) : natS =
   let rid     = !rid_r in
   let old_d   = tree_rr oram rid in
   let (d', _) = mux(ispush, d, old_d) in
   let (id, _) = mux(ispush, rid + 1, rid) in
   tree_add oram id d';
   let (rid', _) = mux(ispush, rid + 1, rid - 1) in
   rid_r := rid'; d'
\end{lstlisting}
\end{boxedminipage}
\precaptionskip
\caption{A deterministic oblivious stack built using a full (recursive) ORAM.}
\postcaptionskip
\label{fig:det-ostack}
\end{figure}

In the code, a stack consists of an ORAM of secret numbers and a
reference storing the index of the root. Function \code{stackop} takes a
stack, a flag indicating whether the operation is a push or pop, and
the value to push, and returns a value. The code reads the value
at the root index (line 9). The next line copies that value to
\code{d'} if the operation is pop, or else puts \code{d} there
if it is a push. Line 11 determines the index of the write it will
perform on line 12: this index (\code{id}) is one more than the root
index if it's a push; it's the current root index if not. As such,
the write on line 12 puts the given value in the next slot in case of a
push, or writes back the value at the current root, if it's a
pop. Finally, line 13 adjusts the root index, and line 14 returns the
result, which is either the popped value or pushed value (if it was a
push).

\begin{figure}
\prefigskip
\centering
\begin{boxedminipage}{0.75\textwidth}
\begin{lstlisting}[escapeinside={([}{])}]
 type block   = { is_dummy : R bit ; idx : R natS ; tag : R natS ; data : natS * (R ([{{\color{\colorMATH}\(\vee \)}}]) R' rnd) }
 type bucket  = block array
 type noram   = bucket array
 type ostack  = noram * natS ref * (R ([{{\color{\colorMATH}\(\vee \)}}]) R' rnd) ref

 let stackop ((noram, rid_r ,pos_r) : ostack) (ispush : bitS) (d : natS) : natS =
 let (rid, pos) = !rid_r, !pos_r in
 let (rid', pos', d') =
   if ispush then
     let (d', _) = noram_rr noram (-1) (castP (rnd (R ([{{\color{\colorMATH}\(\vee \)}}]) R') ())) in
     let b = (d, pos) in
     let pos' = rnd (R ([{{\color{\colorMATH}\(\vee \)}}]) R') () in
     let _ = noram_add noram (rid + 1) (castS pos') b in
     (rid + 1, pos', d')
   else
     let (d', pos') = noram_rr noram rid (castP pos) in
     let b = (d, rnd (R ([{{\color{\colorMATH}\(\vee \)}}]) R') ()) in
     let _ = noram_add noram (-1) 0 b in
     (rid - 1, pos', d') in
 rid_r := rid';
 rpos_r := pos';
 d'
\end{lstlisting}
\end{boxedminipage}
\precaptionskip
\caption{A probabilistic oblivious stack built using a non-recursive ORAM. (Does not use \code{mux}, for simplicity.)}
\postcaptionskip
\label{fig:prob-ostack}
\end{figure}

\begin{figure}
\prefigskip
\centering
\begin{boxedminipage}{0.75\textwidth}
\begin{lstlisting}[escapeinside={([}{])}]
 let stackop ((noram, rid_r, pos_r) : ostack) (ispush : bitS) (d : natS) : natS =
 let (rid, pos) = !rid_r, !pos_r in
 let (rid', pos', d') =
   let (id, new_rid) = mux(ispush, -1, rid + 1) in
   let (to_cast_p, tmp) = mux(ispush, rnd (R ([{{\color{\colorMATH}\(\vee \)}}]) R') (), pos) in
   let (d', pos') = noram_rr noram id (castP to_cast_p) in
   let (pos', _) = mux(ispush, rnd (R ([{{\color{\colorMATH}\(\vee \)}}]) R') (), pos')
   let b = (d, tmp) in
   let (pos_S, _) = mux(ispush, pos', 0)
   let _ = noram_add noram new_rid (castS pos_S) b in
   let (ret_rid, _) = mux(ispush, rid_r - 1, rid_r + 1) in
   (ret_id, pos', d') in rid_r := rid';
 rpos_r := pos';
 d'
\end{lstlisting}
\end{boxedminipage}
\precaptionskip
\caption{A probabilistic oblivious stack built using a non-recursive ORAM using \code{mux}.}
\postcaptionskip
\label{fig:prob-ostack-mux}
\end{figure}

While this code works perfectly well, a \emph{probabilistic} version of the stack, using a
\emph{non-recursive ORAM} would be more space-efficient. In
particular, it will require only $O(L)$ extra space where $L$ is the current
size of the stack, whereas this version requires $O(N)$ extra
space, where $N$ is the size of the ORAM.
To see how, consider that we always access a stack via its head,
using the root index. Thus, in the code in
Figure~\ref{fig:det-ostack}, the non-recursive \texttt{ORAM} internally only
ever uses one slot in its position map. Thus we can do better by using
an NORAM directly, having the stack manage the position tag of the
root. In short, we implement an oblivious stack as a triple comprising
a NORAM, the index of the root element, and its position tag. The latter
two act as a kind of pointer into the NORAM. Each block stored in the
NORAM contains the data and the position tag of the next block in the
stack.

Code implementing the stack following this design is given in
Figure~\ref{fig:prob-ostack}. Note the code branches on the
\code{ispush} variable to make it easier to read; the actual
implementation must use \texttt{mux}s to conditionally execute each
statement in both branches to ensure obliviousness.\footnote{Notice
  that the structure of both branches is roughly parallel, which makes
  converting to the use of \code{mux}es straightforward.}
Line 7 extracts the current root index and position tag. Lines 10--14
handle a push operation. Line 10 first does a ``dummy read'' from the
NORAM; just as we saw with the trivial ORAM \code{add} earlier,
using index \code{-1} results in a dummy
block being returned (the position tag argument is unimportant in this
case). Line 11 constructs a new block \code{b} to push: it consists of
the given data \code{d} paired with the current root's position tag
\code{pos}, thus creating a ``pointer'' to that block. We then
generate a fresh position tag \code{pos'} for this (the new root's)
block, add the block to the \code{noram}. The coercion \code{castP} ascribes
a random number the type \code{natP} (per line 10), while
\code{castS} gives it type \code{natS} (line 13).
The new root index (the old
one plus one), the root's tag, and the dummy block passed in are
returned on line 14. Lines 16--19 handle a pop. Here, the first \code{rr} does
real work, extracting the block that corresponds to the root index and
position tag. We then generate a dummy block to ``add'' to the
ORAM. The updated root index (the old one minus one), its
position tag (returned by the \code{rr}) and the fetched block are
returned. The full mux version is provided in Figure~\ref{fig:prob-ostack-mux}.

This version of an oblivious stack performs better than the
version from Figure~\ref{fig:det-ostack}. The space overhead, due to
the added pointers at the root and within the ORAM, is $O(L)$ where
$L$ is the size of the current stack, not the size $N$ of the
ORAM. The running time is still $O(log\, N)$. Obliviousness is a
direct corollary of implementing our stack on \lang: Because we have
labeled the stack's contents and root as secret, as well as the choice
of operation, nothing can be learned about any of them when observing
the event trace.


\subsection{The Limits of Syntactic Uniformity Enforcement}

Unfortunately, we cannot directly typecheck an implementation of oblivious stacks.
To see why, consider the type of \code{block} of Section~\ref{sec:oram-overview}. In this
type we assert that the position \code{tag} field is in region \code{R} and that
this region is strictly less than \lstinline[mathescape]{R $\vee$ R'}, the region associated with random values stored
inside the NORAM. However, in the code for oblivious stacks, we are storing random numbers in region \lstinline[mathescape]{R $\vee$ R'} that we will later use as the position tag for subsequent operations (which has type \code{R}).
Suppose we change to the type for blocks thus:

\begin{lstlisting}[escapeinside={([}{])},numbers=none]
type block = { is_dummy : R bitS ; idx : R natS ; tag : R ([{{\color{\colorMATH}\(\vee \)}}]) R' natS ; data : (rnd R ([{{\color{\colorMATH}\(\vee \)}}]) R' * rnd R ([{{\color{\colorMATH}\(\vee \)}}]) R') }
\end{lstlisting}

Here, we place the position \code{tag} in region \lstinline[mathescape]{R $\vee$ R'} instead. This almost works---since
the tag argument is public in \code{noram_rr} and \code{noram_add} operations only mux on the index (in the trivial
write to the root bucket). However, the eviction procedure will not typecheck with the position
tag at region \lstinline[mathescape]{R $\vee$ R'}. This is because the eviction procedure performs a \code{mux} on the secret
position tag (in \lstinline[mathescape]{R $\vee$ R'}) to decide where to evict a block (with the data component also in \code{R'}).
Thus, the strict ordering requirement ({{\color{\colorMATH}\(R \vee  R' \sqsubset  R \vee  R'\)}}) for the \code{mux} type rule is not met. The fundamental issue is that we are
storing position tags in the ORAM, so the region associated with position tags \emph{of} the ORAM are the
same as the regions of the data \emph{in} the ORAM.

Our type system rejects the \code{mux} in the eviction procedure (when position tags are typed at region \lstinline[mathescape]{R $\vee$ R'})
because it \emph{does} yield random values which are not uniformly distributed. However, this violation is
actually a false positive. By the time these random values are revealed to the adversary, their uniformity
is re-established. This is obviously the case, because if the ORAM truly implements a map, the result of reading
from the ORAM on line 16 (Figure~\ref{fig:prob-ostack}) of \code{stackop} will yield the same value that was placed into it. This value was a fresh
random value, which is uniformly distributed. The following simple, pathological case illustrates the issue:

\begin{lstlisting}[escapeinside={([}{])}]
let s = I S in (* A secret true bit *)
let (r0, r1) = (flip R0 (), flip R1 ()) in
let r0_s = castS r0 in
let g = s && r0_s in
let (v1, v2) = mux (g, r0, r1) in
let (v, _) = mux (g, v1, v2) in
castP v
\end{lstlisting}

This example flips two coins and uses them as the arguments to the \code{mux} on line 5. Since the guard
of this \code{mux} depends on the value of \code{r0} the resulting values \code{v1, v2} are not uniformly distributed.
Indeed, the type system would reject the \code{mux} on line 5. However, the \code{mux} on line 6 yields another value
\code{v} which is again uniformly distributed and thus safe to reveal. The takeaway here is that a sequence of appropriate
\code{mux}es can temporarily perturb and then re-establish the uniformity of a random value before revealing it to the adversary.
Since our type system forces random values to be uniform \emph{everywhere}, we cannot typecheck instances like this.

We can easily handle this case with the use of unsafe casts---\code{castNU} and \code{castU} for cast ``non-uniform''
and cast ``uniform'' respectively. This allows a value to be labeled as intentionally non-uniform. While a value is marked as non-uniform,
the \code{mux} operations over these values will not be checked for strict ordering. However, it is the programmer's responsibility to ensure
at the point that it is casted back into a random value (using \code{castU}) that it is truly a random value. For example, we could patch
the code above as follows:

\begin{lstlisting}[escapeinside={([}{])}]
let s = I S in (* A secret true bit *)
let (r0, r1) = (flip R0 (), flip R1 ()) in
let r0_s = castS r0 in
let g = s && r0_s in
let (v1, v2) = mux (g, (castNU r0), (castNU r1)) in
let (v, _) = mux (g, v1, v2) in
castP (castU v)
\end{lstlisting}

It is important to note that, if casts are used correctly, then PMTO is preserved. In other words, if all instances of \code{castU} are used
on values which are truly uniformly distributed then the type system ensures that the program is PMTO. These uniformity obligations can be verified
manually, or using external tools~\cite{barthe2018probproglogic}.

Using this simple extension we can modify the \code{stackop} procedure in Figure~\ref{fig:prob-ostack-mux} by inserting a single \code{castU} on \code{pos'} after line 6,
and a \code{castNU} on \code{tmp} after line 7. This has the effect of casting all position tags pushed on the OStack to
``non-uniform'' type, and all position tags popped off the stack to ``uniform'' type again. The safety of these casts relies
on the functional correctness of the non-recursive ORAM. As mentioned above, if the nroram faithfully implements
a map, then we can expect to receive a uniformly distributed value after every pop (since every push generates a fresh random number).



\setlistdepth{9}

\renewlist{enumerate}{enumerate}{9}
\renewlist{itemize}{itemize}{9}

\setlist[enumerate,1]{label={(\arabic*)}} 
\setlist[enumerate,2]{label={(\alph*)}}   
\setlist[enumerate,3]{label={(\roman*)}}  

\raggedbottom

\section{Complete PMTO Proof}
\label{sec:proofs}

In this section we give a complete proof of PMTO. First, in
Section~\ref{sec:proofs:pmto-proof} we present the final proof of PMTO in
top-down breadth-first organization for major lemmas, and depth-first
organization for sublemmas required to prove major lemmas. In many proofs we
abbreviate ``suffices to show'' as ``STS''. Next, in
Section~\ref{sec:proofs:definitions} we show complete definitions for all
semantics, type rules, auxiliary metafunctions, and low-equivalence relations
which are used in the proof. 

The heart of the type system design is typing for flip values:
\begingroup\color{\colorMATH}\begin{mathpar} \fbox{{{\color{\colorMATH}\(
   \inferrule*[vcenter,lab={\mtextsc{ Flip-Value}}
   ]{ {\mtext{Pr}}\left[\hat b \mathrel{\dot =} {\mfootnotesize{{{\color{\colorSYNTAX}\mtexttt{I}}}}} \mathrel{}\middle|\mathrel{} \Phi  \right] = \nicefrac{1}{2} 
   \\ \left[ \hat b \mathrel{\bot \!\!\!\bot } \Psi ^{F},\Psi ^{B}(\{ \rho ^{\prime} \mathrel{|} \rho ^{\prime} \sqsubset  \rho \} ) \mathrel{}\middle|\mathrel{} \Phi  \right]
      }{
      \Psi ^{F},\Psi ^{B},\Phi  \vdash  \hat b \mathrel{:} {\mfootnotesize{{{\color{\colorSYNTAX}\mtexttt{flip}}}}}^{\rho }
   }
   \)}}}
\end{mathpar}\endgroup
This invariant dictates that (1) the distribution is uniform, and (2) that it
is jointly independent of all other flip values in the execution context
{{\color{\colorMATH}\(\Psi ^{F}\)}}, and all other secret bit values in the execution context at strictly
lower region {{\color{\colorMATH}\(\Psi ^{B}\)}}. Joint independence is crucial and strictly stronger than
individual independence; to see this, note that {{\color{\colorMATH}\(A \mathrel{\bot \!\!\!\bot } B\)}} and {{\color{\colorMATH}\(A \mathrel{\bot \!\!\!\bot } C\)}} does {\mtextit{not}}
imply {{\color{\colorMATH}\(A \mathrel{\bot \!\!\!\bot } B,C\)}}, however the converse is true.

The heart of the proof is \nameref{thm:proofs:type-preservation}, and its main
sublemma \nameref{thm:proofs:type-preservation-redex}. The key semantic
property of mux operations used in those lemmas is
\nameref{thm:proofs:cond-stability}.

\subsection{Theorems and Lemmas}\label{sec:proofs:pmto-proof}

The main metatheory result for \lang is PMTO. The proof follows from major
sublemmas.

\noindent
\begin{minipage}{\linewidth}
\begin{theorem}[PMTO]\label{thm:proofs:pmto}\ 
  \begin{itemize}[label={},leftmargin=0pt]\item  {\mtextit{Probabilistic equality modulo adversary observability for source expressions is preserved by the ground truth semantics.}}
  \item  If: \hspace*{0.16em}{{\color{\colorMATH}\(e_{1}\)}}\hspace*{0.16em} and \hspace*{0.16em}{{\color{\colorMATH}\(e_{2}\)}}\hspace*{0.16em} are closed source expressions
  \item  And: \hspace*{0.16em}{{\color{\colorMATH}\(\vdash  e_{1} \mathrel{:} \tau \)}}\hspace*{0.16em} and \hspace*{0.16em}{{\color{\colorMATH}\(\vdash  e_{2} \mathrel{:} \tau \)}}
  \item  And: \hspace*{0.16em}{{\color{\colorMATH}\({\mtext{obs}}(e_{1}) = {\mtext{obs}}(e_{2})\)}}
  \item  Then: 
     \begin{enumerate}[leftmargin=15pt]\item  {{\color{\colorMATH}\({\mtext{nstep}}_{{\mathcal{D}}}(N,\varnothing ,e_{1})\)}}\hspace*{0.16em} and \hspace*{0.16em}{{\color{\colorMATH}\({\mtext{nstep}}_{{\mathcal{D}}}(N,\varnothing ,e_{2})\)}}\hspace*{0.16em} are defined
     \item  {{\color{\colorMATH}\(\widetilde {\mtext{obs}}({\mtext{nstep}}_{{\mathcal{D}}}(N,\varnothing ,e_{1})) = \widetilde {\mtext{obs}}({\mtext{nstep}}_{{\mathcal{D}}}(N,\varnothing ,e_{2}))\)}}
     \end{enumerate}
  \end{itemize}
\end{theorem}
\end{minipage}
\begin{proof}\ 
  \begin{itemize}[label={},leftmargin=0pt]\item  (1) is by \nameref{thm:proofs:progress-ground-truth}
  \item  (2) is by the following:
  \item  {{\color{\colorMATH}\(\begin{array}[t]{rcl
      }   &{}   {}& {\mtext{obs}}(e_{1}) = {\mtext{obs}}(e_{2})
      \cr    &{}\implies   {}& \lbag {{\color{\colorTEXT}\textnormal{\hspace*{0.33em} \nameref{thm:proofs:low-equivalence-completeness-source-expressions} \hspace*{0.33em}}}}\rbag 
      \cr    &{}   {}& e_{1} \sim  e_{2}
      \cr    &{}\implies   {}& \lbag {{\color{\colorTEXT}\textnormal{\hspace*{0.33em} \nameref{thm:proofs:pmto-mixed} \hspace*{0.33em}}}}\rbag 
      \cr    &{}   {}& {\mtext{\underline {nste\hspace{-1pt}}\hspace{1pt}p}}(N,\varnothing ,e_{1}) \approx _{\sim } {\mtext{\underline {nste\hspace{-1pt}}\hspace{1pt}p}}(N,\varnothing ,e_{2})
      \cr    &{}\implies   {}& \lbag {{\color{\colorTEXT}\textnormal{\hspace*{0.33em} \nameref{thm:proofs:low-equivalence-soundness} \hspace*{0.33em}}}}\rbag 
      \cr    &{}   {}& \widehat {\mtext{obs}}(\hat \lceil {\mtext{\underline {nste\hspace{-1pt}}\hspace{1pt}p}}(N,\varnothing ,e_{1})\hat \rceil ) \approx _{=} \widehat {\mtext{obs}}(\hat \lceil {\mtext{\underline {nste\hspace{-1pt}}\hspace{1pt}p}}(N,\varnothing ,e_{2})\hat \rceil )
      \cr    &{}\implies   {}& \lbag {{\color{\colorTEXT}\textnormal{\hspace*{0.33em} \nameref{thm:proofs:simulation-mixed} \hspace*{0.33em}}}}\rbag 
      \cr    &{}   {}& \widehat {\mtext{obs}}({\mtext{nstep}}_{{\mathcal{I}}}(N,\varnothing ,e_{1})) \approx _{=} \widehat {\mtext{obs}}({\mtext{nstep}}_{{\mathcal{I}}}(N,\varnothing ,e_{2}))
      \cr    &{}\implies   {}& \lbag {{\color{\colorTEXT}\textnormal{\hspace*{0.33em} \nameref{thm:proofs:simulation-intensional} \hspace*{0.33em}}}}\rbag 
      \cr    &{}   {}& \widetilde {\mtext{obs}}({\mtext{nstep}}_{{\mathcal{D}}}(N,\varnothing ,e_{1})) = \widetilde {\mtext{obs}}({\mtext{nstep}}_{{\mathcal{D}}}(N,\varnothing ,e_{2}))
      \end{array}\)}}
  \end{itemize}
\end{proof}

\subsubsection{PMTO Proof Key Lemmas}\ \\

\paragraph{\bf Progress (Ground Truth)}\ \\

\noindent
\begin{minipage}{\linewidth}
\begin{lemma}[Progress (Ground Truth)]\label{thm:proofs:progress-ground-truth}\ 
  \begin{itemize}[label={},leftmargin=0pt]\item  {\mtextit{Progress holds for the ground truth semantics.}}
  \item  If: \hspace*{0.16em}{{\color{\colorMATH}\(\vdash  \varsigma \)}}
  \item  Then: \hspace*{0.16em}{{\color{\colorMATH}\({\mtext{nstep}}_{{\mathcal{D}}}(N,\varsigma )\)}}\hspace*{0.16em} is total
  \end{itemize}
\end{lemma}
\end{minipage}
\begin{proof}
  Induction on {{\color{\colorMATH}\(N\)}} and \nameref{thm:proofs:progress-ground-truth-single}
\end{proof}

\noindent
\begin{minipage}{\linewidth}
\begin{lemma}[Progress (Ground Truth) Single]\label{thm:proofs:progress-ground-truth-single}\ 
  \begin{itemize}[label={},leftmargin=0pt]\item  {\mtextit{Progress holds for the ground truth semantics on a single step.}}
  \item  If: \hspace*{0.16em}{{\color{\colorMATH}\(\Sigma  \vdash  \sigma ,e\)}}
  \item  Then either:
     \begin{enumerate}[leftmargin=15pt]\item  {{\color{\colorMATH}\(e = v\)}}\hspace*{0.16em} for \hspace*{0.16em}{{\color{\colorMATH}\(v\)}}\hspace*{0.16em} a value
     \item  {{\color{\colorMATH}\(e = E[e^{\prime}]\)}}\hspace*{0.16em} and \hspace*{0.16em}{{\color{\colorMATH}\(e^{\prime}\)}}\hspace*{0.16em} a redex
     \end{enumerate}
     In both cases \hspace*{0.16em}{{\color{\colorMATH}\({\mtext{step}}_{{\mathcal{D}}}(N,\sigma ,e)\)}}\hspace*{0.16em} is total
  \end{itemize}
\end{lemma}
\end{minipage}
\begin{proof}
  Induction on {{\color{\colorMATH}\(e\)}} and inversion on assumed well-typing
\end{proof}

\paragraph{\bf Low-equivalence Completeness}\ \\

\noindent
\begin{minipage}{\linewidth}
\begin{lemma}[Low-equivalence Completeness (Source Expressions)]\label{thm:proofs:low-equivalence-completeness-source-expressions}\ 
  \begin{itemize}[label={},leftmargin=0pt]\item  {\mtextit{Source expressions which are equal modulo adversary observation are low-equivalent.}}
  \item  If: \hspace*{0.16em}{{\color{\colorMATH}\(e_{1}\)}}\hspace*{0.16em} and \hspace*{0.16em}{{\color{\colorMATH}\(e_{2}\)}}\hspace*{0.16em} are source expressions
  \item  And: \hspace*{0.16em}{{\color{\colorMATH}\({\mtext{obs}}(e_{1}) = {\mtext{obs}}(e_{2})\)}}
  \item  Then: \hspace*{0.16em}{{\color{\colorMATH}\(\lfloor e_{1}\rfloor  \sim  \lfloor e_{2}\rfloor \)}}
  \end{itemize}
\end{lemma}
\end{minipage}
\begin{proof}
  Induction on {{\color{\colorMATH}\(e_{1}\)}} and {{\color{\colorMATH}\(e_{2}\)}}, and discrimination on assumed {{\color{\colorMATH}\({\mtext{obs}}(e_{1}) = {\mtext{obs}}(e_{2})\)}}
\end{proof}

\paragraph{\bf PMTO (Mixed)}\ \\

\noindent
\begin{minipage}{\linewidth}
\begin{lemma}[PMTO (Mixed)]\label{thm:proofs:pmto-mixed}\ 
  \begin{itemize}[label={},leftmargin=0pt]\item  {\mtextit{Probabilistic low-equivalence for source expressions is preserved by the mixed semantics.}}
  \item  If: \hspace*{0.16em}{{\color{\colorMATH}\(e_{1}\)}}\hspace*{0.16em} and \hspace*{0.16em}{{\color{\colorMATH}\(e_{2}\)}}\hspace*{0.16em} are closed source expressions
  \item  And: \hspace*{0.16em}{{\color{\colorMATH}\(\vdash  e_{1} \mathrel{:} \tau \)}}\hspace*{0.16em} and \hspace*{0.16em}{{\color{\colorMATH}\(\vdash  e_{2} \mathrel{:} \tau \)}}
  \item  And: \hspace*{0.16em}{{\color{\colorMATH}\(e_{1} \sim  e_{2}\)}}
  \item  Then: 
     \begin{enumerate}[leftmargin=15pt]\item  {{\color{\colorMATH}\({\mtext{\underline {nste\hspace{-1pt}}\hspace{1pt}p}}(N,\varnothing ,e_{1})\)}}\hspace*{0.16em} and \hspace*{0.16em}{{\color{\colorMATH}\({\mtext{\underline {nste\hspace{-1pt}}\hspace{1pt}p}}(N,\varnothing ,e_{2})\)}}\hspace*{0.16em} are defined
     \item  {{\color{\colorMATH}\({\mtext{\underline {nste\hspace{-1pt}}\hspace{1pt}p}}(N,\varnothing ,e_{1}) \approx _{\sim } {\mtext{\underline {nste\hspace{-1pt}}\hspace{1pt}p}}(N,\varnothing ,e_{2})\)}}
     \end{enumerate}
  \end{itemize}
\end{lemma}
\end{minipage}
\begin{proof}\ 
  \begin{itemize}[label={},leftmargin=0pt]\item  (1) is by \nameref{thm:proofs:progress-mixed}
  \item  (2) is by induction on {{\color{\colorMATH}\(N\)}}
  \item  \begin{itemize}[label=\textbf{-},leftmargin=*]\item  \begin{itemize}[label={},leftmargin=0pt]\item  Case {{\color{\colorMATH}\(N=0\)}}:
        \item  STS: {{\color{\colorMATH}\({\mtext{return}}(e_{1}) \approx _{\sim } {\mtext{return}}(e_{2})\)}}
        \item  By \nameref{thm:proofs:return-equivalence}
        \end{itemize}
     \item  \begin{itemize}[label={},leftmargin=0pt]\item  Case {{\color{\colorMATH}\(N=N+1\)}}:
        \item  {{\color{\colorMATH}\({\mtext{\underline {nste\hspace{-1pt}}\hspace{1pt}p}}(N,\varnothing ,e_{1}) \approx _{\sim } {\mtext{\underline {nste\hspace{-1pt}}\hspace{1pt}p}}(N,\varnothing ,e_{2})\)}} {\mtextit{(IH)}} (by inductive hypothesis)
        \item  STS:
        \item  {{\color{\colorMATH}\(\begin{array}{l
            } {\mtext{do}}\hspace*{0.33em}\begin{array}[t]{l
                    } \underline t\mathord{\cdotp }\underline \varsigma  \leftarrow  {\mtext{\underline {nste\hspace{-1pt}}\hspace{1pt}p}}(N,\varnothing ,e_{1})
                    \cr  \underline \varsigma ^{\prime} \leftarrow  {\mtext{\underline {ste\hspace{-1pt}}\hspace{1pt}p}}(N+1,\underline \varsigma )
                    \cr  {\mtext{return}}(\underline t\mathord{\cdotp }\underline \varsigma ,\underline \varsigma ^{\prime})
                    \end{array}
            \cr  \approx _{\sim }
            \cr  {\mtext{do}}\hspace*{0.33em}\begin{array}[t]{l
                    } \underline t\mathord{\cdotp }\underline \varsigma  \leftarrow  {\mtext{\underline {nste\hspace{-1pt}}\hspace{1pt}p}}(N,\varnothing ,e_{1})
                    \cr  \underline \varsigma ^{\prime} \leftarrow  {\mtext{\underline {ste\hspace{-1pt}}\hspace{1pt}p}}(N+1,\underline \varsigma )
                    \cr  {\mtext{return}}(\underline t\mathord{\cdotp }\underline \varsigma ,\underline \varsigma ^{\prime})
                    \end{array}
            \end{array}\)}}
        \item  By \nameref{thm:proofs:bind-equivalence}, \nameref{thm:proofs:return-equivalence} and {\mtextit{(IH)}}, STS:
        \item  \begin{itemize}[label=\textbf{-},leftmargin=*]\item  \begin{itemize}[label={},leftmargin=0pt]\item  {{\color{\colorMATH}\(\underline t_{1}\mathord{\cdotp }\underline \varsigma _{1} \sim  \underline t_{2}\mathord{\cdotp }\underline \varsigma _{2} \implies   \left[ {\mtext{\underline {ste\hspace{-1pt}}\hspace{1pt}p}}(N+1,\underline \varsigma _{1}) \mathrel{}\middle|\mathrel{} \Phi _{1}\right] \approx _{\sim } \left[{\mtext{\underline {ste\hspace{-1pt}}\hspace{1pt}p}}(N+1,\underline \varsigma _{2}) \mathrel{}\middle|\mathrel{} \Phi _{2}\right]\)}}
              \item  where {{\color{\colorMATH}\(\Phi _{1} \triangleq  [{\mtext{\underline {nste\hspace{-1pt}}\hspace{1pt}p}}(N,\varnothing ,e_{1}) \mathrel{\dot =} \underline t_{1}\mathord{\cdotp }\underline \varsigma _{1}]\)}} and {{\color{\colorMATH}\(\Phi _{2} \triangleq  [{\mtext{\underline {nste\hspace{-1pt}}\hspace{1pt}p}}(N,\varnothing ,e_{2}) \mathrel{\dot =} \underline t_{2}\mathord{\cdotp }\underline \varsigma _{2}]\)}}
              \end{itemize}
           \end{itemize}
        \item  By \nameref{thm:proofs:type-preservation}:
        \item  \begin{itemize}[label=\textbf{-},leftmargin=*]\item  \begin{itemize}[label={},leftmargin=0pt]\item  There exists {{\color{\colorMATH}\(\Sigma _{1}\)}}, {{\color{\colorMATH}\(\Sigma _{2}\)}}, {{\color{\colorMATH}\(\Psi _{1}\)}} and {{\color{\colorMATH}\(\Psi _{2}\)}} 
              \item  {S.t.} {{\color{\colorMATH}\(\Phi _{1},\Sigma _{1} \vdash  \underline \varsigma _{1} \mathrel{;} \Psi _{1}\)}} and {{\color{\colorMATH}\(\Phi _{2},\Sigma _{2} \vdash  \underline \varsigma _{2} \mathrel{;} \Psi _{2}\)}}
              \end{itemize}
           \end{itemize}
        \item  Conclusion is by \nameref{thm:proofs:pmto-mixed-single} applied to premise and the above well-typing
        \end{itemize}
     \end{itemize}
  \end{itemize}
\end{proof}

\noindent
\begin{minipage}{\linewidth}
  \begin{lemma}[PMTO (Mixed) Single]
\label{thm:proofs:pmto-mixed-single}\ 
  \begin{itemize}[label={},leftmargin=0pt]\item  {\mtextit{Probabilistic low-equivalence for source expressions is preserved by the mixed semantics on a single step.}}
  \item  If: \hspace*{0.16em}{{\color{\colorMATH}\(\Phi _{1},\Sigma _{1} \vdash  \underline \varsigma _{1} \mathrel{;} \Psi _{1}\)}}\hspace*{0.16em} and \hspace*{0.16em}{{\color{\colorMATH}\(\Phi _{2},\Sigma _{2} \vdash  \underline \varsigma _{2} \mathrel{;} \Psi _{2}\)}}
  \item  And: \hspace*{0.16em}{{\color{\colorMATH}\(\underline \varsigma _{1} \sim  \underline \varsigma _{2}\)}}
  \item  Then: \hspace*{0.16em}{{\color{\colorMATH}\(\left[ {\mtext{\underline {ste\hspace{-1pt}}\hspace{1pt}p}}(N,\underline \varsigma _{1}) \mathrel{}\middle|\mathrel{} \Phi _{1} \right] \approx _{\sim } \left[{\mtext{\underline {ste\hspace{-1pt}}\hspace{1pt}p}}(N,\underline \varsigma _{2}) \mathrel{}\middle|\mathrel{} \Phi _{2} \right]\)}}
  \end{itemize}
\end{lemma}
\end{minipage}
\begin{proof}
  \begin{itemize}[label={},leftmargin=0pt]\item  By case analysis on {{\color{\colorMATH}\(\underline \varsigma _{1} \sim  \underline \varsigma _{2}\)}} and \nameref{thm:proofs:progress-mixed}; two cases:
  \item  \begin{enumerate}[leftmargin=15pt]\item  \begin{itemize}[label={},leftmargin=0pt]\item  Case {{\color{\colorMATH}\(\underline \varsigma _{1} = \underline \sigma _{1},\underline v_{1}\)}} and {{\color{\colorMATH}\(\underline \varsigma _{2} = \underline \sigma _{2},\underline v_{2}\)}} for {{\color{\colorMATH}\(\underline v_{1}\)}} and {{\color{\colorMATH}\(\underline v_{2}\)}} values
        \item  {{\color{\colorMATH}\({\mtext{\underline {ste\hspace{-1pt}}\hspace{1pt}p}}(N,\underline{\hspace{0.66em}})\)}} is the same as {{\color{\colorMATH}\({\mtext{return}}\)}} on values
        \item  Immediate by \nameref{thm:proofs:return-equivalence}
        \end{itemize}
     \item  \begin{itemize}[label={},leftmargin=0pt]\item  Case {{\color{\colorMATH}\(\underline \varsigma _{1} = \underline \sigma _{1},\underline E_{1}[\underline e_{1}]\)}} and {{\color{\colorMATH}\(\underline \varsigma _{2} = \underline \sigma _{2},\underline E_{2}[\underline e_{2}]\)}} for {{\color{\colorMATH}\(\underline e_{1}\)}} and {{\color{\colorMATH}\(\underline e_{2}\)}} redexes
        \item  {{\color{\colorMATH}\(\underline \sigma _{1} \sim  \underline \sigma _{2}\)}}
        \item  {{\color{\colorMATH}\(\underline e_{1} \sim  \underline e_{2}\)}} (by \nameref{thm:proofs:contexts-preserve-low-equivalence})
        \item  {{\color{\colorMATH}\(\left[ {\mtext{\underline {ste\hspace{-1pt}}\hspace{1pt}p}}(N,\underline \sigma _{1},\underline e_{1}) \mathrel{}\middle|\mathrel{} \Phi _{1} \right] \approx _{\sim } \left[{\mtext{\underline {ste\hspace{-1pt}}\hspace{1pt}p}}(N,\underline \sigma _{2},\underline e_{2}) \mathrel{}\middle|\mathrel{} \Phi _{2}\right]\)}} (by \nameref{thm:proofs:pmto-mixed-redex})
        \item  {{\color{\colorMATH}\(\left[ {\mtext{\underline {ste\hspace{-1pt}}\hspace{1pt}p}}(N,\underline \sigma _{1},\underline E_{1}[\underline e_{1}]) \mathrel{}\middle|\mathrel{} \Phi _{1} \right] \approx _{\sim }
           \left[{\mtext{\underline {ste\hspace{-1pt}}\hspace{1pt}p}}(N,\underline \sigma _{2},\underline E_{2}[\underline e_{2}]) \mathrel{}\middle|\mathrel{} \Phi _{2}\right]\)}} (by
           \nameref{thm:proofs:bind-equivalence},
           \nameref{thm:proofs:return-equivalence} and
           \nameref{thm:proofs:contexts-preserve-low-equivalence})
        \end{itemize}
     \end{enumerate}
  \end{itemize}
\end{proof}

\noindent
\begin{minipage}{\linewidth}
\begin{lemma}[PMTO (Mixed) Redex]\label{thm:proofs:pmto-mixed-redex}\ 
  \begin{itemize}[label={},leftmargin=0pt]\item  {\mtextit{Probabilistic low-equivalence for source expressions is preserved by the mixed semantics on a single step for redex configurations.}}
  \item  If: \hspace*{0.16em}{{\color{\colorMATH}\(\underline \varsigma _{1}\)}}\hspace*{0.16em} and \hspace*{0.16em}{{\color{\colorMATH}\(\underline \varsigma _{2}\)}}\hspace*{0.16em} are redex configurations
  \item  And: \hspace*{0.16em}{{\color{\colorMATH}\(\Phi _{1},\Sigma _{1} \vdash  \underline \varsigma _{1} \mathrel{;} \Psi _{1}\)}}\hspace*{0.16em} and \hspace*{0.16em}{{\color{\colorMATH}\(\Phi _{2},\Sigma _{2} \vdash  \underline \varsigma _{2} \mathrel{;} \Psi _{2}\)}}
  \item  And: \hspace*{0.16em}{{\color{\colorMATH}\(\underline \varsigma _{1} \sim  \underline \varsigma _{2}\)}}
  \item  Then: \hspace*{0.16em}{{\color{\colorMATH}\(\left[ {\mtext{\underline {ste\hspace{-1pt}}\hspace{1pt}p}}(N,\underline \varsigma _{1}) \mathrel{}\middle|\mathrel{} \Phi _{1} \right] \approx _{\sim } \left[{\mtext{\underline {ste\hspace{-1pt}}\hspace{1pt}p}}(N,\underline \varsigma _{2}) \mathrel{}\middle|\mathrel{} \Phi _{2} \right]\)}}
  \end{itemize}
\end{lemma}
\end{minipage}
\begin{proof}
  \begin{itemize}[label={},leftmargin=0pt]\item  By inversion:
  \item  \fbox{{{\color{\colorMATH}\(
     \inferrule*[vcenter,lab=
     ]{ \underline \sigma _{1} \sim  \underline \sigma _{2}
     \\ \underline e_{1} \sim  \underline e_{2}
        }{
        \underline \sigma _{1},\underline e_{1} \sim  \underline \sigma _{2},\underline e_{2}
     }
     \)}}}
  \item  Case analysis on {{\color{\colorMATH}\(\underline e_{1}\)}} and {{\color{\colorMATH}\(\underline e_{2}\)}} and inversion on low-equivalence
     judgment; all cases but two are immediate by
     \nameref{thm:proofs:return-equivalence} because definition of {{\color{\colorMATH}\({\mtext{\underline {ste\hspace{-1pt}}\hspace{1pt}p}}\)}}
     is a {{\color{\colorMATH}\({\mtext{return}}\)}}
  \item  \begin{enumerate}[leftmargin=15pt]\item  \begin{itemize}[label={},leftmargin=0pt]\item  Non-immediate case {{\color{\colorMATH}\(\underline e_{1} = {\mfootnotesize{{{\color{\colorSYNTAX}\mtexttt{cast}}}}}_{P}({\mfootnotesize{{{\color{\colorSYNTAX}\mtexttt{flipv}}}}}(\hat b_{1}))\)}} and {{\color{\colorMATH}\(\underline e_{2} = {\mfootnotesize{{{\color{\colorSYNTAX}\mtexttt{cast}}}}}_{P}({\mfootnotesize{{{\color{\colorSYNTAX}\mtexttt{flipv}}}}}(\hat b_{2}))\)}}:
        \item  By assumed well-typing:
        \item  \begin{itemize}[label=\textbf{-},leftmargin=*]\item  {{\color{\colorMATH}\({\mtext{Pr}}\left[ \hat b_{1} \mathrel{\dot =} {\mfootnotesize{{{\color{\colorSYNTAX}\mtexttt{I}}}}} \mathrel{}\middle|\mathrel{} \Phi _{1} \right] = \nicefrac{1}{2} \)}}
           \item  {{\color{\colorMATH}\({\mtext{Pr}}\left[ \hat b_{2} \mathrel{\dot =} {\mfootnotesize{{{\color{\colorSYNTAX}\mtexttt{I}}}}} \mathrel{}\middle|\mathrel{} \Phi _{2} \right] = \nicefrac{1}{2} \)}}
           \end{itemize}
        \item  By above facts, \nameref{thm:proofs:bind-equivalence} and because {{\color{\colorMATH}\({\mtext{return}}({\mfootnotesize{{{\color{\colorSYNTAX}\mtexttt{bitv}}}}}_{P}({\mfootnotesize{{{\color{\colorSYNTAX}\mtexttt{I}}}}})) \slashedrel\sim  {\mtext{return}}({\mfootnotesize{{{\color{\colorSYNTAX}\mtexttt{bitv}}}}}_{P}({\mfootnotesize{{{\color{\colorSYNTAX}\mtexttt{F}}}}}))\)}}:
        \item  {{\color{\colorMATH}\(\left[ \begin{array}{l@{\hspace*{0.33em}}l
               } {\mtext{do}} & b \leftarrow  \hat b_{1}
               \cr       & {\mtext{return}}({\mfootnotesize{{{\color{\colorSYNTAX}\mtexttt{bitv}}}}}_{P}(b))
               \end{array} \mathrel{}\middle|\mathrel{} \Phi _{1} \right]
            \approx _{\sim }
            \left[ \begin{array}{l@{\hspace*{0.33em}}l
               } {\mtext{do}} & b \leftarrow  \hat b_{2}
               \cr       & {\mtext{return}}({\mfootnotesize{{{\color{\colorSYNTAX}\mtexttt{bitv}}}}}_{P}(b))
               \end{array} \mathrel{}\middle|\mathrel{} \Phi _{2} \right]\)}}
        \end{itemize} 
     \item  \begin{itemize}[label={},leftmargin=0pt]\item  Non-immediate case {{\color{\colorMATH}\(\underline e_{1} = {\mfootnotesize{{{\color{\colorSYNTAX}\mtexttt{if}}}}}({\mfootnotesize{{{\color{\colorSYNTAX}\mtexttt{bitv}}}}}_{P}(\hat b))\{ \underline e_{1 1}\} \{ \underline e_{1 2}\} \)}} and {{\color{\colorMATH}\(\underline e_{2} = {\mfootnotesize{{{\color{\colorSYNTAX}\mtexttt{if}}}}}({\mfootnotesize{{{\color{\colorSYNTAX}\mtexttt{bitv}}}}}_{P}(\hat b))\{ \underline e_{2 1}\} \{ \underline e_{2 2}\} \)}}:
        \item  By assumed well-typing:
        \item  \begin{itemize}[label=\textbf{-},leftmargin=*]\item  {{\color{\colorMATH}\(\hat b = {\mtext{return}}(b)\)}}
           \end{itemize}
        \item  By assumed low-equivalence judgment:
           \begin{itemize}[label=\textbf{-},leftmargin=*]\item  {{\color{\colorMATH}\(\underline e_{1 1} \sim  \underline e_{2 1}\)}} and {{\color{\colorMATH}\(\underline e_{1 2} \sim  \underline e_{2 2}\)}}
           \end{itemize}
        \item  By above facts and \nameref{thm:proofs:monad-laws}:
        \item  {{\color{\colorMATH}\(\left[ \begin{array}{l@{\hspace*{0.33em}}l
               } {\mtext{do}} & b \leftarrow  \hat b
               \cr       & {\mtext{return}}({\mtext{cond}}(b,\underline e_{1 1},\underline e_{1 2}))
               \end{array} \mathrel{}\middle|\mathrel{} \Phi _{1} \right]
            \approx _{\sim }
            \left[ \begin{array}{l@{\hspace*{0.33em}}l
               } {\mtext{do}} & b \leftarrow  \hat b
               \cr       & {\mtext{return}}({\mtext{cond}}(b,\underline e_{2 1},\underline e_{2 2}))
               \end{array} \mathrel{}\middle|\mathrel{} \Phi _{2} \right]\)}}
        \end{itemize}
     \item  \begin{itemize}[label={},leftmargin=0pt]\item  Non-immediate cases {{\color{\colorMATH}\(\underline e_{1}\)}} and {{\color{\colorMATH}\(\underline e_{2}\)}} let-statements or function application
        \item  By \nameref{thm:proofs:pmto-mixed-substitution}
        \end{itemize}
     \end{enumerate}
  \end{itemize}
\end{proof}

\noindent
\begin{minipage}{\linewidth}
\begin{lemma}[PMTO (Mixed) Substitution]\label{thm:proofs:pmto-mixed-substitution}\ 
  \begin{itemize}[label={},leftmargin=0pt]\item  {\mtextit{Low-equivalence is preserved by substitution.}}
  \item  If: \hspace*{0.16em}{{\color{\colorMATH}\(v_{1} \sim  v_{2}\)}}
  \item  And: \hspace*{0.16em}{{\color{\colorMATH}\(e_{1} \sim  e_{2}\)}} 
  \item  And: \hspace*{0.16em}{{\color{\colorMATH}\(x\)}}\hspace*{0.16em} is free in \hspace*{0.16em}{{\color{\colorMATH}\(e_{1}\)}}\hspace*{0.16em} and \hspace*{0.16em}{{\color{\colorMATH}\(e_{2}\)}}
  \item  Then: \hspace*{0.16em}{{\color{\colorMATH}\([v_{1}/x]e_{1} \sim  [v_{2}/x]e_{2}\)}}
  \end{itemize}
\end{lemma}
\end{minipage}
\begin{proof}
  Induction on {{\color{\colorMATH}\(e_{1}\)}} and {{\color{\colorMATH}\(e_{2}\)}} and inversion on assumed low equivalence
\end{proof}

\noindent
\begin{minipage}{\linewidth}
\begin{lemma}[Contexts Preserve Low Equivalence]\label{thm:proofs:contexts-preserve-low-equivalence}\ 
  \begin{itemize}[label={},leftmargin=0pt]\item  {\mtextit{Low-equivalent terms have low-equivalent sub-terms, and contexts respect low-equivalence.}}
  \item  If: \hspace*{0.16em}{{\color{\colorMATH}\(\underline E_{1}[\underline e_{1}] \sim  \underline E_{2}[\underline e_{2}]\)}}
  \item  Then: 
  \item  \begin{enumerate}[leftmargin=15pt]\item  {{\color{\colorMATH}\(\underline e_{1} \sim  \underline e_{2}\)}}
     \item  {{\color{\colorMATH}\(\underline e_{1}^{\prime} \sim  \underline e_{2}^{\prime} \implies   \underline E_{1}[\underline e_{1}] \sim  \underline E_{2}[\underline e_{2}]\)}}
     \end{enumerate}
  \end{itemize}
\end{lemma}
\end{minipage}
\begin{proof}
  \begin{itemize}[label={},leftmargin=0pt]\item  Induction on {{\color{\colorMATH}\(\underline E_{1}\)}} and {{\color{\colorMATH}\(\underline E_{2}\)}} and inversion on assumed low equivalence
  \end{itemize}
\end{proof}

\begin{lemma}[Progress (Mixed)]\label{thm:proofs:progress-mixed}\ 
  \begin{itemize}[label={},leftmargin=0pt]\item  {\mtextit{Progress holds for the mixed semantics.}}
  \item  If: \hspace*{0.16em}{{\color{\colorMATH}\(\Psi _{c},\Phi ,\Sigma  \vdash  \underline \varsigma  \mathrel{:} \tau  \mathrel{;} \Psi \)}}
  \item  Then: {{\color{\colorMATH}\({\mtext{\underline {nste\hspace{-1pt}}\hspace{1pt}p}}(N,\underline \varsigma )\)}} is total
  \end{itemize}
\end{lemma}
\begin{proof}
  Induction on {{\color{\colorMATH}\(N\)}} and \nameref{thm:proofs:progress-mixed-single}
\end{proof}

\noindent
\begin{minipage}{\linewidth}
\begin{lemma}[Progress (Mixed) Single]\label{thm:proofs:progress-mixed-single}\ 
  \begin{itemize}[label={},leftmargin=0pt]\item  {\mtextit{Progress holds for the mixed semantics on a single step.}}
  \item  If: \hspace*{0.16em}{{\color{\colorMATH}\(\Psi _{c},\Phi ,\Sigma  \vdash  \underline \sigma ,\underline e \mathrel{:} \tau  \mathrel{;} \Psi \)}}
  \item  Then either: 
     \begin{enumerate}[leftmargin=15pt]\item  {{\color{\colorMATH}\(\underline e = \underline v\)}} for {{\color{\colorMATH}\(\underline v\)}} a value
     \item  {{\color{\colorMATH}\(\underline e = \underline E[\underline e]\)}} and {{\color{\colorMATH}\(\underline e\)}} a redex
     \end{enumerate}
     In both cases {{\color{\colorMATH}\({\mtext{\underline {ste\hspace{-1pt}}\hspace{1pt}p}}(N,\underline \sigma ,\underline e)\)}} is total.
  \end{itemize}
\end{lemma}
\end{minipage}
\begin{proof}
  Induction on {{\color{\colorMATH}\(\underline e\)}} and inversion on assumed well-typing
\end{proof}

\paragraph{\bf Low-equivalence Soundness}\ \\


\noindent
\begin{minipage}{\linewidth}
\begin{lemma}[Low-equivalence Soundness]\label{thm:proofs:low-equivalence-soundness}\ 
  \begin{itemize}[label={},leftmargin=0pt]\item  {\mtextit{When projected, low-equivalent trace distributions have equal probability
     distributions modulo adversary observation.}}
  \item  If: \hspace*{0.16em}{{\color{\colorMATH}\(\hat {\underline t_{1}} \approx _{\sim } \hat {\underline t_{2}}\)}}
  \item  Then: \hspace*{0.16em}{{\color{\colorMATH}\(\widehat {\mtext{obs}}(\hat \lceil \hat {\underline t_{1}}\hat \rceil ) \approx _{=} \widehat {\mtext{obs}}(\hat \lceil \hat {\underline t_{2}}\hat \rceil )\)}}
  \end{itemize}
\end{lemma}
\end{minipage}
\begin{proof}
  \begin{itemize}[label={},leftmargin=0pt]\item  Rewrite both sides by:
  \item  {{\color{\colorMATH}\(\begin{array}{rcl@{\hspace*{1.00em}}l
      } &{} {}& \widehat {\mtext{obs}}(\hat \lceil \hat {\underline t_{i}}\hat \rceil )
      \cr  &{}={}& \widehat {\mtext{obs}}({\mtext{do}}\hspace*{0.33em}\underline t \leftarrow  \hat {\underline t_{i}} \mathrel{;} \lceil \underline t\rceil )                  & \lbag {{\color{\colorTEXT}\textnormal{\hspace*{0.33em} {defn.} of {{\color{\colorMATH}\(\hat \lceil \underline{\hspace{0.66em}}\hat \rceil \)}} \hspace*{0.33em}}}}\rbag 
      \cr  &{}={}& {\mtext{do}}\hspace*{0.33em}\underline t \leftarrow  \hat {\underline t_{i}} \mathrel{;} \widehat {\mtext{obs}}(\lceil \underline t\rceil )                  & \lbag {{\color{\colorTEXT}\textnormal{\hspace*{0.33em} {defn.} of {{\color{\colorMATH}\(\widehat {\mtext{obs}}\)}} and \nameref{thm:proofs:monad-laws} \hspace*{0.33em}}}}\rbag 
      \end{array}\)}}
  \item  By \nameref{thm:proofs:bind-equivalence} and low-equivalence premise, STS:
  \item  \begin{itemize}[label=\textbf{-},leftmargin=*]\item  {{\color{\colorMATH}\(\underline t_{1} \sim  \underline t_{2} \implies    \widehat {\mtext{obs}}(\lceil \underline t_{1}\rceil ) \approx _{=} \widehat {\mtext{obs}}(\lceil \underline t_{2}\rceil )\)}}
     \end{itemize}
  \item  By \nameref{thm:proofs:low-equivalence-soundness-element}
  \end{itemize}
\end{proof}

\noindent
\begin{minipage}{\linewidth}
\begin{lemma}[Low-equivalence Soundness Element]\label{thm:proofs:low-equivalence-soundness-element}\ 
  \begin{itemize}[label={},leftmargin=0pt]\item  {\mtextit{When projected, low-equivalent traces have equal probability distributions
     modulo adversary observation.}}
  \item  If: \hspace*{0.16em}{{\color{\colorMATH}\(\underline t_{1} \sim  \underline t_{2}\)}}
  \item  Then: \hspace*{0.16em}{{\color{\colorMATH}\(\widehat {\mtext{obs}}(\lceil \underline t_{1}\rceil ) \approx _{=} \widehat {\mtext{obs}}(\lceil \underline t_{2}\rceil )\)}}
  \end{itemize}
\end{lemma}
\end{minipage}
\begin{proof}
  \begin{itemize}[label={},leftmargin=0pt]\item  Induction on traces {{\color{\colorMATH}\(\underline t_{1}\)}} and {{\color{\colorMATH}\(\underline t_{2}\)}} and inversion on assumed low-equivalence
  \item  \begin{enumerate}[leftmargin=15pt]\item  \begin{itemize}[label={},leftmargin=0pt]\item  Case {{\color{\colorMATH}\(\underline t_{1} = \underline t_{2} = {\mfootnotesize{{{\color{\colorSYNTAX}\mtexttt{\epsilon }}}}}\)}}
        \item  Immediate
        \end{itemize}
     \item  \begin{itemize}[label={},leftmargin=0pt]\item  Case {{\color{\colorMATH}\(\underline t_{1} = \underline t_{1}^{\prime}\mathord{\cdotp }\underline \sigma _{1},\underline e_{1}\)}} and {{\color{\colorMATH}\(\underline t_{2} = \underline t_{2}^{\prime}\mathord{\cdotp }\underline \sigma _{2},\underline e_{2}\)}}
        \item  By inversion on assumed low-equivalence:
        \item  \begin{itemize}[label=\textbf{-},leftmargin=*]\item  {{\color{\colorMATH}\(\underline t_{1}^{\prime} \sim  \underline t_{2}^{\prime}\)}}
           \item  {{\color{\colorMATH}\(\underline \sigma _{1} \sim  \underline \sigma _{2}\)}}
           \item  {{\color{\colorMATH}\(\underline e_{1} \sim  \underline e_{2}\)}}
           \end{itemize}
        \item  By induction hypothesis:
        \item  \begin{itemize}[label=\textbf{-},leftmargin=*]\item  {{\color{\colorMATH}\(\widehat {\mtext{obs}}(\lceil \underline t_{1}^{\prime}\rceil ) \approx _{=} \widehat {\mtext{obs}}(\lceil \underline t_{2}^{\prime}\rceil )\)}}
           \end{itemize}
        \item  By \nameref{thm:proofs:low-equivalence-soundness-element-store} and
           \nameref{thm:proofs:low-equivalence-soundness-element-expression}:
           \begin{itemize}[label=\textbf{-},leftmargin=*]\item  {{\color{\colorMATH}\(\widehat {\mtext{obs}}(\lceil \underline \sigma _{1}\rceil ) \approx _{=} \widehat {\mtext{obs}}(\lceil \underline \sigma _{2}\rceil )\)}}
           \item  {{\color{\colorMATH}\(\widehat {\mtext{obs}}(\lceil \underline e_{1}\rceil ) \approx _{=} \widehat {\mtext{obs}}(\lceil \underline e_{2}\rceil )\)}}
           \end{itemize}
       \item  Rewrite both sides by:
       \item  {{\color{\colorMATH}\(\begin{array}{rcl@{\hspace*{1.00em}}l
           } &{} {}& \widehat {\mtext{obs}}(\lceil \underline t_{i}^{\prime}\mathord{\cdotp }\underline \varsigma _{i}\rceil )
           \cr  &{}={}& {\mtext{do}}\hspace*{0.33em}\begin{array}[t]{l
                       } \vphantom{\overset {\mathord{\bullet }}x}\overset {\smash {\mathord{\bullet }}}t \leftarrow  \widehat {\mtext{obs}}(\lceil \underline t_{i}^{\prime}\rceil ) 
                       \cr  \vphantom{\overset {\mathord{\bullet }}x}\overset {\smash {\mathord{\bullet }}}\sigma  \leftarrow  \widehat {\mtext{obs}}(\lceil \underline \sigma _{i}\rceil ) 
                       \cr  \vphantom{\overset {\mathord{\bullet }}x}\overset {\smash {\mathord{\bullet }}}e \leftarrow  \widehat {\mtext{obs}}(\lceil \underline e_{i}\rceil ) 
                       \cr  \underline {\mtext{return}}(\vphantom{\overset {\mathord{\bullet }}x}\overset {\smash {\mathord{\bullet }}}t\mathord{\cdotp }\vphantom{\overset {\mathord{\bullet }}x}\overset {\smash {\mathord{\bullet }}}\sigma ,\vphantom{\overset {\mathord{\bullet }}x}\overset {\smash {\mathord{\bullet }}}e) 
                       \end{array} & \lbag {{\color{\colorTEXT}\textnormal{\hspace*{0.33em} {defn.} of {{\color{\colorMATH}\(\lceil \underline{\hspace{0.66em}}\rceil \)}} and {{\color{\colorMATH}\(\widehat {\mtext{obs}}\)}}, and \nameref{thm:proofs:monad-laws} \hspace*{0.33em}}}}\rbag 
           \end{array}\)}}
       \item  By iterated \nameref{thm:proofs:bind-equivalence}, \nameref{thm:proofs:return-equivalence} and three previously established facts
       \end{itemize}
    \end{enumerate}
  \end{itemize}
\end{proof}

\noindent
\begin{minipage}{\linewidth}
\begin{lemma}[Low-equivalence Soundness Element Store]\label{thm:proofs:low-equivalence-soundness-element-store}\ 
  \begin{itemize}[label={},leftmargin=0pt]\item  {\mtextit{When projected, low-equivalent stores have equal probability
     distributions modulo adversary observation.}}
  \item  If: \hspace*{0.16em}{{\color{\colorMATH}\(\underline \sigma _{1} \sim  \underline \sigma _{2}\)}}
  \item  Then: \hspace*{0.16em}{{\color{\colorMATH}\(\widehat {\mtext{obs}}(\lceil \underline \sigma _{1}\rceil ) \approx _{=} \widehat {\mtext{obs}}(\lceil \underline \sigma _{2}\rceil )\)}}
  \end{itemize}
\end{lemma}
\begin{proof}
  Induction on {{\color{\colorMATH}\(\underline \sigma _{1}\)}} and {{\color{\colorMATH}\(\underline \sigma _{2}\)}}, inversion on assumed low-equivalence,
  \nameref{thm:proofs:monad-laws}, \nameref{thm:proofs:return-equivalence} and
  \nameref{thm:proofs:bind-equivalence}.
\end{proof}
\end{minipage}

\noindent
\begin{minipage}{\linewidth}
\begin{lemma}[Low-equivalence Soundness Element Expression]\label{thm:proofs:low-equivalence-soundness-element-expression}\ 
  \begin{itemize}[label={},leftmargin=0pt]\item  {\mtextit{When projected, low-equivalent expressions have equal probability
     distributions modulo adversary observation.}}
  \item  If: \hspace*{0.16em}{{\color{\colorMATH}\(\underline e_{1} \sim  \underline e_{2}\)}}
  \item  Then: \hspace*{0.16em}{{\color{\colorMATH}\(\widehat {\mtext{obs}}(\lceil \underline e_{1}\rceil ) \approx _{=} \widehat {\mtext{obs}}(\lceil \underline e_{2}\rceil )\)}}
  \end{itemize}
\end{lemma}
\begin{proof}
  Induction on {{\color{\colorMATH}\(\underline e_{1}\)}} and {{\color{\colorMATH}\(\underline e_{2}\)}}, inversion on assumed low-equivalence,
  \nameref{thm:proofs:monad-laws}, \nameref{thm:proofs:return-equivalence} and
  \nameref{thm:proofs:bind-equivalence}.
\end{proof}
\end{minipage}

\paragraph{\bf Simulation (Mixed)}\ \\

\noindent
\begin{minipage}{\linewidth}
\begin{lemma}[Simulation (Mixed)]\label{thm:proofs:simulation-mixed}\ 
 \begin{itemize}[label={},leftmargin=0pt]\item  {\mtextit{When projected, the mixed semantics simulates the intensional standard semantics on source expressions.}}
 \item  If: \hspace*{0.16em}{{\color{\colorMATH}\(e\)}} is a source expression
 \item  Then: \hspace*{0.16em}{{\color{\colorMATH}\(\hat \lceil {\mtext{\underline {nste\hspace{-1pt}}\hspace{1pt}p}}(N,\varnothing ,e)\hat \rceil  = {\mtext{nstep}}_{{\mathcal{I}}}(N,\varnothing ,e)\)}}
 \end{itemize}
\end{lemma}
\end{minipage}
\begin{proof}
  \begin{itemize}[label={},leftmargin=0pt]\item  Induction on {{\color{\colorMATH}\(N\)}}
  \item  \begin{enumerate}[leftmargin=15pt]\item  \begin{itemize}[label={},leftmargin=0pt]\item  Case {{\color{\colorMATH}\(N=0\)}}:
        \item  {{\color{\colorMATH}\(\lceil e\rceil  = e\)}} (by \nameref{thm:proofs:simulation-mixed-zero})
        \end{itemize}
     \item  \begin{itemize}[label={},leftmargin=0pt]\item  Case {{\color{\colorMATH}\(N=N+1\)}}:
        \item  {{\color{\colorMATH}\(\hat \lceil {\mtext{\underline {nste\hspace{-1pt}}\hspace{1pt}p}}(N,\varnothing ,e)\hat \rceil  = {\mtext{nstep}}_{{\mathcal{I}}}(N,\varnothing ,e)\)}} {\mtextit{(IH)}} (by inductive hypothesis)
        \item  By equational reasoning:
        \item  {{\color{\colorMATH}\(\begin{array}[t]{rcl
            } &{} {}& \hat \lceil {\mtext{\underline {nste\hspace{-1pt}}\hspace{1pt}p}}(N+1,\varnothing ,e)\hat \rceil 
            \cr  &{}={}& \lbag {{\color{\colorTEXT}\textnormal{\hspace*{0.33em} {defn.} of {{\color{\colorMATH}\({\mtext{\underline {nste\hspace{-1pt}}\hspace{1pt}p}}\)}} and {{\color{\colorMATH}\(\hat \lceil \underline{\hspace{0.66em}}\hat \rceil \)}}, \nameref{thm:proofs:monad-laws} and \nameref{thm:proofs:monad-commutativity} \hspace*{0.33em}}}}\rbag 
            \cr  &{} {}& {\mtext{do}}\hspace*{0.33em}\begin{array}[t]{l
                        } \underline t\mathord{\cdotp }\underline \varsigma  \leftarrow  {\mtext{\underline {nste\hspace{-1pt}}\hspace{1pt}p}}(N,\varnothing ,e)
                        \cr  t \leftarrow  \lceil \underline t\rceil 
                        \cr  \varsigma  \leftarrow  \lceil \underline \varsigma \rceil 
                        \cr  \underline \varsigma ^{\prime} \leftarrow  {\mtext{\underline {ste\hspace{-1pt}}\hspace{1pt}p}}(N+1,\underline \varsigma )
                        \cr  \varsigma ^{\prime} \leftarrow  \lceil \underline \varsigma ^{\prime}\rceil 
                        \cr  {\mtext{return}}(t\mathord{\cdotp }\varsigma \mathord{\cdotp }\varsigma ^{\prime})
                        \end{array}
            \cr  &{}={}& \lbag {{\color{\colorTEXT}\textnormal{\hspace*{0.33em} \nameref{thm:proofs:simulation-mixed-single}, \nameref{thm:proofs:monad-laws} and \nameref{thm:proofs:monad-idempotence} \hspace*{0.33em}}}}\rbag 
            \cr  &{} {}& {\mtext{do}}\hspace*{0.33em}\begin{array}[t]{l
                        } \underline t\mathord{\cdotp }\underline \varsigma  \leftarrow  {\mtext{\underline {nste\hspace{-1pt}}\hspace{1pt}p}}(N,\varnothing ,e)
                        \cr  t \leftarrow  \lceil \underline t\rceil 
                        \cr  \varsigma  \leftarrow  \lceil \underline \varsigma \rceil 
                        \cr  \varsigma ^{\prime} \leftarrow  {\mtext{step}}_{{\mathcal{I}}}(N+1,\varsigma )
                        \cr  {\mtext{return}}(t\mathord{\cdotp }\varsigma \mathord{\cdotp }\varsigma ^{\prime})
                        \end{array}
            \cr  &{}={}& \lbag {{\color{\colorTEXT}\textnormal{\hspace*{0.33em} {\mtextit{(IH)}}, \nameref{thm:proofs:monad-laws} and {defn.} of {{\color{\colorMATH}\(\lceil \underline{\hspace{0.66em}}\rceil \)}} \hspace*{0.33em}}}}\rbag 
            \cr  &{} {}& {\mtext{do}}\hspace*{0.33em}\begin{array}[t]{l
                        } t\mathord{\cdotp }\varsigma  \leftarrow  {\mtext{nstep}}_{{\mathcal{I}}}(N,\varnothing ,e)
                        \cr  \varsigma ^{\prime} \leftarrow  {\mtext{step}}_{{\mathcal{I}}}(N+1,\varsigma )
                        \cr  {\mtext{return}}(t\mathord{\cdotp }\varsigma \mathord{\cdotp }\varsigma ^{\prime})
                        \end{array}
            \cr  &{}={}& \lbag {{\color{\colorTEXT}\textnormal{\hspace*{0.33em} {defn.} of {{\color{\colorMATH}\({\mtext{nstep}}_{{\mathcal{I}}}\)}} \hspace*{0.33em}}}}\rbag 
            \cr  &{} {}& {\mtext{nstep}}_{{\mathcal{I}}}(N,\varnothing ,e)
            \end{array}\)}}
        \end{itemize}
     \end{enumerate}
  \end{itemize}
\end{proof}

\noindent
\begin{minipage}{\linewidth}
\begin{lemma}[Simulation (Mixed) Zero]\label{thm:proofs:simulation-mixed-zero}\ 
  \begin{itemize}[label={},leftmargin=0pt]\item  {\mtextit{Projection on source expressions is the identity.}}
  \item  If: \hspace*{0.16em}{{\color{\colorMATH}\(e\)}} is a source expression
  \item  Then: \hspace*{0.16em}{{\color{\colorMATH}\(\lceil e\rceil  = e\)}}
  \end{itemize}
\end{lemma}
\end{minipage}
\begin{proof}
  \begin{itemize}[label={},leftmargin=0pt]\item  Induction on {{\color{\colorMATH}\(e\)}}
  \end{itemize}
\end{proof}

\noindent
\begin{minipage}{\linewidth}
\begin{lemma}[Simulation (Mixed) Single]\label{thm:proofs:simulation-mixed-single}\ 
  \begin{itemize}[label={},leftmargin=0pt]\item  {\mtextit{When projected, the mixed semantics simulates the intensional standard
     semantics on source expressions and on a single step.}}
  \item  {{\color{\colorMATH}\(\hat \lceil {\mtext{\underline {ste\hspace{-1pt}}\hspace{1pt}p}}(N,\underline \sigma ,\underline e)\hat \rceil  = \widehat {\mtext{step}}_{{\mathcal{I}}}(N,\lceil \underline \sigma ,\underline e\rceil )\)}}
  \end{itemize}
\end{lemma}
\end{minipage}
\begin{proof}
  \begin{itemize}[label={},leftmargin=0pt]\item  Induction on {{\color{\colorMATH}\(\underline e\)}}; first case is shown as representative trivial case;
     subsequent cases are non-trivial
  \item  \begin{itemize}[label=\textbf{-},leftmargin=*]\item  \begin{itemize}[label={},leftmargin=0pt]\item  Case {{\color{\colorMATH}\(\underline e = b_{\ell }\)}}:
        \item  {{\color{\colorMATH}\(\begin{array}[t]{rcl
            } &{} {}& \hat \lceil {\mtext{\underline {ste\hspace{-1pt}}\hspace{1pt}p}}(N,\underline \sigma ,b_{\ell })\hat \rceil 
            \cr  &{}={}& \lbag {{\color{\colorTEXT}\textnormal{\hspace*{0.33em} {defn.} of {{\color{\colorMATH}\({\mtext{\underline {ste\hspace{-1pt}}\hspace{1pt}p}}\)}} \hspace*{0.33em}}}}\rbag 
            \cr  &{} {}& \hat \lceil {\mtext{return}}(\underline \sigma ,{\mfootnotesize{{{\color{\colorSYNTAX}\mtexttt{bitv}}}}}_{\ell }({\mtext{return}}(b)))\hat \rceil  
            \cr  &{}={}& \lbag {{\color{\colorTEXT}\textnormal{\hspace*{0.33em} {defn.} of {{\color{\colorMATH}\(\hat \lceil \underline{\hspace{0.66em}}\hat \rceil \)}}, {{\color{\colorMATH}\({\mtext{step}}_{{\mathcal{I}}}\)}} and \nameref{thm:proofs:monad-laws} \hspace*{0.33em}}}}\rbag 
            \cr  &{} {}& {\mtext{do}}\hspace*{0.33em}\begin{array}[t]{l
                        } \sigma  \leftarrow  \lceil \underline \sigma \rceil 
                        \cr  e \leftarrow  \lceil b_{\ell }\rceil 
                        \cr  {\mtext{step}}_{{\mathcal{I}}}(\sigma ,e)
                        \end{array}
            \cr  &{}={}& \lbag {{\color{\colorTEXT}\textnormal{\hspace*{0.33em} {defn.} of {{\color{\colorMATH}\(\lceil \underline{\hspace{0.66em}}\rceil \)}}, {{\color{\colorMATH}\(\widehat {\mtext{step}}_{{\mathcal{I}}}\)}} and \nameref{thm:proofs:monad-laws} \hspace*{0.33em}}}}\rbag 
            \cr  &{} {}& \widehat {\mtext{step}}_{{\mathcal{I}}}(N,\lceil \underline \sigma ,b_{\ell }\rceil )
            \end{array}\)}}
        \end{itemize}
     \item  \begin{itemize}[label={},leftmargin=0pt]\item  Case {{\color{\colorMATH}\(\underline e = {\mfootnotesize{{{\color{\colorSYNTAX}\mtexttt{flip}}}}}^{\rho }()\)}}:
        \item  {{\color{\colorMATH}\(\begin{array}[t]{rcl
            } &{} {}& \hat \lceil {\mtext{\underline {ste\hspace{-1pt}}\hspace{1pt}p}}(N,\underline \sigma ,{\mfootnotesize{{{\color{\colorSYNTAX}\mtexttt{flip}}}}}^{\rho }())\hat \rceil 
            \cr  &{}={}& \lbag {{\color{\colorTEXT}\textnormal{\hspace*{0.33em} {defn.} of {{\color{\colorMATH}\({\mtext{\underline {ste\hspace{-1pt}}\hspace{1pt}p}}\)}} \hspace*{0.33em}}}}\rbag 
            \cr  &{} {}& \hat \lceil {\mtext{return}}(\underline \sigma ,{\mfootnotesize{{{\color{\colorSYNTAX}\mtexttt{flipv}}}}}_{\ell }({\mtext{bit}}(N)))\hat \rceil  
            \cr  &{}={}& \lbag {{\color{\colorTEXT}\textnormal{\hspace*{0.33em} {defn.} of {{\color{\colorMATH}\(\hat \lceil \underline{\hspace{0.66em}}\hat \rceil \)}}, {{\color{\colorMATH}\({\mtext{step}}_{{\mathcal{I}}}\)}} and \nameref{thm:proofs:monad-laws} \hspace*{0.33em}}}}\rbag 
            \cr  &{} {}& {\mtext{do}}\hspace*{0.33em}b \leftarrow  {\mtext{bit}}(N) \mathrel{;} {\mtext{return}}(\underline \sigma ,{\mfootnotesize{{{\color{\colorSYNTAX}\mtexttt{flipv}}}}}(b))
            \cr  &{}={}& \lbag {{\color{\colorTEXT}\textnormal{\hspace*{0.33em} {defn.} of {{\color{\colorMATH}\(\lceil \underline{\hspace{0.66em}}\rceil \)}}, {{\color{\colorMATH}\(\widehat {\mtext{step}}_{{\mathcal{I}}}\)}} and \nameref{thm:proofs:monad-laws} \hspace*{0.33em}}}}\rbag 
            \cr  &{} {}& \widehat {\mtext{step}}_{{\mathcal{I}}}(N,\lceil \underline \sigma ,{\mfootnotesize{{{\color{\colorSYNTAX}\mtexttt{flip}}}}}^{\rho }()\rceil )
            \end{array}\)}}
        \end{itemize}
     \item  \begin{itemize}[label={},leftmargin=0pt]\item  Case {{\color{\colorMATH}\(\underline e = {\mfootnotesize{{{\color{\colorSYNTAX}\mtexttt{cast}}}}}_{P}({\mfootnotesize{{{\color{\colorSYNTAX}\mtexttt{flipv}}}}}(\hat b))\)}}:
        \item  {{\color{\colorMATH}\(\begin{array}[t]{rcl
            } &{} {}& \hat \lceil {\mtext{\underline {ste\hspace{-1pt}}\hspace{1pt}p}}(N,\underline \sigma ,{\mfootnotesize{{{\color{\colorSYNTAX}\mtexttt{cast}}}}}_{P}({\mfootnotesize{{{\color{\colorSYNTAX}\mtexttt{flipv}}}}}(\hat b)))\hat \rceil 
            \cr  &{}={}& \lbag {{\color{\colorTEXT}\textnormal{\hspace*{0.33em} {defn.} of {{\color{\colorMATH}\({\mtext{\underline {ste\hspace{-1pt}}\hspace{1pt}p}}\)}} \hspace*{0.33em}}}}\rbag 
            \cr  &{} {}& \hat \lceil {\mtext{do}}\hspace*{0.33em}b \leftarrow  \hat b \mathrel{;} {\mtext{return}}(\underline \sigma ,{\mfootnotesize{{{\color{\colorSYNTAX}\mtexttt{bitv}}}}}_{P}({\mtext{return}}(b)))\hat \rceil  
            \cr  &{}={}& \lbag {{\color{\colorTEXT}\textnormal{\hspace*{0.33em} {defn.} of {{\color{\colorMATH}\(\hat \lceil \underline{\hspace{0.66em}}\hat \rceil \)}}, {{\color{\colorMATH}\({\mtext{step}}_{{\mathcal{I}}}\)}} and \nameref{thm:proofs:monad-laws} \hspace*{0.33em}}}}\rbag 
            \cr  &{} {}& {\mtext{do}}\hspace*{0.33em}b \leftarrow  \hat b \mathrel{;} {\mtext{return}}(\underline \sigma ,{\mfootnotesize{{{\color{\colorSYNTAX}\mtexttt{bitv}}}}}_{P}(b))
            \cr  &{}={}& \lbag {{\color{\colorTEXT}\textnormal{\hspace*{0.33em} {defn.} of {{\color{\colorMATH}\(\lceil \underline{\hspace{0.66em}}\rceil \)}}, {{\color{\colorMATH}\(\widehat {\mtext{step}}_{{\mathcal{I}}}\)}} and \nameref{thm:proofs:monad-laws} \hspace*{0.33em}}}}\rbag 
            \cr  &{} {}& \widehat {\mtext{step}}_{{\mathcal{I}}}(N,\lceil \underline \sigma ,{\mfootnotesize{{{\color{\colorSYNTAX}\mtexttt{cast}}}}}_{P}({\mfootnotesize{{{\color{\colorSYNTAX}\mtexttt{flipv}}}}}(\hat b))\rceil )
            \end{array}\)}}
        \end{itemize}
     \item  {\mtextit{All other cases are analogous to above cases.}}
     \end{itemize}
  \end{itemize}
\end{proof}

\paragraph{\bf Simulation (Intensional)}\ \\

\noindent
\begin{minipage}{\linewidth}
\begin{lemma}[Simulation (Intensional)]
\label{thm:proofs:simulation-intensional}\ 
\begin{itemize}[label={},leftmargin=0pt]\item  {\mtextit{The intensional standard semantics simulates the ground truth semantics on
  source expressions.}}
\item  {{\color{\colorMATH}\({\mtext{Pr}}\left[{\mtext{nstep}}_{{\mathcal{D}}}(N,\varnothing ,e) \mathrel{\dot =} \varsigma \right] = {\mtext{Pr}}\left[{\mtext{nstep}}_{{\mathcal{I}}}(N,\varnothing ,e) \mathrel{\dot =} \varsigma \right]\)}}
\end{itemize}
\end{lemma}
\begin{proof}
  Induction on {{\color{\colorMATH}\(N\)}} and by \nameref{thm:proofs:bind-probability},
  \nameref{thm:proofs:return-probability} and
  \nameref{thm:proofs:simulation-intensional} 
\end{proof}
\end{minipage}

\noindent
\begin{minipage}{\linewidth}
\begin{lemma}[Simulation (Intensional) Single]
\label{thm:proofs:simulation-intensional-single}\ 
\begin{itemize}[label={},leftmargin=0pt]\item  {\mtextit{The intensional standard semantics simulates the ground truth semantics on
  source expressions, and on a single step.}}
\item  {{\color{\colorMATH}\({\mtext{Pr}}\left[{\mtext{step}}_{{\mathcal{D}}}(N,\sigma ,e) \mathrel{\dot =} \varsigma \right] = {\mtext{Pr}}\left[{\mtext{step}}_{{\mathcal{I}}}(N,\sigma ,e) \mathrel{\dot =} \varsigma \right]\)}}
\end{itemize}
\end{lemma}
\begin{proof}
  Induction on {{\color{\colorMATH}\(e\)}} and by \nameref{thm:proofs:bind-probability},
  \nameref{thm:proofs:return-probability} and {{\color{\colorMATH}\({\mtext{bit}}_{{\mathcal{I}}}(N+1) \mathrel{\bot \!\!\!\bot }
  {\mtext{nstep}}_{{\mathcal{I}}}(N,\varsigma )\)}}, which is true by {{\color{\colorMATH}\({\mtext{height}}({\mtext{nstep}}(N,\varsigma )) \leq  N\)}} and
  {{\color{\colorMATH}\({\mtext{bit}}_{{\mathcal{I}}}(N+1) \mathrel{\bot \!\!\!\bot } \hat x\)}} when {{\color{\colorMATH}\({\mtext{height}}(\hat x) \leq  N\)}}
\end{proof}
\end{minipage}

\subsection{Type Preservation}\ \\

\noindent
\begin{minipage}{\linewidth}
\begin{lemma}[Type Preservation]
\label{thm:proofs:type-preservation}\ 
  \begin{itemize}[label={},leftmargin=0pt]\item  {\mtextit{Well-typing is preserved by the mixed semantics {w.r.t.} new trace history.}}
  \item  If: \hspace*{0.16em}{{\color{\colorMATH}\(e\)}} is a closed source expression
  \item  And: {{\color{\colorMATH}\(\vdash  e \mathrel{:} \tau \)}}
  \item  And: \hspace*{0.16em}{{\color{\colorMATH}\(\underline t\mathord{\cdotp }\underline \varsigma  \in  {\mtext{support}}({\mtext{\underline {nste\hspace{-1pt}}\hspace{1pt}p}}(N,\varnothing ,e))\)}}
  \item  Let: \hspace*{0.16em}{{\color{\colorMATH}\(\Phi  \triangleq  [{\mtext{\underline {nste\hspace{-1pt}}\hspace{1pt}p}}(N,\varnothing ,e) \mathrel{\dot =} \underline t\mathord{\cdotp }\underline \varsigma ]\)}}
  \item  Then: there exists \hspace*{0.16em}{{\color{\colorMATH}\(\Sigma \)}}\hspace*{0.16em} and \hspace*{0.16em}{{\color{\colorMATH}\(\Psi \)}}
  \item  {S.t.}: \hspace*{0.16em}{{\color{\colorMATH}\(\Phi ,\Sigma  \vdash  \underline \varsigma  \mathrel{:} \tau ,\Psi \)}}
  \end{itemize}
  \end{lemma}
\end{minipage}
\begin{proof}
  \begin{itemize}[label={},leftmargin=0pt]\item  By \nameref{thm:proofs:type-preservation-strong} which has a stronger conclusion (and therefore induction hypothesis)
  \end{itemize}
\end{proof}

\noindent
\begin{minipage}{\linewidth}
\begin{lemma}[Type Preservation (Strong)]
\label{thm:proofs:type-preservation-strong}\ 
  \begin{itemize}[label={},leftmargin=0pt]\item  {\mtextit{Well-typing is preserved by the mixed semantics {w.r.t.} new trace history.}}
  \item  If: \hspace*{0.16em}{{\color{\colorMATH}\(e\)}} is a closed source expression
  \item  And: {{\color{\colorMATH}\(\vdash  e \mathrel{:} \tau \)}}
  \item  And: \hspace*{0.16em}{{\color{\colorMATH}\(\underline t\mathord{\cdotp }\underline \varsigma  \in  {\mtext{support}}({\mtext{\underline {nste\hspace{-1pt}}\hspace{1pt}p}}(N,\varnothing ,e))\)}}
  \item  Let: \hspace*{0.16em}{{\color{\colorMATH}\(\Phi  \triangleq  [{\mtext{\underline {nste\hspace{-1pt}}\hspace{1pt}p}}(N,\varnothing ,e) \mathrel{\dot =} \underline t\mathord{\cdotp }\underline \varsigma ]\)}}
  \item  Then: there exists \hspace*{0.16em}{{\color{\colorMATH}\(\Sigma \)}}\hspace*{0.16em} and \hspace*{0.16em}{{\color{\colorMATH}\(\Psi \)}}
  \item  {S.t.}: \hspace*{0.16em}{{\color{\colorMATH}\(\Phi ,\Sigma  \vdash  \underline \varsigma  \mathrel{:} \tau ,\Psi \)}}
  \item  And: \hspace*{0.16em}{{\color{\colorMATH}\(\forall  \hat b \in  \Psi .\hspace*{0.33em} \hat b \mathrel{\bot \!\!\!\bot } \{ {\mtext{bit}}(N^{\prime}) \mathrel{|} N^{\prime} \geq  N+1\} \)}}
  \end{itemize}
  \end{lemma}
\end{minipage}
\begin{proof}
  \begin{itemize}[label={},leftmargin=0pt]\item  Induction on {{\color{\colorMATH}\(N\)}}
  \item  \begin{enumerate}[leftmargin=15pt]\item  \begin{itemize}[label={},leftmargin=0pt]\item  Case {{\color{\colorMATH}\(N=0\)}}:
        \item  {{\color{\colorMATH}\(\Phi  = [{\mtext{\underline {nste\hspace{-1pt}}\hspace{1pt}p}}(0,\varnothing ,e) \mathrel{\dot =} {\mtext{return}}(\varnothing ,e)] = [true]\)}}
        \item  {{\color{\colorMATH}\(\Sigma  = \varnothing \)}}
        \item  {{\color{\colorMATH}\(\Psi  = \varnothing \)}}
        \item  {{\color{\colorMATH}\(\varnothing ,\varnothing ,\varnothing  \vdash  \varnothing ,e \mathrel{:} \tau ,\varnothing \)}} (by \nameref{thm:proofs:source-expression-mixed-typing})
        \end{itemize}
     \item  \begin{itemize}[label={},leftmargin=0pt]\item  Case {{\color{\colorMATH}\(N=N+1\)}}:
        \item  By induction hypothesis {\mtextit{(IH)}}:
        \item  \begin{itemize}[label=\textbf{-},leftmargin=*]\item  \begin{itemize}[label={},leftmargin=0pt]\item  {{\color{\colorMATH}\(\Phi ^{\prime},\Sigma ^{\prime} \vdash  \underline \varsigma ^{\prime} \mathrel{:} \tau ,\Psi ^{\prime}\)}} for some {{\color{\colorMATH}\(\Sigma ^{\prime}\)}}, {{\color{\colorMATH}\(\Psi ^{\prime}\)}} 
              \item  and where {{\color{\colorMATH}\(\Phi ^{\prime} \triangleq  [{\mtext{\underline {nste\hspace{-1pt}}\hspace{1pt}p}}(N,\varnothing ,e) \mathrel{\dot =} \underline t^{\prime}\mathord{\cdotp }\underline \varsigma ^{\prime}]\)}}
              \item  and where {{\color{\colorMATH}\(\underline t = \underline t^{\prime}\mathord{\cdotp }\underline \varsigma ^{\prime}\)}}
              \end{itemize}
           \item  {{\color{\colorMATH}\(\forall  \hat b \in  \Psi ^{\prime}.\hspace*{0.33em} \hat b \mathrel{\bot \!\!\!\bot } \{ {\mtext{bit}}(N^{\prime}) \mathrel{|} N^{\prime} \geq  N+1\} \)}}
           \end{itemize}
        \item  By \nameref{thm:proofs:type-preservation-single} and second fact due
           to {\mtextit{(IH)}}:
        \item  \begin{itemize}[label=\textbf{-},leftmargin=*]\item  \begin{itemize}[label={},leftmargin=0pt]\item  {{\color{\colorMATH}\(\Phi ^{\prime \prime},\Sigma ^{\prime \prime} \vdash  \underline \varsigma  \mathrel{:} \tau ,\Psi ^{\prime \prime}\)}} for some {{\color{\colorMATH}\(\Sigma ^{\prime \prime}\)}}, {{\color{\colorMATH}\(\Psi ^{\prime \prime}\)}}
              \item  and where {{\color{\colorMATH}\(\Phi ^{\prime \prime} \triangleq  [\Phi ^{\prime},{\mtext{\underline {ste\hspace{-1pt}}\hspace{1pt}p}}(N,\underline \varsigma ^{\prime}) \mathrel{\dot =} \underline \varsigma ]\)}}
              \end{itemize}
           \item  {{\color{\colorMATH}\(\forall  \hat b \in  \Psi ^{\prime \prime}.\hspace*{0.33em} \hat b \mathrel{\bot \!\!\!\bot } \{ {\mtext{bit}}(N^{\prime}) \mathrel{|} N^{\prime} \geq  N+1+1\} \)}}
           \end{itemize}
        \item  Construct {{\color{\colorMATH}\(\Sigma  \triangleq  \Sigma ^{\prime \prime}\)}} and {{\color{\colorMATH}\(\Psi  \triangleq  \Psi ^{\prime \prime}\)}}; by previous typing and bit
           independence, and {{\color{\colorMATH}\(\Phi ^{\prime \prime} = \Phi \)}} ({b.c.} {{\color{\colorMATH}\(\underline t = \underline t^{\prime}\mathord{\cdotp }\underline \varsigma ^{\prime}\)}})
        \end{itemize} 
     \end{enumerate}
  \end{itemize}
\end{proof}

\noindent
\begin{minipage}{\linewidth}
\begin{lemma}[Source Expression Mixed Typing]
\label{thm:proofs:source-expression-mixed-typing}\ 
  \begin{itemize}[label={},leftmargin=0pt]\item  {\mtextit{Well-typed source expressions are well-typed in the mixed type system.}}
  \item  If: \hspace*{0.16em}{{\color{\colorMATH}\(e\)}}\hspace*{0.16em} is a source expression
  \item  And: \hspace*{0.16em}{{\color{\colorMATH}\(\vdash e \mathrel{:} \tau \)}}\hspace*{0.16em} (via source expression typing)
  \item  Then: \hspace*{0.16em}{{\color{\colorMATH}\(\vdash e \mathrel{:} \tau \)}}\hspace*{0.16em} (via mixed evaluation typing)
  \end{itemize}
\end{lemma}
\end{minipage}
\begin{proof}
  Induction on {{\color{\colorMATH}\(e\)}} and inversion on assumed well-typing 
\end{proof}

\noindent
\begin{minipage}{\linewidth}
\begin{lemma}[Type Preservation Single]
\label{thm:proofs:type-preservation-single}\ 
  \begin{itemize}[label={},leftmargin=0pt]\item  {\mtextit{Well-typing is preserved by the mixed semantics {w.r.t.} new trace
     history on a single step.}}
  \item  If: \hspace*{0.16em}{{\color{\colorMATH}\(\Phi ,\Sigma  \vdash  \underline \varsigma  \mathrel{:} \tau ,\Psi \)}}
  \item  And: \hspace*{0.16em}{{\color{\colorMATH}\(\forall  \hat b \in  \Psi .\hspace*{0.33em}  \hat b \mathrel{\bot \!\!\!\bot } \{ {\mtext{bit}}(N^{\prime}) \mathrel{|} N^{\prime} \geq  N\} \)}}
  \item  And: \hspace*{0.16em}{{\color{\colorMATH}\(\underline \varsigma ^{\prime} \in  {\mtext{support}}({\mtext{step}}_{{\mathcal{I}}}(N,\underline \varsigma ))\)}}
  \item  Let: \hspace*{0.16em}{{\color{\colorMATH}\(\Phi ^{\prime} \triangleq  [\Phi ,{\mtext{step}}_{{\mathcal{I}}}(N,\underline \varsigma ) \mathrel{\dot =} \underline \varsigma ^{\prime}]\)}}
  \item  Then: \hspace*{0.33em}there exists \hspace*{0.16em}{{\color{\colorMATH}\(\Sigma ^{\prime}\)}} and {{\color{\colorMATH}\(\Psi ^{\prime}\)}}
  \item  {S.t.}: \hspace*{0.16em}{{\color{\colorMATH}\(\Phi ^{\prime},\Sigma ^{\prime} \vdash  \underline \varsigma ^{\prime} \mathrel{:} \tau ,\Psi ^{\prime}\)}}
  \item  And: \hspace*{0.16em}{{\color{\colorMATH}\(\forall  \hat b \in  \Psi ^{\prime}.\hspace*{0.33em}  \hat b \mathrel{\bot \!\!\!\bot } \{ {\mtext{bit}}(N^{\prime}) \mathrel{|} N^{\prime} \geq  N+1\} \)}}
  \end{itemize}
\end{lemma}
\end{minipage}
\begin{proof}
  \begin{itemize}[label={},leftmargin=0pt]\item  By \nameref{thm:proofs:progress-mixed} and definition of {{\color{\colorMATH}\({\mtext{step}}_{{\mathcal{I}}}\)}}; two
  cases:
  \item  \begin{enumerate}[leftmargin=15pt]\item  \begin{itemize}[label={},leftmargin=0pt]\item  {{\color{\colorMATH}\(\underline \varsigma  = \underline \sigma ,\underline v\)}}
        \item  {{\color{\colorMATH}\(\underline \varsigma ^{\prime} = \underline \varsigma \)}}\hspace*{0.16em}, \hspace*{0.16em}{{\color{\colorMATH}\(\Phi ^{\prime} = \Phi \)}}\hspace*{0.16em}, \hspace*{0.16em}{{\color{\colorMATH}\(\Sigma ^{\prime} = \Sigma \)}}\hspace*{0.16em} and \hspace*{0.16em}{{\color{\colorMATH}\(\Psi ^{\prime} = \Psi \)}}
        \item  Immediate
        \end{itemize}
     \item  \begin{itemize}[label={},leftmargin=0pt]\item  {{\color{\colorMATH}\(\underline \varsigma  = \underline \sigma ,\underline E[\underline e]\)}}
        \item  {{\color{\colorMATH}\(\underline \varsigma ^{\prime} = \underline \sigma ^{\prime},\underline E[\underline e^{\prime}]\)}}\hspace*{0.16em} for \hspace*{0.16em}{{\color{\colorMATH}\(\underline \sigma ^{\prime},\underline e^{\prime} \in  {\mtext{support}}({\mtext{step}}_{{\mathcal{I}}}(N,\underline \sigma ,\underline e))\)}}
        \item  By~\nameref{thm:proofs:contexts-preserve-typing}:
           \begin{itemize}[label=\textbf{-},leftmargin=*]\item  \begin{itemize}[label={},leftmargin=0pt]\item  There exists \hspace*{0.16em}{{\color{\colorMATH}\(\tau ^{\prime}\)}}\hspace*{0.16em}, \hspace*{0.16em}{{\color{\colorMATH}\(\Psi _{c}\)}}\hspace*{0.16em} and \hspace*{0.16em}{{\color{\colorMATH}\(\Psi ^{\prime}\)}}
              \item  {S.t.}: \hspace*{0.16em}{{\color{\colorMATH}\(\Psi _{c},\Phi ,\Sigma  \vdash  \underline \sigma ,\underline e \mathrel{:} \tau ^{\prime} \mathrel{;} \Psi ^{\prime}\)}}
              \item  And: \hspace*{0.16em}{{\color{\colorMATH}\(\Psi _{c}\uplus \Psi ^{\prime} = \Psi \)}}
              \end{itemize}
           \end{itemize}
        \item  By~\nameref{thm:proofs:type-preservation-redex}:
           \begin{itemize}[label=\textbf{-},leftmargin=*]\item  \begin{itemize}[label={},leftmargin=0pt]\item  There exists \hspace*{0.16em}{{\color{\colorMATH}\(\Sigma ^{\prime}\)}}\hspace*{0.16em} and \hspace*{0.16em}{{\color{\colorMATH}\(\Psi ^{\prime \prime}\)}}
              \item  {S.t.}: \hspace*{0.16em}{{\color{\colorMATH}\(\Sigma ^{\prime} \supseteq  \Sigma \)}}
              \item  And: \hspace*{0.16em}{{\color{\colorMATH}\(\Psi _{c},\Phi ^{\prime},\Sigma ^{\prime} \vdash  \underline \sigma ^{\prime},\underline e^{\prime} \mathrel{:} \tau ^{\prime} \mathrel{;} \Psi ^{\prime \prime}\)}}
              \item  And: \hspace*{0.16em}{{\color{\colorMATH}\(\forall  \rho ,\hat b.\hspace*{0.33em} \Psi _{c}\setminus (\{ \hat b\} ,\varnothing )\uplus \Psi ^{\prime},\Phi  \vdash  \hat b\mathrel{:}{\mfootnotesize{{{\color{\colorSYNTAX}\mtexttt{flip}}}}}^{\rho } \implies   \Psi _{c}\setminus (\{ \hat b\} ,\varnothing )\uplus \Psi ^{\prime \prime},\Phi ^{\prime} \vdash  \hat b\mathrel{:}{\mfootnotesize{{{\color{\colorSYNTAX}\mtexttt{flip}}}}}^{\rho }\)}}
              \item  And: \hspace*{0.16em}{{\color{\colorMATH}\(\forall  \hat b \in  \Psi _{c}\uplus \Psi ^{\prime \prime}.\hspace*{0.33em} \hat b \mathrel{\bot \!\!\!\bot } \{ {\mtext{bit}}(N^{\prime}) \mathrel{|} N^{\prime} \geq  N+1\} \)}}
              \end{itemize}
           \end{itemize}
        \item  By \nameref{thm:proofs:wkn-cxt} and \nameref{thm:proofs:contexts-preserve-typing}:
           \begin{itemize}[label=\textbf{-},leftmargin=*]\item  {{\color{\colorMATH}\(\Phi ^{\prime},\Sigma ^{\prime} \vdash  \underline \sigma ^{\prime},\underline E[\underline e^{\prime}] \mathrel{:} \tau  \mathrel{;} \Psi _{c} \uplus  \Psi ^{\prime \prime}\)}}
           \end{itemize}
        \item  Construct {{\color{\colorMATH}\(\Sigma ^{\prime} \triangleq  \Sigma ^{\prime}\)}} and {{\color{\colorMATH}\(\Psi ^{\prime} = \Psi _{c} \uplus  \Psi ^{\prime \prime}\)}}; by previous typing and bit independence
        \end{itemize}
     \end{enumerate}
  \end{itemize}
\end{proof}

\noindent
\begin{minipage}{\linewidth}
\begin{lemma}[Contexts Preserve Typing]\label{thm:proofs:contexts-preserve-typing}\ 
\begin{itemize}[label={},leftmargin=0pt]\item  If: \hspace*{0.16em}{{\color{\colorMATH}\(\varnothing ,\Phi ,\Sigma ,\varnothing  \vdash  \underline E[\underline e] \mathrel{:} \tau  \mathrel{;} \varnothing  , \Psi \)}}
\item  Then: there exists \hspace*{0.16em}{{\color{\colorMATH}\(\tau ^{\prime},\Psi _{c},\Psi ^{\prime}\)}}\hspace*{0.16em} {s.t.}:
\item  \begin{enumerate}[leftmargin=15pt]\item  {{\color{\colorMATH}\(\Psi _{c},\Phi ,\Sigma ,\varnothing  \vdash  \underline e \mathrel{:} \tau ^{\prime} \mathrel{;} \varnothing  , \Psi ^{\prime}\)}}\hspace*{0.16em} and \hspace*{0.16em}{{\color{\colorMATH}\(\Psi _{c} \uplus  \Psi ^{\prime} = \Psi \)}}
   \item  {{\color{\colorMATH}\(\Psi _{c},\Phi ,\Sigma ,\varnothing  \vdash  \underline e^{\prime} \mathrel{:} \tau ^{\prime} \mathrel{;} \varnothing  , \Psi ^{\prime}\)}}\hspace*{0.16em} and \hspace*{0.16em}{{\color{\colorMATH}\(\Psi _{c} \uplus  \Psi ^{\prime} = \Psi \)}}\hspace*{0.16em} \hspace*{0.16em}{{\color{\colorMATH}\(\implies   \)}}\hspace*{0.16em} {{\color{\colorMATH}\(\Psi _{c},\Phi ,\Sigma ,\varnothing  \vdash  \underline E[\underline e^{\prime}] \mathrel{:} \tau  \mathrel{;} \varnothing  , \Psi \)}}
   \end{enumerate}
\end{itemize}
\end{lemma}
\end{minipage}
\begin{proof}
\begin{itemize}[label={},leftmargin=0pt]\item  Induction on {{\color{\colorMATH}\(E\)}} and inversion on {{\color{\colorMATH}\(\Psi _{c},\Phi ,\Sigma ,\varnothing  \vdash  \underline E[\underline e] \mathrel{:} \tau  \mathrel{;} \varnothing  , \Psi \)}}
\end{itemize}
\end{proof}

\paragraph{\bf Type Preservation Redex}\ \\

\noindent
\begin{minipage}{\linewidth}
\begin{lemma}[Type Preservation Redex]
\label{thm:proofs:type-preservation-redex}\ 
  \begin{itemize}[label={},leftmargin=0pt]\item  If: \hspace*{0.16em}{{\color{\colorMATH}\(\underline e\)}} a redex
  \item  And: \hspace*{0.16em}{{\color{\colorMATH}\(\Psi _{c},\Phi ,\Sigma  \vdash  \underline \sigma ,\underline e \mathrel{:} \tau  \mathrel{;} \Psi \)}}
  \item  And: \hspace*{0.16em}{{\color{\colorMATH}\(\forall  \hat b \in  \Psi _{c}\uplus \Psi .\hspace*{0.33em} \hat b \mathrel{\bot \!\!\!\bot } \{ {\mtext{bit}}(N^{\prime}) \mathrel{|} N^{\prime} \geq  N\} \)}}
  \item  And: \hspace*{0.16em}{{\color{\colorMATH}\(\underline \varsigma  \in  {\mtext{support}}({\mtext{step}}_{{\mathcal{I}}}(N,\underline \sigma ,\underline e))\)}}
  \item  Let: \hspace*{0.16em}{{\color{\colorMATH}\(\Phi ^{\prime} \triangleq  [\Phi ,{\mtext{step}}_{{\mathcal{I}}}(N,\underline \sigma ,\underline e) \mathrel{\dot =} \underline \varsigma ]\)}}
  \item  Then: \hspace*{0.33em}there exists \hspace*{0.16em}{{\color{\colorMATH}\(\Sigma ^{\prime}\)}}\hspace*{0.16em} and \hspace*{0.16em}{{\color{\colorMATH}\(\Psi ^{\prime}\)}}
  \item  {S.t.}: \hspace*{0.16em}{{\color{\colorMATH}\(\Sigma ^{\prime} \supseteq  \Sigma \)}}
  \item  And: \hspace*{0.16em}{{\color{\colorMATH}\(\Psi _{c},\Phi ^{\prime},\Sigma ^{\prime} \vdash  \underline \varsigma  \mathrel{:} \tau  \mathrel{;} \Psi ^{\prime}\)}}
  \item  And: \hspace*{0.16em}{{\color{\colorMATH}\(\forall  \rho ,\hat b.\hspace*{0.33em} \Psi _{c}\setminus (\{ \hat b\} ,\varnothing )\uplus \Psi ,\Phi  \vdash  \hat b\mathrel{:}{\mfootnotesize{{{\color{\colorSYNTAX}\mtexttt{flip}}}}}^{\rho } \implies   \Psi _{c}\setminus (\{ \hat b\} ,\varnothing )\uplus \Psi ^{\prime},\Phi ^{\prime} \vdash  \hat b\mathrel{:}{\mfootnotesize{{{\color{\colorSYNTAX}\mtexttt{flip}}}}}^{\rho }\)}}
  \item  And: \hspace*{0.16em}{{\color{\colorMATH}\(\forall  \hat b \in  \Psi _{c}\uplus \Psi ^{\prime}.\hspace*{0.33em} \hat b \mathrel{\bot \!\!\!\bot } \{ {\mtext{bit}}(N^{\prime}) \mathrel{|} N^{\prime} \geq  N+1\} \)}}
  \end{itemize}
\end{lemma}
\end{minipage}
\begin{proof}
  \begin{itemize}[label={},leftmargin=0pt]\item  By inversion:
  \item  \fbox{{{\color{\colorMATH}\(
     \inferrule*[vcenter,lab=
     ]{ \Psi _{c} \uplus  \Psi _{e},\Phi ,\Sigma  \vdash  \underline \sigma  \mathrel{;} \Psi _{\sigma }
     \\ \Psi _{c} \uplus  \Psi _{\sigma },\Phi ,\Sigma ,\varnothing  \vdash  \underline e \mathrel{:} \tau  \mathrel{;} \varnothing  , \Psi _{e}
        }{
        \Psi _{c} , \Phi  , \Sigma  \vdash  \underline \sigma ,\underline e \mathrel{:} \tau  \mathrel{;} \Psi _{\sigma } \uplus  \Psi _{e}
     }
     \)}}}
  \item  Case analysis on {{\color{\colorMATH}\(\underline e\)}}:
     \begin{enumerate}[leftmargin=15pt]\item  \begin{itemize}[label={},leftmargin=0pt]\item  {{\color{\colorMATH}\(\underline e = {\mfootnotesize{{{\color{\colorSYNTAX}\mtexttt{flip}}}}}^{\rho }()\)}}
        \item  By inversion:
        \item  \fbox{{{\color{\colorMATH}\(
           \inferrule*[vcenter,lab=
           ]{ }{
              \Psi _{c} \uplus  \Psi _{\sigma },\Phi ,\Sigma ,\varnothing  \vdash  {\mfootnotesize{{{\color{\colorSYNTAX}\mtexttt{flip}}}}}^{\rho }() \mathrel{:} {\mfootnotesize{{{\color{\colorSYNTAX}\mtexttt{flip}}}}}^{\rho } \mathrel{;} \oslash  , \varnothing ,\varnothing 
           }
           \)}}}
        \item  {{\color{\colorMATH}\(\tau  = {\mfootnotesize{{{\color{\colorSYNTAX}\mtexttt{flip}}}}}^{\rho }\)}}
        \item  {{\color{\colorMATH}\(\Psi _{e} = \varnothing ,\varnothing \)}}
        \item  {{\color{\colorMATH}\(\underline \varsigma  = \underline \sigma ,{\mfootnotesize{{{\color{\colorSYNTAX}\mtexttt{flipv}}}}}({\mtext{bit}}(N))\)}}
        \item  {{\color{\colorMATH}\(\Phi ^{\prime} = [\Phi ,{\mtext{step}}_{{\mathcal{I}}}(N,\underline \sigma ,{\mfootnotesize{{{\color{\colorSYNTAX}\mtexttt{flip}}}}}^{\rho }() \mathrel{\dot =} \underline \sigma ,{\mfootnotesize{{{\color{\colorSYNTAX}\mtexttt{flipv}}}}}({\mtext{bit}}(N))] = \Phi \)}}
        \item  {{\color{\colorMATH}\(\Psi _{e}^{\prime} \triangleq  \{ {\mtext{bit}}(N)\} ,\varnothing \)}}
        \item  Construct \hspace*{0.16em}{{\color{\colorMATH}\(\Sigma ^{\prime} \triangleq  \Sigma  \supseteq  \Sigma \)}}
        \item  Construct \hspace*{0.16em}{{\color{\colorMATH}\(\Psi ^{\prime} \triangleq  \Psi _{\sigma } \uplus  (\{ {\mtext{bit}}(N)\} ,\varnothing ) = \Psi _{\sigma } \uplus  \Psi _{e}^{\prime}\)}}
        \item  To show: 
           \begin{enumerate}[leftmargin=15pt]\item  {{\color{\colorMATH}\(\Psi _{c}\uplus \{ {\mtext{bit}}(N)\} ,\Phi ,\Sigma  \vdash  \underline \sigma  \mathrel{;} \Psi _{\sigma }\)}}
           \item  {{\color{\colorMATH}\(\Psi _{c}\uplus \Psi _{\sigma },\Phi ,\Sigma  \vdash  {\mfootnotesize{{{\color{\colorSYNTAX}\mtexttt{flipv}}}}}({\mtext{bit}}(N)) \mathrel{:} {\mfootnotesize{{{\color{\colorSYNTAX}\mtexttt{flip}}}}}^{\rho } \mathrel{;} \varnothing \)}}
           \item  {{\color{\colorMATH}\(\begin{array}[t]{l
               } \forall  \rho ^{\prime},\hat b^{\prime}.\hspace*{0.33em} 
               \cr  \Psi _{c}\setminus (\{ \hat b^{\prime}\} ,\varnothing )\uplus \Psi _{\sigma },\Phi ,\Sigma  \vdash  \hat b^{\prime} \mathrel{:} {\mfootnotesize{{{\color{\colorSYNTAX}\mtexttt{flip}}}}}^{\rho ^{\prime}} 
               \cr  \implies   
               \cr  \Psi _{c}\setminus (\{ \hat b^{\prime}\} ,\varnothing )\uplus \Psi _{\sigma }\uplus (\{ {\mtext{bit}}(N)\} ,\varnothing ),\Phi ,\Sigma  \vdash  \hat b^{\prime} \mathrel{:} {\mfootnotesize{{{\color{\colorSYNTAX}\mtexttt{flip}}}}}^{\rho ^{\prime}}
               \end{array}\)}}
           \item  {{\color{\colorMATH}\(\forall  \hat b \in  \Psi _{c}\uplus \Psi _{\sigma }\uplus (\{ {\mtext{bit}}(N)\} ,\varnothing ).\hspace*{0.33em} \hat b \mathrel{\bot \!\!\!\bot } \{ {\mtext{bit}}(N^{\prime}) \mathrel{|} N^{\prime} \geq  N+1\} \)}}
           \end{enumerate}
        \item  (a) is by \nameref{thm:proofs:weaken-store} and \nameref{thm:proofs:type-preservation-flip} applied to {{\color{\colorMATH}\(\Psi _{c}\)}}
        \item  (b-c) are by \nameref{thm:proofs:type-preservation-flip} applied to {{\color{\colorMATH}\(\Psi _{c} \uplus  \Psi _{\sigma }\)}}
        \item  (d) is by assumed bit independence and \nameref{thm:proofs:bit-independence}
        \end{itemize}
     \item  \begin{itemize}[label={},leftmargin=0pt]\item  {{\color{\colorMATH}\(\underline e = {\mfootnotesize{{{\color{\colorSYNTAX}\mtexttt{cast}}}}}_{P}({\mfootnotesize{{{\color{\colorSYNTAX}\mtexttt{flipv}}}}}(\hat b))\)}}
        \item  By inversion:
        \item  \fbox{{{\color{\colorMATH}\(
           \inferrule*[vcenter,lab=
           ]{ \inferrule*[vcenter,lab=
              ]{ \Psi _{c} \uplus  \Psi _{\sigma },\Phi  \vdash  \hat b \mathrel{:} {\mfootnotesize{{{\color{\colorSYNTAX}\mtexttt{flip}}}}}^{\rho }
                 }{
                 \Psi _{c} \uplus  \Psi _{\sigma },\Phi ,\Sigma  \vdash  {\mfootnotesize{{{\color{\colorSYNTAX}\mtexttt{flipv}}}}}(\hat b) \mathrel{:} {\mfootnotesize{{{\color{\colorSYNTAX}\mtexttt{flip}}}}}^{\rho } \mathrel{;} \{ \hat b\} ,\varnothing 
              }
              }{
              \Psi _{c} \uplus  \Psi _{\sigma },\Phi ,\Sigma ,\varnothing  \vdash  {\mfootnotesize{{{\color{\colorSYNTAX}\mtexttt{cast}}}}}_{P}({\mfootnotesize{{{\color{\colorSYNTAX}\mtexttt{flipv}}}}}(\hat b)) \mathrel{:} {\mfootnotesize{{{\color{\colorSYNTAX}\mtexttt{bit}}}}}_{P}^{\bot } \mathrel{;} \oslash  , \{ \hat b\} ,\varnothing 
           }
           \)}}}
        \item  {{\color{\colorMATH}\(\tau  = {\mfootnotesize{{{\color{\colorSYNTAX}\mtexttt{bit}}}}}_{P}^{\bot }\)}}
        \item  {{\color{\colorMATH}\(\Psi _{e} = \{ \hat b\} ,\varnothing \)}}
        \item  {{\color{\colorMATH}\(\underline \varsigma  = \underline \sigma ,{\mfootnotesize{{{\color{\colorSYNTAX}\mtexttt{bitv}}}}}_{P}({\mtext{return}}(b))\)}}\hspace*{0.16em} for \hspace*{0.16em}{{\color{\colorMATH}\(b\in \{ {\mfootnotesize{{{\color{\colorSYNTAX}\mtexttt{O}}}}},{\mfootnotesize{{{\color{\colorSYNTAX}\mtexttt{I}}}}}\} \)}}
        \item  {{\color{\colorMATH}\(\Phi ^{\prime} = [\Phi ,{\mtext{step}}_{{\mathcal{I}}}(N,\underline \sigma ,{\mfootnotesize{{{\color{\colorSYNTAX}\mtexttt{cast}}}}}_{P}({\mfootnotesize{{{\color{\colorSYNTAX}\mtexttt{flipv}}}}}(\hat b))) \mathrel{\dot =} \underline \sigma ,{\mfootnotesize{{{\color{\colorSYNTAX}\mtexttt{bitv}}}}}_{P}({\mtext{return}}(b))] = [\Phi ,\hat b\mathrel{\dot =}b]\)}}
        \item  {{\color{\colorMATH}\(\Psi _{e}^{\prime} \triangleq  \varnothing \)}}
        \item  Construct \hspace*{0.16em}{{\color{\colorMATH}\(\Sigma ^{\prime} \triangleq  \Sigma  \supseteq  \Sigma \)}}
        \item  Construct \hspace*{0.16em}{{\color{\colorMATH}\(\Psi ^{\prime} \triangleq  \Psi _{\sigma } = \Psi _{\sigma } \uplus  \Psi _{e}^{\prime}\)}}
        \item  To show: 
           \begin{enumerate}[leftmargin=15pt]\item  {{\color{\colorMATH}\(\Psi _{c},[\Phi ,\hat b\mathrel{\dot =}b],\Sigma  \vdash  \underline \sigma  \mathrel{;} \Psi _{\sigma }\)}}
           \item  {{\color{\colorMATH}\(\Psi _{c}\uplus \Psi _{\sigma },[\Phi ,\hat b\mathrel{\dot =}b],\Sigma  \vdash  {\mfootnotesize{{{\color{\colorSYNTAX}\mtexttt{bitv}}}}}_{P}({\mtext{return}}(b)) \mathrel{:} {\mfootnotesize{{{\color{\colorSYNTAX}\mtexttt{bit}}}}}_{P}^{\bot } \mathrel{;} \varnothing \)}}
           \item  {{\color{\colorMATH}\(\begin{array}[t]{l
               } \forall  \rho ^{\prime},\hat b^{\prime}.\hspace*{0.33em} 
               \cr  \Psi _{c}\setminus (\{ \hat b^{\prime}\} ,\varnothing )\uplus \Psi _{\sigma }\uplus (\{ \hat b\} ,\varnothing ),\Phi ,\Sigma  \vdash  \hat b^{\prime} \mathrel{:} {\mfootnotesize{{{\color{\colorSYNTAX}\mtexttt{flip}}}}}^{\rho ^{\prime}} 
               \cr  \implies   
               \cr  \Psi _{c}\setminus (\{ \hat b^{\prime}\} ,\varnothing )\uplus \Psi _{\sigma },[\Phi ,\hat b\mathrel{\dot =}b],\Sigma  \vdash  \hat b^{\prime} \mathrel{:} {\mfootnotesize{{{\color{\colorSYNTAX}\mtexttt{flip}}}}}^{\rho ^{\prime}}
               \end{array}\)}}
           \item  {{\color{\colorMATH}\(\forall  \hat b \in  \Psi _{c}\uplus \Psi _{\sigma }.\hspace*{0.33em} \hat b \mathrel{\bot \!\!\!\bot } \{ {\mtext{bit}}(N^{\prime}) \mathrel{|} N^{\prime} \geq  N+1\} \)}}
           \end{enumerate}
        \item  (a) is by \nameref{thm:proofs:weaken-store} and \nameref{thm:proofs:type-preservation-castp} applied to {{\color{\colorMATH}\(\Psi _{c}\)}}
        \item  (b-c) are by \nameref{thm:proofs:type-preservation-castp} applied to {{\color{\colorMATH}\(\Psi _{c} \uplus  \Psi _{\sigma }\)}}
        \item  (d) is by assumption
        \end{itemize}
     \item  \begin{itemize}[label={},leftmargin=0pt]\item  {{\color{\colorMATH}\(\underline e = {\mfootnotesize{{{\color{\colorSYNTAX}\mtexttt{cast}}}}}_{S}({\mfootnotesize{{{\color{\colorSYNTAX}\mtexttt{flipv}}}}}(\hat b))\)}}
        \item  By inversion:
        \item  \fbox{{{\color{\colorMATH}\(
           \inferrule*[vcenter,lab=
           ]{ \inferrule*[vcenter,lab=
              ]{ \Psi _{c} \uplus  \Psi _{\sigma },\Phi  \vdash  \hat b \mathrel{:} {\mfootnotesize{{{\color{\colorSYNTAX}\mtexttt{flip}}}}}^{\rho }
                 }{
                 \Psi _{c} \uplus  \Psi _{\sigma },\Phi ,\Sigma  \vdash  {\mfootnotesize{{{\color{\colorSYNTAX}\mtexttt{flipv}}}}}(\hat b) \mathrel{:} {\mfootnotesize{{{\color{\colorSYNTAX}\mtexttt{flip}}}}}^{\rho } \mathrel{;} \varnothing , \{ \rho  \mapsto  \hat b\} 
              }
              }{
              \Psi _{c} \uplus  \Psi _{\sigma },\Phi ,\Sigma ,\varnothing  \vdash  {\mfootnotesize{{{\color{\colorSYNTAX}\mtexttt{cast}}}}}_{S}({\mfootnotesize{{{\color{\colorSYNTAX}\mtexttt{flipv}}}}}(\hat b)) \mathrel{:} {\mfootnotesize{{{\color{\colorSYNTAX}\mtexttt{bit}}}}}_{S}^{\rho } \mathrel{;} \oslash  , \varnothing  , \{ \rho  \mapsto  \{ \hat b\} \} 
           }
           \)}}}
        \item  {{\color{\colorMATH}\(\tau  = {\mfootnotesize{{{\color{\colorSYNTAX}\mtexttt{bit}}}}}_{S}^{\rho }\)}}
        \item  {{\color{\colorMATH}\(\Psi _{e} = \varnothing ,\{ \rho \mapsto \{ \hat b\} \} \)}}
        \item  {{\color{\colorMATH}\(\underline \varsigma  = \underline \sigma ,{\mfootnotesize{{{\color{\colorSYNTAX}\mtexttt{bitv}}}}}_{S}(\hat b)\)}}
        \item  {{\color{\colorMATH}\(\Phi ^{\prime} = [\Phi ,{\mtext{step}}_{{\mathcal{I}}}(N,\underline \sigma ,{\mfootnotesize{{{\color{\colorSYNTAX}\mtexttt{cast}}}}}_{S}({\mfootnotesize{{{\color{\colorSYNTAX}\mtexttt{flipv}}}}}(\hat b))) \mathrel{\dot =} \underline \sigma ,{\mfootnotesize{{{\color{\colorSYNTAX}\mtexttt{bitv}}}}}_{S}(\hat b)] = \Phi \)}}
        \item  {{\color{\colorMATH}\(\Psi _{e}^{\prime} \triangleq  \varnothing ,\{ \rho \mapsto \{ \hat b\} \}  = \Psi _{e}\)}}
        \item  {{\color{\colorMATH}\(\Sigma  \triangleq  \Sigma  \supseteq  \Sigma \)}}
        \item  {{\color{\colorMATH}\(\Psi ^{\prime} \triangleq  \Psi _{\sigma } \uplus  \Psi _{e} = \Psi _{\sigma } \uplus  \Psi _{e}^{\prime}\)}}
        \item  To show:
           \begin{enumerate}[leftmargin=15pt]\item  {{\color{\colorMATH}\(\Psi _{c}\uplus (\varnothing ,\{ \rho \mapsto \{ \hat b\} \} ),\Phi ,\Sigma  \vdash  \underline \sigma  \mathrel{;} \Psi _{\sigma }\)}}
           \item  {{\color{\colorMATH}\(\Psi _{c}\uplus \Psi _{\sigma },\Phi ,\Sigma  \vdash  {\mfootnotesize{{{\color{\colorSYNTAX}\mtexttt{bitv}}}}}_{S}(\hat b) \mathrel{:} {\mfootnotesize{{{\color{\colorSYNTAX}\mtexttt{bit}}}}}_{S}^{\rho } \mathrel{;} \varnothing ,\{ \rho \mapsto \{ \hat b\} \} \)}}
           \item  {{\color{\colorMATH}\(\begin{array}[t]{l
               } \forall  \rho ^{\prime},\hat b^{\prime}.\hspace*{0.33em} 
               \cr  \Psi _{c}\setminus (\{ \hat b^{\prime}\} ,\varnothing )\uplus \Psi _{\sigma }\uplus (\varnothing ,\{ \rho \mapsto \{ \hat b\} \} ),\Phi ,\Sigma  \vdash  \hat b^{\prime} \mathrel{:} {\mfootnotesize{{{\color{\colorSYNTAX}\mtexttt{flip}}}}}^{\rho ^{\prime}} 
               \cr  \implies   
               \cr  \Psi _{c}\setminus (\{ \hat b^{\prime}\} ,\varnothing )\uplus \Psi _{\sigma }\uplus (\varnothing ,\{ \rho \mapsto \{ \hat b\} \} ),\Phi ,\Sigma  \vdash  \hat b^{\prime} \mathrel{:} {\mfootnotesize{{{\color{\colorSYNTAX}\mtexttt{flip}}}}}^{\rho ^{\prime}}
               \end{array}\)}}
           \item  {{\color{\colorMATH}\(\forall  \hat b \in  \Psi _{c}\uplus \Psi _{\sigma }\uplus (\varnothing ,\{ \rho \mapsto \{ \hat b\} \} ).\hspace*{0.33em} \hat b \mathrel{\bot \!\!\!\bot } \{ {\mtext{bit}}(N^{\prime}) \mathrel{|} N^{\prime} \geq  N+1\} \)}}
           \end{enumerate}
        \item  (a) is by assumption
        \item  (b) is immediate
        \item  (c) is immediate
        \item  (d) is by assumption
        \end{itemize}
     \item  \begin{itemize}[label={},leftmargin=0pt]\item  {{\color{\colorMATH}\(\underline e = {\mfootnotesize{{{\color{\colorSYNTAX}\mtexttt{mux}}}}}({\mfootnotesize{{{\color{\colorSYNTAX}\mtexttt{bitv}}}}}_{S}(\hat b_{1}),{\mfootnotesize{{{\color{\colorSYNTAX}\mtexttt{bitv}}}}}_{S}(\hat b_{2}),{\mfootnotesize{{{\color{\colorSYNTAX}\mtexttt{bitv}}}}}_{S}(\hat b_{3}))\)}}
        \item  By inversion:
        \item  \fbox{\footnotesize{{\color{\colorMATH}\(
           \inferrule*[vcenter,lab=
           ]{ \inferrule*[vcenter,lab=
              ]{ }{
                 \Psi _{c} \uplus  \Psi _{\sigma },\Phi ,\Sigma  \vdash  {\mfootnotesize{{{\color{\colorSYNTAX}\mtexttt{bitv}}}}}_{S}(\hat b_{1}) \mathrel{:} {\mfootnotesize{{{\color{\colorSYNTAX}\mtexttt{bit}}}}}_{S}^{\rho _{1}} \mathrel{;} \varnothing ,\{ \rho _{1}{\mapsto }\{ \hat b_{1}\} \} 
              }
           \\\\
           \\\\ \inferrule*[vcenter,lab=
              ]{ }{
                 \Psi _{c} \uplus  \Psi _{\sigma },\Phi ,\Sigma  \vdash  {\mfootnotesize{{{\color{\colorSYNTAX}\mtexttt{bitv}}}}}_{S}(\hat b_{2}) \mathrel{:} {\mfootnotesize{{{\color{\colorSYNTAX}\mtexttt{bit}}}}}_{S}^{\rho _{2}} \mathrel{;} \varnothing ,\{ \rho _{2}{\mapsto }\{ \hat b_{2}\} \} 
              }
           \\ \inferrule*[vcenter,lab=
              ]{ }{
                 \Psi _{c} \uplus  \Psi _{\sigma },\Phi ,\Sigma  \vdash  {\mfootnotesize{{{\color{\colorSYNTAX}\mtexttt{bitv}}}}}_{S}(\hat b_{3}) \mathrel{:} {\mfootnotesize{{{\color{\colorSYNTAX}\mtexttt{bit}}}}}_{S}^{\rho _{3}} \mathrel{;} \varnothing ,\{ \rho _{3}{\mapsto }\{ \hat b_{3}\} \} 
              }
              }{
              \Psi _{c} \uplus  \Psi _{\sigma },\Phi ,\Sigma ,\varnothing  \vdash  {\mfootnotesize{{{\color{\colorSYNTAX}\mtexttt{mux}}}}}({\mfootnotesize{{{\color{\colorSYNTAX}\mtexttt{bitv}}}}}_{S}(\hat b_{1}),{\mfootnotesize{{{\color{\colorSYNTAX}\mtexttt{bitv}}}}}_{S}(\hat b_{2}),{\mfootnotesize{{{\color{\colorSYNTAX}\mtexttt{bitv}}}}}_{S}(\hat b_{3})) \mathrel{:} {\mfootnotesize{{{\color{\colorSYNTAX}\mtexttt{bit}}}}}_{S}^{\rho _{1}\sqcup \rho _{2}\sqcup \rho _{3}} \times  {\mfootnotesize{{{\color{\colorSYNTAX}\mtexttt{bit}}}}}_{S}^{\rho _{1}\sqcup \rho _{2}\sqcup \rho _{3}}\mathrel{;} \oslash  , \varnothing ,\{ \rho _{i} \mapsto  \{ \hat b_{i}\} \} 
           }
           \)}}}
        \item  {{\color{\colorMATH}\(\tau  = {\mfootnotesize{{{\color{\colorSYNTAX}\mtexttt{bit}}}}}_{S}^{\rho _{1}\sqcup \rho _{2}\sqcup \rho _{3}} \times  {\mfootnotesize{{{\color{\colorSYNTAX}\mtexttt{bit}}}}}_{S}^{\rho _{1}\sqcup \rho _{2}\sqcup \rho _{3}}\)}}
        \item  {{\color{\colorMATH}\(\Psi _{e} = \varnothing ,\{ \rho _{i} \mapsto  \{ \hat b_{i}\} \} \)}}
        \item  {{\color{\colorMATH}\(\underline \varsigma  = \underline \sigma ,\langle {\mfootnotesize{{{\color{\colorSYNTAX}\mtexttt{bitv}}}}}_{S}(\widehat {\mtext{cond}}(\hat b_{1},\hat b_{2},\hat b_{3})),{\mfootnotesize{{{\color{\colorSYNTAX}\mtexttt{bitv}}}}}_{S}(\widehat {\mtext{cond}}(\hat b_{1},\hat b_{3},\hat b_{2}))\rangle \)}}
        \item  {{\color{\colorMATH}\(\Phi ^{\prime} = [\Phi ,\text{\smaller\smaller\({\mtext{step}}_{{\mathcal{I}}}(N,\underline \sigma ,{\mfootnotesize{{{\color{\colorSYNTAX}\mtexttt{mux}}}}}({\mfootnotesize{{{\color{\colorSYNTAX}\mtexttt{bitv}}}}}_{S}(\hat b_{1}),{\mfootnotesize{{{\color{\colorSYNTAX}\mtexttt{bitv}}}}}_{S}(\hat b_{2}),{\mfootnotesize{{{\color{\colorSYNTAX}\mtexttt{bitv}}}}}_{S}(\hat b_{3}))) \mathrel{\dot =} \underline \sigma ,\langle {\mfootnotesize{{{\color{\colorSYNTAX}\mtexttt{bitv}}}}}_{S}(\widehat {\mtext{cond}}(\hat b_{1},\hat b_{2},\hat b_{3})),{\mfootnotesize{{{\color{\colorSYNTAX}\mtexttt{bitv}}}}}_{S}(\widehat {\mtext{cond}}(\hat b_{1},\hat b_{3},\hat b_{2}))\rangle \)}] = \Phi \)}}
        \item  {{\color{\colorMATH}\(\Psi _{e}^{\prime} = \varnothing ,\{ \rho _{1}\sqcup \rho _{2}\sqcup \rho _{3}\mapsto \{ \widehat {\mtext{cond}}(\hat b_{1},\hat b_{2},\hat b_{3}),\widehat {\mtext{cond}}(\hat b_{1},\hat b_{3},\hat b_{2})\} \} \)}}
        \item  {{\color{\colorMATH}\(\Sigma  \triangleq  \Sigma  \supseteq  \Sigma \)}}
        \item  {{\color{\colorMATH}\(\Psi ^{\prime} \triangleq  \Psi _{\sigma } \uplus  (\varnothing ,\{ \rho _{1}\sqcup \rho _{2}\sqcup \rho _{3} \mapsto  \{ \widehat {\mtext{cond}}(\hat b_{1},\hat b_{2},\hat b_{3}),\widehat {\mtext{cond}}(\hat b_{1},\hat b_{3},\hat b_{2})\} \} ) = \Psi _{\sigma } \uplus  \Psi _{e}^{\prime}\)}}
        \item  To show:
           \begin{enumerate}[leftmargin=15pt]\item  {{\color{\colorMATH}\(\Psi _{c}\uplus \Psi _{e}^{\prime},\Phi ,\Sigma  \vdash  \underline \sigma  \mathrel{;} \Psi _{\sigma }\)}}
           \item  {{\color{\colorMATH}\(\Psi _{c}\uplus \Psi _{\sigma },\Phi ,\Sigma  \vdash  \langle {\mfootnotesize{{{\color{\colorSYNTAX}\mtexttt{bitv}}}}}_{S}(\widehat {\mtext{cond}}(\hat b_{1},\hat b_{2},\hat b_{3})),{\mfootnotesize{{{\color{\colorSYNTAX}\mtexttt{bitv}}}}}_{S}(\widehat {\mtext{cond}}(\hat b_{1},\hat b_{3},\hat b_{2}))\rangle  \mathrel{:} {\mfootnotesize{{{\color{\colorSYNTAX}\mtexttt{bit}}}}}_{S}^{\rho _{1}\sqcup \rho _{2}\sqcup \rho _{3}} \times  {\mfootnotesize{{{\color{\colorSYNTAX}\mtexttt{bit}}}}}_{S}^{\rho _{1}\sqcup \rho _{2}\sqcup \rho _{3}} \mathrel{;} \Psi _{e}^{\prime}\)}}
           \item  {{\color{\colorMATH}\(\begin{array}[t]{l
               } \forall  \rho ^{\prime},\hat b^{\prime}.\hspace*{0.33em} 
               \cr  \Psi _{c}\setminus (\{ \hat b^{\prime}\} ,\varnothing )\uplus \Psi _{\sigma }\uplus \Psi _{e},\Phi ,\Sigma  \vdash  \hat b^{\prime} \mathrel{:} {\mfootnotesize{{{\color{\colorSYNTAX}\mtexttt{flip}}}}}^{\rho ^{\prime}} 
               \cr  \implies   
               \cr  \Psi _{c}\setminus (\{ \hat b^{\prime}\} ,\varnothing )\uplus \Psi _{\sigma }\uplus \Psi _{e}^{\prime},\Phi ,\Sigma  \vdash  \hat b^{\prime} \mathrel{:} {\mfootnotesize{{{\color{\colorSYNTAX}\mtexttt{flip}}}}}^{\rho ^{\prime}}
               \end{array}\)}}
           \item  {{\color{\colorMATH}\(\forall  \hat b \in  \Psi _{c}\uplus \Psi _{\sigma }\uplus \Psi _{e}^{\prime}.\hspace*{0.33em} \{ \hat b \mathrel{\bot \!\!\!\bot } {\mtext{bit}}(N^{\prime}) \mathrel{|} N^{\prime} \geq  N+1\} \)}}
           \end{enumerate}
        \item  (a) is by assumption
        \item  (b) is immediate
        \item  (c) is by \nameref{thm:proofs:type-preservation-mux-bits} applied to {{\color{\colorMATH}\(\Psi _{c} \uplus  \Psi _{\sigma }\)}}
        \item  (d) is by assumption and \nameref{thm:proofs:cond-independence}
        \end{itemize}
     \item  \begin{itemize}[label={},leftmargin=0pt]\item  {{\color{\colorMATH}\(\underline e = {\mfootnotesize{{{\color{\colorSYNTAX}\mtexttt{mux}}}}}({\mfootnotesize{{{\color{\colorSYNTAX}\mtexttt{bitv}}}}}_{S}(\hat b_{1}),{\mfootnotesize{{{\color{\colorSYNTAX}\mtexttt{flipv}}}}}(\hat b_{2}),{\mfootnotesize{{{\color{\colorSYNTAX}\mtexttt{flipv}}}}}(\hat b_{3}))\)}}
        \item  By inversion:
        \item  \fbox{\footnotesize{{\color{\colorMATH}\(
           \inferrule*[vcenter,lab=
           ]{ \inferrule*[vcenter,lab=
              ]{ }{
                 \Psi _{c} \uplus  \Psi _{\sigma },\Phi ,\Sigma  \vdash  {\mfootnotesize{{{\color{\colorSYNTAX}\mtexttt{bitv}}}}}_{S}(\hat b_{1}) \mathrel{:} {\mfootnotesize{{{\color{\colorSYNTAX}\mtexttt{bit}}}}}_{S}^{\rho _{1}} \mathrel{;} \varnothing ,\{ \rho _{1}{\mapsto }\{ \hat b_{1}\} \} 
              }
           \\ \rho _{1} \sqsubset  \rho _{2}
           \\ \rho _{1} \sqsubset  \rho _{3}
           \\\\
           \\\\ \inferrule*[vcenter,lab=
              ]{ \Psi _{c} \uplus  \Psi _{\sigma },\Phi  \vdash  \hat b_{2} \mathrel{:} {\mfootnotesize{{{\color{\colorSYNTAX}\mtexttt{flip}}}}}^{\rho _{2}}
                 }{
                 \Psi _{c} \uplus  \Psi _{\sigma },\Phi ,\Sigma  \vdash  {\mfootnotesize{{{\color{\colorSYNTAX}\mtexttt{flipv}}}}}(\hat b_{2}) \mathrel{:} {\mfootnotesize{{{\color{\colorSYNTAX}\mtexttt{flip}}}}}^{\rho _{2}} \mathrel{;} \{ \hat b_{2}\} ,\varnothing 
              }
           \\ \inferrule*[vcenter,lab=
              ]{ \Psi _{c} \uplus  \Psi _{\sigma },\Phi  \vdash  \hat b_{3} \mathrel{:} {\mfootnotesize{{{\color{\colorSYNTAX}\mtexttt{flip}}}}}^{\rho _{3}}
                 }{
                 \Psi _{c} \uplus  \Psi _{\sigma },\Phi ,\Sigma  \vdash  {\mfootnotesize{{{\color{\colorSYNTAX}\mtexttt{flipv}}}}}(\hat b_{3}) \mathrel{:} {\mfootnotesize{{{\color{\colorSYNTAX}\mtexttt{flip}}}}}^{\rho _{3}} \mathrel{;} \{ \hat b_{3}\} ,\varnothing 
              }
              }{
              \Psi _{c} \uplus  \Psi _{\sigma },\Phi ,\Sigma ,\varnothing  \vdash  {\mfootnotesize{{{\color{\colorSYNTAX}\mtexttt{mux}}}}}({\mfootnotesize{{{\color{\colorSYNTAX}\mtexttt{bitv}}}}}_{S}(\hat b_{1}),{\mfootnotesize{{{\color{\colorSYNTAX}\mtexttt{flipv}}}}}(\hat b_{2}),{\mfootnotesize{{{\color{\colorSYNTAX}\mtexttt{flipv}}}}}(\hat b_{3})) \mathrel{:} {\mfootnotesize{{{\color{\colorSYNTAX}\mtexttt{flip}}}}}^{\rho _{1}\sqcup \rho _{2}\sqcup \rho _{3}} \times  {\mfootnotesize{{{\color{\colorSYNTAX}\mtexttt{flip}}}}}^{\rho _{1}\sqcup \rho _{2}\sqcup \rho _{3}}\mathrel{;} \oslash  , \{ \hat b_{2},\hat b_{3}\} ,\{ \rho _{1}{\mapsto }\{ \hat b_{1}\} \} 
           }
           \)}}}
        \item  {{\color{\colorMATH}\(\tau  = {\mfootnotesize{{{\color{\colorSYNTAX}\mtexttt{flip}}}}}^{\rho _{1}\sqcup \rho _{2}\sqcup \rho _{3}} \times  {\mfootnotesize{{{\color{\colorSYNTAX}\mtexttt{flip}}}}}^{\rho _{1}\sqcup \rho _{2}\sqcup \rho _{3}}\)}}
        \item  {{\color{\colorMATH}\(\Psi _{e} = \{ \hat b_{2},\hat b_{3}\} ,\{ \rho _{1}{\mapsto }\{ \hat b_{1}\} \} \)}}
        \item  {{\color{\colorMATH}\(\underline \varsigma  = \underline \sigma ,\langle {\mfootnotesize{{{\color{\colorSYNTAX}\mtexttt{flipv}}}}}(\widehat {\mtext{cond}}(\hat b_{1},\hat b_{2},\hat b_{3})),{\mfootnotesize{{{\color{\colorSYNTAX}\mtexttt{flipv}}}}}(\widehat {\mtext{cond}}(\hat b_{1},\hat b_{3},\hat b_{2}))\rangle \)}}
        \item  {{\color{\colorMATH}\(\Phi ^{\prime} = [\Phi ,\text{\smaller\smaller\({\mtext{step}}_{{\mathcal{I}}}(N,\underline \sigma ,{\mfootnotesize{{{\color{\colorSYNTAX}\mtexttt{mux}}}}}({\mfootnotesize{{{\color{\colorSYNTAX}\mtexttt{bitv}}}}}_{S}(\hat b_{1}),{\mfootnotesize{{{\color{\colorSYNTAX}\mtexttt{flipv}}}}}(\hat b_{2}),{\mfootnotesize{{{\color{\colorSYNTAX}\mtexttt{flipv}}}}}(\hat b_{3}))) \mathrel{\dot =} \underline \sigma ,\langle {\mfootnotesize{{{\color{\colorSYNTAX}\mtexttt{flipv}}}}}(\widehat {\mtext{cond}}(\hat b_{1},\hat b_{2},\hat b_{3})),{\mfootnotesize{{{\color{\colorSYNTAX}\mtexttt{flipv}}}}}(\widehat {\mtext{cond}}(\hat b_{1},\hat b_{3},\hat b_{2}))\rangle \)}] = \Phi \)}}
        \item  {{\color{\colorMATH}\(\Psi _{e}^{\prime} = \{ \widehat {\mtext{cond}}(\hat b_{1},\hat b_{2},\hat b_{3}),\widehat {\mtext{cond}}(\hat b_{1},\hat b_{3},\hat b_{2})\} ,\varnothing \)}}
        \item  {{\color{\colorMATH}\(\Sigma  \triangleq  \Sigma  \supseteq  \Sigma \)}}
        \item  {{\color{\colorMATH}\(\Psi ^{\prime} \triangleq  \Psi _{\sigma } \uplus  (\{ \widehat {\mtext{cond}}(\hat b_{1},\hat b_{2},\hat b_{3}),\widehat {\mtext{cond}}(\hat b_{1},\hat b_{3},\hat b_{2})\} ,\varnothing ) = \Psi _{\sigma } \uplus  \Psi _{e}^{\prime}\)}}
        \item  To show:
           \begin{enumerate}[leftmargin=15pt]\item  {{\color{\colorMATH}\(\Psi _{c}\uplus \Psi _{e}^{\prime},\Phi ,\Sigma  \vdash  \underline \sigma  \mathrel{;} \Psi _{\sigma }\)}}
           \item  {{\color{\colorMATH}\(\Psi _{c}\uplus \Psi _{\sigma },\Phi ,\Sigma  \vdash  \langle {\mfootnotesize{{{\color{\colorSYNTAX}\mtexttt{flipv}}}}}(\widehat {\mtext{cond}}(\hat b_{1},\hat b_{2},\hat b_{3})),{\mfootnotesize{{{\color{\colorSYNTAX}\mtexttt{flipv}}}}}(\widehat {\mtext{cond}}(\hat b_{1},\hat b_{3},\hat b_{2}))\rangle  \mathrel{:} {\mfootnotesize{{{\color{\colorSYNTAX}\mtexttt{flip}}}}}^{\rho _{1}\sqcup \rho _{2}\sqcup \rho _{3}} \times  {\mfootnotesize{{{\color{\colorSYNTAX}\mtexttt{flip}}}}}^{\rho _{1}\sqcup \rho _{2}\sqcup \rho _{3}} \mathrel{;} \Psi _{e}^{\prime}\)}}
           \item  {{\color{\colorMATH}\(\begin{array}[t]{l
               } \forall  \rho ^{\prime},\hat b^{\prime}.\hspace*{0.33em} 
               \cr  \Psi _{c}\setminus (\{ \hat b^{\prime}\} ,\varnothing )\uplus \Psi _{\sigma }\uplus \Psi _{e},\Phi ,\Sigma  \vdash  \hat b^{\prime} \mathrel{:} {\mfootnotesize{{{\color{\colorSYNTAX}\mtexttt{flip}}}}}^{\rho ^{\prime}} 
               \cr  \implies   
               \cr  \Psi _{c}\setminus (\{ \hat b^{\prime}\} ,\varnothing )\uplus \Psi _{\sigma }\uplus \Psi _{e}^{\prime},\Phi ,\Sigma  \vdash  \hat b^{\prime} \mathrel{:} {\mfootnotesize{{{\color{\colorSYNTAX}\mtexttt{flip}}}}}^{\rho ^{\prime}}
               \end{array}\)}}
           \item  {{\color{\colorMATH}\(\forall  \hat b \in  \Psi _{c}\uplus \Psi _{\sigma }\uplus \Psi _{e}^{\prime}.\hspace*{0.33em} \{ \hat b \mathrel{\bot \!\!\!\bot } {\mtext{bit}}(N^{\prime}) \mathrel{|} N^{\prime} \geq  N+1\} \)}}
           \end{enumerate}
        \item  (a) is by \nameref{thm:proofs:weaken-store} and \nameref{thm:proofs:type-preservation-mux-flip} applied to {{\color{\colorMATH}\(\Psi _{c}\)}}
        \item  (b-c) are by \nameref{thm:proofs:type-preservation-mux-flip} applied to {{\color{\colorMATH}\(\Psi _{c} \uplus  \Psi _{\sigma }\)}}
        \item  (d) is by assumption and \nameref{thm:proofs:cond-independence}
        \end{itemize}
     \item  \begin{itemize}[label={},leftmargin=0pt]\item  {{\color{\colorMATH}\(\underline e = {\mfootnotesize{{{\color{\colorSYNTAX}\mtexttt{xor}}}}}({\mfootnotesize{{{\color{\colorSYNTAX}\mtexttt{bitv}}}}}_{S}(\hat b_{1}),{\mfootnotesize{{{\color{\colorSYNTAX}\mtexttt{flipv}}}}}(\hat b_{2}))\)}}
        \item  Analogous to mux-flip case
        \end{itemize}
     \item  \begin{itemize}[label={},leftmargin=0pt]\item  {{\color{\colorMATH}\(\underline e = {\mfootnotesize{{{\color{\colorSYNTAX}\mtexttt{let}}}}}\hspace*{0.33em}x=\underline v\hspace*{0.33em}{\mfootnotesize{{{\color{\colorSYNTAX}\mtexttt{in}}}}}\hspace*{0.33em}\underline e\)}}
        \item  By inversion:
        \item  \fbox{{{\color{\colorMATH}\(
           \inferrule*[vcenter,lab=
           ]{ \Psi _{c} \uplus  \Psi _{\sigma } \uplus  \Psi _{e},\Phi ,\Sigma ,\varnothing  \vdash  \underline v \mathrel{:} \tau ^{\prime} \mathrel{;} \varnothing  , \Psi _{v}
           \\ \Psi _{c} \uplus  \Psi _{\sigma } \uplus  \Psi _{v},\Phi ,\Sigma ,[x\mapsto \tau ^{\prime}] \vdash  \underline e^{\prime} \mathrel{:} \tau  \mathrel{;} \underline{\hspace{0.66em}} , \Psi _{e}^{\prime}
              }{
              \Psi _{c} \uplus  \Psi _{\sigma },\Phi ,\Sigma ,\varnothing  \vdash  {\mfootnotesize{{{\color{\colorSYNTAX}\mtexttt{let}}}}}\hspace*{0.33em}x=\underline v\hspace*{0.33em}{\mfootnotesize{{{\color{\colorSYNTAX}\mtexttt{in}}}}}\hspace*{0.33em}\underline e^{\prime} \mathrel{:} \tau  \mathrel{;} \oslash  , \Psi _{v} \uplus  \Psi _{e}^{\prime}
           }
           \)}}}
        \item  {{\color{\colorMATH}\(\Psi _{e} = \Psi _{v} \uplus  \Psi _{e}^{\prime}\)}}
        \item  {{\color{\colorMATH}\(\underline \varsigma  = \underline \sigma ,[\underline v/x]\underline e\)}}
        \item  {{\color{\colorMATH}\(\Phi ^{\prime} = [\Phi ,{\mtext{step}}_{{\mathcal{I}}}(N,\underline \sigma ,{\mfootnotesize{{{\color{\colorSYNTAX}\mtexttt{let}}}}}\hspace*{0.33em}x=\underline v\hspace*{0.33em}{\mfootnotesize{{{\color{\colorSYNTAX}\mtexttt{in}}}}}\hspace*{0.33em}\underline e^{\prime})) \mathrel{\dot =} \underline \sigma ,[\underline v/x]\underline e] = \Phi \)}}
        \item  By \nameref{thm:proofs:type-preservation-substitution}:
        \item  \begin{itemize}[label=\textbf{-},leftmargin=*]\item  \begin{itemize}[label={},leftmargin=0pt]\item  There exists {{\color{\colorMATH}\(\Psi _{v}^{\prime}\)}} 
              \item  {S.t.}: {{\color{\colorMATH}\(\Psi _{v}^{\prime} \subseteq  \Psi _{v}\)}}
              \item  And: {{\color{\colorMATH}\(\Psi _{c},\Phi ,\Sigma ,\Gamma  \vdash  [\underline v/x]\underline e \mathrel{:} \tau _{2} \mathrel{;} \Gamma ^{\prime},\Psi _{v}^{\prime} \uplus  \Psi _{e}^{\prime}\)}}
              \end{itemize}
           \end{itemize}
        \item  {{\color{\colorMATH}\(\Sigma  \triangleq  \Sigma  \supseteq  \Sigma \)}}
        \item  {{\color{\colorMATH}\(\Psi ^{\prime} \triangleq  \Psi _{\sigma } \uplus  \Psi _{v}^{\prime} \uplus  \Psi _{e}^{\prime}\)}}
        \item  To show:
           \begin{enumerate}[leftmargin=15pt]\item  {{\color{\colorMATH}\(\Psi _{c}\uplus \Psi _{v}^{\prime}\uplus \Psi _{e}^{\prime},\Phi ,\Sigma  \vdash  \underline \sigma  \mathrel{;} \Psi _{\sigma }\)}}
           \item  {{\color{\colorMATH}\(\Psi _{c},\Phi ,\Sigma ,\Gamma  \vdash  [\underline v/x]\underline e \mathrel{:} \tau _{2} \mathrel{;} \Gamma ^{\prime},\Psi _{v}^{\prime} \uplus  \Psi _{e}^{\prime}\)}}
           \item  {{\color{\colorMATH}\(\begin{array}[t]{l
               } \forall  \rho ^{\prime},\hat b^{\prime}.\hspace*{0.33em} 
               \cr  \Psi _{c}\setminus (\{ \hat b^{\prime}\} ,\varnothing )\uplus \Psi _{\sigma }\uplus \Psi _{v}\uplus \Psi _{e}^{\prime},\Phi ,\Sigma  \vdash  \hat b^{\prime} \mathrel{:} {\mfootnotesize{{{\color{\colorSYNTAX}\mtexttt{flip}}}}}^{\rho ^{\prime}} 
               \cr  \implies   
               \cr  \Psi _{c}\setminus (\{ \hat b^{\prime}\} ,\varnothing )\uplus \Psi _{\sigma }\uplus \Psi _{v}^{\prime}\uplus \Psi _{e}^{\prime},\Phi ,\Sigma  \vdash  \hat b^{\prime} \mathrel{:} {\mfootnotesize{{{\color{\colorSYNTAX}\mtexttt{flip}}}}}^{\rho ^{\prime}}
               \end{array}\)}}
           \item  {{\color{\colorMATH}\(\forall  \hat b \in  \Psi _{c}\uplus \Psi _{\sigma }\uplus \Psi _{v}^{\prime}\uplus \Psi _{e}^{\prime}.\hspace*{0.33em} \{ \hat b \mathrel{\bot \!\!\!\bot } {\mtext{bit}}(N^{\prime}) \mathrel{|} N^{\prime} \geq  N+1\} \)}}
           \end{enumerate}
        \item  (a) is by \nameref{thm:proofs:weaken-store} and \nameref{thm:proofs:weaken-flip}
        \item  (b) is by \nameref{thm:proofs:type-preservation-substitution} 
        \item  (c) is by \nameref{thm:proofs:weaken-flip}
        \item  (d) is by assumed bit independence
        \end{itemize}
     \item  \begin{itemize}[label={},leftmargin=0pt]\item  {{\color{\colorMATH}\(\underline e = {\mfootnotesize{{{\color{\colorSYNTAX}\mtexttt{let}}}}}\hspace*{0.33em}x,y=\underline v\hspace*{0.33em}{\mfootnotesize{{{\color{\colorSYNTAX}\mtexttt{in}}}}}\hspace*{0.33em}\underline e\)}} and {{\color{\colorMATH}\(\underline e = ({\mfootnotesize{{{\color{\colorSYNTAX}\mtexttt{fun}}}}}_{y}(x\mathrel{:}\tau ).\hspace*{0.33em}\underline e)(\underline v)\)}}
        \item  Analogous to single-variable let-binding case
        \end{itemize}
     \end{enumerate}
  \end{itemize}
\end{proof}

\noindent
\begin{minipage}{\linewidth}
\begin{lemma}[Type Preservation: Flip]\label{thm:proofs:type-preservation-flip}\ 
  \begin{itemize}[label={},leftmargin=0pt]\item  If: \hspace*{0.16em}{{\color{\colorMATH}\(\forall  \hat b \in  \Psi ^{F},\Psi ^{B}.\hspace*{0.33em} \hat b \mathrel{\bot \!\!\!\bot } \{ {\mtext{bit}}(N^{\prime}) \mathrel{|} N^{\prime} \geq  N\} \)}}
  \item  Then: \hspace*{0.16em}{{\color{\colorMATH}\(\forall  \rho ^{\prime},\hat b^{\prime}.\hspace*{0.33em} \Psi ^{F}\setminus \{ \hat b\} ,\Psi ^{B},\Phi  \vdash  \hat b^{\prime} \mathrel{:} {\mfootnotesize{{{\color{\colorSYNTAX}\mtexttt{flip}}}}}^{\rho ^{\prime}} \implies    \Psi \setminus \{ \hat b\} \uplus \{ {\mtext{bit}}(N)\} ,\Psi ^{B},\Phi  \vdash  \hat b^{\prime} \mathrel{:} {\mfootnotesize{{{\color{\colorSYNTAX}\mtexttt{flip}}}}}^{\rho ^{\prime}}\)}}
  \end{itemize}
\end{lemma}
\end{minipage}
\begin{proof}
  \begin{itemize}[label={},leftmargin=0pt]\item  Assume some {{\color{\colorMATH}\(\rho ^{\prime}\)}}, {{\color{\colorMATH}\(\hat b^{\prime}\)}} where {{\color{\colorMATH}\(\Psi ^{F}\setminus \{ \hat b^{\prime}\} ,\Psi ^{B},\Phi  \vdash  \hat b^{\prime} \mathrel{:} {\mfootnotesize{{{\color{\colorSYNTAX}\mtexttt{flip}}}}}^{\rho ^{\prime}}\)}}
  \item  By inversion:
  \item  \begin{itemize}[label=\textbf{-},leftmargin=*]\item  {{\color{\colorMATH}\({\mtext{Pr}}\left[\hat b^{\prime} \mathrel{\dot =} {\mfootnotesize{{{\color{\colorSYNTAX}\mtexttt{I}}}}} \mathrel{}\middle|\mathrel{} \Phi  \right] = \nicefrac{1}{2} \)}}
     \item  {{\color{\colorMATH}\(\left[\hat b^{\prime} \mathrel{\bot \!\!\!\bot } \Psi ^{F}\setminus \{ \hat b^{\prime}\} ,\Psi ^{B}(\{ \rho ^{\prime \prime} \mathrel{|} \rho ^{\prime \prime} \sqsubset  \rho ^{\prime}\} ) \mathrel{}\middle|\mathrel{} \Phi \right]\)}}
     \end{itemize}
  \item  STS:
     \begin{itemize}[label=\textbf{-},leftmargin=*]\item  {{\color{\colorMATH}\(\left[\hat b^{\prime} \mathrel{\bot \!\!\!\bot } {\mtext{bit}}(N),\Psi ^{F}\setminus \{ \hat b^{\prime}\} ,\Psi ^{B}(\{ \rho ^{\prime \prime} \mathrel{|} \rho ^{\prime \prime} \sqsubset  \rho ^{\prime}\} ) \mathrel{}\middle|\mathrel{} \Phi \right]\)}}
     \end{itemize}
  \item  By assumption of bit independence and second inversion fact
  \end{itemize}
\end{proof}

\noindent
\begin{minipage}{\linewidth}
\begin{lemma}[Type Preservation: CastP]
\label{thm:proofs:type-preservation-castp}\ 
  \begin{itemize}[label={},leftmargin=0pt]\item  If: \hspace*{0.16em}{{\color{\colorMATH}\(\Psi ^{F},\Psi ^{B},\Phi  \vdash  \hat b \mathrel{:} {\mfootnotesize{{{\color{\colorSYNTAX}\mtexttt{flip}}}}}^{\rho }\)}}
  \item  Then:
     \begin{enumerate}[leftmargin=15pt]\item  {{\color{\colorMATH}\(\Psi _{c},[\Phi ,\hat b \mathrel{\dot =} b] \vdash  {\mtext{return}}(b) \mathrel{:} {\mfootnotesize{{{\color{\colorSYNTAX}\mtexttt{bit}}}}}_{P}^{\bot } \mathrel{;} \varnothing \)}}
     \item  {{\color{\colorMATH}\(\forall  \rho ^{\prime},\hat b^{\prime}.\hspace*{0.33em} \Psi ^{F}\setminus \{ \hat b^{\prime}\} \uplus \{ \hat b\} ,\Psi ^{B},\Phi  \vdash  \hat b^{\prime} \mathrel{:} {\mfootnotesize{{{\color{\colorSYNTAX}\mtexttt{flip}}}}}^{\rho ^{\prime}} \implies    \Psi ^{F}\setminus \{ \hat b^{\prime}\} ,\Psi ^{B},[\Phi ,\hat b\mathrel{\dot =}b] \vdash  \hat b^{\prime} \mathrel{:} {\mfootnotesize{{{\color{\colorSYNTAX}\mtexttt{flip}}}}}^{\rho ^{\prime}}\)}}
     \end{enumerate}
  \end{itemize}
\end{lemma}
\end{minipage}
\begin{proof}
\begin{enumerate}[leftmargin=15pt]\item  Immediate by constructing type derivation
\item  \begin{itemize}[label={},leftmargin=0pt]\item  Assume some {{\color{\colorMATH}\(\rho ^{\prime}\)}} and {{\color{\colorMATH}\(\hat b^{\prime}\)}} where {{\color{\colorMATH}\(\Psi ^{F}\setminus \{ \hat b^{\prime}\} \uplus \{ \hat b\} ,\Psi ^{B},\Phi  \vdash  \hat b^{\prime} \mathrel{:} {\mfootnotesize{{{\color{\colorSYNTAX}\mtexttt{flip}}}}}^{\rho ^{\prime}}\)}}
   \item  By inversion: 
   \item  \begin{itemize}[label=\textbf{-},leftmargin=*]\item  {{\color{\colorMATH}\({\mtext{Pr}}\left[\hat b^{\prime} \mathrel{\dot =} {\mfootnotesize{{{\color{\colorSYNTAX}\mtexttt{I}}}}} \mathrel{}\middle|\mathrel{} \Phi  \right] = \nicefrac{1}{2} \)}} {\mtextit{(H1)}}
      \item  {{\color{\colorMATH}\(\left[\hat b^{\prime} \mathrel{\bot \!\!\!\bot } \hat b,\Psi ^{F}\setminus \{ \hat b^{\prime}\} ,\Psi ^{B}(\{ \rho ^{\prime \prime} \mathrel{|} \rho ^{\prime \prime} \sqsubset  \rho ^{\prime}\} ) \mathrel{}\middle|\mathrel{} \Phi  \right]\)}} {\mtextit{(H2)}}
      \end{itemize}
   \item  STS:
      \begin{enumerate}[leftmargin=15pt]\item  {{\color{\colorMATH}\({\mtext{Pr}}\left[\hat b^{\prime} \mathrel{\dot =} {\mfootnotesize{{{\color{\colorSYNTAX}\mtexttt{I}}}}} \mathrel{}\middle|\mathrel{} \Phi ,\hat b\mathrel{\dot =}b \right] = \nicefrac{1}{2} \)}} 
      \item  {{\color{\colorMATH}\(\left[\hat b^{\prime} \mathrel{\bot \!\!\!\bot } \Psi ^{F}\setminus \{ \hat b^{\prime}\} ,\Psi ^{B}(\{ \rho ^{\prime \prime} \mathrel{|} \rho ^{\prime \prime} \sqsubset  \rho ^{\prime}\} ) \mathrel{}\middle|\mathrel{} \Phi ,\hat b\mathrel{\dot =}b \right]\)}}
      \end{enumerate}
   \item  {\mtextit{(a)}} is by \nameref{thm:proofs:decomposition} applied {\mtextit{(H2)}} to establish {{\color{\colorMATH}\(\left[\hat b^{\prime} \mathrel{\bot \!\!\!\bot } \hat b \mathrel{}\middle|\mathrel{} \Phi  \right]\)}}, which is then applied to {\mtextit{(H1)}}
   \item  {\mtextit{(b)}} is by \nameref{thm:proofs:weak-union} applied to {\mtextit{(H2)}}, moving {{\color{\colorMATH}\(\hat b\)}} from the RHS of independence into the condition
   \end{itemize}
\end{enumerate}
\end{proof}

\noindent
\begin{minipage}{\linewidth}
\begin{lemma}[Type Preservation: Mux BitS]\label{thm:proofs:type-preservation-mux-bits}\ 
  \begin{itemize}[label={},leftmargin=0pt]\item  If: \hspace*{0.16em}{{\color{\colorMATH}\(\Psi ^{F}\setminus \{ \hat b^{\prime}\} ,\Psi ^{B}\cup \{ \rho _{1}{\mapsto }\{ \hat b_{1}\} ,\rho _{2}{\mapsto }\{ \hat b_{2}\} ,\rho _{3}{\mapsto }\{ \hat b_{3}\} \} ,\Phi  \vdash  \hat b^{\prime} \mathrel{:} {\mfootnotesize{{{\color{\colorSYNTAX}\mtexttt{flip}}}}}^{\rho ^{\prime}}\)}}
  \item  Then: \hspace*{0.16em}{{\color{\colorMATH}\(\Psi ^{F}\setminus \{ \hat b^{\prime}\} ,\Psi ^{B}\cup \{ \rho _{1}\sqcup \rho _{2}\sqcup \rho _{3}{\mapsto }\{ \widehat {\mtext{cond}}(\hat b_{1},\hat b_{2},\hat b_{3})\} \} ,\Phi  \vdash  \hat b^{\prime} \mathrel{:} {\mfootnotesize{{{\color{\colorSYNTAX}\mtexttt{flip}}}}}^{\rho ^{\prime}}\)}}
  \end{itemize}
\end{lemma}
\end{minipage}
\begin{proof}
  \begin{itemize}[label={},leftmargin=0pt]\item  By inversion:
  \item  \begin{itemize}[label=\textbf{-},leftmargin=*]\item  {{\color{\colorMATH}\({\mtext{Pr}}\left[\hat b^{\prime} \mathrel{\dot =} {\mfootnotesize{{{\color{\colorSYNTAX}\mtexttt{I}}}}} \mathrel{}\middle|\mathrel{} \Phi  \right] = \nicefrac{1}{2} \)}}
     \item  {{\color{\colorMATH}\(\left[\hat b^{\prime} \mathrel{\bot \!\!\!\bot } \hat b,\Psi ^{F}\setminus \{ \hat b^{\prime}\} ,(\Psi ^{B}\cup \{ \rho _{1}{\mapsto }\{ \hat b_{1}\} ,\rho _{2}{\mapsto }\{ \hat b_{2}\} ,\rho _{3}{\mapsto }\{ \hat b_{3}\} \} )(\{ \rho ^{\prime \prime} \mathrel{|} \rho ^{\prime \prime} \sqsubset  \rho ^{\prime}\} ) \mathrel{}\middle|\mathrel{} \Phi  \right]\)}} {\mtextit{(H)}}
     \end{itemize}
  \item  STS:
  \item  \begin{itemize}[label=\textbf{-},leftmargin=*]\item  {{\color{\colorMATH}\(\left[\hat b^{\prime} \mathrel{\bot \!\!\!\bot } {\mtext{bit}}(N),\Psi ^{F}\setminus \{ \hat b^{\prime}\} ,(\Psi ^{B}\cup \{ \rho _{1}\sqcup \rho _{2}\sqcup \rho _{3}{\mapsto }\{ \widehat {\mtext{cond}}(\hat b_{1},\hat b_{2},\hat b_{3})\} \} )(\{ \rho ^{\prime \prime} \mathrel{|} \rho ^{\prime \prime} \sqsubset  \rho ^{\prime}\} ) \mathrel{}\middle|\mathrel{} \Phi \right]\)}}
     \end{itemize}
  \item  \begin{enumerate}[leftmargin=15pt]\item  \begin{itemize}[label={},leftmargin=0pt]\item  Case {{\color{\colorMATH}\(\rho _{1}\sqcup \rho _{2}\sqcup \rho _{3} \slashedrel\sqsubset  \rho ^{\prime}\)}}:
        \item  {{\color{\colorMATH}\((\Psi ^{B}\cup \{ \rho _{1}\sqcup \rho _{2}\sqcup \rho _{3}{\mapsto }\{ \widehat {\mtext{cond}}(\hat b_{1},\hat b_{2},\hat b_{3})\} \} )(\{ \rho ^{\prime \prime} \mathrel{|} \rho ^{\prime \prime} \sqsubset  \rho ^{\prime}\} ) = \Psi ^{B}\)}}
        \item  By {\mtextit{(H)}} and \nameref{thm:proofs:decomposition}
        \end{itemize}
     \item  \begin{itemize}[label={},leftmargin=0pt]\item  Case {{\color{\colorMATH}\(\rho _{1}\sqcup \rho _{2}\sqcup \rho _{3} \sqsubset  \rho ^{\prime}\)}}:
        \item  {{\color{\colorMATH}\((\Psi ^{B}\cup \{ \rho _{1}\sqcup \rho _{2}\sqcup \rho _{3}{\mapsto }\{ \widehat {\mtext{cond}}(\hat b_{1},\hat b_{2},\hat b_{3})\} \} )(\{ \rho ^{\prime \prime} \mathrel{|} \rho ^{\prime \prime} \sqsubset  \rho ^{\prime}\} ) = \Psi ^{B}(\{ \rho ^{\prime \prime} \mathrel{|} \rho ^{\prime \prime} \sqsubset  \rho ^{\prime}\} )\cup \{ \widehat {\mtext{cond}}(\hat b_{1},\hat b_{2},\hat b_{3})\} \)}}
        \item  {{\color{\colorMATH}\((\Psi ^{B}\cup \{ \rho _{1}{\mapsto }\{ \hat b_{1}\} ,\rho _{2}{\mapsto }\{ \hat b_{2}\} ,\rho _{3}{\mapsto }\{ \hat b_{3}\} \} )(\{ \rho ^{\prime \prime} \mathrel{|} \rho ^{\prime \prime} \sqsubset  \rho ^{\prime}\} ) = \Psi ^{B}(\{ \rho ^{\prime \prime} \mathrel{|} \rho ^{\prime \prime} \sqsubset  \rho ^{\prime}\} )\cup \{ \hat b_{1},\hat b_{2},\hat b_{3}\} \)}}
        \item  By {\mtextit{(H)}} and \nameref{thm:proofs:cond-independence}
        \end{itemize}
     \end{enumerate}
  \end{itemize}
\end{proof}

\noindent
\begin{minipage}{\linewidth}
\begin{lemma}[Type Preservation: Flip]\label{thm:proofs:type-preservation-mux-flip}\ 
  \begin{itemize}[label={},leftmargin=0pt]\item  If: \hspace*{0.16em}{{\color{\colorMATH}\(\Psi ^{F}\uplus \{ \hat b_{3}\} ,\Psi ^{B}\uplus \{ \rho _{1}\mapsto \{ \hat b_{1}\} \} ,\Phi  \vdash  \hat b_{2} \mathrel{:} {\mfootnotesize{{{\color{\colorSYNTAX}\mtexttt{flip}}}}}^{\rho _{2}}\)}}
  \item  And: \hspace*{0.16em}{{\color{\colorMATH}\(\Psi ^{F}\uplus \{ \hat b_{2}\} ,\Psi ^{B}\uplus \{ \rho _{1}\mapsto \{ \hat b_{1}\} \} ,\Phi  \vdash  \hat b_{3} \mathrel{:} {\mfootnotesize{{{\color{\colorSYNTAX}\mtexttt{flip}}}}}^{\rho _{3}}\)}}
  \item  And: \hspace*{0.16em}{{\color{\colorMATH}\(\rho _{1} \sqsubset  \rho _{2}\)}}\hspace*{0.16em} and \hspace*{0.16em}{{\color{\colorMATH}\(\rho _{1} \sqsubset  \rho _{3}\)}}
  \item  Then: 
  \item  \begin{enumerate}[leftmargin=15pt]\item  {{\color{\colorMATH}\(\Psi ^{F}\uplus \{ \widehat {\mtext{cond}}(\hat b_{1},\hat b_{3},\hat b_{2})\} ,\Psi ^{B},\Phi  \vdash  \widehat {\mtext{cond}}(\hat b_{1},\hat b_{2},\hat b_{3}) \mathrel{:} {\mfootnotesize{{{\color{\colorSYNTAX}\mtexttt{flip}}}}}^{\rho _{1}\sqcup \rho _{2}\sqcup \rho _{3}}\)}}
     \item  {{\color{\colorMATH}\(\Psi ^{F}\uplus \{ \widehat {\mtext{cond}}(\hat b_{1},\hat b_{2},\hat b_{3})\} ,\Psi ^{B},\Phi  \vdash  \widehat {\mtext{cond}}(\hat b_{1},\hat b_{3},\hat b_{2}) \mathrel{:} {\mfootnotesize{{{\color{\colorSYNTAX}\mtexttt{flip}}}}}^{\rho _{1}\sqcup \rho _{2}\sqcup \rho _{3}}\)}}
     \item  {{\color{\colorMATH}\(\begin{array}[t]{l
         } \forall  \rho ^{\prime},\hat b^{\prime}.\hspace*{0.33em} 
         \cr  \Psi ^{F}\setminus \{ \hat b^{\prime}\} \uplus \{ \hat b_{2},\hat b_{3}\} ,\Psi ^{B}\cup \{ \rho _{1}\mapsto \{ \hat b_{1}\} \} ,\Phi  \vdash  \hat b^{\prime} \mathrel{:} {\mfootnotesize{{{\color{\colorSYNTAX}\mtexttt{flip}}}}}^{\rho ^{\prime}} 
         \cr  \implies    
         \cr  \Psi ^{F}\setminus \{ \hat b^{\prime}\} \uplus \{ \widehat {\mtext{cond}}(\hat b_{1},\hat b_{2},\hat b_{3}),\widehat {\mtext{cond}}(\hat b_{1},\hat b_{3},\hat b_{2})\} ,\Psi ^{B},\Phi  \vdash  \hat b^{\prime} \mathrel{:} {\mfootnotesize{{{\color{\colorSYNTAX}\mtexttt{flip}}}}}^{\rho ^{\prime}}
         \end{array}\)}}
     \end{enumerate}
  \end{itemize}
\end{lemma}
\end{minipage}
\begin{proof}
  \begin{itemize}[label={},leftmargin=0pt]\item  By inversion:
  \item  \begin{itemize}[label=\textbf{-},leftmargin=*]\item  {{\color{\colorMATH}\({\mtext{Pr}}\left[ \hat b_{2} \mathrel{\dot =} {\mfootnotesize{{{\color{\colorSYNTAX}\mtexttt{I}}}}} \mathrel{}\middle|\mathrel{} \Phi  \right] = \nicefrac{1}{2} \)}} {\mtextit{(H11)}}
     \item  {{\color{\colorMATH}\(\left[ \hat b_{2} \mathrel{\bot \!\!\!\bot } \Psi ^{F}\uplus \{ \hat b_{3}\} ,(\Psi ^{B}\uplus \{ \rho _{1}\mapsto \{ \hat b_{1}\} \} )(\{ \rho ^{\prime} \mathrel{|} \rho ^{\prime} \sqsubset  \rho _{2}\} ) \mathrel{}\middle|\mathrel{} \Phi  \right]\)}} {\mtextit{(H12)}}
     \item  {{\color{\colorMATH}\({\mtext{Pr}}\left[ \hat b_{3} \mathrel{\dot =} {\mfootnotesize{{{\color{\colorSYNTAX}\mtexttt{I}}}}} \mathrel{}\middle|\mathrel{} \Phi  \right] = \nicefrac{1}{2} \)}} {\mtextit{(H21)}}
     \item  {{\color{\colorMATH}\(\left[ \hat b_{3} \mathrel{\bot \!\!\!\bot } \Psi ^{F}\uplus \{ \hat b_{2}\} ,(\Psi ^{B}\uplus \{ \rho _{1}\mapsto \{ \hat b_{1}\} \} )(\{ \rho ^{\prime} \mathrel{|} \rho ^{\prime} \sqsubset  \rho _{3}\} ) \mathrel{}\middle|\mathrel{} \Phi  \right]\)}} {\mtextit{(H22)}}
     \end{itemize}
  \item  By {{\color{\colorMATH}\(\rho _{1} \sqsubset  \rho _{2}\)}} and {{\color{\colorMATH}\(\rho _{1} \sqsubset  \rho _{3}\)}}:
  \item  \begin{itemize}[label=\textbf{-},leftmargin=*]\item  {{\color{\colorMATH}\(\left[ \hat b_{2} \mathrel{\bot \!\!\!\bot } \Psi ^{F}\uplus \{ \hat b_{3}\} ,\Psi ^{B}(\{ \rho ^{\prime} \mathrel{|} \rho ^{\prime} \sqsubset  \rho _{2}\} ),\hat b_{1} \mathrel{}\middle|\mathrel{} \Phi  \right]\)}} {\mtextit{(H13)}}
     \item  {{\color{\colorMATH}\(\left[ \hat b_{3} \mathrel{\bot \!\!\!\bot } \Psi ^{F}\uplus \{ \hat b_{2}\} ,\Psi ^{B}(\{ \rho ^{\prime} \mathrel{|} \rho ^{\prime} \sqsubset  \rho _{2}\} ),\hat b_{1} \mathrel{}\middle|\mathrel{} \Phi  \right]\)}} {\mtextit{(H23)}}
     \end{itemize}
  \item  \begin{enumerate}[leftmargin=15pt]\item  \begin{itemize}[label={},leftmargin=0pt]\item  STS:
        \item  \begin{enumerate}[leftmargin=15pt]\item  {{\color{\colorMATH}\({\mtext{Pr}}\left[ \widehat {\mtext{cond}}(\hat b_{1},\hat b_{2},\hat b_{3}) \mathrel{\dot =} {\mfootnotesize{{{\color{\colorSYNTAX}\mtexttt{I}}}}}\right] = \nicefrac{1}{2} \)}} {\mtextit{(i)}}
           \item  {{\color{\colorMATH}\(\left[ \widehat {\mtext{cond}}(\hat b_{1},\hat b_{2},\hat b_{3}) \mathrel{\bot \!\!\!\bot }  \Psi ^{F}\uplus \{ \widehat {\mtext{cond}}(\hat b_{1},\hat b_{3},\hat b_{2})\} ,\Psi ^{B}(\{ \rho ^{\prime} \mathrel{|} \rho ^{\prime} \sqsubset  \rho _{1}\sqcup \rho _{2}\sqcup \rho _{3}\} ) \mathrel{}\middle|\mathrel{} \Phi  \right]\)}} {\mtextit{(ii)}}
           \end{enumerate}
        \item  {\mtextit{(i)}} is by \nameref{thm:proofs:cond-stability} applied to {\mtextit{(H13)}} and \nameref{thm:proofs:decomposition} 
           (to achieve {{\color{\colorMATH}\(\left[ \hat b_{2} \mathrel{\bot \!\!\!\bot } \hat b_{1} \mathrel{}\middle|\mathrel{} \Phi  \right]\)}}), {\mtextit{(H23)}} and
           \nameref{thm:proofs:decomposition} (to achieve {{\color{\colorMATH}\(\left[ \hat b_{3} \mathrel{\bot \!\!\!\bot } \hat b_{1} \mathrel{}\middle|\mathrel{} \Phi  \right]\)}}), {\mtextit{(H11)}} and {\mtextit{(H21)}}
        \item  {\mtextit{(ii)}} is by:
        \item  {{\color{\colorMATH}\(\begin{array}[t]{rcl
            } &{} {}& {\mtext{Pr}}\left[ \widehat {\mtext{cond}}(\hat b_{1},\hat b_{2},\hat b_{3})  \mathrel{}\middle|\mathrel{} \Psi ^{F}\uplus \{ \widehat {\mtext{cond}}(\hat b_{1},\hat b_{3},\hat b_{2})\} ,\Psi ^{B}(\{ \rho ^{\prime} \mathrel{|} \rho ^{\prime} \sqsubset  \rho _{1}\sqcup \rho _{2}\sqcup \rho _{3}\} ) , \Phi  \right]
            \cr  &{}={}& \lbag {{\color{\colorTEXT}\textnormal{\hspace*{0.33em} \nameref{thm:proofs:total-probability} \hspace*{0.33em}}}}\rbag 
            \cr  &{} {}& {\mtext{Pr}}\left[\hat b_{2}  \mathrel{}\middle|\mathrel{} \hat b_{1}\mathrel{\dot =}{\mfootnotesize{{{\color{\colorSYNTAX}\mtexttt{I}}}}},\Psi ^{F}\uplus \{ \widehat {\mtext{cond}}(\hat b_{1},\hat b_{3},\hat b_{2})\} ,\Psi ^{B}(\{ \rho ^{\prime} \mathrel{|} \rho ^{\prime} \sqsubset  \rho _{1}\sqcup \rho _{2}\sqcup \rho _{3}\} ) , \Phi \right]{\mtext{Pr}}\left[\hat b_{1}\mathrel{\dot =}{\mfootnotesize{{{\color{\colorSYNTAX}\mtexttt{I}}}}}\right]
            \cr  &{} {}& +
            \cr  &{} {}& {\mtext{Pr}}\left[\hat b_{3}  \mathrel{}\middle|\mathrel{} \hat b_{1}\mathrel{\dot =}{\mfootnotesize{{{\color{\colorSYNTAX}\mtexttt{I}}}}},\Psi ^{F}\uplus \{ \widehat {\mtext{cond}}(\hat b_{1},\hat b_{3},\hat b_{2})\} ,\Psi ^{B}(\{ \rho ^{\prime} \mathrel{|} \rho ^{\prime} \sqsubset  \rho _{1}\sqcup \rho _{2}\sqcup \rho _{3}\} ) , \Phi \right]{\mtext{Pr}}\left[\hat b_{1}\mathrel{\dot =}{\mfootnotesize{{{\color{\colorSYNTAX}\mtexttt{I}}}}}\right]
            \cr  &{}={}& \lbag {{\color{\colorTEXT}\textnormal{\hspace*{0.33em} {\mtextit{(H13)}}, {\mtextit{(H23)}} and \nameref{thm:proofs:decomposition} \hspace*{0.33em}}}}\rbag 
            \cr  &{} {}& {\mtext{Pr}}\left[\hat b_{2}  \mathrel{}\middle|\mathrel{} \hat b_{1}\mathrel{\dot =}{\mfootnotesize{{{\color{\colorSYNTAX}\mtexttt{I}}}}}, \Phi \right]{\mtext{Pr}}\left[\hat b_{1}\mathrel{\dot =}{\mfootnotesize{{{\color{\colorSYNTAX}\mtexttt{I}}}}}\right] + {\mtext{Pr}}\left[\hat b_{3}  \mathrel{}\middle|\mathrel{} \hat b_{1}\mathrel{\dot =}{\mfootnotesize{{{\color{\colorSYNTAX}\mtexttt{I}}}}}, \Phi \right]{\mtext{Pr}}\left[\hat b_{1}\mathrel{\dot =}{\mfootnotesize{{{\color{\colorSYNTAX}\mtexttt{I}}}}}\right]
            \cr  &{}={}& \lbag {{\color{\colorTEXT}\textnormal{\hspace*{0.33em} \nameref{thm:proofs:total-probability} \hspace*{0.33em}}}}\rbag 
            \cr  &{} {}& {\mtext{Pr}}\left[ \widehat {\mtext{cond}}(\hat b_{1},\hat b_{2},\hat b_{3})  \mathrel{}\middle|\mathrel{}  \Phi  \right]
            \end{array}\)}}
        \end{itemize}
     \item  \begin{itemize}[label={},leftmargin=0pt]\item  STS:
        \item  \begin{enumerate}[leftmargin=15pt]\item  {{\color{\colorMATH}\({\mtext{Pr}}\left[ \widehat {\mtext{cond}}(\hat b_{1},\hat b_{3},\hat b_{2}) \mathrel{\dot =} {\mfootnotesize{{{\color{\colorSYNTAX}\mtexttt{I}}}}}\right] = \nicefrac{1}{2} \)}}
           \item  {{\color{\colorMATH}\(\left[ \widehat {\mtext{cond}}(\hat b_{1},\hat b_{3},\hat b_{2}) \mathrel{\bot \!\!\!\bot }  \Psi ^{F}\uplus \{ \widehat {\mtext{cond}}(\hat b_{1},\hat b_{2},\hat b_{3})\} ,\Psi ^{B}(\{ \rho ^{\prime} \mathrel{|} \rho ^{\prime} \sqsubset  \rho _{1}\sqcup \rho _{2}\sqcup \rho _{3}\} ) \mathrel{}\middle|\mathrel{} \Phi  \right]\)}}
           \end{enumerate}
        \item  Analogous to previous cases
        \end{itemize}
     \item  \begin{itemize}[label={},leftmargin=0pt]\item  Assume {{\color{\colorMATH}\(\rho ^{\prime}\)}} and {{\color{\colorMATH}\(\hat b^{\prime}\)}} where {{\color{\colorMATH}\(\Psi ^{F}\setminus \{ \hat b^{\prime}\} \uplus \{ \hat b_{2},\hat b_{3}\} ,\Psi ^{B}\cup \{ \rho _{1}\mapsto \{ \hat b_{1}\} \} ,\Phi  \vdash  \hat b^{\prime} \mathrel{:} {\mfootnotesize{{{\color{\colorSYNTAX}\mtexttt{flip}}}}}^{\rho ^{\prime}}\)}}
        \item  By inversion:
        \item  \begin{itemize}[label=\textbf{-},leftmargin=*]\item  {{\color{\colorMATH}\({\mtext{Pr}}\left[\hat b^{\prime} \mathrel{\dot =} {\mfootnotesize{{{\color{\colorSYNTAX}\mtexttt{I}}}}}\right] = \nicefrac{1}{2} \)}}
           \item  {{\color{\colorMATH}\(\left[\hat b^{\prime} \mathrel{\bot \!\!\!\bot } \Psi ^{F}\setminus \{ \hat b^{\prime}\} \uplus \{ \hat b_{2},\hat b_{3}\} ,(\Psi ^{B}\cup \{ \rho _{1}\mapsto \{ \hat b_{1}\} \} )(\{ \rho ^{\prime \prime} \mathrel{|} \rho ^{\prime \prime} \sqsubset  \rho ^{\prime}\} ) \mathrel{}\middle|\mathrel{} \Phi  \right]\)}} {\mtextit{(i)}}
           \end{itemize}
        \item  STS:
           \begin{itemize}[label=\textbf{-},leftmargin=*]\item  {{\color{\colorMATH}\(\left[\hat b^{\prime} \mathrel{\bot \!\!\!\bot } \Psi ^{F}\setminus \{ \hat b^{\prime}\} \uplus \{ \widehat {\mtext{cond}}(\hat b_{1},\hat b_{2},\hat b_{3}),\widehat {\mtext{cond}}(\hat b_{1},\hat b_{3},\hat b_{2})\} ,\Psi ^{B}(\{ \rho ^{\prime \prime} \mathrel{|} \rho ^{\prime \prime} \sqsubset  \rho ^{\prime}\} ) \mathrel{}\middle|\mathrel{} \Phi \right]\)}}
           \end{itemize}
        \item  {{\color{\colorMATH}\(\begin{array}[t]{rcl
            } &{} {}& {\mtext{Pr}}\left[\hat b^{\prime} \mathrel{}\middle|\mathrel{} \Psi ^{F}\setminus \{ \hat b^{\prime}\} \uplus \{ \widehat {\mtext{cond}}(\hat b_{1},\hat b_{2},\hat b_{3}),\widehat {\mtext{cond}}(\hat b_{1},\hat b_{3},\hat b_{2})\} ,\Psi ^{B}(\{ \rho ^{\prime \prime} \mathrel{|} \rho ^{\prime \prime} \sqsubset  \rho ^{\prime}\} ) , \Phi \right]
            \cr  &{}={}& \lbag {{\color{\colorTEXT}\textnormal{\hspace*{0.33em} \nameref{thm:proofs:total-probability} \hspace*{0.33em}}}}\rbag 
            \cr  &{} {}& {\mtext{Pr}}\left[\hat b^{\prime} \mathrel{}\middle|\mathrel{} \hat b_{1}\mathrel{\dot =}{\mfootnotesize{{{\color{\colorSYNTAX}\mtexttt{I}}}}},\Psi ^{F}\setminus \{ \hat b^{\prime}\} \uplus \{ \hat b_{2},\hat b_{3}\} ,\Psi ^{B}(\{ \rho ^{\prime \prime} \mathrel{|} \rho ^{\prime \prime} \sqsubset  \rho ^{\prime}\} ) , \Phi \right]{\mtext{Pr}}\left[\hat b_{1} \mathrel{\dot =} {\mfootnotesize{{{\color{\colorSYNTAX}\mtexttt{I}}}}}\right]
            \cr  &{} {}& +
            \cr  &{} {}& {\mtext{Pr}}\left[\hat b^{\prime} \mathrel{}\middle|\mathrel{} \hat b_{1}\mathrel{\dot =}{\mfootnotesize{{{\color{\colorSYNTAX}\mtexttt{O}}}}},\Psi ^{F}\setminus \{ \hat b^{\prime}\} \uplus \{ \hat b_{2},\hat b_{3}\} ,\Psi ^{B}(\{ \rho ^{\prime \prime} \mathrel{|} \rho ^{\prime \prime} \sqsubset  \rho ^{\prime}\} ) , \Phi \right]{\mtext{Pr}}\left[\hat b_{1} \mathrel{\dot =} {\mfootnotesize{{{\color{\colorSYNTAX}\mtexttt{O}}}}}\right]
            \cr  &{}={}& \lbag {{\color{\colorTEXT}\textnormal{\hspace*{0.33em} {\mtextit{(i)}}, {\mtextit{(H13)}} and {\mtextit{(H23)}} \hspace*{0.33em}}}}\rbag 
            \cr  &{} {}& {\mtext{Pr}}\left[\hat b^{\prime} \mathrel{}\middle|\mathrel{} \hat b_{1}\mathrel{\dot =}{\mfootnotesize{{{\color{\colorSYNTAX}\mtexttt{I}}}}} , \Phi \right]{\mtext{Pr}}\left[\hat b_{1} \mathrel{\dot =} {\mfootnotesize{{{\color{\colorSYNTAX}\mtexttt{I}}}}}\right] + {\mtext{Pr}}\left[\hat b^{\prime} \mathrel{}\middle|\mathrel{} \hat b_{1}\mathrel{\dot =}{\mfootnotesize{{{\color{\colorSYNTAX}\mtexttt{O}}}}} , \Phi \right]{\mtext{Pr}}\left[\hat b_{1} \mathrel{\dot =} {\mfootnotesize{{{\color{\colorSYNTAX}\mtexttt{O}}}}}\right]
            \cr  &{}={}& \lbag {{\color{\colorTEXT}\textnormal{\hspace*{0.33em} \nameref{thm:proofs:total-probability} \hspace*{0.33em}}}}\rbag 
            \cr  &{} {}& {\mtext{Pr}}\left[\hat b^{\prime} \mathrel{}\middle|\mathrel{} \Phi \right]
            \end{array}\)}}
        \end{itemize}
     \end{enumerate}
  \end{itemize}
\end{proof}

\noindent
\begin{minipage}{\linewidth}
\begin{lemma}[Type Preservation: Substitution]
\label{thm:proofs:type-preservation-substitution}\ 
  \begin{itemize}[label={},leftmargin=0pt]\item  If: \hspace*{0.16em}{{\color{\colorMATH}\({\mathcal{K}}(\tau _{1}) = {\mfootnotesize{{{\color{\colorSYNTAX}\mtexttt{A}}}}}\)}}
  \item  And: \hspace*{0.16em}{{\color{\colorMATH}\(\Psi _{c}\uplus \Psi _{2},\Phi ,\Sigma ,\varnothing  \vdash  \underline v \mathrel{:} \tau _{1},\Psi _{1}\)}}
  \item  And: \hspace*{0.16em}{{\color{\colorMATH}\(\Psi _{c}\uplus \Psi _{1},\Phi ,\Sigma ,\Gamma \uplus [x\mapsto \tau _{1}] \vdash  \underline e \mathrel{:} \tau _{2} \mathrel{;} \Gamma ^{\prime}\uplus [x\mapsto \vphantom{\overset {\mathord{\bullet }}x}\overset {\smash {\mathord{\bullet }}}\tau _{1}^{\prime}],\Psi _{2}\)}}
  \item  Then: \hspace*{0.33em}there exists \hspace*{0.16em}{{\color{\colorMATH}\(\Psi _{1}^{\prime}\)}}
  \item  {S.t.}: \hspace*{0.16em}{{\color{\colorMATH}\(\Psi _{1}^{\prime} \subseteq  \Psi _{1}\)}} 
  \item  And: \hspace*{0.16em}{{\color{\colorMATH}\(\Psi _{c},\Phi ,\Sigma ,\Gamma  \vdash  [\underline v/x]\underline e \mathrel{:} \tau _{2} \mathrel{;} \Gamma ^{\prime},\Psi _{1}^{\prime} \uplus  \Psi _{2}\)}}
  \end{itemize} 
\end{lemma}
\end{minipage}
\begin{proof}\ 
  \begin{itemize}[label={},leftmargin=0pt]\item  Case analysis on {{\color{\colorMATH}\(\vphantom{\overset {\mathord{\bullet }}x}\overset {\smash {\mathord{\bullet }}}\tau _{1}^{\prime}\)}}:
     \begin{enumerate}[leftmargin=15pt]\item  \begin{itemize}[label={},leftmargin=0pt]\item  {{\color{\colorMATH}\(\vphantom{\overset {\mathord{\bullet }}x}\overset {\smash {\mathord{\bullet }}}\tau _{1}^{\prime} = \tau _{1}\)}}
        \item  {{\color{\colorMATH}\(\Psi _{1}^{\prime} \triangleq  \varnothing \)}}
        \item  By {{\color{\colorMATH}\(\varnothing  \subseteq  \Psi _{1}\)}}, \nameref{thm:proofs:aff-sub-unu} and \nameref{thm:proofs:wkn-exp} applied to 
           \nameref{thm:proofs:wkn-bit} and \nameref{thm:proofs:weaken-flip}
        \end{itemize}
     \item  \begin{itemize}[label={},leftmargin=0pt]\item  {{\color{\colorMATH}\(\vphantom{\overset {\mathord{\bullet }}x}\overset {\smash {\mathord{\bullet }}}\tau _{1}^{\prime} = {\mathord{\bullet }}\)}}
        \item  {{\color{\colorMATH}\(\Psi _{1}^{\prime} \triangleq  \Psi _{1}\)}}
        \item  By {{\color{\colorMATH}\(\Psi _{1} \subseteq  \Psi _{1}\)}} and \nameref{thm:proofs:aff-sub-use}
        \end{itemize}
     \end{enumerate}
  \end{itemize}
\end{proof}

\noindent
\begin{minipage}{\linewidth}
\begin{lemma}[Affine Substitution Used]
\label{thm:proofs:aff-sub-use}\ 
  \begin{itemize}[label={},leftmargin=0pt]\item  If: \hspace*{0.16em}{{\color{\colorMATH}\({\mathcal{K}}(\tau _{1}) = {\mfootnotesize{{{\color{\colorSYNTAX}\mtexttt{A}}}}}\)}}
  \item  And: \hspace*{0.16em}{{\color{\colorMATH}\(\Psi _{c}\uplus \Psi _{2},\Phi ,\Sigma ,\varnothing  \vdash  \underline v \mathrel{:} \tau _{1} \mathrel{;} \varnothing  , \Psi _{1}\)}}
  \item  And: \hspace*{0.16em}{{\color{\colorMATH}\(\Psi _{c}\uplus \Psi _{1},\Phi ,\Sigma ,\Gamma \uplus [x\mapsto \tau _{1}] \vdash  \underline e \mathrel{:} \tau _{2} \mathrel{;} \Gamma ^{\prime}\uplus [x\mapsto {\mathord{\bullet }}] , \Psi _{2}\)}}
  \item  Then: \hspace*{0.16em}{{\color{\colorMATH}\(\Psi _{c},\Phi ,\Sigma ,\Gamma  \vdash  [\underline v/x]\underline e \mathrel{:} \tau _{2} \mathrel{;} \Gamma ^{\prime} , \Psi _{1} \uplus  \Psi _{2}\)}}
  \end{itemize}
\end{lemma}
\end{minipage}
\begin{proof}\ 
  \begin{itemize}[label={},leftmargin=0pt]\item  Induction on {{\color{\colorMATH}\(\underline e\)}}, \nameref{thm:proofs:cxt-mon} and \nameref{thm:proofs:aff-sub-unu}
  \item  Representative inductive case:
     \begin{itemize}[label={},leftmargin=0pt]\item  {{\color{\colorMATH}\(\underline e = \langle \underline e_{1},\underline e_{2}\rangle \)}}
     \item  Must be one of the following (by \nameref{thm:proofs:cxt-mon}):
     \item  \begin{enumerate}[leftmargin=15pt]\item  \begin{itemize}[label={},leftmargin=0pt]\item  \fbox{{{\color{\colorMATH}\(
              \inferrule*[vcenter,lab=
              ]{ \Psi _{c}\uplus \Psi _{1}\uplus \Psi _{2 2},\Phi ,\Sigma ,\Gamma \uplus [x\mapsto \tau _{1}] \vdash  \underline e_{2 1} \mathrel{:} \tau _{2 1} \mathrel{;} \Gamma ^{\prime \prime}\uplus [x\mapsto \tau _{1}] , \Psi _{2 1}
              \\ \Psi _{c}\uplus \Psi _{1}\uplus \Psi _{2 1},\Phi ,\Sigma ,\Gamma ^{\prime \prime}\uplus [x\mapsto \tau _{1}] \vdash  \underline e_{2 2} \mathrel{:} \tau _{2 2} \mathrel{;} \Gamma ^{\prime}\uplus [x\mapsto {\mathord{\bullet }}] , \Psi _{2 2}
                 }{
                 \Psi _{c}\uplus \Psi _{1},\Phi ,\Sigma ,\Gamma \uplus [x\mapsto \tau _{1}] \vdash  \langle \underline e_{2 1},\underline e_{2 2}\rangle  \mathrel{:} \tau _{2 1} \times  \tau _{2 2} \mathrel{;} \Gamma ^{\prime}\uplus [x\mapsto {\mathord{\bullet }}] , \Psi _{2 1} \uplus  \Psi _{2 2}
              }
              \)}}}
           \item  Goal: {{\color{\colorMATH}\(\Psi _{c},\Phi ,\Sigma ,\Gamma \uplus [x\mapsto \tau _{1}] \vdash  [\underline v/x]\langle \underline e_{2 1},\underline e_{2 2}\rangle  \mathrel{:} \tau _{2 1} \times  \tau _{2 2} \mathrel{;} \Gamma ^{\prime}\uplus [x\mapsto {\mathord{\bullet }}] , \Psi _{1} \uplus  \Psi _{2 1} \uplus  \Psi _{2 2}\)}}
           \item  {{\color{\colorMATH}\([\underline v/x]\langle \underline e_{2 1},\underline e_{2 2}\rangle  = \langle \underline e_{2 1},[\underline v/x]\underline e_{2 2}\rangle \)}} (by \nameref{thm:proofs:aff-sub-unu})
           \item  STS:
              \begin{enumerate}[leftmargin=15pt]\item  {{\color{\colorMATH}\(\Psi _{c}\uplus \Psi _{2 2},\Phi ,\Sigma ,\Gamma  \vdash  \underline e_{2 1} \mathrel{:} \tau _{2 1} \mathrel{;} \Gamma ^{\prime \prime} , \Psi _{2 1}\)}} (by \nameref{thm:proofs:wkn-exp})
              \item  {{\color{\colorMATH}\(\Psi _{c}\uplus \Psi _{2 1},\Phi ,\Sigma ,\Gamma ^{\prime \prime} \vdash  [\underline v/x]\underline e_{2 2} \mathrel{:} \tau  \mathrel{;} \Gamma ^{\prime} , \Psi _{1} \uplus  \Psi _{2 2}\)}} (by Inductive Hypothesis)
              \end{enumerate}
           \end{itemize}
        \item  \begin{itemize}[label={},leftmargin=0pt]\item  \fbox{{{\color{\colorMATH}\(
              \inferrule*[vcenter,lab=
              ]{ \Psi _{c}\uplus \Psi _{1}\uplus \Psi _{2 2},\Phi ,\Sigma ,\Gamma \uplus [x\mapsto \tau _{1}] \vdash  \underline e_{2 1} \mathrel{:} \tau _{2 1} \mathrel{;} \Gamma ^{\prime \prime}\uplus [x\mapsto {\mathord{\bullet }}] , \Psi _{2 1}
              \\ \Psi _{c}\uplus \Psi _{1}\uplus \Psi _{2 1},\Phi ,\Sigma ,\Gamma ^{\prime \prime}\uplus [x\mapsto {\mathord{\bullet }}] \vdash  \underline e_{2 2} \mathrel{:} \tau _{2 2} \mathrel{;} \Gamma ^{\prime}\uplus [x\mapsto {\mathord{\bullet }}] , \Psi _{2 2}
                 }{
                 \Psi _{c},\Phi ,\Sigma ,\Gamma \uplus [x\mapsto \tau _{1}] \vdash  \langle \underline e_{2 1},\underline e_{2 2}\rangle  \mathrel{:} \tau _{2 1} \times  \tau _{2 2} \mathrel{;} \Gamma ^{\prime}\uplus [x\mapsto {\mathord{\bullet }}] , \Psi _{2 1} \uplus  \Psi _{2 2}
              }
              \)}}}
           \item  Analogous to (1) where {{\color{\colorMATH}\([\underline v/x]\langle \underline e_{2 1},\underline e_{2 2}\rangle  = \langle [\underline v/x]\underline e_{2 1},\underline e_{2 2}\rangle \)}}
           \end{itemize}
        \end{enumerate}
     \end{itemize}
  \end{itemize}
\end{proof}

\noindent
\begin{minipage}{\linewidth}
\begin{lemma}[Affine Substitution Unused]
\label{thm:proofs:aff-sub-unu}\ 
  \begin{itemize}[label={},leftmargin=0pt]\item  If: \hspace*{0.16em}{{\color{\colorMATH}\(\Psi _{c},\Phi ,\Sigma ,\Gamma \uplus [x\mapsto {\mathord{\bullet }}] \vdash  \underline e \mathrel{:} \tau _{2} \mathrel{;} \Gamma ^{\prime}\uplus [x\mapsto {\mathord{\bullet }}] , \Psi \)}}
  \item  Or: \hspace*{0.16em}{{\color{\colorMATH}\({\mathcal{K}}(\tau _{1}) = {\mfootnotesize{{{\color{\colorSYNTAX}\mtexttt{A}}}}}\)}}\hspace*{0.16em} and \hspace*{0.16em}{{\color{\colorMATH}\(\Psi _{c},\Phi ,\Sigma ,\Gamma \uplus [x\mapsto \tau _{1}] \vdash  \underline e \mathrel{:} \tau _{2} \mathrel{;} \Gamma ^{\prime}\uplus [x\mapsto \tau _{1}] , \Psi \)}}
  \item  Then: \hspace*{0.16em}{{\color{\colorMATH}\([\underline v/x]\underline e = \underline e\)}}
  \end{itemize}
\end{lemma}
\end{minipage}
\begin{proof}\ 
  \begin{itemize}[label={},leftmargin=0pt]\item  Induction on {{\color{\colorMATH}\(\underline e\)}} and \nameref{thm:proofs:cxt-mon}
  \end{itemize}
\end{proof}

\noindent
\begin{minipage}{\linewidth}
\begin{lemma}[Context Monotonicity]
\label{thm:proofs:cxt-mon}\ 
  \begin{itemize}[label={},leftmargin=0pt]\item  If: \hspace*{0.16em}{{\color{\colorMATH}\(\Psi _{c},\Phi ,\Sigma ,\Gamma  \vdash  \underline e \mathrel{:} \tau _{2} \mathrel{;} \Gamma ^{\prime} , \Psi \)}}
  \item  Then: \hspace*{0.16em}{{\color{\colorMATH}\(\Gamma (x) \sqsubseteq  \Gamma ^{\prime}(x)\)}}
  \end{itemize}
\end{lemma}
\end{minipage}
\begin{proof}\ 
  \begin{itemize}[label={},leftmargin=0pt]\item  Induction on {{\color{\colorMATH}\(\underline e\)}} and partial order properties
  \end{itemize}
\end{proof}

\paragraph{\bf Weakening}\ \\

\noindent
\begin{minipage}{\linewidth}
\begin{lemma}[Weaken Context]
\label{thm:proofs:wkn-cxt}\ 
\begin{itemize}[label={},leftmargin=0pt]\item  If: \hspace*{0.16em}{{\color{\colorMATH}\(\Psi _{c},\Phi ,\Sigma  \vdash  \underline \sigma ,\underline e \mathrel{:} \tau ,\Psi \)}}
\item  And: \hspace*{0.16em}{{\color{\colorMATH}\(\Psi _{c},\Phi ^{\prime},\Sigma ^{\prime} \vdash  \underline \sigma ^{\prime},\underline e^{\prime} \mathrel{:} \tau ,\Psi ^{\prime}\)}}
\item  And: \hspace*{0.16em}{{\color{\colorMATH}\(\Sigma ^{\prime} \supseteq  \Sigma \)}}
\item  And: \hspace*{0.16em}{{\color{\colorMATH}\(\forall  \hat b,\rho .\hspace*{0.33em} \Psi \uplus \Psi _{c},\Phi  \vdash  \hat b\mathrel{:}{\mfootnotesize{{{\color{\colorSYNTAX}\mtexttt{bit}}}}}_{S}^{\rho } \implies   \Psi ^{\prime}\uplus \Psi _{c},\Phi ^{\prime} \vdash  \hat b\mathrel{:}{\mfootnotesize{{{\color{\colorSYNTAX}\mtexttt{bit}}}}}_{S}^{\rho }\)}}
\item  And: \hspace*{0.16em}{{\color{\colorMATH}\(\forall  \hat b,\rho .\hspace*{0.33em} \Psi /\Psi _{c},\Phi  \vdash  \hat b\mathrel{:}{\mfootnotesize{{{\color{\colorSYNTAX}\mtexttt{flip}}}}}^{\rho } \implies   \Psi ^{\prime}/\Psi _{c},\Phi ^{\prime} \vdash  \hat b\mathrel{:}{\mfootnotesize{{{\color{\colorSYNTAX}\mtexttt{flip}}}}}^{\rho }\)}}
\item  And: \hspace*{0.16em}{{\color{\colorMATH}\(\Phi ,\Sigma  \vdash  \underline \sigma ,\underline E[\underline e] \mathrel{:} \tau ^{\prime},\Psi _{c} \uplus  \Psi \)}}
\item  Then: \hspace*{0.16em}{{\color{\colorMATH}\(\Phi ^{\prime},\Sigma ^{\prime} \vdash  \underline \sigma ^{\prime},\underline E[\underline e^{\prime}] \mathrel{:} \tau ^{\prime} , \Psi _{c} \uplus  \Psi ^{\prime}\)}}
\end{itemize}
\end{lemma}
\end{minipage}
\begin{proof}\ 
\begin{itemize}[label={},leftmargin=0pt]\item  Induction on {{\color{\colorMATH}\(\underline E\)}} and \nameref{thm:proofs:wkn-exp}
\end{itemize}
\end{proof}

\noindent
\begin{minipage}{\linewidth}
\begin{lemma}[Weaken Store]
\label{thm:proofs:weaken-store}\ 
\begin{itemize}[label={},leftmargin=0pt]\item  If: \hspace*{0.16em}{{\color{\colorMATH}\(\Psi _{c},\Phi ,\Sigma  \vdash  \underline \sigma  \mathrel{;}  \Psi \)}}
\item  And: \hspace*{0.16em}{{\color{\colorMATH}\(\Sigma ^{\prime} \supseteq  \Sigma \)}}
\item  And: \hspace*{0.16em}{{\color{\colorMATH}\(\forall  \hat b,\rho .\hspace*{0.33em} \Psi _{c}/\Psi ,\Phi  \vdash  \hat b\mathrel{:}{\mfootnotesize{{{\color{\colorSYNTAX}\mtexttt{flip}}}}}^{\rho } \implies   \Psi _{c}^{\prime}/\Psi ,\Phi ^{\prime} \vdash  \hat b\mathrel{:}{\mfootnotesize{{{\color{\colorSYNTAX}\mtexttt{flip}}}}}^{\rho }\)}}
\item  Then: \hspace*{0.16em}{{\color{\colorMATH}\(\Psi _{c}^{\prime},\Phi ^{\prime},\Sigma ^{\prime} \vdash  \underline \sigma  \mathrel{;} \Psi \)}}
\end{itemize}
\end{lemma}
\end{minipage}
\begin{proof}\ 
\begin{itemize}[label={},leftmargin=0pt]\item  Induction on {{\color{\colorMATH}\(\underline \sigma \)}}, \nameref{thm:proofs:wkn-exp} and {{\color{\colorMATH}\(\Sigma (\iota ) = \tau  \implies   \Sigma ^{\prime}(\iota ) = \tau \)}}
\end{itemize}
\end{proof}

\noindent
\begin{minipage}{\linewidth}
\begin{lemma}[Weaken Expression]
\label{thm:proofs:wkn-exp}\ 
\begin{itemize}[label={},leftmargin=0pt]\item  If: \hspace*{0.16em}{{\color{\colorMATH}\(\Psi _{c},\Phi ,\Sigma ,\Gamma  \vdash  \underline e \mathrel{:} \tau  \mathrel{;} \Gamma ^{\prime} , \Psi \)}}
\item  And: \hspace*{0.16em}{{\color{\colorMATH}\(\Sigma ^{\prime} \supseteq  \Sigma \)}}
  \item  And: \hspace*{0.16em}{{\color{\colorMATH}\(\forall  \hat b,\rho .\hspace*{0.33em} \Psi _{c}/\Psi ,\Phi  \vdash  \hat b\mathrel{:}{\mfootnotesize{{{\color{\colorSYNTAX}\mtexttt{flip}}}}}^{\rho } \implies   \Psi _{c}^{\prime}/\Psi ,\Phi ^{\prime} \vdash  \hat b\mathrel{:}{\mfootnotesize{{{\color{\colorSYNTAX}\mtexttt{flip}}}}}^{\rho }\)}} {\mtextit{(H)}}
\item  Then: \hspace*{0.16em}{{\color{\colorMATH}\(\Psi _{c}^{\prime},\Phi ^{\prime},\Sigma ^{\prime},\Gamma  \vdash  \underline e \mathrel{:} \tau  \mathrel{;} \Gamma ^{\prime} , \Psi \)}}
\end{itemize}
\end{lemma}
\begin{itemize}[label={},leftmargin=0pt]\item  Induction on {{\color{\colorMATH}\(\underline e\)}}, \nameref{thm:proofs:wkn-bit} and application of {\mtextit{(H)}} on flip values
\end{itemize}
\end{minipage}

\noindent
\begin{minipage}{\linewidth}
\begin{lemma}[Weaken Bit Value]
\label{thm:proofs:wkn-bit}\ 
\begin{itemize}[label={},leftmargin=0pt]\item  If: \hspace*{0.16em}{{\color{\colorMATH}\(\Psi _{c},\Phi  \vdash  \hat b \mathrel{:} {\mfootnotesize{{{\color{\colorSYNTAX}\mtexttt{bitv}}}}}_{\ell }^{\rho }\)}}
\item  Then: \hspace*{0.16em}{{\color{\colorMATH}\(\Psi _{c}^{\prime},\Phi ^{\prime} \vdash  \hat b \mathrel{:} {\mfootnotesize{{{\color{\colorSYNTAX}\mtexttt{bit}}}}}_{\ell }^{\rho }\)}}
\end{itemize}
\end{lemma}
\end{minipage}
\begin{proof}\ 
\begin{itemize}[label={},leftmargin=0pt]\item  Immediate by inversion and re-construction of the type derivation
\end{itemize}
\end{proof}

\noindent
\begin{minipage}{\linewidth}
\begin{lemma}[Weaken Flip]
\label{thm:proofs:weaken-flip}\ 
\begin{itemize}[label={},leftmargin=0pt]\item  If: \hspace*{0.16em}{{\color{\colorMATH}\(\Psi _{c}^{F},\Psi _{c}^{B},\Phi  \vdash  \hat b \mathrel{:} {\mfootnotesize{{{\color{\colorSYNTAX}\mtexttt{flipv}}}}}^{\rho }\)}}
\item  And: \hspace*{0.16em}{{\color{\colorMATH}\(\Psi _{c}^{F \prime},\Psi _{c}^{B \prime} \subseteq  \Psi _{c}^{F},\Psi _{c}^{B}\)}}
\item  Then: \hspace*{0.16em}{{\color{\colorMATH}\(\Psi _{c}^{F \prime},\Psi _{c}^{B},\Phi  \vdash  \hat b \mathrel{:} {\mfootnotesize{{{\color{\colorSYNTAX}\mtexttt{flipv}}}}}^{\rho }\)}}
\end{itemize}
\end{lemma}
\end{minipage}
\begin{proof}\ 
\begin{itemize}[label={},leftmargin=0pt]\item  By inversion:
\item  \fbox{{{\color{\colorMATH}\(
   \inferrule*[vcenter,lab=
   ]{ {\mtext{Pr}}\left[\hat b \mathrel{\dot =} {\mfootnotesize{{{\color{\colorSYNTAX}\mtexttt{I}}}}} \mathrel{}\middle|\mathrel{} \Phi  \right] = \nicefrac{1}{2} 
   \\ \left[ \hat b \mathrel{\bot \!\!\!\bot } \Psi _{c}^{F},\Psi _{c}^{B}(\{ \rho ^{\prime} \mathrel{|} \rho ^{\prime} \sqsubset  \rho \} ) \mathrel{}\middle|\mathrel{} \Phi  \right] \hspace*{0.33em}{{\color{\colorTEXT}\textnormal{{\mtextit{(H)}}}}}
      }{
      \Psi _{c}^{F},\Psi _{c}^{B},\Phi ,\Sigma  \vdash  \hat b \mathrel{:} {\mfootnotesize{{{\color{\colorSYNTAX}\mtexttt{flipv}}}}}^{\rho }
   }
   \)}}}
\item  STS: \hspace*{0.16em}{{\color{\colorMATH}\(\left[ \hat b \mathrel{\bot \!\!\!\bot } \Psi _{c}^{F \prime},\Psi _{c}^{B \prime}(\{ \rho ^{\prime} \mathrel{|} \rho ^{\prime} \sqsubset  \rho \} ) \mathrel{}\middle|\mathrel{} \Phi  \right]\)}}
\item  By {\mtextit{(H)}} and \nameref{thm:proofs:decomposition} with {{\color{\colorMATH}\(\Psi _{c}^{F \prime} \subseteq  \Psi _{c}^{F}\)}} and {{\color{\colorMATH}\(\Psi _{c}^{B \prime}(\{ \rho ^{\prime} \mathrel{|} \rho ^{\prime} \sqsubset  \rho \} ) \subseteq  \Psi _{c}^{B}(\{ \rho ^{\prime} \mathrel{|} \rho ^{\prime} \sqsubset  \rho \} )\)}}
\end{itemize}
\end{proof}

\subsubsection{Intensional Distribution Lemmas}

All of the following lemmas are proved for intensional distributions {{\color{\colorMATH}\(\hat x \in 
{\mathcal{I}}(A)\)}}, however except for \nameref{thm:proofs:monad-idempotence}, each of
the properties are also true of denotational distributions {{\color{\colorMATH}\(\tilde x \in  {\mathcal{D}}(A)\)}}
(although the proof given only applies to intensional distributions). Recall that 
trees are considered equal {{\color{\colorMATH}\(=\)}} when they are syntactically equal modulo
height extension, {i.e.}, {{\color{\colorMATH}\(\hat x = \text{\guilsinglleft}\hat x \hat x\text{\guilsinglright}\)}}.

\noindent
\begin{minipage}{\linewidth}
\begin{lemma}[Proper Distribution]\label{thm:proofs:proper-distribution}\ 
\begin{enumerate}[leftmargin=15pt]\item  {{\color{\colorMATH}\(\sum \limits_{x\in {\mtext{support}}(\hat x)}{\mtext{Pr}}\left[\hat x \mathrel{\dot =} x\right] = 1\)}}
\item  \begin{itemize}[label={},leftmargin=0pt]\item  If: \hspace*{0.16em}{{\color{\colorMATH}\({\mtext{Pr}}\left[\hat y \mathrel{\dot =} y\right] > 0\)}}
   \item  Then: \hspace*{0.16em}{{\color{\colorMATH}\({\mtext{Pr}}\left[\hat x \mathrel{\dot =} x \mathrel{}\middle|\mathrel{} \hat y \mathrel{\dot =} y\right]\)}} is defined
   \item  And: \hspace*{0.16em}{{\color{\colorMATH}\(\sum \limits_{x\in {\mtext{support}}(\hat x)}{\mtext{Pr}}\left[\hat x \mathrel{\dot =} x \mathrel{}\middle|\mathrel{} \hat y \mathrel{\dot =} y\right] = 1\)}}
   \end{itemize}
\end{enumerate}
\end{lemma}
\end{minipage}
\begin{proof}
  Induction on the tree-structure of {{\color{\colorMATH}\(\hat x\)}}
\end{proof}

\noindent
\begin{minipage}{\linewidth}
\begin{lemma}[Return Probability]\label{thm:proofs:return-probability}\ 
  \begin{enumerate}[leftmargin=15pt]\item  {{\color{\colorMATH}\({\mtext{Pr}}\left[{\mtext{return}}_{{\mathcal{I}}}(x) \mathrel{\dot =} x\right] = 1\)}}
  \item  {{\color{\colorMATH}\({\mtext{Pr}}\left[{\mtext{return}}_{{\mathcal{I}}}(x) \mathrel{\dot =} y\right] = 0\)}} when {{\color{\colorMATH}\(x \neq  y\)}}
  \end{enumerate}
\end{lemma}
\end{minipage}
\begin{proof}
  \begin{itemize}[label={},leftmargin=0pt]\item  Immediate by definition of {{\color{\colorMATH}\({\mtext{return}}\)}} and {{\color{\colorMATH}\({\mtext{Pr}}\)}}
  \end{itemize}
\end{proof}

\noindent
\begin{minipage}{\linewidth}
\begin{lemma}[Bind Probability]\label{thm:proofs:bind-probability}\ 
  \begin{itemize}[label={},leftmargin=0pt]\item  {{\color{\colorMATH}\({\mtext{Pr}}\left[{\mtext{do}}\hspace*{0.33em}x \leftarrow  \hat x \mathrel{;} f(x) \mathrel{\dot =} y\right] = \sum \limits_{x}{\mtext{Pr}}\left[f(x) \mathrel{\dot =} y \mathrel{}\middle|\mathrel{} \hat x \mathrel{\dot =} x\right]{\mtext{Pr}}\left[\hat x \mathrel{\dot =} x\right]\)}}
  \end{itemize}
\end{lemma}
\end{minipage}
\begin{proof}
  \begin{itemize}[label={},leftmargin=0pt]\item  Induction on the tree-structure of {{\color{\colorMATH}\(\hat x\)}}
  \end{itemize}
\end{proof}

\noindent
\begin{minipage}{\linewidth}
\begin{lemma}[Monad Laws]\label{thm:proofs:monad-laws}\ 
\begin{itemize}[label={},leftmargin=0pt]\item  {{\color{\colorMATH}\(\begin{array}[t]{l@{\hspace*{1.00em}\hspace*{1.00em}\hspace*{1.00em}}r
    } ({\mtext{do}}\hspace*{0.33em}x \leftarrow  {\mtext{return}}_{{\mathcal{I}}}(y) \mathrel{;} f(x)) = f(y)                                    & {{\color{\colorTEXT}\textnormal{{\mtextit{(left-unit)}}}}}
    \cr  ({\mtext{do}}\hspace*{0.33em}x \leftarrow  \hat x \mathrel{;} {\mtext{return}}(x)) = \hat x                                         & {{\color{\colorTEXT}\textnormal{{\mtextit{(right-unit)}}}}}
    \cr  ({\mtext{do}}\hspace*{0.33em}y \leftarrow  ({\mtext{do}}\hspace*{0.33em}x \leftarrow  \hat x \mathrel{;} f(x)) \mathrel{;} g(y)) = ({\mtext{do}}\hspace*{0.33em}x \leftarrow  \hat x \mathrel{;} y \leftarrow  f(x) \mathrel{;} g(y)) & {{\color{\colorTEXT}\textnormal{{\mtextit{(associativity)}}}}}
    \end{array}\)}}
\end{itemize}
\end{lemma}
\end{minipage}
\begin{proof}
  \begin{enumerate}[leftmargin=15pt]\item  \begin{itemize}[label={},leftmargin=0pt]\item  (left-unit) 
     \item  immediate from definitions
     \end{itemize}
  \item  \begin{itemize}[label={},leftmargin=0pt]\item  (right-unit) 
     \item  Case analysis on {{\color{\colorMATH}\(\hat x\)}}
     \item  \begin{itemize}[label=\textbf{-},leftmargin=*]\item  \begin{itemize}[label={},leftmargin=0pt]\item  Case {{\color{\colorMATH}\(\hat x=x\)}}:
           \item  {{\color{\colorMATH}\(x = x\)}}; immediate
           \end{itemize}
        \item  \begin{itemize}[label={},leftmargin=0pt]\item  Case {{\color{\colorMATH}\(\hat x = \text{\guilsinglleft}\hat x_{1} \hat x_{2}\text{\guilsinglright}\)}}:
           \item  {{\color{\colorMATH}\(\text{\guilsinglleft}\pi _{1}(\text{\guilsinglleft}\hat x_{1} \hat x_{2}\text{\guilsinglright}) \pi _{2}(\text{\guilsinglleft}\hat x_{1} \hat x_{2}\text{\guilsinglright})\text{\guilsinglright} = \text{\guilsinglleft}\hat x_{1} \hat x_{2}\text{\guilsinglright}\)}}; immediate
           \end{itemize}
        \end{itemize}
     \end{itemize}
  \item  \begin{itemize}[label={},leftmargin=0pt]\item  (associativity)
     \item  Case analysis on {{\color{\colorMATH}\(\hat x\)}}:
     \item  \begin{itemize}[label=\textbf{-},leftmargin=*]\item  \begin{itemize}[label={},leftmargin=0pt]\item  Case {{\color{\colorMATH}\(\hat x=x\)}}:
           \item {{\color{\colorMATH}\(({\mtext{do}}\hspace*{0.33em}y \leftarrow  f(x) \mathrel{;} g(y)) = ({\mtext{do}}\hspace*{0.33em}y \leftarrow  f(x) \mathrel{;} g(y))\)}}; immediate
           \end{itemize}
        \item  \begin{itemize}[label={},leftmargin=0pt]\item  Case {{\color{\colorMATH}\(\hat x=\text{\guilsinglleft}\hat x_{1} \hat x_{2}\text{\guilsinglright}\)}}:
           \item {{\color{\colorMATH}\(\text{\guilsinglleft}\pi _{1}(g(\pi _{1}(f(\hat x_{1})))) \pi _{2}(g(\pi _{2}(f(\hat x_{2}))))\text{\guilsinglright} = \text{\guilsinglleft}\pi _{1}(g(\pi _{1}(f(\hat x_{1})))) \pi _{2}(g(\pi _{2}(f(\hat x_{2}))))\text{\guilsinglright}\)}}; immediate
           \end{itemize}
        \end{itemize}
     \end{itemize}
  \end{enumerate}
\end{proof}

\noindent
\begin{minipage}{\linewidth}
\begin{lemma}[Monad Commutativity]\label{thm:proofs:monad-commutativity}\ 
  \begin{itemize}[label={},leftmargin=0pt]\item  {{\color{\colorMATH}\(({\mtext{do}}\hspace*{0.33em}x \leftarrow  \hat x \mathrel{;} y \leftarrow  \hat y \mathrel{;} f(x,y)) = ({\mtext{do}}\hspace*{0.33em}y \leftarrow  \hat y \mathrel{;} x \leftarrow  \hat x \mathrel{;} f(x,y))\)}}
  \end{itemize}
\end{lemma}
\end{minipage}
\begin{proof}
  \begin{itemize}[label={},leftmargin=0pt]\item  Case analysis on {{\color{\colorMATH}\(\hat x\)}}:
  \item  \begin{itemize}[label=\textbf{-},leftmargin=*]\item  \begin{itemize}[label={},leftmargin=0pt]\item  Case {{\color{\colorMATH}\(\hat x=x\)}}:
        \item  {{\color{\colorMATH}\(({\mtext{do}}\hspace*{0.33em}y \leftarrow  \hat y \mathrel{;} f(x,y)) = ({\mtext{do}}\hspace*{0.33em}y \leftarrow  \hat y \mathrel{;} f(x,y))\)}}; immediate
        \end{itemize}
     \item  \begin{itemize}[label={},leftmargin=0pt]\item  Case {{\color{\colorMATH}\(\hat x=\text{\guilsinglleft}\hat x_{1} \hat x_{2}\text{\guilsinglright}\)}}:
        \item  {{\color{\colorMATH}\(\begin{array}[t]{l
            } \text{\guilsinglleft}\pi _{1}({\mtext{do}}\hspace*{0.33em}y \leftarrow  \hat y \mathrel{;} f(\hat x_{1},y)) \pi _{2}({\mtext{do}}\hspace*{0.33em}y \leftarrow  \hat y \mathrel{;} f(\hat x_{2},y))\text{\guilsinglright}
            \cr  =
            \cr  {\mtext{do}}\hspace*{0.33em}y \leftarrow  \hat y \mathrel{;} \text{\guilsinglleft}\pi _{1}(f(\hat x_{1},y)) \pi _{2}(f(\hat x_{2},y))\text{\guilsinglright}
            \end{array}\)}}
        \item  Finally by case analysis on {{\color{\colorMATH}\(\hat y\)}}
        \end{itemize}
     \end{itemize}
  \end{itemize}
\end{proof}

\noindent
\begin{minipage}{\linewidth}
\begin{lemma}[Monad Idempotence (Intensional Only)]\label{thm:proofs:monad-idempotence}\ 
  \begin{itemize}[label={},leftmargin=0pt]\item  {\mtextit{The intensional distribution monad {{\color{\colorMATH}\({\mathcal{I}}\)}} is idempotent.}}
  \item  {\mtextit{NOTE: this is in contrast with the denotational distribution monad {{\color{\colorMATH}\({\mathcal{D}}\)}} which is {\mtextit{not}} idempotent.}}
  \item  {{\color{\colorMATH}\(({\mtext{do}}\hspace*{0.33em}x_{1} \leftarrow  \hat x \mathrel{;} x_{2} \leftarrow  \hat x \mathrel{;} f(x_{1},x_{2})) = ({\mtext{do}}\hspace*{0.33em}x \leftarrow  \hat x \mathrel{;} f(x,x))\)}}
  \end{itemize}
\end{lemma}
\end{minipage}
\begin{proof}
  \begin{itemize}[label={},leftmargin=0pt]\item  Case analysis on {{\color{\colorMATH}\(\hat x\)}} (analogous to monad laws and commutativity proofs)
  \end{itemize}
\end{proof}

\noindent
\begin{minipage}{\linewidth}
\begin{lemma}[Bit Independence]\label{thm:proofs:bit-independence}\ 
  \begin{itemize}[label={},leftmargin=0pt]\item  {\mtextit{A particular random bit is independent of all other random bits.}}
  \item  {{\color{\colorMATH}\({\mtext{bit}}(N) \mathrel{\bot \!\!\!\bot } {\mtext{bit}}(N^{\prime})\)}} for {{\color{\colorMATH}\(N \neq  N^{\prime}\)}}
  \end{itemize}
\end{lemma}
\end{minipage}
\begin{proof}
  \begin{itemize}[label={},leftmargin=0pt]\item  Induction on {{\color{\colorMATH}\(N\)}} and {{\color{\colorMATH}\(N^{\prime}\)}}
  \end{itemize}
\end{proof}

\noindent
\begin{minipage}{\linewidth}
\begin{lemma}[Cond Independence]\label{thm:proofs:cond-independence}\ 
  \begin{itemize}[label={},leftmargin=0pt]\item  {\mtextit{A conditional is independent when its inputs are jointly indpendent.}}
  \item  {{\color{\colorMATH}\(\hat x \mathrel{\bot \!\!\!\bot } \hat b,\hat y,\hat z \implies   \hat x \mathrel{\bot \!\!\!\bot } {\mtext{cond}}(\hat b,\hat y,\hat z)\)}}
  \end{itemize}
\end{lemma}
\end{minipage}
\begin{proof}
  \begin{itemize}[label={},leftmargin=0pt]\item  By \nameref{thm:proofs:total-probability} on {{\color{\colorMATH}\(\hat b\)}} and unfolding definition
     of {{\color{\colorMATH}\(\hat b\)}}
  \end{itemize}
\end{proof}

\noindent
\begin{minipage}{\linewidth}
\begin{lemma}[Cond Stability]\label{thm:proofs:cond-stability}\ 
  \begin{itemize}[label={},leftmargin=0pt]\item  {\mtextit{A conditional is stable when the guard is independent of branches, and
     branches have equal distributions.}}
  \item  If: \hspace*{0.16em}{{\color{\colorMATH}\(\hat b \mathrel{\bot \!\!\!\bot } \hat x_{1}\)}}
  \item  And: \hspace*{0.16em}{{\color{\colorMATH}\(\hat b \mathrel{\bot \!\!\!\bot } \hat x_{2}\)}}
  \item  And: \hspace*{0.16em}{{\color{\colorMATH}\({\mtext{Pr}}\left[\hat x_{1} \mathrel{\dot =} x\right] = {\mtext{Pr}}\left[\hat x_{2} \mathrel{\dot =} x \right]\)}}
  \item  Then: 
  \item  \begin{enumerate}[leftmargin=15pt]\item  {{\color{\colorMATH}\({\mtext{Pr}}\left[{\mtext{cond}}(\hat b,\hat x_{1},\hat x_{2}) \mathrel{\dot =} x\right] = {\mtext{Pr}}\left[\hat x_{1} \mathrel{\dot =} x\right] = {\mtext{Pr}}\left[\hat x_{2} \mathrel{\dot =} x \right]\)}}
     \item  {{\color{\colorMATH}\(\hat b \mathrel{\bot \!\!\!\bot } {\mtext{cond}}(\hat b,\hat x_{1},\hat x_{2})\)}}
     \end{enumerate}
  \end{itemize}
\end{lemma}
\end{minipage}
\begin{proof}
  \begin{itemize}[label={},leftmargin=0pt]\item  \begin{enumerate}[leftmargin=15pt]\item  \begin{itemize}[label={},leftmargin=0pt]\item  {{\color{\colorMATH}\(\begin{array}[t]{rcl
            } &{} {}& {\mtext{Pr}}\left[{\mtext{cond}}(\hat b,\hat x_{1},\hat x_{2}) \mathrel{\dot =} x \mathrel{}\middle|\mathrel{} \hat b \mathrel{\dot =} b \right]
            \cr  &{}={}& \lbag {{\color{\colorTEXT}\textnormal{\hspace*{0.33em} \nameref{thm:proofs:total-probability} \hspace*{0.33em}}}}\rbag 
            \cr  &{} {}& {\mtext{Pr}}\left[\hat x_{1} \mathrel{\dot =} x \mathrel{}\middle|\mathrel{} \hat b \mathrel{\dot =} {\mfootnotesize{{{\color{\colorSYNTAX}\mtexttt{I}}}}} \right]{\mtext{Pr}}\left[ \hat b \mathrel{\dot =} {\mfootnotesize{{{\color{\colorSYNTAX}\mtexttt{I}}}}} \right] + {\mtext{Pr}}\left[\hat x_{2} \mathrel{\dot =} x \mathrel{}\middle|\mathrel{} \hat b \mathrel{\dot =} {\mfootnotesize{{{\color{\colorSYNTAX}\mtexttt{O}}}}} \right]{\mtext{Pr}}\left[ \hat b \mathrel{\dot =} {\mfootnotesize{{{\color{\colorSYNTAX}\mtexttt{O}}}}} \right]
            \cr  &{}={}& \lbag {{\color{\colorTEXT}\textnormal{\hspace*{0.33em} {{\color{\colorMATH}\(\hat b \mathrel{\bot \!\!\!\bot } \hat x_{i}\)}} \hspace*{0.33em}}}}\rbag 
            \cr  &{} {}& {\mtext{Pr}}\left[\hat x_{1} \mathrel{\dot =} x \right]{\mtext{Pr}}\left[ \hat b \mathrel{\dot =} {\mfootnotesize{{{\color{\colorSYNTAX}\mtexttt{I}}}}} \right] + {\mtext{Pr}}\left[\hat x_{2} \mathrel{\dot =} x \right]{\mtext{Pr}}\left[ \hat b \mathrel{\dot =} {\mfootnotesize{{{\color{\colorSYNTAX}\mtexttt{O}}}}} \right]
            \cr  &{}={}& \lbag {{\color{\colorTEXT}\textnormal{\hspace*{0.33em} {{\color{\colorMATH}\({\mtext{Pr}}\left[\hat x_{1} \mathrel{\dot =} x\right] = {\mtext{Pr}}\left[\hat x_{2} \mathrel{\dot =} x \right]\)}} \hspace*{0.33em}}}}\rbag 
            \cr  &{} {}& {\mtext{Pr}}\left[\hat x_{1} \mathrel{\dot =} x \right]({\mtext{Pr}}\left[ \hat b \mathrel{\dot =} {\mfootnotesize{{{\color{\colorSYNTAX}\mtexttt{I}}}}} \right] + {\mtext{Pr}}\left[ \hat b \mathrel{\dot =} {\mfootnotesize{{{\color{\colorSYNTAX}\mtexttt{O}}}}} \right])
            \cr  &{}={}& \lbag {{\color{\colorTEXT}\textnormal{\hspace*{0.33em} \nameref{thm:proofs:proper-distribution} \hspace*{0.33em}}}}\rbag 
            \cr  &{} {}& {\mtext{Pr}}\left[\hat x_{1} \mathrel{\dot =} x \right]
            \cr  &{}={}& {\mtext{Pr}}\left[\hat x_{2} \mathrel{\dot =} x \right]
            \end{array}\)}}
         \end{itemize}
      \item  Follows direction from (1)
      \end{enumerate}
  \end{itemize}
\end{proof}

\subsubsection{Probability Facts}

All of the following facts are stated using intensional distribution notation
{{\color{\colorMATH}\(\hat x \in  {\mathcal{I}}(A)\)}}, however they are true of any model which supports joint
probabilities, including {{\color{\colorMATH}\(\tilde x \in  {\mathcal{D}}(A)\)}}. Proofs are not given because they are
standard properties {w.r.t.} standard definitions.

\noindent
\begin{minipage}{\linewidth}
\begin{fact}[Conditional Decomposition]\label{thm:proofs:conditional-decomposition}\ 
  \begin{itemize}[label={},leftmargin=0pt]\item  {{\color{\colorMATH}\({\mtext{Pr}}\left[\hat x\mathrel{\dot =}x\mathrel{}\middle|\mathrel{}\hat y\mathrel{\dot =}y\right] = \frac{{\mtext{Pr}}\left[\hat x\mathrel{\dot =}x,\hat y\mathrel{\dot =}y\right]}{{\mtext{Pr}}\left[\hat y\mathrel{\dot =}y\right]}\)}}
  \end{itemize}
\end{fact}
\end{minipage}

\noindent
\begin{minipage}{\linewidth}
\begin{fact}[Bayes' Rule]\label{thm:proofs:bayes-rule}\ 
  \begin{itemize}[label={},leftmargin=0pt]\item  {{\color{\colorMATH}\({\mtext{Pr}}\left[\overline {\hat x \mathrel{\dot =} x} \mathrel{}\middle|\mathrel{} \overline {\hat y \mathrel{\dot =} y},\overline {\hat z \mathrel{\dot =} z}\right] = \frac{{\mtext{Pr}}\left[ \overline {\hat y \mathrel{\dot =} y} \mathrel{}\middle|\mathrel{} \overline {\hat x \mathrel{\dot =} x},\overline {\hat z\mathrel{\dot =}z}\right]{\mtext{Pr}}\left[\overline {\hat x \mathrel{\dot =} x} \mathrel{}\middle|\mathrel{} \overline {\hat z\mathrel{\dot =}z}\right]}{{\mtext{Pr}}\left[\overline {\hat y \mathrel{\dot =} y} \mathrel{}\middle|\mathrel{} \overline {\hat z\mathrel{\dot =}z}\right]}\)}}
  \end{itemize}
\end{fact}
\end{minipage}

\noindent
\begin{minipage}{\linewidth}
\begin{fact}[Total Probability]
\label{thm:proofs:total-probability}\ 
\begin{itemize}[label={},leftmargin=0pt]\item  {{\color{\colorMATH}\({\mtext{Pr}}\left[\hat x \mathrel{\dot =} x\right] = \sum \limits_{y\in {\mtext{support}}(\hat y)}{\mtext{Pr}}\left[\hat x \mathrel{\dot =} x \mathrel{}\middle|\mathrel{} \hat y\mathrel{\dot =}y\right]{\mtext{Pr}}\left[\hat y\mathrel{\dot =}y\right]\)}}
\end{itemize}
\end{fact}
\end{minipage}
\begin{proof}
  Induction on the tree-structure of {{\color{\colorMATH}\(\hat x\)}} and {{\color{\colorMATH}\(\hat y\)}}
\end{proof}

\noindent
\begin{minipage}{\linewidth}
\begin{fact}[Decomposition]\label{thm:proofs:decomposition}\ 
  \begin{enumerate}[leftmargin=15pt]\item  {{\color{\colorMATH}\(\hat x \mathrel{\bot \!\!\!\bot } \hat y,\hat z \implies    \hat x \mathrel{\bot \!\!\!\bot } \hat y\)}}
  \item  {{\color{\colorMATH}\(\hat x \mathrel{\bot \!\!\!\bot } \hat y,\hat z \implies    \hat x \mathrel{\bot \!\!\!\bot } \hat z\)}}
  \end{enumerate}
\end{fact}
\end{minipage}

\noindent
\begin{minipage}{\linewidth}
\begin{fact}[Decomposition]\label{thm:proofs:weak-union}\ 
  \begin{enumerate}[leftmargin=15pt]\item  {{\color{\colorMATH}\(\hat x \mathrel{\bot \!\!\!\bot } \hat y,\hat z \implies    \left[\hat x \mathrel{\bot \!\!\!\bot } \hat y \mathrel{}\middle|\mathrel{} \hat z\right]\)}}
  \item  {{\color{\colorMATH}\(\hat x \mathrel{\bot \!\!\!\bot } \hat y,\hat z \implies    \left[\hat x \mathrel{\bot \!\!\!\bot } \hat z \mathrel{}\middle|\mathrel{} \hat y\right]\)}}
  \end{enumerate}
\end{fact}
\end{minipage}

\noindent
\begin{minipage}{\linewidth}
\begin{fact}[Decomposition]\label{thm:proofs:contraction}\ 
  \begin{itemize}[label={},leftmargin=0pt]\item  If: \hspace*{0.16em}{{\color{\colorMATH}\(\hat x \mathrel{\bot \!\!\!\bot } \hat y\)}}
  \item  And: \hspace*{0.16em}{{\color{\colorMATH}\(\left[\hat x \mathrel{\bot \!\!\!\bot } \hat y \mathrel{}\middle|\mathrel{} \hat z\right]\)}}
  \item  Then: \hspace*{0.16em}{{\color{\colorMATH}\(\hat x \mathrel{\bot \!\!\!\bot } \hat y,\hat z\)}}
  \end{itemize}
\end{fact}
\end{minipage}

\noindent
\begin{minipage}{\linewidth}
\begin{fact}[Independence Equivalences]\label{thm:proofs:independence-equivalences}\ 
  \begin{enumerate}[leftmargin=15pt]\item  {{\color{\colorMATH}\(\begin{array}[t]{l
      } \left[ \overline {\hat x} \mathrel{\bot \!\!\!\bot } \overline {\hat y}  \mathrel{}\middle|\mathrel{} \overline {\hat z \mathrel{\dot =} z} \right]
      \cr  \overset \vartriangle \iff    \forall \overline x,\overline y.\hspace*{0.33em} {\mtext{Pr}}\left[\overline {\hat x \mathrel{\dot =} x},\overline {\hat y \mathrel{\dot =} y},\overline {\hat z\mathrel{\dot =}z}\right] = {\mtext{Pr}}\left[\overline {\hat x\mathrel{\dot =}x}\mathrel{}\middle|\mathrel{}\overline {\hat z\mathrel{\dot =}z}\right]{\mtext{Pr}}\left[\overline {\hat y\mathrel{\dot =}y}\mathrel{}\middle|\mathrel{}\overline {\hat z\mathrel{\dot =}z}\right] 
      \cr  \iff      \forall \overline x,\overline y.\hspace*{0.33em} {\mtext{Pr}}\left[\overline {\hat x\mathrel{\dot =}x}\mathrel{}\middle|\mathrel{}\overline {\hat y\mathrel{\dot =}y},\overline {\hat z\mathrel{\dot =}z}\right]={\mtext{Pr}}\left[\overline {\hat x\mathrel{\dot =}x}\mathrel{}\middle|\mathrel{}\overline {\hat z\mathrel{\dot =}z}\right] 
      \cr  \iff      \forall \overline x,\overline y.\hspace*{0.33em} {\mtext{Pr}}\left[\overline {\hat y \mathrel{\dot =} y}\mathrel{}\middle|\mathrel{}\overline {\hat x\mathrel{\dot =}x},\overline {\hat z\mathrel{\dot =}z}\right]={\mtext{Pr}}\left[\overline {\hat y\mathrel{\dot =}y}\mathrel{}\middle|\mathrel{}\overline {\hat z\mathrel{\dot =}z}\right]
      \cr  \iff      \left[ \overline {\hat y} \mathrel{\bot \!\!\!\bot } \overline {\hat x}  \mathrel{}\middle|\mathrel{} \overline {\hat z \mathrel{\dot =} z} \right]
      \end{array}\)}}
  \end{enumerate}
\end{fact}
\end{minipage}

\noindent
\begin{minipage}{\linewidth}
\begin{fact}[Return Equivalence]\label{thm:proofs:return-equivalence}\ 
  \begin{itemize}[label={},leftmargin=0pt]\item  If: \hspace*{0.16em}{{\color{\colorMATH}\(x_{1} \sim _{A} x_{2}\)}}
  \item  Then: \hspace*{0.16em}{{\color{\colorMATH}\(\left[{\mtext{return}}_{{\mathcal{I}}}(x_{1}) \mathrel{}\middle|\mathrel{} \overline {\hat y \mathrel{\dot =} y}\right] \approx _{\sim _{A}} \left[{\mtext{return}}_{{\mathcal{I}}}(x_{2}) \mathrel{}\middle|\mathrel{} \overline {\hat z \mathrel{\dot =} z} \right]\)}}
  \end{itemize}
\end{fact}
\end{minipage}

\noindent
\begin{minipage}{\linewidth}
\begin{fact}[Bind Equivalence]\label{thm:proofs:bind-equivalence}\ 
  \begin{itemize}[label={},leftmargin=0pt]\item  For: \hspace*{0.16em}{{\color{\colorMATH}\(f_{1},f_{2} \in  A \rightarrow  {\mathcal{I}}(B)\)}}
  \item  If: \hspace*{0.16em}{{\color{\colorMATH}\(\hat x_{1} \approx _{\sim _{A}} \hat x_{2}\)}}
  \item  And: 
  \item  {{\color{\colorMATH}\(\forall  (x_{1} \in  {\mtext{support}}(\hat x_{1})),(x_{2} \in  {\mtext{support}}(\hat x_{2})).\hspace*{0.33em} x_{1} \sim _{A} x_{2} \implies    \left[f_{1}(x_{1}) \mathrel{}\middle|\mathrel{} \hat x_{1} \mathrel{\dot =} x_{1}\right] \approx _{\sim _{B}} \left[f_{2}(x_{2}) \mathrel{}\middle|\mathrel{} \hat x_{2} \mathrel{\dot =} x_{2} \right]\)}}
  \item  Then: \hspace*{0.16em}{{\color{\colorMATH}\((\hat x_{1} \gg = f_{1}) \approx _{\sim _{B}} (\hat x_{2} \gg = f_{2})\)}}
  \end{itemize}
\end{fact}
\end{minipage}

\noindent
\begin{minipage}{\linewidth}
\begin{fact}[Extensional Equivalence]\label{thm:proofs:extensional-equivalence}\ 
  \begin{itemize}[label={},leftmargin=0pt]\item  {{\color{\colorMATH}\(\begin{array}{c
      } \left[ \overline {\hat x_{1}} \mathrel{}\middle|\mathrel{} \overline {\hat y \mathrel{\dot =} y}\right] \approx _{=} \left[\overline {\hat x_{2}} \mathrel{}\middle|\mathrel{} \overline {\hat z \mathrel{\dot =} z} \right]
      \cr  \iff 
      \cr  \forall  \overline x,\overline y.\hspace*{0.33em} {\mtext{Pr}}\left[\overline {\hat x_{1} \mathrel{\dot =} x} \mathrel{}\middle|\mathrel{} \overline {\hat y \mathrel{\dot =} y} \right] = {\mtext{Pr}}\left[\overline {\hat x_{2} \mathrel{\dot =} x} \mathrel{}\middle|\mathrel{} \overline {\hat z \mathrel{\dot =} z}\right]
      \end{array}\)}}
  \end{itemize}
\end{fact}
\end{minipage}

\noindent
\begin{minipage}{\linewidth}
\begin{fact}[Distribution Equality Injective Function]\label{thm:proofs:distribution-equality-injective-function}\ 
  \begin{itemize}[label={},leftmargin=0pt]\item  If: \hspace*{0.16em}{{\color{\colorMATH}\(\hat x_{1} \approx _{=} \hat x_{2}\)}}
  \item  And: \hspace*{0.16em}{{\color{\colorMATH}\(f\)}} is injective
  \item  Then: \hspace*{0.16em}{{\color{\colorMATH}\(({\mtext{do}}\hspace*{0.33em}x \leftarrow  \hat x_{1} \mathrel{;} {\mtext{return}}(f(x))) \approx _{=} ({\mtext{do}}\hspace*{0.33em}x \leftarrow  \hat x_{2} \mathrel{;} {\mtext{return}}(f(x)))\)}}
  \end{itemize}
\end{fact}
\end{minipage}

\clearpage
\subsection{Definitions}\label{sec:proofs:definitions}

\begin{figure}[H]\small 
\begingroup\color{\colorMATH}\begin{gather*}
\begin{tabularx}{\linewidth}{>{\centering\arraybackslash\(}X<{\)}}\hfill\hspace{0pt} \begingroup\color{\colorTEXT}\boxed{\begingroup\color{\colorMATH} {\mathcal{D}}(A) \in  {\mtext{set}} \endgroup}\endgroup
  \\[-1.75ex]
  \cr \hfill\hspace{0pt}  x \in  A
    \hfill\hspace{0pt}  \tilde x \in  {\mathcal{D}}(A) \triangleq  \left\{  f \in  A \rightarrow  {\mathbb{R}} \mathrel{}\middle|\mathrel{} \sum \limits_{x\in A} f(x) = 1 \right\} 
    \hfill\hspace{0pt}  {\mtext{Pr}}\left[\tilde x \mathrel{\dot =} x \right] \triangleq  \tilde x(x)
    \hfill\hspace{0pt}
  \\[-1.75ex]
  \cr \begin{array}{rcl@{\hspace*{1.00em}\hspace*{1.00em}\hspace*{1.00em}}rcl
    } {\mtext{return}}    &{}\in {}& {\mathcal{D}}(A)
     & {\mtext{return}}(x) &{}\triangleq {}& \lambda x^{\prime}.\hspace*{0.33em} \left\{ \begin{array}{l@{\hspace*{1.00em}}c@{\hspace*{1.00em}}l
                             } 1 &{}{{\color{\colorTEXT}\textnormal{{\mtextit{if}}}}}{}& x = x^{\prime}
                             \cr  0 &{}{{\color{\colorTEXT}\textnormal{{\mtextit{if}}}}}{}& x \neq  x^{\prime}
                             \end{array}\right.
    \cr  {\mtext{bind}}         &{}\in {}& {\mathcal{D}}(A) \times  (A \rightarrow  {\mathcal{D}}(B)) \rightarrow  {\mathcal{D}}(B)
     & {\mtext{bind}}(\tilde x,f)  &{}\triangleq {}& \lambda y.\hspace*{0.33em} \sum \limits_{x}f(x)(y)\tilde x(x)
    \cr  {\mtext{bit}}          &{}\in {}& {\mathbb{N}} \rightarrow  {\mathcal{D}}({\mathbb{B}})
     & {\mtext{bit}}(N)       &{}\triangleq {}& \lambda b.\hspace*{0.33em} \nicefrac{1}{2} 
    \end{array}
  \\[-1.75ex]
  \cr  {\mtext{Pr}}\left[\overline {\tilde x \mathrel{\dot =} x} \mathrel{}\middle|\mathrel{} \overline {\tilde y \mathrel{\dot =} y} \right] \triangleq  \frac{{\mtext{Pr}}\left[\overline {\tilde x\mathrel{\dot =}x},\overline {\tilde y\mathrel{\dot =}y}\right]} {{\mtext{Pr}}\left[\tilde y\mathrel{\dot =}y\right]}
  \\[-1.75ex]
  \cr
\end{tabularx}
\end{gather*}\endgroup
\caption{(1) Denotational Distribution Monad; (2) Intensional Distribution Monad}
\label{fig:proofs:denotational-distributions}
\end{figure}

\begin{figure}[H]\small
\begingroup\color{\colorMATH}\begin{gather*}\begin{tabularx}{\linewidth}{>{\centering\arraybackslash\(}X<{\)}}\hfill\hspace{0pt} \begin{array}{rcrcl@{\hspace*{1.00em}}l
      } \ell    &{}\in {}& {\mtext{label}}   &{}\mathrel{\Coloneqq }{}& {\mtext{P}} \mathrel{|} {\mtext{S}}                       & {{\color{\colorTEXT}\textnormal{public and secret}}}
      \cr      &{} {}& \multicolumn{3}{c}{{{\color{\colorTEXT}\textnormal{({\mtextit{where}} {{\color{\colorMATH}\({\mtext{P}}\sqsubset {\mtext{S}}\)}})}}}}                   & {{\color{\colorTEXT}\textnormal{security labels}}}
      \cr  \rho    &{}\in {}& R         &{} {}&                                 & {{\color{\colorTEXT}\textnormal{probability region}}}
      \cr  b   &{}\in {}& {\mathbb{B}}         &{}\mathrel{\Coloneqq }{}& {\mfootnotesize{{{\color{\colorSYNTAX}\mtexttt{O}}}}} \mathrel{|} {\mfootnotesize{{{\color{\colorSYNTAX}\mtexttt{I}}}}}                       & {{\color{\colorTEXT}\textnormal{bits}}}
      \cr  x,y &{}\in {}& {\mtext{var}}     &{} {}&                                 & {{\color{\colorTEXT}\textnormal{variables}}}
      \cr  v   &{}\in {}& {\mtext{val}}     &{}\mathrel{\Coloneqq }{}& x                               & {{\color{\colorTEXT}\textnormal{variable values}}}
      \cr      &{} {}&           &{}\mathrel{|}{}& \langle v,v\rangle                            & {{\color{\colorTEXT}\textnormal{tuple values}}}
      \cr      &{} {}&           &{}\mathrel{|}{}& {\mfootnotesize{{{\color{\colorSYNTAX}\mtexttt{fun}}}}}_{y}(x{\mathrel{:}}\tau ).e               & {{\color{\colorTEXT}\textnormal{function values}}}
      \cr  \tau    &{}\in {}& {\mtext{type}}    &{}\mathrel{\Coloneqq }{}& {\mfootnotesize{{{\color{\colorSYNTAX}\mtexttt{bit}}}}}_{\ell }^{\rho }                     & {{\color{\colorTEXT}\textnormal{non-random bit}}}
      \cr      &{} {}&           &{}\mathrel{|}{}& {\mfootnotesize{{{\color{\colorSYNTAX}\mtexttt{flip}}}}}^{\rho }                       & {{\color{\colorTEXT}\textnormal{secret uniform bit}}}
      \cr      &{} {}&           &{}\mathrel{|}{}& {\mfootnotesize{{{\color{\colorSYNTAX}\mtexttt{ref}}}}}(\tau )                        & {{\color{\colorTEXT}\textnormal{reference}}}
      \cr      &{} {}&           &{}\mathrel{|}{}& \tau  \times  \tau                            & {{\color{\colorTEXT}\textnormal{tuple}}}
      \cr      &{} {}&           &{}\mathrel{|}{}& \tau  \rightarrow  \tau                            & {{\color{\colorTEXT}\textnormal{function}}}
      \end{array}
    \hfill\hspace{0pt} \begin{array}{rcl@{\hspace*{1.00em}}l
      } e   &{}\in {}& {\mtext{exp}}
      \cr  e   &{}\mathrel{\Coloneqq }{}& v                               & {{\color{\colorTEXT}\textnormal{value expressions}}}
      \cr      &{}\mathrel{|}{}& b_{\ell }                            & {{\color{\colorTEXT}\textnormal{bit literal}}}
      \cr      &{}\mathrel{|}{}& {\mfootnotesize{{{\color{\colorSYNTAX}\mtexttt{flip}}}}}^{\rho }()                     & {{\color{\colorTEXT}\textnormal{coin flip in region}}}
      \cr      &{}\mathrel{|}{}& {\mfootnotesize{{{\color{\colorSYNTAX}\mtexttt{cast}}}}}_{\ell }(v)                    & {{\color{\colorTEXT}\textnormal{cast flip to bit}}}
      \cr      &{}\mathrel{|}{}& {\mfootnotesize{{{\color{\colorSYNTAX}\mtexttt{mux}}}}}(e,e,e)                    & {{\color{\colorTEXT}\textnormal{atomic conditional}}}
      \cr      &{}\mathrel{|}{}& {\mfootnotesize{{{\color{\colorSYNTAX}\mtexttt{xor}}}}}(e,e)                      & {{\color{\colorTEXT}\textnormal{bit xor}}}
      \cr      &{}\mathrel{|}{}& {\mfootnotesize{{{\color{\colorSYNTAX}\mtexttt{if}}}}}(e)\{ e\} \{ e\}                    & {{\color{\colorTEXT}\textnormal{branch conditional}}}
      \cr      &{}\mathrel{|}{}& {\mfootnotesize{{{\color{\colorSYNTAX}\mtexttt{ref}}}}}(e)                        & {{\color{\colorTEXT}\textnormal{reference creation}}}
      \cr      &{}\mathrel{|}{}& {\mfootnotesize{{{\color{\colorSYNTAX}\mtexttt{read}}}}}(e)                       & {{\color{\colorTEXT}\textnormal{reference read}}}
      \cr      &{}\mathrel{|}{}& {\mfootnotesize{{{\color{\colorSYNTAX}\mtexttt{write}}}}}(e,e)                    & {{\color{\colorTEXT}\textnormal{reference write}}}
      \cr      &{}\mathrel{|}{}& \langle e,e\rangle                            & {{\color{\colorTEXT}\textnormal{tuple creation}}}
      \cr      &{}\mathrel{|}{}& {\mfootnotesize{{{\color{\colorSYNTAX}\mtexttt{let}}}}}\hspace*{0.33em}x = e\hspace*{0.33em}{\mfootnotesize{{{\color{\colorSYNTAX}\mtexttt{in}}}}}\hspace*{0.33em}e              & {{\color{\colorTEXT}\textnormal{variable binding}}}
      \cr      &{}\mathrel{|}{}& {\mfootnotesize{{{\color{\colorSYNTAX}\mtexttt{let}}}}}\hspace*{0.33em}x,y = e\hspace*{0.33em}{\mfootnotesize{{{\color{\colorSYNTAX}\mtexttt{in}}}}}\hspace*{0.33em}e            & {{\color{\colorTEXT}\textnormal{tuple elimination}}}
      \cr      &{}\mathrel{|}{}& e(e)                            & {{\color{\colorTEXT}\textnormal{fun. application}}}
      \end{array}
    \hfill\hspace{0pt}
  \cr
\end{tabularx}
\end{gather*}\endgroup
\caption{(1) Source Syntax; and (2) Runtime syntax: standard, mixed and adversary}
\end{figure}

\begin{figure}[H]\small
\begingroup\color{\colorMATH}\begin{gather*} 
\begin{tabularx}{\linewidth}{>{\centering\arraybackslash\(}X<{\)}}\hfill\hspace{0pt} \begin{array}{rcrcl@{\hspace*{1.00em}}l
    } \vphantom{\overset {\mathord{\bullet }}x}\overset {\smash {\mathord{\bullet }}}\tau  &{}\in {}& {\mtext{t\vphantom{\overset {\mathord{\bullet }}x}\overset {\smash {\mathord{\bullet }}}ype}} &{}\mathrel{\Coloneqq }{}& \tau    \mathrel{|} {\mathord{\bullet }}  \hspace*{0.33em}\hspace*{0.33em}{{\color{\colorTEXT}\textnormal{({\mtextit{where}} {{\color{\colorMATH}\(\tau  \sqsubset  {\mathord{\bullet }}\)}})}}}
    \cr  \kappa    &{}\in {}& {\mtext{kind}}     &{}\mathrel{\Coloneqq }{}& {\mfootnotesize{{{\color{\colorSYNTAX}\mtexttt{U}}}}} \mathrel{|} {\mfootnotesize{{{\color{\colorSYNTAX}\mtexttt{A}}}}}\hspace*{0.33em}\hspace*{0.33em}{{\color{\colorTEXT}\textnormal{({\mtextit{where}} {{\color{\colorMATH}\({\mfootnotesize{{{\color{\colorSYNTAX}\mtexttt{U}}}}}\sqsubset {\mfootnotesize{{{\color{\colorSYNTAX}\mtexttt{A}}}}}\)}})}}}
    \end{array}
  \hfill\hspace{0pt} \begin{array}{c
    } \Gamma  \in  {\mtext{tcxt}} \triangleq  var \rightharpoonup  {\mtext{t\vphantom{\overset {\mathord{\bullet }}x}\overset {\smash {\mathord{\bullet }}}ype}}
    \cr  (\Gamma _{1}\sqcup \Gamma _{2})(x)\triangleq \Gamma _{1}(x)\sqcup \Gamma _{2}(x)
    \end{array}
  \hfill\hspace{0pt}
\\[-1.75ex]
\cr \hfill\hspace{0pt} \begingroup\color{\colorTEXT}\boxed{\begingroup\color{\colorMATH} {\mathcal{K}} \in  {\mtext{type}} \rightarrow  {\mtext{kind}} \endgroup}\endgroup
\\[-1.75ex]
\cr
\end{tabularx}
\end{gather*}\endgroup
\caption{Source Typing}
\end{figure}

\begin{figure}[H]\small
\begingroup\color{\colorMATH}\begin{gather*}\begin{tabularx}{\linewidth}{>{\centering\arraybackslash\(}X<{\)}}\hfill\hspace{0pt} \begingroup\color{\colorTEXT}\boxed{\begingroup\color{\colorMATH} {\mtext{step}}_{{\mathcal{M}}} \in  {\mathbb{N}} \times  {\mtext{config}} \rightharpoonup  {\mathcal{M}}({\mtext{config}}) \endgroup}\endgroup
  \\[-1.75ex]
  \cr
\end{tabularx}
\end{gather*}\endgroup
\caption{\lang Semantics Standard and Mixed}
\end{figure}

\begin{figure}[H]\small
\begingroup\color{\colorMATH}\begin{gather*}\begin{tabularx}{\linewidth}{>{\centering\arraybackslash\(}X<{\)}}\hfill\hspace{0pt} \begingroup\color{\colorTEXT}\boxed{\begingroup\color{\colorMATH} {\mtext{obs}} \in  ({\mtext{exp}} \rightarrow  {\mtext{e\vphantom{\overset {\mathord{\bullet }}x}\overset {\smash {\mathord{\bullet }}}xp}}) \uplus  ({\mtext{store}} \rightarrow  {\mtext{st\vphantom{\overset {\mathord{\bullet }}x}\overset {\smash {\mathord{\bullet }}}ore}}) \uplus  ({\mtext{config}} \rightarrow  {\mtext{co\vphantom{\overset {\mathord{\bullet }}x}\overset {\smash {\mathord{\bullet }}}nfig}}) \uplus  ({\mtext{trace}} \rightarrow  {\mtext{tr\vphantom{\overset {\mathord{\bullet }}x}\overset {\smash {\mathord{\bullet }}}ace}}) \endgroup}\endgroup
  \\[-1.75ex]
  \cr
\end{tabularx}
\end{gather*}\endgroup
\caption{Adversary Observation}
\end{figure}

\begin{figure}[H]\small
\begingroup\color{\colorMATH}\begin{gather*}\begin{tabularx}{\linewidth}{>{\centering\arraybackslash\(}X<{\)}}\hfill\hspace{0pt} \begingroup\color{\colorTEXT}\boxed{\begingroup\color{\colorMATH} \lceil \underline{\hspace{0.66em}}\rceil  \in  {\mtext{\underline {ex\hspace{-1pt}}\hspace{1pt}p}} \rightarrow  {\mathcal{I}}({\mtext{exp}}) \endgroup}\endgroup
  \\[-1.75ex]
  \cr
\end{tabularx}
\end{gather*}\endgroup
\caption{Projection}
\end{figure}

\begin{figure}[H]\small
\begingroup\color{\colorMATH}\begin{gather*}
\begin{tabularx}{\linewidth}{>{\centering\arraybackslash\(}X<{\)}}\hfill\hspace{0pt} \begingroup\color{\colorTEXT}\boxed{\begingroup\color{\colorMATH} \Sigma ,\Gamma  \vdash  e \mathrel{:} \tau  \mathrel{;} \Gamma  \endgroup}\endgroup
\\
\end{tabularx}
\end{gather*}\endgroup
\caption{\lang Type System Evaluation Standard Expressions}
\end{figure} 

\begin{figure}[H]\small
\begingroup\color{\colorMATH}\begin{gather*}
\begin{tabularx}{\linewidth}{>{\centering\arraybackslash\(}X<{\)}}\hfill\hspace{0pt} \Psi ^{F} {\in } {\mtext{flipset}} \triangleq  \wp ({\mathcal{I}}({\mathbb{B}}))
  \hfill\hspace{0pt} \Psi ^{B} {\in } {\mtext{bitset}}  \triangleq  {\mathbb{R}} {\rightarrow } \wp ({\mathcal{I}}({\mathbb{B}}))
  \hfill\hspace{0pt} \Psi     {\in } {\mtext{fbset}}   \mathrel{\Coloneqq } \Psi ^{F}{,}\Psi ^{B}
  \hfill\hspace{0pt} \Phi     {\in } {\mtext{history}}  \mathrel{\Coloneqq } \overline {\hat {\underline \varsigma } \mathrel{\dot =} \underline \varsigma }
  \hfill\hspace{0pt} 
\\[-1.75ex]
\cr (\Psi _{1}^{F},\Psi _{1}^{B}) \uplus  (\Psi _{2}^{F},\Psi _{2}^{B}) \triangleq  (\Psi _{1}^{F} \uplus  \Psi _{2}^{F}),(\Psi _{1}^{B} \cup  \Psi _{2}^{B})
\\[-1.75ex]\hfill\hspace{0pt} \begingroup\color{\colorTEXT}\boxed{\begingroup\color{\colorMATH} \Psi ,\Phi ,\Sigma ,\Gamma  \vdash  \underline e \mathrel{:} \tau  \mathrel{;} \Gamma ,\Psi  \endgroup}\endgroup
\\
\end{tabularx}
\end{gather*}\endgroup
\caption{\lang Type System Evaluation Mixed Expressions}
\end{figure} 

\begin{figure}[H]\small
\begingroup\color{\colorMATH}\begin{gather*}
\begin{tabularx}{\linewidth}{>{\centering\arraybackslash\(}X<{\)}}\hfill\hspace{0pt} \begingroup\color{\colorTEXT}\boxed{\begingroup\color{\colorMATH}\Psi ,\Phi \vdash \hat b\mathrel{:}{\mfootnotesize{{{\color{\colorSYNTAX}\mtexttt{flip}}}}}^{\rho }\endgroup}\endgroup
\\[-1.75ex]
\parbox{\linewidth}{ \def\MathparLineskip{\lineskip=4pt}
\begingroup\color{\colorMATH}\begin{mathpar} \inferrule*[vcenter,lab={\mtextsc{ Flip-Value}}
   ]{ {\mtext{Pr}}\left[\hat b \mathrel{\dot =} {\mfootnotesize{{{\color{\colorSYNTAX}\mtexttt{I}}}}} \mathrel{}\middle|\mathrel{} \Phi  \right] = \nicefrac{1}{2} 
   \\ \left[ \hat b \mathrel{\bot \!\!\!\bot } \Psi ^{F},\Psi ^{B}(\{ \rho ^{\prime} \mathrel{|} \rho ^{\prime} \sqsubset  \rho \} ) \mathrel{}\middle|\mathrel{} \Phi  \right]
      }{
      \Psi ^{F},\Psi ^{B},\Phi  \vdash  \hat b \mathrel{:} {\mfootnotesize{{{\color{\colorSYNTAX}\mtexttt{flip}}}}}^{\rho }
   }
\end{mathpar}\endgroup
}
\cr \hfill\hspace{0pt} \begingroup\color{\colorTEXT}\boxed{\begingroup\color{\colorMATH}\Psi ,\Phi ,\Sigma ,\Gamma \vdash v\mathrel{:}\tau \mathrel{;}\Gamma ,\Psi \endgroup}\endgroup
\\
\end{tabularx}
\end{gather*}\endgroup
\caption{\lang Type System Evaluation Mixed Values, Store and Configurations}
\end{figure}

\begin{figure}[H]\small
\begingroup\color{\colorMATH}\begin{gather*}
\begin{tabularx}{\linewidth}{>{\centering\arraybackslash\(}X<{\)}}\hfill\hspace{0pt} \begingroup\color{\colorTEXT}\boxed{\begingroup\color{\colorMATH} \underline e_{1} \sim  \underline e_{2} \endgroup}\endgroup
\cr
\end{tabularx}
\end{gather*}\endgroup
\caption{Low Equivalence Relation}
\end{figure}

\else
\fi

\end{document}
\endinput